%% file: lusiani-phipsi17-procs.tex
\pdfoutput=1

\documentclass{webofc}
\usepackage[varg]{txfonts}   

\usepackage{overpic}
\usepackage{relsize}
\usepackage{alumisc}
\usepackage{ifthen}
\usepackage{hyperref}
\usepackage{booktabs}
\usepackage[svgnames,x11names]{xcolor}
\usepackage{tikz}
\usepackage{amsmath}
\usepackage{amssymb}
\usepackage{amstext}

\newenvironment{ensuredisplaymath}%
{\(\displaystyle}
{\)}

\newcommand{\EE}[1]{\ensuremath{{\cdot}10^{#1}}}

\newcommand{\gev}{\ensuremath{\,\text{Ge\kern -0.1em V}}\xspace}
\newcommand{\gevc}{\ensuremath{\,\text{Ge\kern -0.1em V}\!/c}\xspace}
\newcommand{\gevcc}{\ensuremath{\,\text{Ge\kern -0.1em V}\!/c^2}\xspace}
\newcommand{\mev}{\ensuremath{\,\text{Me\kern -0.1em V}}\xspace}
\newcommand{\mevc}{\ensuremath{\,\text{Me\kern -0.1em V}\!/c}\xspace}
\newcommand{\mevcc}{\ensuremath{\,\text{Me\kern -0.1em V}\!/c^2}\xspace}

\newcommand{\babar}{\mbox{%
    \slshape B\kern-0.1em{\smaller A}\kern-0.1em
    B\kern-0.1em{\smaller A\kern-0.2em R}}\xspace}

\newcommand{\mtau}      {\ensuremath{\tau}\xspace}

\newcommand{\nub}       {\ensuremath{\overline{\nu}}\xspace}

\newcommand{\Bbar}   {\kern 0.18em\overline{\kern -0.18em B}{}\xspace}

\newcommand{\Bu}     {\ensuremath{B^+}\xspace}
\newcommand{\Bub}    {\ensuremath{B^-}\xspace}

\newcommand{\BpBm}   {\ensuremath{\Bu {\kern -0.16em \Bub}}\xspace}
\newcommand{\Bz}     {\ensuremath{B^0}\xspace}
\newcommand{\Bzb}    {\ensuremath{\Bbar^0}\xspace}
\newcommand{\BzBzb}  {\ensuremath{\Bz {\kern -0.16em \Bzb}}\xspace}

\newcommand{\BR}{{\cal B}\xspace}

\def\Y#1S{\ensuremath{\Upsilon{(#1S)}}\xspace}

\newcommand{\Vud}{\ensuremath{\left|V_{ud}\right|}\xspace}
\newcommand{\Vus}{\ensuremath{\left|V_{us}\right|}\xspace}
\newcommand{\Vub}{\ensuremath{\left|V_{ub}\right|}\xspace}

\newcommand{\leptrad}{\ensuremath{r}\xspace}%

\makeatletter
\newcommand{\htdef}[2]{%
  \@namedef{hfagtau@#1}{#2}%
}

\newcommand{\htuse}[1]{%
  \ifcsname hfagtau@#1\endcsname
  \@nameuse{hfagtau@#1}%
  \else
  \@latex@error{Undefined name hfagtau@#1}\@eha
  \fi
}

\newcommand{\htquantdef}[6]{%
  \ifx&#2&\else
  \@namedef{hfagtau@#1.gn}{\ensuremath{#2}}%
  \fi
  \ifx&#3&\else
  \@namedef{hfagtau@#1.td}{\ensuremath{#3}}%
  \fi
  \ifx&#6&%
    \@namedef{hfagtau@#1}{\ensuremath{#5}}%
  \else
    \ifthenelse{\equal{#6}{0}}{%
      \@namedef{hfagtau@#1}{\ensuremath{#5}}%
    }{%
      \@namedef{hfagtau@#1}{\ensuremath{#4}}%
      \@namedef{hfagtau@#1.v}{\ensuremath{#5}}%
      \@namedef{hfagtau@#1.e}{\ensuremath{#6}}%
    }%
  \fi
}

\newcommand{\htmeasdef}[8]{%
  \@namedef{hfagtau@#1,quant}{\ensuremath{#2}}%
  \@namedef{hfagtau@#1,exp}{#3}%
  \@namedef{hfagtau@#1,ref}{\cite{#4}}%
  \@namedef{hfagtau@#1}{\ensuremath{#5}}%
  \@namedef{hfagtau@#1,val}{\ensuremath{#6}}%
  \@namedef{hfagtau@#1,stat}{\ensuremath{#7}}%
  \@namedef{hfagtau@#1,syst}{\ensuremath{#8}}%
}

\newcommand{\htconstrdef}[4]{%
  \@namedef{hfagtau@#1.left}{\ensuremath{#2}}%
  \@namedef{hfagtau@#1.right}{\ensuremath{#3}}%
  \@namedef{hfagtau@#1.right.split}{\ensuremath{#4}}%
  \@namedef{hfagtau@#1.constr.eq}{\htuse{#1.left} ={}& \htuse{#1.right}}%
}

\newcommand{\htQuantLine}[3]{\ensuremath{\htuse{#1.td}}&\ensuremath{#2}\\}

\makeatother

\newif\ifhevea\heveafalse

\input{tau-defs}
\input{tau-br-fit-hfag-2016}
\input{tau-br-fit-hfag-2016-elab-vadirect}

\begin{document}
\title{Status and progress of the HFLAV-Tau group activities}
%
%

\author{Alberto Lusiani\inst{1,2}\fnsep\thanks{\email{alberto.lusiani@pi.infn.it}}}

\institute{%
  Scuola Normale Superiore, Pisa, Italy\and
  INFN sezione di Pisa, Pisa, Italy}

\abstract{%
  We report the status and progress of the Heavy Flavour Averaging Group
  work on the
  global fit of $\tau$ tau lepton branching fractions, on the lepton
  universality tests and on the measurement of the
  Cabibbo-Kobayashi-Maskawa matrix element \Vus using $\tau$ lepton
  measurements. We also review the prospects for improving the precision
  of the \Vus measurement.
}
\maketitle

\section{Introduction}
\label{sec:intro}

To best exploit the many experimental measurements on the $\tau$ lepton
branching fractions, it is convenient to perform a global fit. The
results are useful to compute the lepton universality tests, the strong
coupling constant at the $\tau$ mass, the Cabibbo-Kobayashi-Maskawa
(CKM) matrix element \Vus and the muon $g-2$ hadronic vacuum
polarization contribution.

The $\tau$ subgroup of the Heavy Flavour Averaging Group (HFLAV)
has updated in 2016 the global fit of the $\tau$
branching fractions, the  lepton universality tests and the \Vus
determination based on $\tau$ measurements. A version of the global fit with
unitarity constraint has been
published in the  Review of
Particle Physics~\cite{Patrignani:2016xqp} (RPP).
All the 2016 HFLAV-Tau results have
been submitted for publication as part of the 2016 HFLAV
report~\cite{Amhis:2016xyh}. These results are labelled ``Spring
2017'' and have only minimal differences to formerly reported
preliminary results~\cite{Lusiani:2017spn}.

In the following, we summarize the latest HFLAV-Tau results,
briefly mentioning the changes since the HFLAV 2014
report~\cite{Amhis:2014hma} and the minor modifications
since the Tau 2016 conference update~\cite{Lusiani:2017spn}. 
Furthermore, we review the prospects for
improving the \Vus determination based on $\tau$ measurements.

\section{Branching fractions fits}
\label{sec:tau-br-fit}

A global fit of the available experimental measurements is used to
determine the $\tau$ branching fractions, together with their
uncertainties and statistical correlations.
According to the HFLAV
methodology~\cite{Asner:2010qj}, we considered all significant
correlations induced by common systematic effects across different
measurements, and we updated the measurements to take into account
updates on other Physics quantities from which each specific
measurement has documented dependencies. After doing that, we get a
statistically consistent global fit by applying just a single scale
factor of 5.44 to the published
uncertainties of the measurements of
\(\Gamma_{96} = \tau \to KKK\nu\) by \babar{} and
Belle{}, which would be
otherwise severely statistically inconsistent.

The HFLAV Spring 2017 $\tau$ branching fractions fit has $\chi^2/\text{d.o.f.} =
\htuse{Chisq}/\htuse{Dof}$, corresponding to a confidence level
$\text{CL} = \htuse{ChisqProb}$. The procedure uses a total of \htuse{MeasNum}
measurements to fit \htuse{QuantNum} quantities subjected to
\htuse{ConstrNum} constraints. The fit is
statistically consistent with unitarity, and the unitarity residual is
$\htuse{Gamma998.td} = \htuse{Gamma998}$.


With respect to the 2014 HFLAV report, we removed the Belle result on
$\tau^- \to \htuse{Gamma33.td}$~\cite{Ryu:2014vpc} because the published
information does not permit a reliable determination of the correlations
with the other results in the same paper.  The inclusion of that result
made the correlation matrix of the 2014 fit results non
positive-definite.
Furthermore, as detailed
elsewhere~\cite{Amhis:2016xyh,Lusiani:2017spn}, we discarded some
old preliminary results that are not
expected to be published, we removed a result that was
effectively 100\% correlated with other measurements by the same
experiment, and applied a few minor corrections to the constraints that
relate  different branching fractions.
We have found that all these modifications have negligible impact on
the lepton universality and the \Vus measurements.


As is standard for the RPP branching fraction fits, the RPP 2016 $\tau$ branching
fraction fit provided by HFLAV-Tau is unitarity constrained, while the
HFLAV Spring 2017 fit is not. In the RPP 2016, the ALEPH experiment measurements are used in a
slightly different way compared to the HFLAV fit, as documented
elsewhere~\cite{Amhis:2016xyh,Lusiani:2017spn}.
The two fits also have a slightly different treatment of the
experimental results related to $\htuse{Gamma805.gn} =
\BR(\tau\to\htuse{Gamma805.td})$~\cite{Amhis:2016xyh,Lusiani:2017spn}.

\section{Tests of lepton universality}
\label{sec:tau:leptonuniv}

According to the Standard Model (SM), the partial width of a heavier lepton $\lepth$
decaying to a lighter lepton $\leptl$~\cite{Marciano:1988vm} is,
\begin{align*}
  \Gamma(\lepth {\to} \nu_{\lepth} \leptl \nub_{\leptl} (\gamma)) =
  \frac{\BR(\lepth {\to} \nu_{\lepth} \leptl \nub_{\leptl})}{\tau_{\lepth}} =
  \frac {G_{\lepth} G_{\leptl} m^5_{\lepth}}{192 \pi^3}\, f\left(\frac {m^2_{\leptl}}{m^2_{\lepth}}\right)
  \radRatio^{\lepth}_W \radRatio^{\lepth}_\gamma~,
\end{align*}
where
\begin{alignat*}{3}
 G_{\leptl} &= \frac {g^2_{\leptl}}{4 \sqrt{2} M^2_W}~, &\quad&&
 f(x) &= 1 -8x +8x^3 -x^4 -12x^2 \text{ln}x~, \\
 \radRatio^{\lepth}_W &= 1 + \frac {3}{5} \frac {m^2_{\lepth}}{M^2_W}  +
 \frac {9}{5} \frac {m^2_{\leptl}}{M^2_W}~, & \quad\quad&&
 \radRatio^{\lepth}_\gamma &= 1+\frac {\alpha(m_{\lepth})}{2\pi} \left(\frac {25}{4}-\pi^2\right)~.
\end{alignat*}
With respect to the recently reported preliminary
results~\cite{Lusiani:2017spn}, we include now the subleading term to
$\radRatio^{\lepth}_W$~\cite{Ferroglia:2013dga,Fael:2013pja}, which does
not change the reported numbers.
We use $\radRatio^\tau_\gamma=1-43.2\cdot 10^{-4}$ and
$\radRatio^\mu_\gamma=1-42.4\cdot 10^{-4}$~\cite{Marciano:1988vm} and $M_W$
from PDG 2015~\cite{PDG_2014}.
We use HFLAV Spring 2017 averages and PDG 2015 for the other quantities.
Using leptonic processes we obtain
\begin{align*}
  &\left( {g_\tau}/{g_\mu} \right) = \htuse{gtaubygmu_tau}~, \quad
  \left( {g_\tau}/{g_e} \right) = \htuse{gtaubyge_tau}~, \quad
  \left( {g_\mu}/{g_e} \right) = \htuse{gmubyge_tau}~.
\end{align*}
\begin{figure}[tb]
  \begin{center}
    \sidecaption
    \fbox{%
      \begin{overpic}[trim=5 10 5 5,width=0.6\linewidth-2\fboxsep-2\fboxrule,clip]{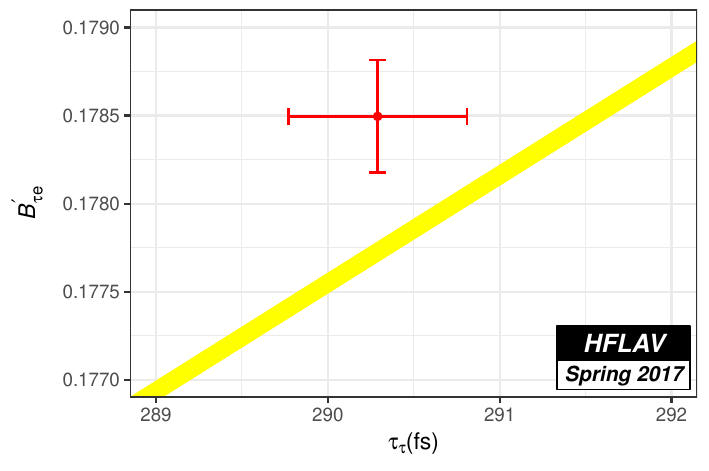}
    \end{overpic}}
    \caption{%
      Test of the SM prediction of the relation between the $\tau$
      leptonic branching fractions and the $\tau$ lifetime and
      mass. $B^\prime_{\tau e}$ denotes the statistical average of
      $B_e = \BR(\tau \to e \bar{\nu}_e \nu_\tau)$ and the $B_e$ SM
      prediction from the $B_\mu$ measurement
      $B_e(B_\mu) = B_\mu\cdot(f_{\tau e}/f_{\tau\mu})$. The yellow band
      represents the uncertainty from the $\tau$ lifetime.
      \label{fig:tau:leptunivploty}%
    }
  \end{center}
\end{figure}
Figure~\ref{fig:tau:leptunivploty} shows the test of the SM prediction
of the relation between the $\tau$ leptonic branching fractions
$B_\ell = \BR(\tau \to \ell \bar{\nu}_\ell \nu_\tau)$, with $\ell=e,\mu$, the
$\tau$ lifetime $\tau_\tau$, the $\tau$ mass $m_\tau$, and the respective muon parameters,
\begin{align*}
  \BR^{\text{SM}}_{\tau \ell} =
  \BR_{\mu e}
  \frac{\tau_\tau}{\tau_\mu}
  \frac{m_\tau^5}{m_\mu^5}
  \frac{%
  f_{\tau \ell}\leptrad^{\tau}_W\leptrad^{\tau}_\gamma}{%
  f_{\mu e} \leptrad^{\mu}_W \leptrad^{\mu}_\gamma},\quad
  \text{with {}} f_{\lepth \leptl} = f\left(\frac {m^2_{\leptl}}{m^2_{\lepth}}\right)~.
\end{align*}
Using semi-hadronic processes
\begin{align*}
  \left( \frac{g_\tau}{g_\mu} \right)^2 =
  \frac{\BR({\tau \to h \nu_\tau})}{\BR({h \to \mu \bar{\nu}_\mu})}
  \frac{2m_h m^2_{\mu}\tau_h}{(1+\delta_{h})m^3_{\tau}\tau_{\tau}}
  \left( \frac{1-m^2_{\mu}/m^2_h}{1-m^2_h/m^2_{\tau}} \right)^2~,
\end{align*}
where $h$ = $\pi$ or $K$ and the radiative corrections are
$\delta_{\pi} = (\htuse{dRrad_taupi_by_pimu})\%$ and
$\delta_{K} = (\htuse{dRrad_tauK_by_Kmu})\%$~\cite{Decker:1994dd}.
We measure:
\begin{align*}
  \left( {g_\tau}/{g_\mu} \right)_\pi &= \htuse{gtaubygmu_pi}~,
  & \left( {g_\tau}/{g_\mu} \right)_K = \htuse{gtaubygmu_K}~.
\end{align*}
Similar tests could be performed with decays to electrons, however they are
less precise because the hadron two body decays to electrons are
helicity-suppressed.
Averaging the three \(g_\tau/g_\mu\) ratios we obtain
\begin{align*}
  \left( {g_\tau}/{g_\mu} \right)_{\tau{+}\pi{+}K} &= \htuse{gtaubygmu_fit}~,
\end{align*}
accounting for all correlations.

\section{$\Vus$ measurement}
\label{sec:tau:vus}

The measurements of the kaon branching fractions are used in conjunction
with lattice QCD estimates of hadronic form factors to provide the
most precise determinations of \Vus~\cite{Patrignani:2016xqp}. The $\tau$
exclusive branching fractions to strange final states can be used in a
similar way to obtain additional less precise \Vus determinations. Furthermore, the
inclusive branching fraction of the $\tau$ to all strange final states,
$\BR(\tau \to X_s\nu)$, can be used to compute \Vus with a procedure
that does not require lattice QCD estimates and has an independent and
small theory uncertainty~\cite{Gamiz:2006xx}:
\begin{align*}
  \VusTauIncl &= \sqrt{\Rstrange/\left[\frac{\Rnonstrange}{\Vud^2} -  \delta R_{\text{theory}}\right]}~.
\end{align*}
\Rstrange and \Rnonstrange are the $\tau$ hadronic
partial widths to
strange and to non-strange hadronic final states (\Gammastrange and
\Gammahad) divided by the universality-improved
branching fraction $\BR(\tau \to e \nu \bar{\nu}) = \BR_e^{\text{uni}} =
(\htuse{Be_univ})\%$~\cite{Amhis:2016xyh,Lusiani:2017spn}.
We compute $\delta R_{\text{theory}} = \htuse{deltaR_su3break}$ using
inputs from Ref.~\cite{Gamiz:2006xx} and \mbox{$m_s =
  (\htuse{m_s})\,\text{MeV}$}~\cite{PDG_2014}.
Alternative determinations
exist~\cite{Gamiz:2006xx,Gamiz:2007qs,Maltman:2010hb}.
We use $\Vud = \htuse{Vud}$~\cite{Hardy:2014qxa}.
We obtain $\VusTauIncl = \htuse{Vus}$, which
is $\htuse{Vus_mism_sigma_abs}\sigma$ lower than the unitarity CKM
prediction $\VusUni = \htuse{Vus_uni}$, from $(\VusUni)^2 = 1 -
\Vud^2$ (\Vub is negligible). The \VusTauIncl uncertainty includes a systematic uncertainty
contribution of \htuse{Vus_err_th_perc}\% from the theory uncertainty on
$\delta R_{\text{theory}}$.

We follow the same procedure of the HFLAV 2012 report~\cite{Amhis:2012bh} to compute \Vus from
the ratio of branching fractions $\BR(\tau \to \htuse{Gamma10.td}) / \BR(\tau \to \htuse{Gamma9.td}) =
\htuse{Gamma10by9}$
from the equation
\begin{align*}
\frac{\BR(\tau \to \htuse{Gamma10.td})}{\BR(\tau \to \htuse{Gamma9.td})} &=
\frac{f_K^2 \Vus^2}{f_\pi^2 \Vud^2} \frac{\left( 1 - m_K^2/m_\tau^2 \right)^2}{\left( 1 -  m_\pi^2/m_\tau^2 \right)^2}
\radRatio_{\tau K/\tau\pi}~.
\end{align*}
We use $f_K/f_\pi = \htuse{f_K_by_f_pi}$ from the
FLAG 2016 Lattice averages with $N_f=2+1+1$~\cite{Aoki:2016frl}. 
The HFLAV Spring 2017 report~\cite{Amhis:2016xyh} contains more
details on the radiative
correction term $\radRatio_{\tau K/\tau\pi} = \htuse{Rrad_kmunu_by_pimunu}$.
We obtain $\VusTauKpi = \htuse{Vus_tauKpi}$,
$\htuse{Vus_tauKpi_mism_sigma_abs}\sigma$ below the CKM unitarity prediction.

Until the most recent HFLAV-Tau update~\cite{Lusiani:2017spn}, we
reported a third \Vus measurement using $\BR(\tau \to K\nu)$.
The HFLAV Spring 2017 update does not include any more that last
determination, because it did not include the long-distance radiative
corrections in addition to the short-distance contribution. This change
has a negligible effect on the overall precision of \Vus.


\begin{figure}[tb]
  \begin{center}
    \sidecaption
    \fbox{\begin{overpic}[trim=0 0 0 0,width=0.6\linewidth-2\fboxsep-2\fboxrule,clip]{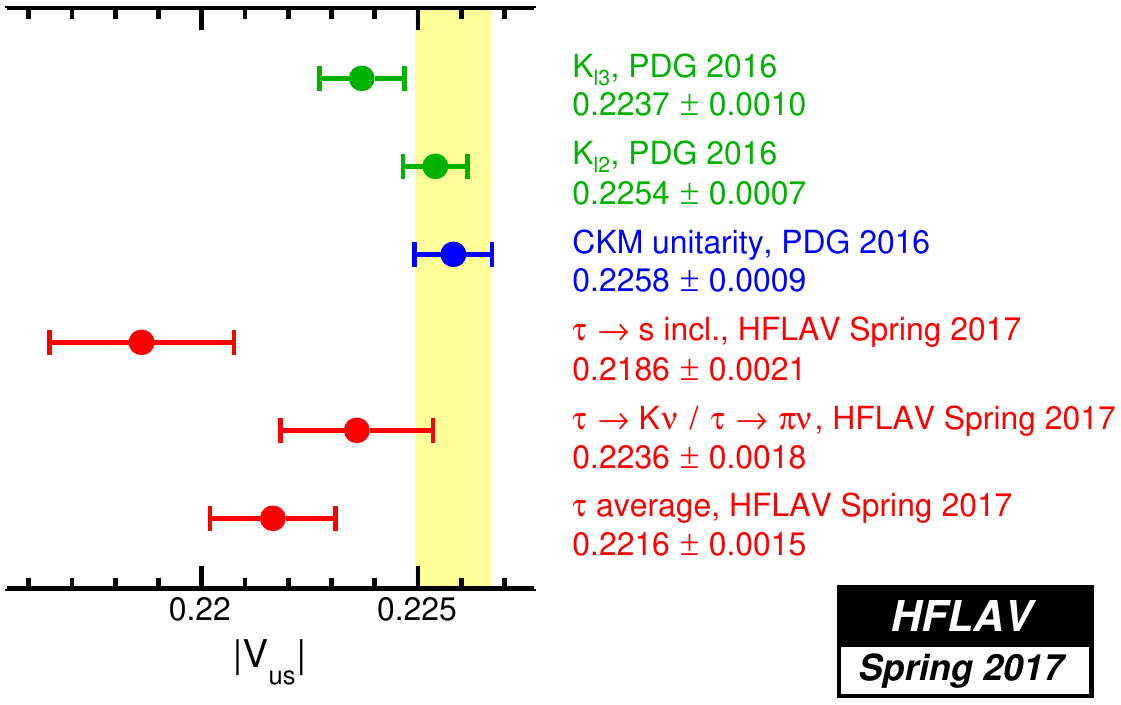}
      \end{overpic}}
    \caption{\Vus computed using the $\tau$ measurements compared to the
      determinations based on the CKM matrix unitarity and on the kaon
      branching fractions.
      \label{fig:tau:vus-summary}%
    }
  \end{center}
\end{figure}

Averaging the two \Vus determinations, we obtain
$\Vus_\tau = \htuse{Vus_tau}$, which is $\htuse{Vus_tau_mism_sigma_abs}\sigma$
below the CKM unitarity determination.
Figure~\ref{fig:tau:vus-summary} summarizes the
\Vus results, reporting also
\Vus computed from kaon measurements~\cite{Patrignani:2016xqp} and \Vus obtained
from \Vud and the CKM matrix unitarity~\cite{Patrignani:2016xqp}.

\section{Studies on \Vus from $\tau$ data}


\newcommand{\valstatsyst}[3]{%
  \ensuremath{#1 \pm #2\,\text{(stat.)} \pm #3\,\text{(syst.)}}\xspace
}

\colorlet{xbarcolor}{RoyalBlue}
\newlength\xbarBaseWidth
\newlength\xbarHeight
\setlength{\xbarHeight}{1.5ex}%
\newcommand{\xbarline}[2]{%
  #1 & #2 & \color{xbarcolor}\rule{#2\xbarBaseWidth}{\xbarHeight}
}
\newcommand{\xbarlineEmph}[2]{%
  \magentaEm #1 & \magentaEm #2 & \color{magentaEm!80!white}\rule{#2\xbarBaseWidth}{\xbarHeight}
}
\colorlet{magentaEm}{DarkMagenta}
\DeclareRobustCommand*{\magentaEm}{\color{magentaEm}}


Assuming the SM, the following $\tau$ branching fractions have been
computed using the precisely measured $K_{\ell2}$ and $K_{\ell3}$
branching fractions and the measured $\tau^- \to (K\pi)^- \nu_\tau$
spectra~\cite{Antonelli:2013usa}:
\begin{align*}
  & \BR(\tau^-\to K^- \nu_\tau) &&= (0.713 \pm 0.003)\%~, \\[-0.5ex]
  & \BR(\tau^-\to K^- \pi^0 \nu_\tau) &&= (0.471 \pm 0.018)\%~, \\[-0.5ex]
  & \BR(\tau^-\to K^0 \pi^- \nu_\tau) &&= (0.857 \pm 0.030)\%~.
\end{align*}
It has been observed~\cite{Antonelli:2013usa, Pich:2013lsa} that all the
above indirect values are higher than the corresponding directly measured $\tau$
branching fractions. If the indirect values replace the direct ones,
\mbox{$\Vus = 0.2207 \pm 0.027$}~\cite{Antonelli:2013usa}. If the indirect values are instead
statistically combined with the direct measurements used in the HFLAV
Spring 2017 global fit according to their
respective uncertainties, we obtain \mbox{$\Vus = 0.2195 \pm
0.021$}. Both \Vus values are close to the one computed using just the
HFLAV Spring 2017 fit outputs.

An alternative determination of $\Vus =
\valstatsyst{0.2231}{0.0027}{0.0004}$~\cite{Hudspith:2017vew}
relies on  the \mtau
spectral functions in addition to the inclusive $\tau \to X_s \nu$
branching fraction, for the purpose of reducing the QCD-related
uncertainties. This study replaces the HFLAV fit values for
$\BR(\tau^-\to K^- \nu_\tau)$, $\BR(\tau^-\to K^- \pi^0 \nu_\tau)$ and
$\BR(\tau^-\to K^0 \pi^- \nu_\tau)$ with the indirectly values in
Ref.~\cite{Antonelli:2013usa}. This last \Vus value is entirely
consistent with CKM unitarity.

A PHD thesis published in 2011~\cite{Adametz:2011vla} reports
measurements of 4 \mtau branching fractions with a kaon in the finel
state, performed in a not-yet-released \babar analysis:
\begin{align*}
  & \BR(\tau\to K\nu) &&= (\valstatsyst{0.710}{0.003}{0.016})\%~, \\[-0.5ex]
  & \BR(\tau\to K \pi^0 \nu) &&= (\valstatsyst{0.500}{0.002}{0.014})\%~, \\[-0.5ex]
  & \BR(\tau\to K 2\pi^0 \nu) &&= (\valstatsyst{5.66}{0.14}{0.32})\EE{-4}~, \\[-0.5ex]
  & \BR(\tau\to K 3\pi^0 \nu) &&= (\valstatsyst{1.64}{0.28}{0.38})\EE{-4}~.
\end{align*}
The 3 measurements with neutral pions in the final state correspond to
3 out of the 5 largest contributions to the uncertainty on \Vus from the
$\tau$ branching fractions. If we add to the HFLAV
Spring 2017 inputs the above 4 results and the
additional measurements from the same reference on $\BR(\tau\to \pi
3\pi^0 \nu)$ and $\BR(\tau\to \pi 4\pi^0 \nu)$, we perform the global
$\tau$ branching fraction fit and we use the fit results,
we obtain $\Vus = 0.2192 \pm 0.019$, close to the HFLAV Spring 2017 value.

The main contributions to the HFLAV-Tau \Vus determination uncertainty come
from the $\tau$ branching fractions to strange final states and from the
theory. After we add to the HFLAV-Tau $\tau$ measurements both the above
mentioned in-progress \babar results and the indirect measurements from
the kaon branching fractions, the following list shows the
resulting individual contributions to the \Vus uncertainty.

\begin{center}
{%
  \setlength{\xbarBaseWidth}{100ex}%
  \begin{tabular}{lrl}
    \xbarline{$\pi^- \bar{K}^0 \pi^0 \pi^0 \nu_\tau ~(\text{ex.~}K^0)$}{0.3908} \\
    \xbarline{$\bar{K}^0 h^- h^- h^+ \nu_\tau$}{0.3430} \\
    \xbarline{$K^- \pi^- \pi^+ \pi^0 \nu_\tau ~(\text{ex.~}K^0,\omega,\eta)$}{0.2422} \\
    \xbarline{$\pi^- \bar{K}^0 \pi^0 \nu_\tau$}{0.2178} \\
    \xbarline{$K^- \omega \nu_\tau$}{0.1563} \\
    \xbarline{$\pi^- \bar{K}^0 \nu_\tau$}{0.1501} \\
    \xbarline{$K^- \pi^- \pi^+ \nu_\tau ~(\text{ex.~}K^0,\omega)$}{0.1140} \\
    \xbarline{$K^- \pi^0 \nu_\tau$}{0.0881} \\
    \xbarline{$K^- 3\pi^0 \nu_\tau ~(\text{ex.~}K^0,\eta)$}{0.0703} \\
    \xbarline{$K^- \nu_\tau$}{0.0473} \\
    \xbarline{$K^- 2\pi^0 \nu_\tau ~(\text{ex.~}K^0)$}{0.0414} \\
    \xbarline{$\pi^- \bar{K}^0 \eta \nu_\tau$}{0.0253} \\
    \xbarline{$K^- \pi^0 \eta \nu_\tau$}{0.0197} \\
    \xbarline{$K^- \eta \nu_\tau$}{0.0136} \\
    \xbarline{$K^- \phi \nu_\tau ~(\phi \to K^+ K^-)$}{0.0136} \\
    \xbarline{$K^- \phi \nu_\tau ~(\phi \to K_S^0 K_L^0)$}{0.0095} \\
    \xbarline{$K^- 2\pi^- 2\pi^+ \nu_\tau ~(\text{ex.~}K^0)$}{0.0020} \\
    \xbarline{$K^- 2\pi^- 2\pi^+ \pi^0 \nu_\tau ~(\text{ex.~}K^0)$}{0.0010} \\
    \xbarline{$\tau\to\text{non-strange}$}{0.0853} \\
    \xbarline{theory}{0.4724}
  \end{tabular}}
\end{center}

The Belle II super flavour factory may provide the required more precise
measurements of the small $\tau$ strange branching fractions that
presently limit the \Vus precision. Furthermore, precision $\tau$
branching fractions and spectral function measurements can contribute to
improving the theory uncertainty.

\section{Conclusions}
\label{sec:conclusion}

The Standard Model lepton universality is confirmed by the updated HFLAV
Spring 2017 tests. The \Vus determination from
inclusive $\tau \to X_s \nu$ decays is
$\htuse{Vus_mism_sigma_abs}\sigma$ lower than the unitarity CKM
prediction. The discrepancy persists also when considering indirect
determinations of some $\tau$ branching fractions from kaon
measurements and additional experimental measurements from an
in-progress \babar study.
However, an alternative \Vus computation relying also on
the \mtau spectral functions obtains a \Vus value that is entirely consistent
with CKM unitarity. More precise
$\tau$ branching fractions and spectral functions measurements are
desirable to improve our understanding of the observed discrepancy on
\Vus from inclusive $\tau \to X_s \nu$ decays.

\ifdefined\bibtexflag
\bibliography{%
  additional%
  ,bibtex/pub-2017%
  ,bibtex/pub-2016%
  ,bibtex/tau-lepton%
  ,bibtex/pub-extra%
  ,lattice%
  ,tau-refs%
  ,tau-refs-pdg%
  ,Summer16%
}
\else

\input{lusiani-phipsi17-procs-bt.bbl}
\fi
\end{document}

%% file: tau-defs.tex
\newcommand{\BRF}[2]{#2}
\renewcommand{\BR}{\ensuremath{B}\xspace}
\ifhevea
\renewcommand{\babar}{BaBar\xspace}
\else
\providecommand{\babar}{\mbox{\slshape B\kern-0.1em{\smaller A}\kern-0.1em
    B\kern-0.1em{\smaller A\kern-0.2em R}}\xspace}
\fi

\newcommand{\lepth}{\lambda}
\newcommand{\leptl}{\rho}

\newcommand{\radRatio}{R}

\newcommand{\Rhad}{\ensuremath{R_{\text{had}}}\xspace}

\newcommand{\Gammahad}{\ensuremath{\Gamma_{\text{had}}}\xspace}
\newcommand{\Rstrange}{\ensuremath{R_s}\xspace}
\newcommand{\BRstrange}{\ensuremath{\BR_s}\xspace}
\newcommand{\Gammastrange}{\ensuremath{\Gamma_s}\xspace}
\newcommand{\Rnonstrange}{\ensuremath{R_{\text{VA}}}\xspace}
\newcommand{\BRnonstrange}{\ensuremath{\BR_{\text{VA}}}\xspace}
\newcommand{\Gammanonstrange}{\ensuremath{\Gamma_{\text{VA}}}\xspace}

\newcommand{\VusUni}{\ensuremath{\Vus_{\text{uni}}}\xspace}
\newcommand{\VusTauIncl}{\ensuremath{\Vus_{\tau s}}\xspace}
\newcommand{\VusTauKpi}{\ensuremath{\Vus_{\tau K/\pi}}\xspace}
\newcommand{\VusTauKnu}{\ensuremath{\Vus_{\tau K}}\xspace}

\providecommand{\slashLikeH}{%
  \raisebox{.9ex}{%
    \scalebox{.7}{%
      \rotatebox[origin=c]{18}{$-$}%
    }%
  }%
}
\providecommand{\hslash}{%
  {%
   \vphantom{h}%
   \ooalign{\kern.05em\smash{\slashLikeH}\hidewidth\cr$h$\cr}%
   \kern.05em
  }%
}

%% file: tau-br-fit-hfag-2016.tex
\htdef{UnitarityResid}{(0.0355 \pm 0.1031)\%}%
\htdef{MeasNum}{170}%
\htdef{QuantNum}{135}%
\htdef{QuantNumNonRatio}{118}%
\htdef{QuantNumRatio}{17}%
\htdef{QuantNumWithMeas}{84}%
\htdef{QuantNumNonRatioWithMeas}{71}%
\htdef{QuantNumRatioWithMeas}{13}%
\htdef{QuantNumPdg}{129}%
\htdef{QuantNumNonRatioPdg}{112}%
\htdef{QuantNumRatioPdg}{17}%
\htdef{QuantNumWithMeasPdg}{82}%
\htdef{QuantNumNonRatioWithMeasPdg}{69}%
\htdef{QuantNumRatioWithMeasPdg}{13}%
\htdef{IndepQuantNum}{47}%
\htdef{BaseQuantNum}{47}%
\htdef{UnitarityQuantNum}{48}%
\htdef{ConstrNum}{88}%
\htdef{ConstrNumPdg}{82}%
\htdef{Chisq}{137}%
\htdef{Dof}{123}%
\htdef{ChisqProb}{17.84\%}%
\htdef{ChisqProbRound}{18\%}%
\htmeasdef{ALEPH.Gamma10.pub.BARATE.99K}{Gamma10}{ALEPH}{Barate:1999hi}{0.00696 \pm 0.0002865}{0.00696}{\pm 0.0002865}{0}%
\htmeasdef{ALEPH.Gamma103.pub.SCHAEL.05C}{Gamma103}{ALEPH}{Schael:2005am}{0.00072 \pm 0.00015}{0.00072}{\pm 0.00015}{0}%
\htmeasdef{ALEPH.Gamma104.pub.SCHAEL.05C}{Gamma104}{ALEPH}{Schael:2005am}{( 0.021 \pm 0.007 \pm 0.009 ) \cdot 10^{ -2 }}{0.021e-2}{\pm 0.007e-2}{0.009e-2}%
\htmeasdef{ALEPH.Gamma126.pub.BUSKULIC.97C}{Gamma126}{ALEPH}{Buskulic:1996qs}{0.0018 \pm 0.0004472}{0.0018}{\pm 0.0004472}{0}%
\htmeasdef{ALEPH.Gamma128.pub.BUSKULIC.97C}{Gamma128}{ALEPH}{Buskulic:1996qs}{( 2.9 {}^{+1.3\cdot 10^{-4}}_{-1.2} \pm 0.7 ) \cdot 10^{ -4 }}{2.9e-4}{{}^{+1.3e-4}_{-1.2e-4}}{0.7e-4}%
\htmeasdef{ALEPH.Gamma13.pub.SCHAEL.05C}{Gamma13}{ALEPH}{Schael:2005am}{0.25924 \pm 0.00128973}{0.25924}{\pm 0.00128973}{0}%
\htmeasdef{ALEPH.Gamma150.pub.BUSKULIC.97C}{Gamma150}{ALEPH}{Buskulic:1996qs}{0.0191 \pm 0.000922}{0.0191}{\pm 0.000922}{0}%
\htmeasdef{ALEPH.Gamma150by66.pub.BUSKULIC.96}{Gamma150by66}{ALEPH}{Buskulic:1995ty}{0.431 \pm 0.033}{0.431}{\pm 0.033}{0}%
\htmeasdef{ALEPH.Gamma152.pub.BUSKULIC.97C}{Gamma152}{ALEPH}{Buskulic:1996qs}{0.0043 \pm 0.000781}{0.0043}{\pm 0.000781}{0}%
\htmeasdef{ALEPH.Gamma16.pub.BARATE.99K}{Gamma16}{ALEPH}{Barate:1999hi}{0.00444 \pm 0.0003538}{0.00444}{\pm 0.0003538}{0}%
\htmeasdef{ALEPH.Gamma19.pub.SCHAEL.05C}{Gamma19}{ALEPH}{Schael:2005am}{0.09295 \pm 0.00121655}{0.09295}{\pm 0.00121655}{0}%
\htmeasdef{ALEPH.Gamma23.pub.BARATE.99K}{Gamma23}{ALEPH}{Barate:1999hi}{0.00056 \pm 0.00025}{0.00056}{\pm 0.00025}{0}%
\htmeasdef{ALEPH.Gamma26.pub.SCHAEL.05C}{Gamma26}{ALEPH}{Schael:2005am}{0.01082 \pm 0.000925581}{0.01082}{\pm 0.000925581}{0}%
\htmeasdef{ALEPH.Gamma28.pub.BARATE.99K}{Gamma28}{ALEPH}{Barate:1999hi}{0.00037 \pm 0.0002371}{0.00037}{\pm 0.0002371}{0}%
\htmeasdef{ALEPH.Gamma3.pub.SCHAEL.05C}{Gamma3}{ALEPH}{Schael:2005am}{0.17319 \pm 0.000769675}{0.17319}{\pm 0.000769675}{0}%
\htmeasdef{ALEPH.Gamma30.pub.SCHAEL.05C}{Gamma30}{ALEPH}{Schael:2005am}{0.00112 \pm 0.000509313}{0.00112}{\pm 0.000509313}{0}%
\htmeasdef{ALEPH.Gamma33.pub.BARATE.98E}{Gamma33}{ALEPH}{Barate:1997tt}{0.0097 \pm 0.000849}{0.0097}{\pm 0.000849}{0}%
\htmeasdef{ALEPH.Gamma35.pub.BARATE.99K}{Gamma35}{ALEPH}{Barate:1999hi}{0.00928 \pm 0.000564}{0.00928}{\pm 0.000564}{0}%
\htmeasdef{ALEPH.Gamma37.pub.BARATE.98E}{Gamma37}{ALEPH}{Barate:1997tt}{0.00158 \pm 0.0004531}{0.00158}{\pm 0.0004531}{0}%
\htmeasdef{ALEPH.Gamma37.pub.BARATE.99K}{Gamma37}{ALEPH}{Barate:1999hi}{0.00162 \pm 0.0002371}{0.00162}{\pm 0.0002371}{0}%
\htmeasdef{ALEPH.Gamma40.pub.BARATE.98E}{Gamma40}{ALEPH}{Barate:1997tt}{0.00294 \pm 0.0008184}{0.00294}{\pm 0.0008184}{0}%
\htmeasdef{ALEPH.Gamma40.pub.BARATE.99K}{Gamma40}{ALEPH}{Barate:1999hi}{0.00347 \pm 0.0006464}{0.00347}{\pm 0.0006464}{0}%
\htmeasdef{ALEPH.Gamma42.pub.BARATE.98E}{Gamma42}{ALEPH}{Barate:1997tt}{0.00152 \pm 0.0007885}{0.00152}{\pm 0.0007885}{0}%
\htmeasdef{ALEPH.Gamma42.pub.BARATE.99K}{Gamma42}{ALEPH}{Barate:1999hi}{0.00143 \pm 0.0002915}{0.00143}{\pm 0.0002915}{0}%
\htmeasdef{ALEPH.Gamma44.pub.BARATE.99R}{Gamma44}{ALEPH}{Barate:1999hj}{0.00026 \pm 0.00024}{0.00026}{\pm 0.00024}{0}%
\htmeasdef{ALEPH.Gamma47.pub.BARATE.98E}{Gamma47}{ALEPH}{Barate:1997tt}{0.00026 \pm 0.0001118}{0.00026}{\pm 0.0001118}{0}%
\htmeasdef{ALEPH.Gamma48.pub.BARATE.98E}{Gamma48}{ALEPH}{Barate:1997tt}{0.00101 \pm 0.0002642}{0.00101}{\pm 0.0002642}{0}%
\htmeasdef{ALEPH.Gamma5.pub.SCHAEL.05C}{Gamma5}{ALEPH}{Schael:2005am}{0.17837 \pm 0.000804984}{0.17837}{\pm 0.000804984}{0}%
\htmeasdef{ALEPH.Gamma51.pub.BARATE.98E}{Gamma51}{ALEPH}{Barate:1997tt}{( 3.1 \pm 1.1 \pm 0.5 ) \cdot 10^{ -4 }}{3.1e-4}{\pm 1.1e-4}{0.5e-4}%
\htmeasdef{ALEPH.Gamma53.pub.BARATE.98E}{Gamma53}{ALEPH}{Barate:1997tt}{0.00023 \pm 0.000202485}{0.00023}{\pm 0.000202485}{0}%
\htmeasdef{ALEPH.Gamma58.pub.SCHAEL.05C}{Gamma58}{ALEPH}{Schael:2005am}{0.09469 \pm 0.000957758}{0.09469}{\pm 0.000957758}{0}%
\htmeasdef{ALEPH.Gamma66.pub.SCHAEL.05C}{Gamma66}{ALEPH}{Schael:2005am}{0.04734 \pm 0.000766942}{0.04734}{\pm 0.000766942}{0}%
\htmeasdef{ALEPH.Gamma76.pub.SCHAEL.05C}{Gamma76}{ALEPH}{Schael:2005am}{0.00435 \pm 0.000460977}{0.00435}{\pm 0.000460977}{0}%
\htmeasdef{ALEPH.Gamma8.pub.SCHAEL.05C}{Gamma8}{ALEPH}{Schael:2005am}{0.11524 \pm 0.00104805}{0.11524}{\pm 0.00104805}{0}%
\htmeasdef{ALEPH.Gamma805.pub.SCHAEL.05C}{Gamma805}{ALEPH}{Schael:2005am}{( 4 \pm 2 ) \cdot 10^{ -4 }}{4e-04}{\pm 2e-04}{0}%
\htmeasdef{ALEPH.Gamma85.pub.BARATE.98}{Gamma85}{ALEPH}{Barate:1997ma}{0.00214 \pm 0.0004701}{0.00214}{\pm 0.0004701}{0}%
\htmeasdef{ALEPH.Gamma88.pub.BARATE.98}{Gamma88}{ALEPH}{Barate:1997ma}{0.00061 \pm 0.0004295}{0.00061}{\pm 0.0004295}{0}%
\htmeasdef{ALEPH.Gamma93.pub.BARATE.98}{Gamma93}{ALEPH}{Barate:1997ma}{0.00163 \pm 0.0002702}{0.00163}{\pm 0.0002702}{0}%
\htmeasdef{ALEPH.Gamma94.pub.BARATE.98}{Gamma94}{ALEPH}{Barate:1997ma}{0.00075 \pm 0.0003265}{0.00075}{\pm 0.0003265}{0}%
\htmeasdef{ARGUS.Gamma103.pub.ALBRECHT.88B}{Gamma103}{ARGUS}{Albrecht:1987zf}{0.00064 \pm 0.00023 \pm 0.0001}{0.00064}{\pm 0.00023}{0.0001}%
\htmeasdef{ARGUS.Gamma3by5.pub.ALBRECHT.92D}{Gamma3by5}{ARGUS}{Albrecht:1991rh}{0.997 \pm 0.035 \pm 0.04}{0.997}{\pm 0.035}{0.04}%
\htmeasdef{BaBar.Gamma10by5.pub.AUBERT.10F}{Gamma10by5}{\babar}{Aubert:2009qj}{0.03882 \pm 0.00032 \pm 0.00057}{0.03882}{\pm 0.00032}{0.00057}%
\htmeasdef{BaBar.Gamma128.pub.DEL-AMO-SANCHEZ.11E}{Gamma128}{\babar}{delAmoSanchez:2010pc}{0.000142 \pm 1.1\cdot 10^{-5} \pm 7\cdot 10^{-6}}{0.000142}{\pm 1.1e-05}{7e-06}%
\htmeasdef{BaBar.Gamma16.pub.AUBERT.07AP}{Gamma16}{\babar}{Aubert:2007jh}{0.00416 \pm 3\cdot 10^{-5} \pm 0.00018}{0.00416}{\pm 3e-05}{0.00018}%
\htmeasdef{BaBar.Gamma3by5.pub.AUBERT.10F}{Gamma3by5}{\babar}{Aubert:2009qj}{0.9796 \pm 0.0016 \pm 0.0036}{0.9796}{\pm 0.0016}{0.0036}%
\htmeasdef{BaBar.Gamma47.pub.LEES.12Y}{Gamma47}{\babar}{Lees:2012de}{( 2.31 \pm 0.04 \pm 0.08 ) \cdot 10^{ -4 }}{2.31e-4}{\pm 0.04e-4}{0.08e-4}%
\htmeasdef{BaBar.Gamma50.pub.LEES.12Y}{Gamma50}{\babar}{Lees:2012de}{( 1.60 \pm 0.20 \pm 0.22 ) \cdot 10^{ -5 }}{1.60e-5}{\pm 0.20e-5}{0.22e-5}%
\htmeasdef{BaBar.Gamma60.pub.AUBERT.08}{Gamma60}{\babar}{Aubert:2007mh}{0.0883 \pm 0.0001 \pm 0.0013}{0.0883}{\pm 0.0001}{0.0013}%
\htmeasdef{BaBar.Gamma811.pub.LEES.12X}{Gamma811}{\babar}{Lees:2012ks}{( 7.3 \pm 1.2 \pm 1.2 ) \cdot 10^{ -5 }}{7.3e-5}{\pm 1.2e-5}{1.2e-5}%
\htmeasdef{BaBar.Gamma812.pub.LEES.12X}{Gamma812}{\babar}{Lees:2012ks}{( 0.1 \pm 0.08 \pm 0.30 ) \cdot 10^{ -4 }}{0.1e-4}{\pm 0.08e-4}{0.30e-4}%
\htmeasdef{BaBar.Gamma821.pub.LEES.12X}{Gamma821}{\babar}{Lees:2012ks}{( 7.68 \pm 0.04 \pm 0.40 ) \cdot 10^{ -4 }}{7.68e-4}{\pm 0.04e-4}{0.40e-4}%
\htmeasdef{BaBar.Gamma822.pub.LEES.12X}{Gamma822}{\babar}{Lees:2012ks}{( 0.6 \pm 0.5 \pm 1.1 ) \cdot 10^{ -6 }}{0.6e-06}{\pm 0.5e-06}{1.1e-06}%
\htmeasdef{BaBar.Gamma831.pub.LEES.12X}{Gamma831}{\babar}{Lees:2012ks}{( 8.4 \pm 0.4 \pm 0.6 ) \cdot 10^{ -5 }}{8.4e-5}{\pm 0.4e-5}{0.6e-5}%
\htmeasdef{BaBar.Gamma832.pub.LEES.12X}{Gamma832}{\babar}{Lees:2012ks}{( 0.36 \pm 0.03 \pm 0.09 ) \cdot 10^{ -4 }}{0.36e-4}{\pm 0.03e-4}{0.09e-4}%
\htmeasdef{BaBar.Gamma833.pub.LEES.12X}{Gamma833}{\babar}{Lees:2012ks}{( 1.1 \pm 0.4 \pm 0.4 ) \cdot 10^{ -6 }}{1.1e-6}{\pm 0.4e-6}{0.4e-6}%
\htmeasdef{BaBar.Gamma85.pub.AUBERT.08}{Gamma85}{\babar}{Aubert:2007mh}{0.00273 \pm 2\cdot 10^{-5} \pm 9\cdot 10^{-5}}{0.00273}{\pm 2e-05}{9e-05}%
\htmeasdef{BaBar.Gamma910.pub.LEES.12X}{Gamma910}{\babar}{Lees:2012ks}{( 8.27 \pm 0.88 \pm 0.81 ) \cdot 10^{ -5 }}{8.27e-5}{\pm 0.88e-5}{0.81e-5}%
\htmeasdef{BaBar.Gamma911.pub.LEES.12X}{Gamma911}{\babar}{Lees:2012ks}{( 4.57 \pm 0.77 \pm 0.50 ) \cdot 10^{ -5 }}{4.57e-5}{\pm 0.77e-5}{0.50e-5}%
\htmeasdef{BaBar.Gamma920.pub.LEES.12X}{Gamma920}{\babar}{Lees:2012ks}{( 5.20 \pm 0.31 \pm 0.37 ) \cdot 10^{ -5 }}{5.20e-5}{\pm 0.31e-5}{0.37e-5}%
\htmeasdef{BaBar.Gamma93.pub.AUBERT.08}{Gamma93}{\babar}{Aubert:2007mh}{0.001346 \pm 1\cdot 10^{-5} \pm 3.6\cdot 10^{-5}}{0.001346}{\pm 1e-05}{3.6e-05}%
\htmeasdef{BaBar.Gamma930.pub.LEES.12X}{Gamma930}{\babar}{Lees:2012ks}{( 5.39 \pm 0.27 \pm 0.41 ) \cdot 10^{ -5 }}{5.39e-5}{\pm 0.27e-5}{0.41e-5}%
\htmeasdef{BaBar.Gamma944.pub.LEES.12X}{Gamma944}{\babar}{Lees:2012ks}{( 8.26 \pm 0.35 \pm 0.51 ) \cdot 10^{ -5 }}{8.26e-5}{\pm 0.35e-5}{0.51e-5}%
\htmeasdef{BaBar.Gamma96.pub.AUBERT.08}{Gamma96}{\babar}{Aubert:2007mh}{1.5777\cdot 10^{-5} \pm 1.3\cdot 10^{-6} \pm 1.2308\cdot 10^{-6}}{1.5777e-05}{\pm 1.3e-06}{1.2308e-06}%
\htmeasdef{BaBar.Gamma9by5.pub.AUBERT.10F}{Gamma9by5}{\babar}{Aubert:2009qj}{0.5945 \pm 0.0014 \pm 0.0061}{0.5945}{\pm 0.0014}{0.0061}%
\htmeasdef{Belle.Gamma126.pub.INAMI.09}{Gamma126}{Belle}{Inami:2008ar}{0.00135 \pm 3\cdot 10^{-5} \pm 7\cdot 10^{-5}}{0.00135}{\pm 3e-05}{7e-05}%
\htmeasdef{Belle.Gamma128.pub.INAMI.09}{Gamma128}{Belle}{Inami:2008ar}{0.000158 \pm 5\cdot 10^{-6} \pm 9\cdot 10^{-6}}{0.000158}{\pm 5e-06}{9e-06}%
\htmeasdef{Belle.Gamma13.pub.FUJIKAWA.08}{Gamma13}{Belle}{Fujikawa:2008ma}{0.2567 \pm 1\cdot 10^{-4} \pm 0.0039}{0.2567}{\pm 1e-04}{0.0039}%
\htmeasdef{Belle.Gamma130.pub.INAMI.09}{Gamma130}{Belle}{Inami:2008ar}{4.6\cdot 10^{-5} \pm 1.1\cdot 10^{-5} \pm 4\cdot 10^{-6}}{4.6e-05}{\pm 1.1e-05}{4e-06}%
\htmeasdef{Belle.Gamma132.pub.INAMI.09}{Gamma132}{Belle}{Inami:2008ar}{8.8\cdot 10^{-5} \pm 1.4\cdot 10^{-5} \pm 6\cdot 10^{-6}}{8.8e-05}{\pm 1.4e-05}{6e-06}%
\htmeasdef{Belle.Gamma35.pub.RYU.14vpc}{Gamma35}{Belle}{Ryu:2014vpc}{8.32\cdot 10^{-3} \pm 0.3\% \pm 1.8\%}{8.32e-03}{\pm 0.3\%}{1.8\%}%
\htmeasdef{Belle.Gamma37.pub.RYU.14vpc}{Gamma37}{Belle}{Ryu:2014vpc}{14.8\cdot 10^{-4} \pm 0.9\% \pm 3.7\%}{14.8e-04}{\pm 0.9\%}{3.7\%}%
\htmeasdef{Belle.Gamma40.pub.RYU.14vpc}{Gamma40}{Belle}{Ryu:2014vpc}{3.86\cdot 10^{-3} \pm 0.8\% \pm 3.5\%}{3.86e-03}{\pm 0.8\%}{3.5\%}%
\htmeasdef{Belle.Gamma42.pub.RYU.14vpc}{Gamma42}{Belle}{Ryu:2014vpc}{14.96\cdot 10^{-4} \pm 1.3\% \pm 4.9\%}{14.96e-04}{\pm 1.3\%}{4.9\%}%
\htmeasdef{Belle.Gamma47.pub.RYU.14vpc}{Gamma47}{Belle}{Ryu:2014vpc}{2.33\cdot 10^{-4} \pm 1.4\% \pm 4.0\%}{2.33e-04}{\pm 1.4\%}{4.0\%}%
\htmeasdef{Belle.Gamma50.pub.RYU.14vpc}{Gamma50}{Belle}{Ryu:2014vpc}{2.00\cdot 10^{-5} \pm 10.8\% \pm 10.1\%}{2.00e-05}{\pm 10.8\%}{10.1\%}%
\htmeasdef{Belle.Gamma60.pub.LEE.10}{Gamma60}{Belle}{Lee:2010tc}{0.0842 \pm 0 {}^{+0.0026}_{-0.0025}}{0.0842}{\pm 0}{{}^{+0.0026}_{-0.0025}}%
\htmeasdef{Belle.Gamma85.pub.LEE.10}{Gamma85}{Belle}{Lee:2010tc}{0.0033 \pm 1\cdot 10^{-5} {}^{+0.00016}_{-0.00017}}{0.0033}{\pm 1e-05}{{}^{+0.00016}_{-0.00017}}%
\htmeasdef{Belle.Gamma93.pub.LEE.10}{Gamma93}{Belle}{Lee:2010tc}{0.00155 \pm 1\cdot 10^{-5} {}^{+6\cdot 10^{-5}}_{-5\cdot 10^{-5}}}{0.00155}{\pm 1e-05}{{}^{+6e-05}_{-5e-05}}%
\htmeasdef{Belle.Gamma96.pub.LEE.10}{Gamma96}{Belle}{Lee:2010tc}{3.29\cdot 10^{-5} \pm 1.7\cdot 10^{-6} {}^{+1.9\cdot 10^{-6}}_{-2.0\cdot 10^{-6}}}{3.29e-05}{\pm 1.7e-06}{{}^{+1.9e-06}_{-2.0e-06}}%
\htmeasdef{CELLO.Gamma54.pub.BEHREND.89B}{Gamma54}{CELLO}{Behrend:1989wc}{0.15 \pm 0.004 \pm 0.003}{0.15}{\pm 0.004}{0.003}%
\htmeasdef{CLEO.Gamma10.pub.BATTLE.94}{Gamma10}{CLEO}{Battle:1994by}{0.0066 \pm 0.0007 \pm 0.0009}{0.0066}{\pm 0.0007}{0.0009}%
\htmeasdef{CLEO.Gamma102.pub.GIBAUT.94B}{Gamma102}{CLEO}{Gibaut:1994ik}{0.00097 \pm 5\cdot 10^{-5} \pm 0.00011}{0.00097}{\pm 5e-05}{0.00011}%
\htmeasdef{CLEO.Gamma103.pub.GIBAUT.94B}{Gamma103}{CLEO}{Gibaut:1994ik}{0.00077 \pm 5\cdot 10^{-5} \pm 9\cdot 10^{-5}}{0.00077}{\pm 5e-05}{9e-05}%
\htmeasdef{CLEO.Gamma104.pub.ANASTASSOV.01}{Gamma104}{CLEO}{Anastassov:2000xu}{0.00017 \pm 2\cdot 10^{-5} \pm 2\cdot 10^{-5}}{0.00017}{\pm 2e-05}{2e-05}%
\htmeasdef{CLEO.Gamma126.pub.ARTUSO.92}{Gamma126}{CLEO}{Artuso:1992qu}{0.0017 \pm 0.0002 \pm 0.0002}{0.0017}{\pm 0.0002}{0.0002}%
\htmeasdef{CLEO.Gamma128.pub.BARTELT.96}{Gamma128}{CLEO}{Bartelt:1996iv}{( 2.6 \pm 0.5 \pm 0.5 ) \cdot 10^{ -4 }}{2.6e-4}{\pm 0.5e-4}{0.5e-4}%
\htmeasdef{CLEO.Gamma13.pub.ARTUSO.94}{Gamma13}{CLEO}{Artuso:1994ii}{0.2587 \pm 0.0012 \pm 0.0042}{0.2587}{\pm 0.0012}{0.0042}%
\htmeasdef{CLEO.Gamma130.pub.BISHAI.99}{Gamma130}{CLEO}{Bishai:1998gf}{( 1.77 \pm 0.56 \pm 0.71 ) \cdot 10^{ -4 }}{1.77e-4}{\pm 0.56e-4}{0.71e-4}%
\htmeasdef{CLEO.Gamma132.pub.BISHAI.99}{Gamma132}{CLEO}{Bishai:1998gf}{( 2.2 \pm 0.70 \pm 0.22 ) \cdot 10^{ -4 }}{2.2e-4}{\pm 0.70e-4}{0.22e-4}%
\htmeasdef{CLEO.Gamma150.pub.BARINGER.87}{Gamma150}{CLEO}{Baringer:1987tr}{0.016 \pm 0.0027 \pm 0.0041}{0.016}{\pm 0.0027}{0.0041}%
\htmeasdef{CLEO.Gamma150by66.pub.BALEST.95C}{Gamma150by66}{CLEO}{Balest:1995kq}{0.464 \pm 0.016 \pm 0.017}{0.464}{\pm 0.016}{0.017}%
\htmeasdef{CLEO.Gamma152by76.pub.BORTOLETTO.93}{Gamma152by76}{CLEO}{Bortoletto:1993px}{0.81 \pm 0.06 \pm 0.06}{0.81}{\pm 0.06}{0.06}%
\htmeasdef{CLEO.Gamma16.pub.BATTLE.94}{Gamma16}{CLEO}{Battle:1994by}{0.0051 \pm 0.001 \pm 0.0007}{0.0051}{\pm 0.001}{0.0007}%
\htmeasdef{CLEO.Gamma19by13.pub.PROCARIO.93}{Gamma19by13}{CLEO}{Procario:1992hd}{0.342 \pm 0.006 \pm 0.016}{0.342}{\pm 0.006}{0.016}%
\htmeasdef{CLEO.Gamma23.pub.BATTLE.94}{Gamma23}{CLEO}{Battle:1994by}{0.0009 \pm 0.001 \pm 0.0003}{0.0009}{\pm 0.001}{0.0003}%
\htmeasdef{CLEO.Gamma26by13.pub.PROCARIO.93}{Gamma26by13}{CLEO}{Procario:1992hd}{0.044 \pm 0.003 \pm 0.005}{0.044}{\pm 0.003}{0.005}%
\htmeasdef{CLEO.Gamma29.pub.PROCARIO.93}{Gamma29}{CLEO}{Procario:1992hd}{0.0016 \pm 0.0005 \pm 0.0005}{0.0016}{\pm 0.0005}{0.0005}%
\htmeasdef{CLEO.Gamma31.pub.BATTLE.94}{Gamma31}{CLEO}{Battle:1994by}{0.017 \pm 0.0012 \pm 0.0019}{0.017}{\pm 0.0012}{0.0019}%
\htmeasdef{CLEO.Gamma34.pub.COAN.96}{Gamma34}{CLEO}{Coan:1996iu}{0.00855 \pm 0.00036 \pm 0.00073}{0.00855}{\pm 0.00036}{0.00073}%
\htmeasdef{CLEO.Gamma37.pub.COAN.96}{Gamma37}{CLEO}{Coan:1996iu}{0.00151 \pm 0.00021 \pm 0.00022}{0.00151}{\pm 0.00021}{0.00022}%
\htmeasdef{CLEO.Gamma39.pub.COAN.96}{Gamma39}{CLEO}{Coan:1996iu}{0.00562 \pm 0.0005 \pm 0.00048}{0.00562}{\pm 0.0005}{0.00048}%
\htmeasdef{CLEO.Gamma3by5.pub.ANASTASSOV.97}{Gamma3by5}{CLEO}{Anastassov:1996tc}{0.9777 \pm 0.0063 \pm 0.0087}{0.9777}{\pm 0.0063}{0.0087}%
\htmeasdef{CLEO.Gamma42.pub.COAN.96}{Gamma42}{CLEO}{Coan:1996iu}{0.00145 \pm 0.00036 \pm 0.0002}{0.00145}{\pm 0.00036}{0.0002}%
\htmeasdef{CLEO.Gamma47.pub.COAN.96}{Gamma47}{CLEO}{Coan:1996iu}{0.00023 \pm 5\cdot 10^{-5} \pm 3\cdot 10^{-5}}{0.00023}{\pm 5e-05}{3e-05}%
\htmeasdef{CLEO.Gamma5.pub.ANASTASSOV.97}{Gamma5}{CLEO}{Anastassov:1996tc}{0.1776 \pm 0.0006 \pm 0.0017}{0.1776}{\pm 0.0006}{0.0017}%
\htmeasdef{CLEO.Gamma57.pub.BALEST.95C}{Gamma57}{CLEO}{Balest:1995kq}{0.0951 \pm 0.0007 \pm 0.002}{0.0951}{\pm 0.0007}{0.002}%
\htmeasdef{CLEO.Gamma66.pub.BALEST.95C}{Gamma66}{CLEO}{Balest:1995kq}{0.0423 \pm 0.0006 \pm 0.0022}{0.0423}{\pm 0.0006}{0.0022}%
\htmeasdef{CLEO.Gamma69.pub.EDWARDS.00A}{Gamma69}{CLEO}{Edwards:1999fj}{0.0419 \pm 0.001 \pm 0.0021}{0.0419}{\pm 0.001}{0.0021}%
\htmeasdef{CLEO.Gamma76by54.pub.BORTOLETTO.93}{Gamma76by54}{CLEO}{Bortoletto:1993px}{0.034 \pm 0.002 \pm 0.003}{0.034}{\pm 0.002}{0.003}%
\htmeasdef{CLEO.Gamma78.pub.ANASTASSOV.01}{Gamma78}{CLEO}{Anastassov:2000xu}{0.00022 \pm 3\cdot 10^{-5} \pm 4\cdot 10^{-5}}{0.00022}{\pm 3e-05}{4e-05}%
\htmeasdef{CLEO.Gamma8.pub.ANASTASSOV.97}{Gamma8}{CLEO}{Anastassov:1996tc}{0.1152 \pm 0.0005 \pm 0.0012}{0.1152}{\pm 0.0005}{0.0012}%
\htmeasdef{CLEO.Gamma80by60.pub.RICHICHI.99}{Gamma80by60}{CLEO}{Richichi:1998bc}{0.0544 \pm 0.0021 \pm 0.0053}{0.0544}{\pm 0.0021}{0.0053}%
\htmeasdef{CLEO.Gamma81by69.pub.RICHICHI.99}{Gamma81by69}{CLEO}{Richichi:1998bc}{0.0261 \pm 0.0045 \pm 0.0042}{0.0261}{\pm 0.0045}{0.0042}%
\htmeasdef{CLEO.Gamma93by60.pub.RICHICHI.99}{Gamma93by60}{CLEO}{Richichi:1998bc}{0.016 \pm 0.0015 \pm 0.003}{0.016}{\pm 0.0015}{0.003}%
\htmeasdef{CLEO.Gamma94by69.pub.RICHICHI.99}{Gamma94by69}{CLEO}{Richichi:1998bc}{0.0079 \pm 0.0044 \pm 0.0016}{0.0079}{\pm 0.0044}{0.0016}%
\htmeasdef{CLEO3.Gamma151.pub.ARMS.05}{Gamma151}{CLEO3}{Arms:2005qg}{( 4.1 \pm 0.6 \pm 0.7 ) \cdot 10^{ -4 }}{4.1e-4}{\pm 0.6e-4}{0.7e-4}%
\htmeasdef{CLEO3.Gamma60.pub.BRIERE.03}{Gamma60}{CLEO3}{Briere:2003fr}{0.0913 \pm 0.0005 \pm 0.0046}{0.0913}{\pm 0.0005}{0.0046}%
\htmeasdef{CLEO3.Gamma85.pub.BRIERE.03}{Gamma85}{CLEO3}{Briere:2003fr}{0.00384 \pm 0.00014 \pm 0.00038}{0.00384}{\pm 0.00014}{0.00038}%
\htmeasdef{CLEO3.Gamma88.pub.ARMS.05}{Gamma88}{CLEO3}{Arms:2005qg}{0.00074 \pm 8\cdot 10^{-5} \pm 0.00011}{0.00074}{\pm 8e-05}{0.00011}%
\htmeasdef{CLEO3.Gamma93.pub.BRIERE.03}{Gamma93}{CLEO3}{Briere:2003fr}{0.00155 \pm 6\cdot 10^{-5} \pm 9\cdot 10^{-5}}{0.00155}{\pm 6e-05}{9e-05}%
\htmeasdef{CLEO3.Gamma94.pub.ARMS.05}{Gamma94}{CLEO3}{Arms:2005qg}{( 5.5 \pm 1.4 \pm 1.2 ) \cdot 10^{ -5 }}{5.5e-05}{\pm 1.4e-05}{1.2e-05}%
\htmeasdef{DELPHI.Gamma10.pub.ABREU.94K}{Gamma10}{DELPHI}{Abreu:1994fi}{0.0085 \pm 0.0018}{0.0085}{\pm 0.0018}{0}%
\htmeasdef{DELPHI.Gamma103.pub.ABDALLAH.06A}{Gamma103}{DELPHI}{Abdallah:2003cw}{0.00097 \pm 0.00015 \pm 5\cdot 10^{-5}}{0.00097}{\pm 0.00015}{5e-05}%
\htmeasdef{DELPHI.Gamma104.pub.ABDALLAH.06A}{Gamma104}{DELPHI}{Abdallah:2003cw}{0.00016 \pm 0.00012 \pm 6\cdot 10^{-5}}{0.00016}{\pm 0.00012}{6e-05}%
\htmeasdef{DELPHI.Gamma13.pub.ABDALLAH.06A}{Gamma13}{DELPHI}{Abdallah:2003cw}{0.2574 \pm 0.00201 \pm 0.00138}{0.2574}{\pm 0.00201}{0.00138}%
\htmeasdef{DELPHI.Gamma19.pub.ABDALLAH.06A}{Gamma19}{DELPHI}{Abdallah:2003cw}{0.09498 \pm 0.0032 \pm 0.00275}{0.09498}{\pm 0.0032}{0.00275}%
\htmeasdef{DELPHI.Gamma25.pub.ABDALLAH.06A}{Gamma25}{DELPHI}{Abdallah:2003cw}{0.01403 \pm 0.00214 \pm 0.00224}{0.01403}{\pm 0.00214}{0.00224}%
\htmeasdef{DELPHI.Gamma3.pub.ABREU.99X}{Gamma3}{DELPHI}{Abreu:1999rb}{0.17325 \pm 0.00095 \pm 0.00077}{0.17325}{\pm 0.00095}{0.00077}%
\htmeasdef{DELPHI.Gamma31.pub.ABREU.94K}{Gamma31}{DELPHI}{Abreu:1994fi}{0.0154 \pm 0.0024}{0.0154}{\pm 0.0024}{0}%
\htmeasdef{DELPHI.Gamma5.pub.ABREU.99X}{Gamma5}{DELPHI}{Abreu:1999rb}{0.17877 \pm 0.00109 \pm 0.0011}{0.17877}{\pm 0.00109}{0.0011}%
\htmeasdef{DELPHI.Gamma57.pub.ABDALLAH.06A}{Gamma57}{DELPHI}{Abdallah:2003cw}{0.09317 \pm 0.0009 \pm 0.00082}{0.09317}{\pm 0.0009}{0.00082}%
\htmeasdef{DELPHI.Gamma66.pub.ABDALLAH.06A}{Gamma66}{DELPHI}{Abdallah:2003cw}{0.04545 \pm 0.00106 \pm 0.00103}{0.04545}{\pm 0.00106}{0.00103}%
\htmeasdef{DELPHI.Gamma7.pub.ABREU.92N}{Gamma7}{DELPHI}{Abreu:1992gn}{0.124 \pm 0.007 \pm 0.007}{0.124}{\pm 0.007}{0.007}%
\htmeasdef{DELPHI.Gamma74.pub.ABDALLAH.06A}{Gamma74}{DELPHI}{Abdallah:2003cw}{0.00561 \pm 0.00068 \pm 0.00095}{0.00561}{\pm 0.00068}{0.00095}%
\htmeasdef{DELPHI.Gamma8.pub.ABDALLAH.06A}{Gamma8}{DELPHI}{Abdallah:2003cw}{0.11571 \pm 0.0012 \pm 0.00114}{0.11571}{\pm 0.0012}{0.00114}%
\htmeasdef{HRS.Gamma102.pub.BYLSMA.87}{Gamma102}{HRS}{Bylsma:1986zy}{0.00102 \pm 0.00029}{0.00102}{\pm 0.00029}{0}%
\htmeasdef{HRS.Gamma103.pub.BYLSMA.87}{Gamma103}{HRS}{Bylsma:1986zy}{0.00051 \pm 0.0002}{0.00051}{\pm 0.0002}{0}%
\htmeasdef{L3.Gamma102.pub.ACHARD.01D}{Gamma102}{L3}{Achard:2001pk}{0.0017 \pm 0.00022 \pm 0.00026}{0.0017}{\pm 0.00022}{0.00026}%
\htmeasdef{L3.Gamma13.pub.ACCIARRI.95}{Gamma13}{L3}{Acciarri:1994vr}{0.2505 \pm 0.0035 \pm 0.005}{0.2505}{\pm 0.0035}{0.005}%
\htmeasdef{L3.Gamma19.pub.ACCIARRI.95}{Gamma19}{L3}{Acciarri:1994vr}{0.0888 \pm 0.0037 \pm 0.0042}{0.0888}{\pm 0.0037}{0.0042}%
\htmeasdef{L3.Gamma26.pub.ACCIARRI.95}{Gamma26}{L3}{Acciarri:1994vr}{0.017 \pm 0.0024 \pm 0.0038}{0.017}{\pm 0.0024}{0.0038}%
\htmeasdef{L3.Gamma3.pub.ACCIARRI.01F}{Gamma3}{L3}{Acciarri:2001sg}{0.17342 \pm 0.0011 \pm 0.00067}{0.17342}{\pm 0.0011}{0.00067}%
\htmeasdef{L3.Gamma35.pub.ACCIARRI.95F}{Gamma35}{L3}{Acciarri:1995kx}{0.0095 \pm 0.0015 \pm 0.0006}{0.0095}{\pm 0.0015}{0.0006}%
\htmeasdef{L3.Gamma40.pub.ACCIARRI.95F}{Gamma40}{L3}{Acciarri:1995kx}{0.0041 \pm 0.0012 \pm 0.0003}{0.0041}{\pm 0.0012}{0.0003}%
\htmeasdef{L3.Gamma5.pub.ACCIARRI.01F}{Gamma5}{L3}{Acciarri:2001sg}{0.17806 \pm 0.00104 \pm 0.00076}{0.17806}{\pm 0.00104}{0.00076}%
\htmeasdef{L3.Gamma54.pub.ADEVA.91F}{Gamma54}{L3}{Adeva:1991qq}{0.144 \pm 0.006 \pm 0.003}{0.144}{\pm 0.006}{0.003}%
\htmeasdef{L3.Gamma55.pub.ACHARD.01D}{Gamma55}{L3}{Achard:2001pk}{0.14556 \pm 0.00105 \pm 0.00076}{0.14556}{\pm 0.00105}{0.00076}%
\htmeasdef{L3.Gamma7.pub.ACCIARRI.95}{Gamma7}{L3}{Acciarri:1994vr}{0.1247 \pm 0.0026 \pm 0.0043}{0.1247}{\pm 0.0026}{0.0043}%
\htmeasdef{OPAL.Gamma10.pub.ABBIENDI.01J}{Gamma10}{OPAL}{Abbiendi:2000ee}{0.00658 \pm 0.00027 \pm 0.00029}{0.00658}{\pm 0.00027}{0.00029}%
\htmeasdef{OPAL.Gamma103.pub.ACKERSTAFF.99E}{Gamma103}{OPAL}{Ackerstaff:1998ia}{0.00091 \pm 0.00014 \pm 6\cdot 10^{-5}}{0.00091}{\pm 0.00014}{6e-05}%
\htmeasdef{OPAL.Gamma104.pub.ACKERSTAFF.99E}{Gamma104}{OPAL}{Ackerstaff:1998ia}{0.00027 \pm 0.00018 \pm 9\cdot 10^{-5}}{0.00027}{\pm 0.00018}{9e-05}%
\htmeasdef{OPAL.Gamma13.pub.ACKERSTAFF.98M}{Gamma13}{OPAL}{Ackerstaff:1997tx}{0.2589 \pm 0.0017 \pm 0.0029}{0.2589}{\pm 0.0017}{0.0029}%
\htmeasdef{OPAL.Gamma16.pub.ABBIENDI.04J}{Gamma16}{OPAL}{Abbiendi:2004xa}{0.00471 \pm 0.00059 \pm 0.00023}{0.00471}{\pm 0.00059}{0.00023}%
\htmeasdef{OPAL.Gamma17.pub.ACKERSTAFF.98M}{Gamma17}{OPAL}{Ackerstaff:1997tx}{0.0991 \pm 0.0031 \pm 0.0027}{0.0991}{\pm 0.0031}{0.0027}%
\htmeasdef{OPAL.Gamma3.pub.ABBIENDI.03}{Gamma3}{OPAL}{Abbiendi:2002jw}{0.1734 \pm 0.0009 \pm 0.0006}{0.1734}{\pm 0.0009}{0.0006}%
\htmeasdef{OPAL.Gamma31.pub.ABBIENDI.01J}{Gamma31}{OPAL}{Abbiendi:2000ee}{0.01528 \pm 0.00039 \pm 0.0004}{0.01528}{\pm 0.00039}{0.0004}%
\htmeasdef{OPAL.Gamma33.pub.AKERS.94G}{Gamma33}{OPAL}{Akers:1994td}{0.0097 \pm 0.0009 \pm 0.0006}{0.0097}{\pm 0.0009}{0.0006}%
\htmeasdef{OPAL.Gamma35.pub.ABBIENDI.00C}{Gamma35}{OPAL}{Abbiendi:1999pm}{0.00933 \pm 0.00068 \pm 0.00049}{0.00933}{\pm 0.00068}{0.00049}%
\htmeasdef{OPAL.Gamma38.pub.ABBIENDI.00C}{Gamma38}{OPAL}{Abbiendi:1999pm}{0.0033 \pm 0.00055 \pm 0.00039}{0.0033}{\pm 0.00055}{0.00039}%
\htmeasdef{OPAL.Gamma43.pub.ABBIENDI.00C}{Gamma43}{OPAL}{Abbiendi:1999pm}{0.00324 \pm 0.00074 \pm 0.00066}{0.00324}{\pm 0.00074}{0.00066}%
\htmeasdef{OPAL.Gamma5.pub.ABBIENDI.99H}{Gamma5}{OPAL}{Abbiendi:1998cx}{0.1781 \pm 0.0009 \pm 0.0006}{0.1781}{\pm 0.0009}{0.0006}%
\htmeasdef{OPAL.Gamma55.pub.AKERS.95Y}{Gamma55}{OPAL}{Akers:1995ry}{0.1496 \pm 0.0009 \pm 0.0022}{0.1496}{\pm 0.0009}{0.0022}%
\htmeasdef{OPAL.Gamma57by55.pub.AKERS.95Y}{Gamma57by55}{OPAL}{Akers:1995ry}{0.66 \pm 0.004 \pm 0.014}{0.66}{\pm 0.004}{0.014}%
\htmeasdef{OPAL.Gamma7.pub.ALEXANDER.91D}{Gamma7}{OPAL}{Alexander:1991am}{0.121 \pm 0.007 \pm 0.005}{0.121}{\pm 0.007}{0.005}%
\htmeasdef{OPAL.Gamma8.pub.ACKERSTAFF.98M}{Gamma8}{OPAL}{Ackerstaff:1997tx}{0.1198 \pm 0.0013 \pm 0.0016}{0.1198}{\pm 0.0013}{0.0016}%
\htmeasdef{OPAL.Gamma85.pub.ABBIENDI.04J}{Gamma85}{OPAL}{Abbiendi:2004xa}{0.00415 \pm 0.00053 \pm 0.0004}{0.00415}{\pm 0.00053}{0.0004}%
\htmeasdef{OPAL.Gamma92.pub.ABBIENDI.00D}{Gamma92}{OPAL}{Abbiendi:1999cq}{0.00159 \pm 0.00053 \pm 0.0002}{0.00159}{\pm 0.00053}{0.0002}%
\htmeasdef{TPC.Gamma54.pub.AIHARA.87B}{Gamma54}{TPC}{Aihara:1986mw}{0.151 \pm 0.008 \pm 0.006}{0.151}{\pm 0.008}{0.006}%
\htmeasdef{TPC.Gamma82.pub.BAUER.94}{Gamma82}{TPC}{Bauer:1993wn}{0.0058 {}^{+0.0015}_{-0.0013} \pm 0.0012}{0.0058}{{}^{+0.0015}_{-0.0013}}{0.0012}%
\htmeasdef{TPC.Gamma92.pub.BAUER.94}{Gamma92}{TPC}{Bauer:1993wn}{0.0015 {}^{+0.0009}_{-0.0007} \pm 0.0003}{0.0015}{{}^{+0.0009}_{-0.0007}}{0.0003}%
\htdef{Gamma1.qt}{\ensuremath{0.8519 \pm 0.0011}}%
\htdef{Gamma2.qt}{\ensuremath{0.8453 \pm 0.0010}}%
\htdef{Gamma3.qt}{\ensuremath{0.17392 \pm 0.00040}}%
\htdef{ALEPH.Gamma3.pub.SCHAEL.05C,qt}{\ensuremath{0.17319 \pm 0.00077 \pm 0.00000}}%
\htdef{DELPHI.Gamma3.pub.ABREU.99X,qt}{\ensuremath{0.17325 \pm 0.00095 \pm 0.00077}}%
\htdef{L3.Gamma3.pub.ACCIARRI.01F,qt}{\ensuremath{0.17342 \pm 0.00110 \pm 0.00067}}%
\htdef{OPAL.Gamma3.pub.ABBIENDI.03,qt}{\ensuremath{0.17340 \pm 0.00090 \pm 0.00060}}%
\htdef{Gamma3by5.qt}{\ensuremath{0.9762 \pm 0.0028}}%
\htdef{ARGUS.Gamma3by5.pub.ALBRECHT.92D,qt}{\ensuremath{0.9970 \pm 0.0350 \pm 0.0400}}%
\htdef{BaBar.Gamma3by5.pub.AUBERT.10F,qt}{\ensuremath{0.9796 \pm 0.0016 \pm 0.0036}}%
\htdef{CLEO.Gamma3by5.pub.ANASTASSOV.97,qt}{\ensuremath{0.9777 \pm 0.0063 \pm 0.0087}}%
\htdef{Gamma5.qt}{\ensuremath{0.17816 \pm 0.00041}}%
\htdef{ALEPH.Gamma5.pub.SCHAEL.05C,qt}{\ensuremath{0.17837 \pm 0.00080 \pm 0.00000}}%
\htdef{CLEO.Gamma5.pub.ANASTASSOV.97,qt}{\ensuremath{0.17760 \pm 0.00060 \pm 0.00170}}%
\htdef{DELPHI.Gamma5.pub.ABREU.99X,qt}{\ensuremath{0.17877 \pm 0.00109 \pm 0.00110}}%
\htdef{L3.Gamma5.pub.ACCIARRI.01F,qt}{\ensuremath{0.17806 \pm 0.00104 \pm 0.00076}}%
\htdef{OPAL.Gamma5.pub.ABBIENDI.99H,qt}{\ensuremath{0.17810 \pm 0.00090 \pm 0.00060}}%
\htdef{Gamma7.qt}{\ensuremath{0.12023 \pm 0.00054}}%
\htdef{DELPHI.Gamma7.pub.ABREU.92N,qt}{\ensuremath{0.12400 \pm 0.00700 \pm 0.00700}}%
\htdef{L3.Gamma7.pub.ACCIARRI.95,qt}{\ensuremath{0.12470 \pm 0.00260 \pm 0.00430}}%
\htdef{OPAL.Gamma7.pub.ALEXANDER.91D,qt}{\ensuremath{0.12100 \pm 0.00700 \pm 0.00500}}%
\htdef{Gamma8.qt}{\ensuremath{0.11506 \pm 0.00054}}%
\htdef{ALEPH.Gamma8.pub.SCHAEL.05C,qt}{\ensuremath{0.11524 \pm 0.00105 \pm 0.00000}}%
\htdef{CLEO.Gamma8.pub.ANASTASSOV.97,qt}{\ensuremath{0.11520 \pm 0.00050 \pm 0.00120}}%
\htdef{DELPHI.Gamma8.pub.ABDALLAH.06A,qt}{\ensuremath{0.11571 \pm 0.00120 \pm 0.00114}}%
\htdef{OPAL.Gamma8.pub.ACKERSTAFF.98M,qt}{\ensuremath{0.11980 \pm 0.00130 \pm 0.00160}}%
\htdef{Gamma8by5.qt}{\ensuremath{0.6458 \pm 0.0033}}%
\htdef{Gamma9.qt}{\ensuremath{0.10810 \pm 0.00053}}%
\htdef{Gamma9by5.qt}{\ensuremath{0.6068 \pm 0.0032}}%
\htdef{BaBar.Gamma9by5.pub.AUBERT.10F,qt}{\ensuremath{0.5945 \pm 0.0014 \pm 0.0061}}%
\htdef{Gamma10.qt}{\ensuremath{(0.6960 \pm 0.0096) \cdot 10^{-2}}}%
\htdef{ALEPH.Gamma10.pub.BARATE.99K,qt}{\ensuremath{(0.6960 \pm 0.0287 \pm 0.0000) \cdot 10^{-2} }}%
\htdef{CLEO.Gamma10.pub.BATTLE.94,qt}{\ensuremath{(0.6600 \pm 0.0700 \pm 0.0900) \cdot 10^{-2} }}%
\htdef{DELPHI.Gamma10.pub.ABREU.94K,qt}{\ensuremath{(0.8500 \pm 0.1800 \pm 0.0000) \cdot 10^{-2} }}%
\htdef{OPAL.Gamma10.pub.ABBIENDI.01J,qt}{\ensuremath{(0.6580 \pm 0.0270 \pm 0.0290) \cdot 10^{-2} }}%
\htdef{Gamma10by5.qt}{\ensuremath{(3.906 \pm 0.054) \cdot 10^{-2}}}%
\htdef{BaBar.Gamma10by5.pub.AUBERT.10F,qt}{\ensuremath{(3.882 \pm 0.032 \pm 0.057) \cdot 10^{-2} }}%
\htdef{Gamma10by9.qt}{\ensuremath{(6.438 \pm 0.094) \cdot 10^{-2}}}%
\htdef{Gamma11.qt}{\ensuremath{0.36973 \pm 0.00097}}%
\htdef{Gamma12.qt}{\ensuremath{0.36475 \pm 0.00097}}%
\htdef{Gamma13.qt}{\ensuremath{0.25935 \pm 0.00091}}%
\htdef{ALEPH.Gamma13.pub.SCHAEL.05C,qt}{\ensuremath{0.25924 \pm 0.00129 \pm 0.00000}}%
\htdef{Belle.Gamma13.pub.FUJIKAWA.08,qt}{\ensuremath{0.25670 \pm 0.00010 \pm 0.00390}}%
\htdef{CLEO.Gamma13.pub.ARTUSO.94,qt}{\ensuremath{0.25870 \pm 0.00120 \pm 0.00420}}%
\htdef{DELPHI.Gamma13.pub.ABDALLAH.06A,qt}{\ensuremath{0.25740 \pm 0.00201 \pm 0.00138}}%
\htdef{L3.Gamma13.pub.ACCIARRI.95,qt}{\ensuremath{0.25050 \pm 0.00350 \pm 0.00500}}%
\htdef{OPAL.Gamma13.pub.ACKERSTAFF.98M,qt}{\ensuremath{0.25890 \pm 0.00170 \pm 0.00290}}%
\htdef{Gamma14.qt}{\ensuremath{0.25502 \pm 0.00092}}%
\htdef{Gamma16.qt}{\ensuremath{(0.4327 \pm 0.0149) \cdot 10^{-2}}}%
\htdef{ALEPH.Gamma16.pub.BARATE.99K,qt}{\ensuremath{(0.4440 \pm 0.0354 \pm 0.0000) \cdot 10^{-2} }}%
\htdef{BaBar.Gamma16.pub.AUBERT.07AP,qt}{\ensuremath{(0.4160 \pm 0.0030 \pm 0.0180) \cdot 10^{-2} }}%
\htdef{CLEO.Gamma16.pub.BATTLE.94,qt}{\ensuremath{(0.5100 \pm 0.1000 \pm 0.0700) \cdot 10^{-2} }}%
\htdef{OPAL.Gamma16.pub.ABBIENDI.04J,qt}{\ensuremath{(0.4710 \pm 0.0590 \pm 0.0230) \cdot 10^{-2} }}%
\htdef{Gamma17.qt}{\ensuremath{0.10775 \pm 0.00095}}%
\htdef{OPAL.Gamma17.pub.ACKERSTAFF.98M,qt}{\ensuremath{0.09910 \pm 0.00310 \pm 0.00270}}%
\htdef{Gamma18.qt}{\ensuremath{(9.458 \pm 0.097) \cdot 10^{-2}}}%
\htdef{Gamma19.qt}{\ensuremath{(9.306 \pm 0.097) \cdot 10^{-2}}}%
\htdef{ALEPH.Gamma19.pub.SCHAEL.05C,qt}{\ensuremath{(9.295 \pm 0.122 \pm 0.000) \cdot 10^{-2} }}%
\htdef{DELPHI.Gamma19.pub.ABDALLAH.06A,qt}{\ensuremath{(9.498 \pm 0.320 \pm 0.275) \cdot 10^{-2} }}%
\htdef{L3.Gamma19.pub.ACCIARRI.95,qt}{\ensuremath{(8.880 \pm 0.370 \pm 0.420) \cdot 10^{-2} }}%
\htdef{Gamma19by13.qt}{\ensuremath{0.3588 \pm 0.0044}}%
\htdef{CLEO.Gamma19by13.pub.PROCARIO.93,qt}{\ensuremath{0.3420 \pm 0.0060 \pm 0.0160}}%
\htdef{Gamma20.qt}{\ensuremath{(9.242 \pm 0.100) \cdot 10^{-2}}}%
\htdef{Gamma23.qt}{\ensuremath{(0.0640 \pm 0.0220) \cdot 10^{-2}}}%
\htdef{ALEPH.Gamma23.pub.BARATE.99K,qt}{\ensuremath{(0.0560 \pm 0.0250 \pm 0.0000) \cdot 10^{-2} }}%
\htdef{CLEO.Gamma23.pub.BATTLE.94,qt}{\ensuremath{(0.0900 \pm 0.1000 \pm 0.0300) \cdot 10^{-2} }}%
\htdef{Gamma24.qt}{\ensuremath{(1.318 \pm 0.065) \cdot 10^{-2}}}%
\htdef{Gamma25.qt}{\ensuremath{(1.233 \pm 0.065) \cdot 10^{-2}}}%
\htdef{DELPHI.Gamma25.pub.ABDALLAH.06A,qt}{\ensuremath{(1.403 \pm 0.214 \pm 0.224) \cdot 10^{-2} }}%
\htdef{Gamma26.qt}{\ensuremath{(1.158 \pm 0.072) \cdot 10^{-2}}}%
\htdef{ALEPH.Gamma26.pub.SCHAEL.05C,qt}{\ensuremath{(1.082 \pm 0.093 \pm 0.000) \cdot 10^{-2} }}%
\htdef{L3.Gamma26.pub.ACCIARRI.95,qt}{\ensuremath{(1.700 \pm 0.240 \pm 0.380) \cdot 10^{-2} }}%
\htdef{Gamma26by13.qt}{\ensuremath{(4.465 \pm 0.277) \cdot 10^{-2}}}%
\htdef{CLEO.Gamma26by13.pub.PROCARIO.93,qt}{\ensuremath{(4.400 \pm 0.300 \pm 0.500) \cdot 10^{-2} }}%
\htdef{Gamma27.qt}{\ensuremath{(1.029 \pm 0.075) \cdot 10^{-2}}}%
\htdef{Gamma28.qt}{\ensuremath{(4.283 \pm 2.161) \cdot 10^{-4}}}%
\htdef{ALEPH.Gamma28.pub.BARATE.99K,qt}{\ensuremath{(3.700 \pm 2.371 \pm 0.000) \cdot 10^{-4} }}%
\htdef{Gamma29.qt}{\ensuremath{(0.1568 \pm 0.0391) \cdot 10^{-2}}}%
\htdef{CLEO.Gamma29.pub.PROCARIO.93,qt}{\ensuremath{(0.1600 \pm 0.0500 \pm 0.0500) \cdot 10^{-2} }}%
\htdef{Gamma30.qt}{\ensuremath{(0.1099 \pm 0.0391) \cdot 10^{-2}}}%
\htdef{ALEPH.Gamma30.pub.SCHAEL.05C,qt}{\ensuremath{(0.1120 \pm 0.0509 \pm 0.0000) \cdot 10^{-2} }}%
\htdef{Gamma31.qt}{\ensuremath{(1.545 \pm 0.030) \cdot 10^{-2}}}%
\htdef{CLEO.Gamma31.pub.BATTLE.94,qt}{\ensuremath{(1.700 \pm 0.120 \pm 0.190) \cdot 10^{-2} }}%
\htdef{DELPHI.Gamma31.pub.ABREU.94K,qt}{\ensuremath{(1.540 \pm 0.240 \pm 0.000) \cdot 10^{-2} }}%
\htdef{OPAL.Gamma31.pub.ABBIENDI.01J,qt}{\ensuremath{(1.528 \pm 0.039 \pm 0.040) \cdot 10^{-2} }}%
\htdef{Gamma32.qt}{\ensuremath{(0.8528 \pm 0.0286) \cdot 10^{-2}}}%
\htdef{Gamma33.qt}{\ensuremath{(0.9372 \pm 0.0292) \cdot 10^{-2}}}%
\htdef{ALEPH.Gamma33.pub.BARATE.98E,qt}{\ensuremath{(0.9700 \pm 0.0849 \pm 0.0000) \cdot 10^{-2} }}%
\htdef{OPAL.Gamma33.pub.AKERS.94G,qt}{\ensuremath{(0.9700 \pm 0.0900 \pm 0.0600) \cdot 10^{-2} }}%
\htdef{Gamma34.qt}{\ensuremath{(0.9865 \pm 0.0139) \cdot 10^{-2}}}%
\htdef{CLEO.Gamma34.pub.COAN.96,qt}{\ensuremath{(0.8550 \pm 0.0360 \pm 0.0730) \cdot 10^{-2} }}%
\htdef{Gamma35.qt}{\ensuremath{(0.8386 \pm 0.0141) \cdot 10^{-2}}}%
\htdef{ALEPH.Gamma35.pub.BARATE.99K,qt}{\ensuremath{(0.9280 \pm 0.0564 \pm 0.0000) \cdot 10^{-2} }}%
\htdef{Belle.Gamma35.pub.RYU.14vpc,qt}{\ensuremath{(0.8320 \pm 0.0025 \pm 0.0150) \cdot 10^{-2} }}%
\htdef{L3.Gamma35.pub.ACCIARRI.95F,qt}{\ensuremath{(0.9500 \pm 0.1500 \pm 0.0600) \cdot 10^{-2} }}%
\htdef{OPAL.Gamma35.pub.ABBIENDI.00C,qt}{\ensuremath{(0.9330 \pm 0.0680 \pm 0.0490) \cdot 10^{-2} }}%
\htdef{Gamma37.qt}{\ensuremath{(0.1479 \pm 0.0053) \cdot 10^{-2}}}%
\htdef{ALEPH.Gamma37.pub.BARATE.98E,qt}{\ensuremath{(0.1580 \pm 0.0453 \pm 0.0000) \cdot 10^{-2} }}%
\htdef{ALEPH.Gamma37.pub.BARATE.99K,qt}{\ensuremath{(0.1620 \pm 0.0237 \pm 0.0000) \cdot 10^{-2} }}%
\htdef{Belle.Gamma37.pub.RYU.14vpc,qt}{\ensuremath{(0.1480 \pm 0.0013 \pm 0.0055) \cdot 10^{-2} }}%
\htdef{CLEO.Gamma37.pub.COAN.96,qt}{\ensuremath{(0.1510 \pm 0.0210 \pm 0.0220) \cdot 10^{-2} }}%
\htdef{Gamma38.qt}{\ensuremath{(0.2982 \pm 0.0079) \cdot 10^{-2}}}%
\htdef{OPAL.Gamma38.pub.ABBIENDI.00C,qt}{\ensuremath{(0.3300 \pm 0.0550 \pm 0.0390) \cdot 10^{-2} }}%
\htdef{Gamma39.qt}{\ensuremath{(0.5314 \pm 0.0134) \cdot 10^{-2}}}%
\htdef{CLEO.Gamma39.pub.COAN.96,qt}{\ensuremath{(0.5620 \pm 0.0500 \pm 0.0480) \cdot 10^{-2} }}%
\htdef{Gamma40.qt}{\ensuremath{(0.3812 \pm 0.0129) \cdot 10^{-2}}}%
\htdef{ALEPH.Gamma40.pub.BARATE.98E,qt}{\ensuremath{(0.2940 \pm 0.0818 \pm 0.0000) \cdot 10^{-2} }}%
\htdef{ALEPH.Gamma40.pub.BARATE.99K,qt}{\ensuremath{(0.3470 \pm 0.0646 \pm 0.0000) \cdot 10^{-2} }}%
\htdef{Belle.Gamma40.pub.RYU.14vpc,qt}{\ensuremath{(0.3860 \pm 0.0031 \pm 0.0135) \cdot 10^{-2} }}%
\htdef{L3.Gamma40.pub.ACCIARRI.95F,qt}{\ensuremath{(0.4100 \pm 0.1200 \pm 0.0300) \cdot 10^{-2} }}%
\htdef{Gamma42.qt}{\ensuremath{(0.1502 \pm 0.0071) \cdot 10^{-2}}}%
\htdef{ALEPH.Gamma42.pub.BARATE.98E,qt}{\ensuremath{(0.1520 \pm 0.0789 \pm 0.0000) \cdot 10^{-2} }}%
\htdef{ALEPH.Gamma42.pub.BARATE.99K,qt}{\ensuremath{(0.1430 \pm 0.0291 \pm 0.0000) \cdot 10^{-2} }}%
\htdef{Belle.Gamma42.pub.RYU.14vpc,qt}{\ensuremath{(0.1496 \pm 0.0019 \pm 0.0073) \cdot 10^{-2} }}%
\htdef{CLEO.Gamma42.pub.COAN.96,qt}{\ensuremath{(0.1450 \pm 0.0360 \pm 0.0200) \cdot 10^{-2} }}%
\htdef{Gamma43.qt}{\ensuremath{(0.4046 \pm 0.0260) \cdot 10^{-2}}}%
\htdef{OPAL.Gamma43.pub.ABBIENDI.00C,qt}{\ensuremath{(0.3240 \pm 0.0740 \pm 0.0660) \cdot 10^{-2} }}%
\htdef{Gamma44.qt}{\ensuremath{(2.340 \pm 2.306) \cdot 10^{-4}}}%
\htdef{ALEPH.Gamma44.pub.BARATE.99R,qt}{\ensuremath{(2.600 \pm 2.400 \pm 0.000) \cdot 10^{-4} }}%
\htdef{Gamma46.qt}{\ensuremath{(0.1513 \pm 0.0247) \cdot 10^{-2}}}%
\htdef{Gamma47.qt}{\ensuremath{(2.332 \pm 0.065) \cdot 10^{-4}}}%
\htdef{ALEPH.Gamma47.pub.BARATE.98E,qt}{\ensuremath{(2.600 \pm 1.118 \pm 0.000) \cdot 10^{-4} }}%
\htdef{BaBar.Gamma47.pub.LEES.12Y,qt}{\ensuremath{(2.310 \pm 0.040 \pm 0.080) \cdot 10^{-4} }}%
\htdef{Belle.Gamma47.pub.RYU.14vpc,qt}{\ensuremath{(2.330 \pm 0.033 \pm 0.093) \cdot 10^{-4} }}%
\htdef{CLEO.Gamma47.pub.COAN.96,qt}{\ensuremath{(2.300 \pm 0.500 \pm 0.300) \cdot 10^{-4} }}%
\htdef{Gamma48.qt}{\ensuremath{(0.1047 \pm 0.0247) \cdot 10^{-2}}}%
\htdef{ALEPH.Gamma48.pub.BARATE.98E,qt}{\ensuremath{(0.1010 \pm 0.0264 \pm 0.0000) \cdot 10^{-2} }}%
\htdef{Gamma49.qt}{\ensuremath{(3.540 \pm 1.193) \cdot 10^{-4}}}%
\htdef{Gamma50.qt}{\ensuremath{(1.815 \pm 0.207) \cdot 10^{-5}}}%
\htdef{BaBar.Gamma50.pub.LEES.12Y,qt}{\ensuremath{(1.600 \pm 0.200 \pm 0.220) \cdot 10^{-5} }}%
\htdef{Belle.Gamma50.pub.RYU.14vpc,qt}{\ensuremath{(2.000 \pm 0.216 \pm 0.202) \cdot 10^{-5} }}%
\htdef{Gamma51.qt}{\ensuremath{(3.177 \pm 1.192) \cdot 10^{-4}}}%
\htdef{ALEPH.Gamma51.pub.BARATE.98E,qt}{\ensuremath{(3.100 \pm 1.100 \pm 0.500) \cdot 10^{-4} }}%
\htdef{Gamma53.qt}{\ensuremath{(2.218 \pm 2.024) \cdot 10^{-4}}}%
\htdef{ALEPH.Gamma53.pub.BARATE.98E,qt}{\ensuremath{(2.300 \pm 2.025 \pm 0.000) \cdot 10^{-4} }}%
\htdef{Gamma54.qt}{\ensuremath{0.15215 \pm 0.00061}}%
\htdef{CELLO.Gamma54.pub.BEHREND.89B,qt}{\ensuremath{0.15000 \pm 0.00400 \pm 0.00300}}%
\htdef{L3.Gamma54.pub.ADEVA.91F,qt}{\ensuremath{0.14400 \pm 0.00600 \pm 0.00300}}%
\htdef{TPC.Gamma54.pub.AIHARA.87B,qt}{\ensuremath{0.15100 \pm 0.00800 \pm 0.00600}}%
\htdef{Gamma55.qt}{\ensuremath{0.14567 \pm 0.00057}}%
\htdef{L3.Gamma55.pub.ACHARD.01D,qt}{\ensuremath{0.14556 \pm 0.00105 \pm 0.00076}}%
\htdef{OPAL.Gamma55.pub.AKERS.95Y,qt}{\ensuremath{0.14960 \pm 0.00090 \pm 0.00220}}%
\htdef{Gamma56.qt}{\ensuremath{(9.780 \pm 0.054) \cdot 10^{-2}}}%
\htdef{Gamma57.qt}{\ensuremath{(9.439 \pm 0.053) \cdot 10^{-2}}}%
\htdef{CLEO.Gamma57.pub.BALEST.95C,qt}{\ensuremath{(9.510 \pm 0.070 \pm 0.200) \cdot 10^{-2} }}%
\htdef{DELPHI.Gamma57.pub.ABDALLAH.06A,qt}{\ensuremath{(9.317 \pm 0.090 \pm 0.082) \cdot 10^{-2} }}%
\htdef{Gamma57by55.qt}{\ensuremath{0.6480 \pm 0.0030}}%
\htdef{OPAL.Gamma57by55.pub.AKERS.95Y,qt}{\ensuremath{0.6600 \pm 0.0040 \pm 0.0140}}%
\htdef{Gamma58.qt}{\ensuremath{(9.408 \pm 0.053) \cdot 10^{-2}}}%
\htdef{ALEPH.Gamma58.pub.SCHAEL.05C,qt}{\ensuremath{(9.469 \pm 0.096 \pm 0.000) \cdot 10^{-2} }}%
\htdef{Gamma59.qt}{\ensuremath{(9.290 \pm 0.052) \cdot 10^{-2}}}%
\htdef{Gamma60.qt}{\ensuremath{(9.000 \pm 0.051) \cdot 10^{-2}}}%
\htdef{BaBar.Gamma60.pub.AUBERT.08,qt}{\ensuremath{(8.830 \pm 0.010 \pm 0.130) \cdot 10^{-2} }}%
\htdef{Belle.Gamma60.pub.LEE.10,qt}{\ensuremath{(8.420 \pm 0.000 {}^{+0.260}_{-0.250}) \cdot 10^{-2} }}%
\htdef{CLEO3.Gamma60.pub.BRIERE.03,qt}{\ensuremath{(9.130 \pm 0.050 \pm 0.460) \cdot 10^{-2} }}%
\htdef{Gamma62.qt}{\ensuremath{(8.970 \pm 0.052) \cdot 10^{-2}}}%
\htdef{Gamma63.qt}{\ensuremath{(5.325 \pm 0.050) \cdot 10^{-2}}}%
\htdef{Gamma64.qt}{\ensuremath{(5.120 \pm 0.049) \cdot 10^{-2}}}%
\htdef{Gamma65.qt}{\ensuremath{(4.790 \pm 0.052) \cdot 10^{-2}}}%
\htdef{Gamma66.qt}{\ensuremath{(4.606 \pm 0.051) \cdot 10^{-2}}}%
\htdef{ALEPH.Gamma66.pub.SCHAEL.05C,qt}{\ensuremath{(4.734 \pm 0.077 \pm 0.000) \cdot 10^{-2} }}%
\htdef{CLEO.Gamma66.pub.BALEST.95C,qt}{\ensuremath{(4.230 \pm 0.060 \pm 0.220) \cdot 10^{-2} }}%
\htdef{DELPHI.Gamma66.pub.ABDALLAH.06A,qt}{\ensuremath{(4.545 \pm 0.106 \pm 0.103) \cdot 10^{-2} }}%
\htdef{Gamma67.qt}{\ensuremath{(2.820 \pm 0.070) \cdot 10^{-2}}}%
\htdef{Gamma68.qt}{\ensuremath{(4.651 \pm 0.053) \cdot 10^{-2}}}%
\htdef{Gamma69.qt}{\ensuremath{(4.519 \pm 0.052) \cdot 10^{-2}}}%
\htdef{CLEO.Gamma69.pub.EDWARDS.00A,qt}{\ensuremath{(4.190 \pm 0.100 \pm 0.210) \cdot 10^{-2} }}%
\htdef{Gamma70.qt}{\ensuremath{(2.769 \pm 0.071) \cdot 10^{-2}}}%
\htdef{Gamma74.qt}{\ensuremath{(0.5135 \pm 0.0312) \cdot 10^{-2}}}%
\htdef{DELPHI.Gamma74.pub.ABDALLAH.06A,qt}{\ensuremath{(0.5610 \pm 0.0680 \pm 0.0950) \cdot 10^{-2} }}%
\htdef{Gamma75.qt}{\ensuremath{(0.5024 \pm 0.0310) \cdot 10^{-2}}}%
\htdef{Gamma76.qt}{\ensuremath{(0.4925 \pm 0.0310) \cdot 10^{-2}}}%
\htdef{ALEPH.Gamma76.pub.SCHAEL.05C,qt}{\ensuremath{(0.4350 \pm 0.0461 \pm 0.0000) \cdot 10^{-2} }}%
\htdef{Gamma76by54.qt}{\ensuremath{(3.237 \pm 0.202) \cdot 10^{-2}}}%
\htdef{CLEO.Gamma76by54.pub.BORTOLETTO.93,qt}{\ensuremath{(3.400 \pm 0.200 \pm 0.300) \cdot 10^{-2} }}%
\htdef{Gamma77.qt}{\ensuremath{(9.759 \pm 3.550) \cdot 10^{-4}}}%
\htdef{Gamma78.qt}{\ensuremath{(2.107 \pm 0.299) \cdot 10^{-4}}}%
\htdef{CLEO.Gamma78.pub.ANASTASSOV.01,qt}{\ensuremath{(2.200 \pm 0.300 \pm 0.400) \cdot 10^{-4} }}%
\htdef{Gamma79.qt}{\ensuremath{(0.6297 \pm 0.0141) \cdot 10^{-2}}}%
\htdef{Gamma80.qt}{\ensuremath{(0.4363 \pm 0.0073) \cdot 10^{-2}}}%
\htdef{Gamma80by60.qt}{\ensuremath{(4.847 \pm 0.080) \cdot 10^{-2}}}%
\htdef{CLEO.Gamma80by60.pub.RICHICHI.99,qt}{\ensuremath{(5.440 \pm 0.210 \pm 0.530) \cdot 10^{-2} }}%
\htdef{Gamma81.qt}{\ensuremath{(8.726 \pm 1.177) \cdot 10^{-4}}}%
\htdef{Gamma81by69.qt}{\ensuremath{(1.931 \pm 0.266) \cdot 10^{-2}}}%
\htdef{CLEO.Gamma81by69.pub.RICHICHI.99,qt}{\ensuremath{(2.610 \pm 0.450 \pm 0.420) \cdot 10^{-2} }}%
\htdef{Gamma82.qt}{\ensuremath{(0.4780 \pm 0.0137) \cdot 10^{-2}}}%
\htdef{TPC.Gamma82.pub.BAUER.94,qt}{\ensuremath{(0.5800 {}^{+0.1500}_{-0.1300} \pm 0.1200) \cdot 10^{-2} }}%
\htdef{Gamma83.qt}{\ensuremath{(0.3741 \pm 0.0135) \cdot 10^{-2}}}%
\htdef{Gamma84.qt}{\ensuremath{(0.3441 \pm 0.0070) \cdot 10^{-2}}}%
\htdef{Gamma85.qt}{\ensuremath{(0.2929 \pm 0.0067) \cdot 10^{-2}}}%
\htdef{ALEPH.Gamma85.pub.BARATE.98,qt}{\ensuremath{(0.2140 \pm 0.0470 \pm 0.0000) \cdot 10^{-2} }}%
\htdef{BaBar.Gamma85.pub.AUBERT.08,qt}{\ensuremath{(0.2730 \pm 0.0020 \pm 0.0090) \cdot 10^{-2} }}%
\htdef{Belle.Gamma85.pub.LEE.10,qt}{\ensuremath{(0.3300 \pm 0.0010 {}^{+0.0160}_{-0.0170}) \cdot 10^{-2} }}%
\htdef{CLEO3.Gamma85.pub.BRIERE.03,qt}{\ensuremath{(0.3840 \pm 0.0140 \pm 0.0380) \cdot 10^{-2} }}%
\htdef{OPAL.Gamma85.pub.ABBIENDI.04J,qt}{\ensuremath{(0.4150 \pm 0.0530 \pm 0.0400) \cdot 10^{-2} }}%
\htdef{Gamma85by60.qt}{\ensuremath{(3.254 \pm 0.074) \cdot 10^{-2}}}%
\htdef{Gamma87.qt}{\ensuremath{(0.1331 \pm 0.0119) \cdot 10^{-2}}}%
\htdef{Gamma88.qt}{\ensuremath{(8.115 \pm 1.168) \cdot 10^{-4}}}%
\htdef{ALEPH.Gamma88.pub.BARATE.98,qt}{\ensuremath{(6.100 \pm 4.295 \pm 0.000) \cdot 10^{-4} }}%
\htdef{CLEO3.Gamma88.pub.ARMS.05,qt}{\ensuremath{(7.400 \pm 0.800 \pm 1.100) \cdot 10^{-4} }}%
\htdef{Gamma89.qt}{\ensuremath{(7.761 \pm 1.168) \cdot 10^{-4}}}%
\htdef{Gamma92.qt}{\ensuremath{(0.1495 \pm 0.0033) \cdot 10^{-2}}}%
\htdef{OPAL.Gamma92.pub.ABBIENDI.00D,qt}{\ensuremath{(0.1590 \pm 0.0530 \pm 0.0200) \cdot 10^{-2} }}%
\htdef{TPC.Gamma92.pub.BAUER.94,qt}{\ensuremath{(0.1500 {}^{+0.0900}_{-0.0700} \pm 0.0300) \cdot 10^{-2} }}%
\htdef{Gamma93.qt}{\ensuremath{(0.1434 \pm 0.0027) \cdot 10^{-2}}}%
\htdef{ALEPH.Gamma93.pub.BARATE.98,qt}{\ensuremath{(0.1630 \pm 0.0270 \pm 0.0000) \cdot 10^{-2} }}%
\htdef{BaBar.Gamma93.pub.AUBERT.08,qt}{\ensuremath{(0.1346 \pm 0.0010 \pm 0.0036) \cdot 10^{-2} }}%
\htdef{Belle.Gamma93.pub.LEE.10,qt}{\ensuremath{(0.1550 \pm 0.0010 {}^{+0.0060}_{-0.0050}) \cdot 10^{-2} }}%
\htdef{CLEO3.Gamma93.pub.BRIERE.03,qt}{\ensuremath{(0.1550 \pm 0.0060 \pm 0.0090) \cdot 10^{-2} }}%
\htdef{Gamma93by60.qt}{\ensuremath{(1.593 \pm 0.030) \cdot 10^{-2}}}%
\htdef{CLEO.Gamma93by60.pub.RICHICHI.99,qt}{\ensuremath{(1.600 \pm 0.150 \pm 0.300) \cdot 10^{-2} }}%
\htdef{Gamma94.qt}{\ensuremath{(0.611 \pm 0.183) \cdot 10^{-4}}}%
\htdef{ALEPH.Gamma94.pub.BARATE.98,qt}{\ensuremath{(7.500 \pm 3.265 \pm 0.000) \cdot 10^{-4} }}%
\htdef{CLEO3.Gamma94.pub.ARMS.05,qt}{\ensuremath{(0.550 \pm 0.140 \pm 0.120) \cdot 10^{-4} }}%
\htdef{Gamma94by69.qt}{\ensuremath{(0.1353 \pm 0.0405) \cdot 10^{-2}}}%
\htdef{CLEO.Gamma94by69.pub.RICHICHI.99,qt}{\ensuremath{(0.7900 \pm 0.4400 \pm 0.1600) \cdot 10^{-2} }}%
\htdef{Gamma96.qt}{\ensuremath{(2.174 \pm 0.800) \cdot 10^{-5}}}%
\htdef{BaBar.Gamma96.pub.AUBERT.08,qt}{\ensuremath{(1.578 \pm 0.130 \pm 0.123) \cdot 10^{-5} }}%
\htdef{Belle.Gamma96.pub.LEE.10,qt}{\ensuremath{(3.290 \pm 0.170 {}^{+0.190}_{-0.200}) \cdot 10^{-5} }}%
\htdef{Gamma102.qt}{\ensuremath{(0.0985 \pm 0.0037) \cdot 10^{-2}}}%
\htdef{CLEO.Gamma102.pub.GIBAUT.94B,qt}{\ensuremath{(0.0970 \pm 0.0050 \pm 0.0110) \cdot 10^{-2} }}%
\htdef{HRS.Gamma102.pub.BYLSMA.87,qt}{\ensuremath{(0.1020 \pm 0.0290 \pm 0.0000) \cdot 10^{-2} }}%
\htdef{L3.Gamma102.pub.ACHARD.01D,qt}{\ensuremath{(0.1700 \pm 0.0220 \pm 0.0260) \cdot 10^{-2} }}%
\htdef{Gamma103.qt}{\ensuremath{(8.216 \pm 0.316) \cdot 10^{-4}}}%
\htdef{ALEPH.Gamma103.pub.SCHAEL.05C,qt}{\ensuremath{(7.200 \pm 1.500 \pm 0.000) \cdot 10^{-4} }}%
\htdef{ARGUS.Gamma103.pub.ALBRECHT.88B,qt}{\ensuremath{(6.400 \pm 2.300 \pm 1.000) \cdot 10^{-4} }}%
\htdef{CLEO.Gamma103.pub.GIBAUT.94B,qt}{\ensuremath{(7.700 \pm 0.500 \pm 0.900) \cdot 10^{-4} }}%
\htdef{DELPHI.Gamma103.pub.ABDALLAH.06A,qt}{\ensuremath{(9.700 \pm 1.500 \pm 0.500) \cdot 10^{-4} }}%
\htdef{HRS.Gamma103.pub.BYLSMA.87,qt}{\ensuremath{(5.100 \pm 2.000 \pm 0.000) \cdot 10^{-4} }}%
\htdef{OPAL.Gamma103.pub.ACKERSTAFF.99E,qt}{\ensuremath{(9.100 \pm 1.400 \pm 0.600) \cdot 10^{-4} }}%
\htdef{Gamma104.qt}{\ensuremath{(1.634 \pm 0.114) \cdot 10^{-4}}}%
\htdef{ALEPH.Gamma104.pub.SCHAEL.05C,qt}{\ensuremath{(2.100 \pm 0.700 \pm 0.900) \cdot 10^{-4} }}%
\htdef{CLEO.Gamma104.pub.ANASTASSOV.01,qt}{\ensuremath{(1.700 \pm 0.200 \pm 0.200) \cdot 10^{-4} }}%
\htdef{DELPHI.Gamma104.pub.ABDALLAH.06A,qt}{\ensuremath{(1.600 \pm 1.200 \pm 0.600) \cdot 10^{-4} }}%
\htdef{OPAL.Gamma104.pub.ACKERSTAFF.99E,qt}{\ensuremath{(2.700 \pm 1.800 \pm 0.900) \cdot 10^{-4} }}%
\htdef{Gamma106.qt}{\ensuremath{(0.7748 \pm 0.0534) \cdot 10^{-2}}}%
\htdef{Gamma110.qt}{\ensuremath{(2.909 \pm 0.048) \cdot 10^{-2}}}%
\htdef{Gamma126.qt}{\ensuremath{(0.1386 \pm 0.0072) \cdot 10^{-2}}}%
\htdef{ALEPH.Gamma126.pub.BUSKULIC.97C,qt}{\ensuremath{(0.1800 \pm 0.0447 \pm 0.0000) \cdot 10^{-2} }}%
\htdef{Belle.Gamma126.pub.INAMI.09,qt}{\ensuremath{(0.1350 \pm 0.0030 \pm 0.0070) \cdot 10^{-2} }}%
\htdef{CLEO.Gamma126.pub.ARTUSO.92,qt}{\ensuremath{(0.1700 \pm 0.0200 \pm 0.0200) \cdot 10^{-2} }}%
\htdef{Gamma128.qt}{\ensuremath{(1.547 \pm 0.080) \cdot 10^{-4}}}%
\htdef{ALEPH.Gamma128.pub.BUSKULIC.97C,qt}{\ensuremath{(2.900 {}^{+1.300}_{-1.200} \pm 0.700) \cdot 10^{-4} }}%
\htdef{BaBar.Gamma128.pub.DEL-AMO-SANCHEZ.11E,qt}{\ensuremath{(1.420 \pm 0.110 \pm 0.070) \cdot 10^{-4} }}%
\htdef{Belle.Gamma128.pub.INAMI.09,qt}{\ensuremath{(1.580 \pm 0.050 \pm 0.090) \cdot 10^{-4} }}%
\htdef{CLEO.Gamma128.pub.BARTELT.96,qt}{\ensuremath{(2.600 \pm 0.500 \pm 0.500) \cdot 10^{-4} }}%
\htdef{Gamma130.qt}{\ensuremath{(0.483 \pm 0.116) \cdot 10^{-4}}}%
\htdef{Belle.Gamma130.pub.INAMI.09,qt}{\ensuremath{(0.460 \pm 0.110 \pm 0.040) \cdot 10^{-4} }}%
\htdef{CLEO.Gamma130.pub.BISHAI.99,qt}{\ensuremath{(1.770 \pm 0.560 \pm 0.710) \cdot 10^{-4} }}%
\htdef{Gamma132.qt}{\ensuremath{(0.937 \pm 0.149) \cdot 10^{-4}}}%
\htdef{Belle.Gamma132.pub.INAMI.09,qt}{\ensuremath{(0.880 \pm 0.140 \pm 0.060) \cdot 10^{-4} }}%
\htdef{CLEO.Gamma132.pub.BISHAI.99,qt}{\ensuremath{(2.200 \pm 0.700 \pm 0.220) \cdot 10^{-4} }}%
\htdef{Gamma136.qt}{\ensuremath{(2.184 \pm 0.130) \cdot 10^{-4}}}%
\htdef{Gamma149.qt}{\ensuremath{(2.401 \pm 0.075) \cdot 10^{-2}}}%
\htdef{Gamma150.qt}{\ensuremath{(1.995 \pm 0.064) \cdot 10^{-2}}}%
\htdef{ALEPH.Gamma150.pub.BUSKULIC.97C,qt}{\ensuremath{(1.910 \pm 0.092 \pm 0.000) \cdot 10^{-2} }}%
\htdef{CLEO.Gamma150.pub.BARINGER.87,qt}{\ensuremath{(1.600 \pm 0.270 \pm 0.410) \cdot 10^{-2} }}%
\htdef{Gamma150by66.qt}{\ensuremath{0.4332 \pm 0.0139}}%
\htdef{ALEPH.Gamma150by66.pub.BUSKULIC.96,qt}{\ensuremath{0.4310 \pm 0.0330 \pm 0.0000}}%
\htdef{CLEO.Gamma150by66.pub.BALEST.95C,qt}{\ensuremath{0.4640 \pm 0.0160 \pm 0.0170}}%
\htdef{Gamma151.qt}{\ensuremath{(4.100 \pm 0.922) \cdot 10^{-4}}}%
\htdef{CLEO3.Gamma151.pub.ARMS.05,qt}{\ensuremath{(4.100 \pm 0.600 \pm 0.700) \cdot 10^{-4} }}%
\htdef{Gamma152.qt}{\ensuremath{(0.4058 \pm 0.0419) \cdot 10^{-2}}}%
\htdef{ALEPH.Gamma152.pub.BUSKULIC.97C,qt}{\ensuremath{(0.4300 \pm 0.0781 \pm 0.0000) \cdot 10^{-2} }}%
\htdef{Gamma152by54.qt}{\ensuremath{(2.667 \pm 0.275) \cdot 10^{-2}}}%
\htdef{Gamma152by76.qt}{\ensuremath{0.8241 \pm 0.0757}}%
\htdef{CLEO.Gamma152by76.pub.BORTOLETTO.93,qt}{\ensuremath{0.8100 \pm 0.0600 \pm 0.0600}}%
\htdef{Gamma167.qt}{\ensuremath{(4.445 \pm 1.636) \cdot 10^{-5}}}%
\htdef{Gamma168.qt}{\ensuremath{(2.174 \pm 0.800) \cdot 10^{-5}}}%
\htdef{Gamma169.qt}{\ensuremath{(1.520 \pm 0.560) \cdot 10^{-5}}}%
\htdef{Gamma800.qt}{\ensuremath{(1.954 \pm 0.065) \cdot 10^{-2}}}%
\htdef{Gamma802.qt}{\ensuremath{(0.2923 \pm 0.0067) \cdot 10^{-2}}}%
\htdef{Gamma803.qt}{\ensuremath{(4.103 \pm 1.429) \cdot 10^{-4}}}%
\htdef{Gamma804.qt}{\ensuremath{(2.332 \pm 0.065) \cdot 10^{-4}}}%
\htdef{Gamma805.qt}{\ensuremath{(4.000 \pm 2.000) \cdot 10^{-4}}}%
\htdef{ALEPH.Gamma805.pub.SCHAEL.05C,qt}{\ensuremath{(4.000 \pm 2.000 \pm 0.000) \cdot 10^{-4} }}%
\htdef{Gamma806.qt}{\ensuremath{(1.815 \pm 0.207) \cdot 10^{-5}}}%
\htdef{Gamma810.qt}{\ensuremath{(1.924 \pm 0.298) \cdot 10^{-4}}}%
\htdef{Gamma811.qt}{\ensuremath{(7.105 \pm 1.586) \cdot 10^{-5}}}%
\htdef{BaBar.Gamma811.pub.LEES.12X,qt}{\ensuremath{(7.300 \pm 1.200 \pm 1.200) \cdot 10^{-5} }}%
\htdef{Gamma812.qt}{\ensuremath{(1.344 \pm 2.683) \cdot 10^{-5}}}%
\htdef{BaBar.Gamma812.pub.LEES.12X,qt}{\ensuremath{(1.000 \pm 0.800 \pm 3.000) \cdot 10^{-5} }}%
\htdef{Gamma820.qt}{\ensuremath{(8.197 \pm 0.315) \cdot 10^{-4}}}%
\htdef{Gamma821.qt}{\ensuremath{(7.677 \pm 0.297) \cdot 10^{-4}}}%
\htdef{BaBar.Gamma821.pub.LEES.12X,qt}{\ensuremath{(7.680 \pm 0.040 \pm 0.400) \cdot 10^{-4} }}%
\htdef{Gamma822.qt}{\ensuremath{(0.596 \pm 1.208) \cdot 10^{-6}}}%
\htdef{BaBar.Gamma822.pub.LEES.12X,qt}{\ensuremath{(0.600 \pm 0.500 \pm 1.100) \cdot 10^{-6} }}%
\htdef{Gamma830.qt}{\ensuremath{(1.623 \pm 0.114) \cdot 10^{-4}}}%
\htdef{Gamma831.qt}{\ensuremath{(8.359 \pm 0.626) \cdot 10^{-5}}}%
\htdef{BaBar.Gamma831.pub.LEES.12X,qt}{\ensuremath{(8.400 \pm 0.400 \pm 0.600) \cdot 10^{-5} }}%
\htdef{Gamma832.qt}{\ensuremath{(3.771 \pm 0.875) \cdot 10^{-5}}}%
\htdef{BaBar.Gamma832.pub.LEES.12X,qt}{\ensuremath{(3.600 \pm 0.300 \pm 0.900) \cdot 10^{-5} }}%
\htdef{Gamma833.qt}{\ensuremath{(1.108 \pm 0.566) \cdot 10^{-6}}}%
\htdef{BaBar.Gamma833.pub.LEES.12X,qt}{\ensuremath{(1.100 \pm 0.400 \pm 0.400) \cdot 10^{-6} }}%
\htdef{Gamma910.qt}{\ensuremath{(7.136 \pm 0.424) \cdot 10^{-5}}}%
\htdef{BaBar.Gamma910.pub.LEES.12X,qt}{\ensuremath{(8.270 \pm 0.880 \pm 0.810) \cdot 10^{-5} }}%
\htdef{Gamma911.qt}{\ensuremath{(4.420 \pm 0.867) \cdot 10^{-5}}}%
\htdef{BaBar.Gamma911.pub.LEES.12X,qt}{\ensuremath{(4.570 \pm 0.770 \pm 0.500) \cdot 10^{-5} }}%
\htdef{Gamma920.qt}{\ensuremath{(5.197 \pm 0.444) \cdot 10^{-5}}}%
\htdef{BaBar.Gamma920.pub.LEES.12X,qt}{\ensuremath{(5.200 \pm 0.310 \pm 0.370) \cdot 10^{-5} }}%
\htdef{Gamma930.qt}{\ensuremath{(5.005 \pm 0.297) \cdot 10^{-5}}}%
\htdef{BaBar.Gamma930.pub.LEES.12X,qt}{\ensuremath{(5.390 \pm 0.270 \pm 0.410) \cdot 10^{-5} }}%
\htdef{Gamma944.qt}{\ensuremath{(8.606 \pm 0.511) \cdot 10^{-5}}}%
\htdef{BaBar.Gamma944.pub.LEES.12X,qt}{\ensuremath{(8.260 \pm 0.350 \pm 0.510) \cdot 10^{-5} }}%
\htdef{Gamma945.qt}{\ensuremath{(1.929 \pm 0.378) \cdot 10^{-4}}}%
\htdef{Gamma998.qt}{\ensuremath{(0.0355 \pm 0.1031) \cdot 10^{-2}}}%
\htdef{Gamma1.qm}{%
\begin{ensuredisplaymath}
\htuse{Gamma1.gn} = \htuse{Gamma1.td}
\end{ensuredisplaymath}
 & \htuse{Gamma1.qt} & \hfagFitLabel}%
\htdef{Gamma2.qm}{%
\begin{ensuredisplaymath}
\htuse{Gamma2.gn} = \htuse{Gamma2.td}
\end{ensuredisplaymath}
 & \htuse{Gamma2.qt} & \hfagFitLabel}%
\htdef{Gamma3.qm}{%
\begin{ensuredisplaymath}
\htuse{Gamma3.gn} = \htuse{Gamma3.td}
\end{ensuredisplaymath}
 & \htuse{Gamma3.qt} & \hfagFitLabel\\
\htuse{ALEPH.Gamma3.pub.SCHAEL.05C,qt} & \htuse{ALEPH.Gamma3.pub.SCHAEL.05C,exp} & \htuse{ALEPH.Gamma3.pub.SCHAEL.05C,ref} \\
\htuse{DELPHI.Gamma3.pub.ABREU.99X,qt} & \htuse{DELPHI.Gamma3.pub.ABREU.99X,exp} & \htuse{DELPHI.Gamma3.pub.ABREU.99X,ref} \\
\htuse{L3.Gamma3.pub.ACCIARRI.01F,qt} & \htuse{L3.Gamma3.pub.ACCIARRI.01F,exp} & \htuse{L3.Gamma3.pub.ACCIARRI.01F,ref} \\
\htuse{OPAL.Gamma3.pub.ABBIENDI.03,qt} & \htuse{OPAL.Gamma3.pub.ABBIENDI.03,exp} & \htuse{OPAL.Gamma3.pub.ABBIENDI.03,ref}
}%
\htdef{Gamma3by5.qm}{%
\begin{ensuredisplaymath}
\htuse{Gamma3by5.gn} = \htuse{Gamma3by5.td}
\end{ensuredisplaymath}
 & \htuse{Gamma3by5.qt} & \hfagFitLabel\\
\htuse{ARGUS.Gamma3by5.pub.ALBRECHT.92D,qt} & \htuse{ARGUS.Gamma3by5.pub.ALBRECHT.92D,exp} & \htuse{ARGUS.Gamma3by5.pub.ALBRECHT.92D,ref} \\
\htuse{BaBar.Gamma3by5.pub.AUBERT.10F,qt} & \htuse{BaBar.Gamma3by5.pub.AUBERT.10F,exp} & \htuse{BaBar.Gamma3by5.pub.AUBERT.10F,ref} \\
\htuse{CLEO.Gamma3by5.pub.ANASTASSOV.97,qt} & \htuse{CLEO.Gamma3by5.pub.ANASTASSOV.97,exp} & \htuse{CLEO.Gamma3by5.pub.ANASTASSOV.97,ref}
}%
\htdef{Gamma5.qm}{%
\begin{ensuredisplaymath}
\htuse{Gamma5.gn} = \htuse{Gamma5.td}
\end{ensuredisplaymath}
 & \htuse{Gamma5.qt} & \hfagFitLabel\\
\htuse{ALEPH.Gamma5.pub.SCHAEL.05C,qt} & \htuse{ALEPH.Gamma5.pub.SCHAEL.05C,exp} & \htuse{ALEPH.Gamma5.pub.SCHAEL.05C,ref} \\
\htuse{CLEO.Gamma5.pub.ANASTASSOV.97,qt} & \htuse{CLEO.Gamma5.pub.ANASTASSOV.97,exp} & \htuse{CLEO.Gamma5.pub.ANASTASSOV.97,ref} \\
\htuse{DELPHI.Gamma5.pub.ABREU.99X,qt} & \htuse{DELPHI.Gamma5.pub.ABREU.99X,exp} & \htuse{DELPHI.Gamma5.pub.ABREU.99X,ref} \\
\htuse{L3.Gamma5.pub.ACCIARRI.01F,qt} & \htuse{L3.Gamma5.pub.ACCIARRI.01F,exp} & \htuse{L3.Gamma5.pub.ACCIARRI.01F,ref} \\
\htuse{OPAL.Gamma5.pub.ABBIENDI.99H,qt} & \htuse{OPAL.Gamma5.pub.ABBIENDI.99H,exp} & \htuse{OPAL.Gamma5.pub.ABBIENDI.99H,ref}
}%
\htdef{Gamma7.qm}{%
\begin{ensuredisplaymath}
\htuse{Gamma7.gn} = \htuse{Gamma7.td}
\end{ensuredisplaymath}
 & \htuse{Gamma7.qt} & \hfagFitLabel\\
\htuse{DELPHI.Gamma7.pub.ABREU.92N,qt} & \htuse{DELPHI.Gamma7.pub.ABREU.92N,exp} & \htuse{DELPHI.Gamma7.pub.ABREU.92N,ref} \\
\htuse{L3.Gamma7.pub.ACCIARRI.95,qt} & \htuse{L3.Gamma7.pub.ACCIARRI.95,exp} & \htuse{L3.Gamma7.pub.ACCIARRI.95,ref} \\
\htuse{OPAL.Gamma7.pub.ALEXANDER.91D,qt} & \htuse{OPAL.Gamma7.pub.ALEXANDER.91D,exp} & \htuse{OPAL.Gamma7.pub.ALEXANDER.91D,ref}
}%
\htdef{Gamma8.qm}{%
\begin{ensuredisplaymath}
\htuse{Gamma8.gn} = \htuse{Gamma8.td}
\end{ensuredisplaymath}
 & \htuse{Gamma8.qt} & \hfagFitLabel\\
\htuse{ALEPH.Gamma8.pub.SCHAEL.05C,qt} & \htuse{ALEPH.Gamma8.pub.SCHAEL.05C,exp} & \htuse{ALEPH.Gamma8.pub.SCHAEL.05C,ref} \\
\htuse{CLEO.Gamma8.pub.ANASTASSOV.97,qt} & \htuse{CLEO.Gamma8.pub.ANASTASSOV.97,exp} & \htuse{CLEO.Gamma8.pub.ANASTASSOV.97,ref} \\
\htuse{DELPHI.Gamma8.pub.ABDALLAH.06A,qt} & \htuse{DELPHI.Gamma8.pub.ABDALLAH.06A,exp} & \htuse{DELPHI.Gamma8.pub.ABDALLAH.06A,ref} \\
\htuse{OPAL.Gamma8.pub.ACKERSTAFF.98M,qt} & \htuse{OPAL.Gamma8.pub.ACKERSTAFF.98M,exp} & \htuse{OPAL.Gamma8.pub.ACKERSTAFF.98M,ref}
}%
\htdef{Gamma8by5.qm}{%
\begin{ensuredisplaymath}
\htuse{Gamma8by5.gn} = \htuse{Gamma8by5.td}
\end{ensuredisplaymath}
 & \htuse{Gamma8by5.qt} & \hfagFitLabel}%
\htdef{Gamma9.qm}{%
\begin{ensuredisplaymath}
\htuse{Gamma9.gn} = \htuse{Gamma9.td}
\end{ensuredisplaymath}
 & \htuse{Gamma9.qt} & \hfagFitLabel}%
\htdef{Gamma9by5.qm}{%
\begin{ensuredisplaymath}
\htuse{Gamma9by5.gn} = \htuse{Gamma9by5.td}
\end{ensuredisplaymath}
 & \htuse{Gamma9by5.qt} & \hfagFitLabel\\
\htuse{BaBar.Gamma9by5.pub.AUBERT.10F,qt} & \htuse{BaBar.Gamma9by5.pub.AUBERT.10F,exp} & \htuse{BaBar.Gamma9by5.pub.AUBERT.10F,ref}
}%
\htdef{Gamma10.qm}{%
\begin{ensuredisplaymath}
\htuse{Gamma10.gn} = \htuse{Gamma10.td}
\end{ensuredisplaymath}
 & \htuse{Gamma10.qt} & \hfagFitLabel\\
\htuse{ALEPH.Gamma10.pub.BARATE.99K,qt} & \htuse{ALEPH.Gamma10.pub.BARATE.99K,exp} & \htuse{ALEPH.Gamma10.pub.BARATE.99K,ref} \\
\htuse{CLEO.Gamma10.pub.BATTLE.94,qt} & \htuse{CLEO.Gamma10.pub.BATTLE.94,exp} & \htuse{CLEO.Gamma10.pub.BATTLE.94,ref} \\
\htuse{DELPHI.Gamma10.pub.ABREU.94K,qt} & \htuse{DELPHI.Gamma10.pub.ABREU.94K,exp} & \htuse{DELPHI.Gamma10.pub.ABREU.94K,ref} \\
\htuse{OPAL.Gamma10.pub.ABBIENDI.01J,qt} & \htuse{OPAL.Gamma10.pub.ABBIENDI.01J,exp} & \htuse{OPAL.Gamma10.pub.ABBIENDI.01J,ref}
}%
\htdef{Gamma10by5.qm}{%
\begin{ensuredisplaymath}
\htuse{Gamma10by5.gn} = \htuse{Gamma10by5.td}
\end{ensuredisplaymath}
 & \htuse{Gamma10by5.qt} & \hfagFitLabel\\
\htuse{BaBar.Gamma10by5.pub.AUBERT.10F,qt} & \htuse{BaBar.Gamma10by5.pub.AUBERT.10F,exp} & \htuse{BaBar.Gamma10by5.pub.AUBERT.10F,ref}
}%
\htdef{Gamma10by9.qm}{%
\begin{ensuredisplaymath}
\htuse{Gamma10by9.gn} = \htuse{Gamma10by9.td}
\end{ensuredisplaymath}
 & \htuse{Gamma10by9.qt} & \hfagFitLabel}%
\htdef{Gamma11.qm}{%
\begin{ensuredisplaymath}
\htuse{Gamma11.gn} = \htuse{Gamma11.td}
\end{ensuredisplaymath}
 & \htuse{Gamma11.qt} & \hfagFitLabel}%
\htdef{Gamma12.qm}{%
\begin{ensuredisplaymath}
\htuse{Gamma12.gn} = \htuse{Gamma12.td}
\end{ensuredisplaymath}
 & \htuse{Gamma12.qt} & \hfagFitLabel}%
\htdef{Gamma13.qm}{%
\begin{ensuredisplaymath}
\htuse{Gamma13.gn} = \htuse{Gamma13.td}
\end{ensuredisplaymath}
 & \htuse{Gamma13.qt} & \hfagFitLabel\\
\htuse{ALEPH.Gamma13.pub.SCHAEL.05C,qt} & \htuse{ALEPH.Gamma13.pub.SCHAEL.05C,exp} & \htuse{ALEPH.Gamma13.pub.SCHAEL.05C,ref} \\
\htuse{Belle.Gamma13.pub.FUJIKAWA.08,qt} & \htuse{Belle.Gamma13.pub.FUJIKAWA.08,exp} & \htuse{Belle.Gamma13.pub.FUJIKAWA.08,ref} \\
\htuse{CLEO.Gamma13.pub.ARTUSO.94,qt} & \htuse{CLEO.Gamma13.pub.ARTUSO.94,exp} & \htuse{CLEO.Gamma13.pub.ARTUSO.94,ref} \\
\htuse{DELPHI.Gamma13.pub.ABDALLAH.06A,qt} & \htuse{DELPHI.Gamma13.pub.ABDALLAH.06A,exp} & \htuse{DELPHI.Gamma13.pub.ABDALLAH.06A,ref} \\
\htuse{L3.Gamma13.pub.ACCIARRI.95,qt} & \htuse{L3.Gamma13.pub.ACCIARRI.95,exp} & \htuse{L3.Gamma13.pub.ACCIARRI.95,ref} \\
\htuse{OPAL.Gamma13.pub.ACKERSTAFF.98M,qt} & \htuse{OPAL.Gamma13.pub.ACKERSTAFF.98M,exp} & \htuse{OPAL.Gamma13.pub.ACKERSTAFF.98M,ref}
}%
\htdef{Gamma14.qm}{%
\begin{ensuredisplaymath}
\htuse{Gamma14.gn} = \htuse{Gamma14.td}
\end{ensuredisplaymath}
 & \htuse{Gamma14.qt} & \hfagFitLabel}%
\htdef{Gamma16.qm}{%
\begin{ensuredisplaymath}
\htuse{Gamma16.gn} = \htuse{Gamma16.td}
\end{ensuredisplaymath}
 & \htuse{Gamma16.qt} & \hfagFitLabel\\
\htuse{ALEPH.Gamma16.pub.BARATE.99K,qt} & \htuse{ALEPH.Gamma16.pub.BARATE.99K,exp} & \htuse{ALEPH.Gamma16.pub.BARATE.99K,ref} \\
\htuse{BaBar.Gamma16.pub.AUBERT.07AP,qt} & \htuse{BaBar.Gamma16.pub.AUBERT.07AP,exp} & \htuse{BaBar.Gamma16.pub.AUBERT.07AP,ref} \\
\htuse{CLEO.Gamma16.pub.BATTLE.94,qt} & \htuse{CLEO.Gamma16.pub.BATTLE.94,exp} & \htuse{CLEO.Gamma16.pub.BATTLE.94,ref} \\
\htuse{OPAL.Gamma16.pub.ABBIENDI.04J,qt} & \htuse{OPAL.Gamma16.pub.ABBIENDI.04J,exp} & \htuse{OPAL.Gamma16.pub.ABBIENDI.04J,ref}
}%
\htdef{Gamma17.qm}{%
\begin{ensuredisplaymath}
\htuse{Gamma17.gn} = \htuse{Gamma17.td}
\end{ensuredisplaymath}
 & \htuse{Gamma17.qt} & \hfagFitLabel\\
\htuse{OPAL.Gamma17.pub.ACKERSTAFF.98M,qt} & \htuse{OPAL.Gamma17.pub.ACKERSTAFF.98M,exp} & \htuse{OPAL.Gamma17.pub.ACKERSTAFF.98M,ref}
}%
\htdef{Gamma18.qm}{%
\begin{ensuredisplaymath}
\htuse{Gamma18.gn} = \htuse{Gamma18.td}
\end{ensuredisplaymath}
 & \htuse{Gamma18.qt} & \hfagFitLabel}%
\htdef{Gamma19.qm}{%
\begin{ensuredisplaymath}
\htuse{Gamma19.gn} = \htuse{Gamma19.td}
\end{ensuredisplaymath}
 & \htuse{Gamma19.qt} & \hfagFitLabel\\
\htuse{ALEPH.Gamma19.pub.SCHAEL.05C,qt} & \htuse{ALEPH.Gamma19.pub.SCHAEL.05C,exp} & \htuse{ALEPH.Gamma19.pub.SCHAEL.05C,ref} \\
\htuse{DELPHI.Gamma19.pub.ABDALLAH.06A,qt} & \htuse{DELPHI.Gamma19.pub.ABDALLAH.06A,exp} & \htuse{DELPHI.Gamma19.pub.ABDALLAH.06A,ref} \\
\htuse{L3.Gamma19.pub.ACCIARRI.95,qt} & \htuse{L3.Gamma19.pub.ACCIARRI.95,exp} & \htuse{L3.Gamma19.pub.ACCIARRI.95,ref}
}%
\htdef{Gamma19by13.qm}{%
\begin{ensuredisplaymath}
\htuse{Gamma19by13.gn} = \htuse{Gamma19by13.td}
\end{ensuredisplaymath}
 & \htuse{Gamma19by13.qt} & \hfagFitLabel\\
\htuse{CLEO.Gamma19by13.pub.PROCARIO.93,qt} & \htuse{CLEO.Gamma19by13.pub.PROCARIO.93,exp} & \htuse{CLEO.Gamma19by13.pub.PROCARIO.93,ref}
}%
\htdef{Gamma20.qm}{%
\begin{ensuredisplaymath}
\htuse{Gamma20.gn} = \htuse{Gamma20.td}
\end{ensuredisplaymath}
 & \htuse{Gamma20.qt} & \hfagFitLabel}%
\htdef{Gamma23.qm}{%
\begin{ensuredisplaymath}
\htuse{Gamma23.gn} = \htuse{Gamma23.td}
\end{ensuredisplaymath}
 & \htuse{Gamma23.qt} & \hfagFitLabel\\
\htuse{ALEPH.Gamma23.pub.BARATE.99K,qt} & \htuse{ALEPH.Gamma23.pub.BARATE.99K,exp} & \htuse{ALEPH.Gamma23.pub.BARATE.99K,ref} \\
\htuse{CLEO.Gamma23.pub.BATTLE.94,qt} & \htuse{CLEO.Gamma23.pub.BATTLE.94,exp} & \htuse{CLEO.Gamma23.pub.BATTLE.94,ref}
}%
\htdef{Gamma24.qm}{%
\begin{ensuredisplaymath}
\htuse{Gamma24.gn} = \htuse{Gamma24.td}
\end{ensuredisplaymath}
 & \htuse{Gamma24.qt} & \hfagFitLabel}%
\htdef{Gamma25.qm}{%
\begin{ensuredisplaymath}
\htuse{Gamma25.gn} = \htuse{Gamma25.td}
\end{ensuredisplaymath}
 & \htuse{Gamma25.qt} & \hfagFitLabel\\
\htuse{DELPHI.Gamma25.pub.ABDALLAH.06A,qt} & \htuse{DELPHI.Gamma25.pub.ABDALLAH.06A,exp} & \htuse{DELPHI.Gamma25.pub.ABDALLAH.06A,ref}
}%
\htdef{Gamma26.qm}{%
\begin{ensuredisplaymath}
\htuse{Gamma26.gn} = \htuse{Gamma26.td}
\end{ensuredisplaymath}
 & \htuse{Gamma26.qt} & \hfagFitLabel\\
\htuse{ALEPH.Gamma26.pub.SCHAEL.05C,qt} & \htuse{ALEPH.Gamma26.pub.SCHAEL.05C,exp} & \htuse{ALEPH.Gamma26.pub.SCHAEL.05C,ref} \\
\htuse{L3.Gamma26.pub.ACCIARRI.95,qt} & \htuse{L3.Gamma26.pub.ACCIARRI.95,exp} & \htuse{L3.Gamma26.pub.ACCIARRI.95,ref}
}%
\htdef{Gamma26by13.qm}{%
\begin{ensuredisplaymath}
\htuse{Gamma26by13.gn} = \htuse{Gamma26by13.td}
\end{ensuredisplaymath}
 & \htuse{Gamma26by13.qt} & \hfagFitLabel\\
\htuse{CLEO.Gamma26by13.pub.PROCARIO.93,qt} & \htuse{CLEO.Gamma26by13.pub.PROCARIO.93,exp} & \htuse{CLEO.Gamma26by13.pub.PROCARIO.93,ref}
}%
\htdef{Gamma27.qm}{%
\begin{ensuredisplaymath}
\htuse{Gamma27.gn} = \htuse{Gamma27.td}
\end{ensuredisplaymath}
 & \htuse{Gamma27.qt} & \hfagFitLabel}%
\htdef{Gamma28.qm}{%
\begin{ensuredisplaymath}
\htuse{Gamma28.gn} = \htuse{Gamma28.td}
\end{ensuredisplaymath}
 & \htuse{Gamma28.qt} & \hfagFitLabel\\
\htuse{ALEPH.Gamma28.pub.BARATE.99K,qt} & \htuse{ALEPH.Gamma28.pub.BARATE.99K,exp} & \htuse{ALEPH.Gamma28.pub.BARATE.99K,ref}
}%
\htdef{Gamma29.qm}{%
\begin{ensuredisplaymath}
\htuse{Gamma29.gn} = \htuse{Gamma29.td}
\end{ensuredisplaymath}
 & \htuse{Gamma29.qt} & \hfagFitLabel\\
\htuse{CLEO.Gamma29.pub.PROCARIO.93,qt} & \htuse{CLEO.Gamma29.pub.PROCARIO.93,exp} & \htuse{CLEO.Gamma29.pub.PROCARIO.93,ref}
}%
\htdef{Gamma30.qm}{%
\begin{ensuredisplaymath}
\htuse{Gamma30.gn} = \htuse{Gamma30.td}
\end{ensuredisplaymath}
 & \htuse{Gamma30.qt} & \hfagFitLabel\\
\htuse{ALEPH.Gamma30.pub.SCHAEL.05C,qt} & \htuse{ALEPH.Gamma30.pub.SCHAEL.05C,exp} & \htuse{ALEPH.Gamma30.pub.SCHAEL.05C,ref}
}%
\htdef{Gamma31.qm}{%
\begin{ensuredisplaymath}
\htuse{Gamma31.gn} = \htuse{Gamma31.td}
\end{ensuredisplaymath}
 & \htuse{Gamma31.qt} & \hfagFitLabel\\
\htuse{CLEO.Gamma31.pub.BATTLE.94,qt} & \htuse{CLEO.Gamma31.pub.BATTLE.94,exp} & \htuse{CLEO.Gamma31.pub.BATTLE.94,ref} \\
\htuse{DELPHI.Gamma31.pub.ABREU.94K,qt} & \htuse{DELPHI.Gamma31.pub.ABREU.94K,exp} & \htuse{DELPHI.Gamma31.pub.ABREU.94K,ref} \\
\htuse{OPAL.Gamma31.pub.ABBIENDI.01J,qt} & \htuse{OPAL.Gamma31.pub.ABBIENDI.01J,exp} & \htuse{OPAL.Gamma31.pub.ABBIENDI.01J,ref}
}%
\htdef{Gamma32.qm}{%
\begin{ensuredisplaymath}
\htuse{Gamma32.gn} = \htuse{Gamma32.td}
\end{ensuredisplaymath}
 & \htuse{Gamma32.qt} & \hfagFitLabel}%
\htdef{Gamma33.qm}{%
\begin{ensuredisplaymath}
\htuse{Gamma33.gn} = \htuse{Gamma33.td}
\end{ensuredisplaymath}
 & \htuse{Gamma33.qt} & \hfagFitLabel\\
\htuse{ALEPH.Gamma33.pub.BARATE.98E,qt} & \htuse{ALEPH.Gamma33.pub.BARATE.98E,exp} & \htuse{ALEPH.Gamma33.pub.BARATE.98E,ref} \\
\htuse{OPAL.Gamma33.pub.AKERS.94G,qt} & \htuse{OPAL.Gamma33.pub.AKERS.94G,exp} & \htuse{OPAL.Gamma33.pub.AKERS.94G,ref}
}%
\htdef{Gamma34.qm}{%
\begin{ensuredisplaymath}
\htuse{Gamma34.gn} = \htuse{Gamma34.td}
\end{ensuredisplaymath}
 & \htuse{Gamma34.qt} & \hfagFitLabel\\
\htuse{CLEO.Gamma34.pub.COAN.96,qt} & \htuse{CLEO.Gamma34.pub.COAN.96,exp} & \htuse{CLEO.Gamma34.pub.COAN.96,ref}
}%
\htdef{Gamma35.qm}{%
\begin{ensuredisplaymath}
\htuse{Gamma35.gn} = \htuse{Gamma35.td}
\end{ensuredisplaymath}
 & \htuse{Gamma35.qt} & \hfagFitLabel\\
\htuse{ALEPH.Gamma35.pub.BARATE.99K,qt} & \htuse{ALEPH.Gamma35.pub.BARATE.99K,exp} & \htuse{ALEPH.Gamma35.pub.BARATE.99K,ref} \\
\htuse{Belle.Gamma35.pub.RYU.14vpc,qt} & \htuse{Belle.Gamma35.pub.RYU.14vpc,exp} & \htuse{Belle.Gamma35.pub.RYU.14vpc,ref} \\
\htuse{L3.Gamma35.pub.ACCIARRI.95F,qt} & \htuse{L3.Gamma35.pub.ACCIARRI.95F,exp} & \htuse{L3.Gamma35.pub.ACCIARRI.95F,ref} \\
\htuse{OPAL.Gamma35.pub.ABBIENDI.00C,qt} & \htuse{OPAL.Gamma35.pub.ABBIENDI.00C,exp} & \htuse{OPAL.Gamma35.pub.ABBIENDI.00C,ref}
}%
\htdef{Gamma37.qm}{%
\begin{ensuredisplaymath}
\htuse{Gamma37.gn} = \htuse{Gamma37.td}
\end{ensuredisplaymath}
 & \htuse{Gamma37.qt} & \hfagFitLabel\\
\htuse{ALEPH.Gamma37.pub.BARATE.98E,qt} & \htuse{ALEPH.Gamma37.pub.BARATE.98E,exp} & \htuse{ALEPH.Gamma37.pub.BARATE.98E,ref} \\
\htuse{ALEPH.Gamma37.pub.BARATE.99K,qt} & \htuse{ALEPH.Gamma37.pub.BARATE.99K,exp} & \htuse{ALEPH.Gamma37.pub.BARATE.99K,ref} \\
\htuse{Belle.Gamma37.pub.RYU.14vpc,qt} & \htuse{Belle.Gamma37.pub.RYU.14vpc,exp} & \htuse{Belle.Gamma37.pub.RYU.14vpc,ref} \\
\htuse{CLEO.Gamma37.pub.COAN.96,qt} & \htuse{CLEO.Gamma37.pub.COAN.96,exp} & \htuse{CLEO.Gamma37.pub.COAN.96,ref}
}%
\htdef{Gamma38.qm}{%
\begin{ensuredisplaymath}
\htuse{Gamma38.gn} = \htuse{Gamma38.td}
\end{ensuredisplaymath}
 & \htuse{Gamma38.qt} & \hfagFitLabel\\
\htuse{OPAL.Gamma38.pub.ABBIENDI.00C,qt} & \htuse{OPAL.Gamma38.pub.ABBIENDI.00C,exp} & \htuse{OPAL.Gamma38.pub.ABBIENDI.00C,ref}
}%
\htdef{Gamma39.qm}{%
\begin{ensuredisplaymath}
\htuse{Gamma39.gn} = \htuse{Gamma39.td}
\end{ensuredisplaymath}
 & \htuse{Gamma39.qt} & \hfagFitLabel\\
\htuse{CLEO.Gamma39.pub.COAN.96,qt} & \htuse{CLEO.Gamma39.pub.COAN.96,exp} & \htuse{CLEO.Gamma39.pub.COAN.96,ref}
}%
\htdef{Gamma40.qm}{%
\begin{ensuredisplaymath}
\htuse{Gamma40.gn} = \htuse{Gamma40.td}
\end{ensuredisplaymath}
 & \htuse{Gamma40.qt} & \hfagFitLabel\\
\htuse{ALEPH.Gamma40.pub.BARATE.98E,qt} & \htuse{ALEPH.Gamma40.pub.BARATE.98E,exp} & \htuse{ALEPH.Gamma40.pub.BARATE.98E,ref} \\
\htuse{ALEPH.Gamma40.pub.BARATE.99K,qt} & \htuse{ALEPH.Gamma40.pub.BARATE.99K,exp} & \htuse{ALEPH.Gamma40.pub.BARATE.99K,ref} \\
\htuse{Belle.Gamma40.pub.RYU.14vpc,qt} & \htuse{Belle.Gamma40.pub.RYU.14vpc,exp} & \htuse{Belle.Gamma40.pub.RYU.14vpc,ref} \\
\htuse{L3.Gamma40.pub.ACCIARRI.95F,qt} & \htuse{L3.Gamma40.pub.ACCIARRI.95F,exp} & \htuse{L3.Gamma40.pub.ACCIARRI.95F,ref}
}%
\htdef{Gamma42.qm}{%
\begin{ensuredisplaymath}
\htuse{Gamma42.gn} = \htuse{Gamma42.td}
\end{ensuredisplaymath}
 & \htuse{Gamma42.qt} & \hfagFitLabel\\
\htuse{ALEPH.Gamma42.pub.BARATE.98E,qt} & \htuse{ALEPH.Gamma42.pub.BARATE.98E,exp} & \htuse{ALEPH.Gamma42.pub.BARATE.98E,ref} \\
\htuse{ALEPH.Gamma42.pub.BARATE.99K,qt} & \htuse{ALEPH.Gamma42.pub.BARATE.99K,exp} & \htuse{ALEPH.Gamma42.pub.BARATE.99K,ref} \\
\htuse{Belle.Gamma42.pub.RYU.14vpc,qt} & \htuse{Belle.Gamma42.pub.RYU.14vpc,exp} & \htuse{Belle.Gamma42.pub.RYU.14vpc,ref} \\
\htuse{CLEO.Gamma42.pub.COAN.96,qt} & \htuse{CLEO.Gamma42.pub.COAN.96,exp} & \htuse{CLEO.Gamma42.pub.COAN.96,ref}
}%
\htdef{Gamma43.qm}{%
\begin{ensuredisplaymath}
\htuse{Gamma43.gn} = \htuse{Gamma43.td}
\end{ensuredisplaymath}
 & \htuse{Gamma43.qt} & \hfagFitLabel\\
\htuse{OPAL.Gamma43.pub.ABBIENDI.00C,qt} & \htuse{OPAL.Gamma43.pub.ABBIENDI.00C,exp} & \htuse{OPAL.Gamma43.pub.ABBIENDI.00C,ref}
}%
\htdef{Gamma44.qm}{%
\begin{ensuredisplaymath}
\htuse{Gamma44.gn} = \htuse{Gamma44.td}
\end{ensuredisplaymath}
 & \htuse{Gamma44.qt} & \hfagFitLabel\\
\htuse{ALEPH.Gamma44.pub.BARATE.99R,qt} & \htuse{ALEPH.Gamma44.pub.BARATE.99R,exp} & \htuse{ALEPH.Gamma44.pub.BARATE.99R,ref}
}%
\htdef{Gamma46.qm}{%
\begin{ensuredisplaymath}
\htuse{Gamma46.gn} = \htuse{Gamma46.td}
\end{ensuredisplaymath}
 & \htuse{Gamma46.qt} & \hfagFitLabel}%
\htdef{Gamma47.qm}{%
\begin{ensuredisplaymath}
\htuse{Gamma47.gn} = \htuse{Gamma47.td}
\end{ensuredisplaymath}
 & \htuse{Gamma47.qt} & \hfagFitLabel\\
\htuse{ALEPH.Gamma47.pub.BARATE.98E,qt} & \htuse{ALEPH.Gamma47.pub.BARATE.98E,exp} & \htuse{ALEPH.Gamma47.pub.BARATE.98E,ref} \\
\htuse{BaBar.Gamma47.pub.LEES.12Y,qt} & \htuse{BaBar.Gamma47.pub.LEES.12Y,exp} & \htuse{BaBar.Gamma47.pub.LEES.12Y,ref} \\
\htuse{Belle.Gamma47.pub.RYU.14vpc,qt} & \htuse{Belle.Gamma47.pub.RYU.14vpc,exp} & \htuse{Belle.Gamma47.pub.RYU.14vpc,ref} \\
\htuse{CLEO.Gamma47.pub.COAN.96,qt} & \htuse{CLEO.Gamma47.pub.COAN.96,exp} & \htuse{CLEO.Gamma47.pub.COAN.96,ref}
}%
\htdef{Gamma48.qm}{%
\begin{ensuredisplaymath}
\htuse{Gamma48.gn} = \htuse{Gamma48.td}
\end{ensuredisplaymath}
 & \htuse{Gamma48.qt} & \hfagFitLabel\\
\htuse{ALEPH.Gamma48.pub.BARATE.98E,qt} & \htuse{ALEPH.Gamma48.pub.BARATE.98E,exp} & \htuse{ALEPH.Gamma48.pub.BARATE.98E,ref}
}%
\htdef{Gamma49.qm}{%
\begin{ensuredisplaymath}
\htuse{Gamma49.gn} = \htuse{Gamma49.td}
\end{ensuredisplaymath}
 & \htuse{Gamma49.qt} & \hfagFitLabel}%
\htdef{Gamma50.qm}{%
\begin{ensuredisplaymath}
\htuse{Gamma50.gn} = \htuse{Gamma50.td}
\end{ensuredisplaymath}
 & \htuse{Gamma50.qt} & \hfagFitLabel\\
\htuse{BaBar.Gamma50.pub.LEES.12Y,qt} & \htuse{BaBar.Gamma50.pub.LEES.12Y,exp} & \htuse{BaBar.Gamma50.pub.LEES.12Y,ref} \\
\htuse{Belle.Gamma50.pub.RYU.14vpc,qt} & \htuse{Belle.Gamma50.pub.RYU.14vpc,exp} & \htuse{Belle.Gamma50.pub.RYU.14vpc,ref}
}%
\htdef{Gamma51.qm}{%
\begin{ensuredisplaymath}
\htuse{Gamma51.gn} = \htuse{Gamma51.td}
\end{ensuredisplaymath}
 & \htuse{Gamma51.qt} & \hfagFitLabel\\
\htuse{ALEPH.Gamma51.pub.BARATE.98E,qt} & \htuse{ALEPH.Gamma51.pub.BARATE.98E,exp} & \htuse{ALEPH.Gamma51.pub.BARATE.98E,ref}
}%
\htdef{Gamma53.qm}{%
\begin{ensuredisplaymath}
\htuse{Gamma53.gn} = \htuse{Gamma53.td}
\end{ensuredisplaymath}
 & \htuse{Gamma53.qt} & \hfagFitLabel\\
\htuse{ALEPH.Gamma53.pub.BARATE.98E,qt} & \htuse{ALEPH.Gamma53.pub.BARATE.98E,exp} & \htuse{ALEPH.Gamma53.pub.BARATE.98E,ref}
}%
\htdef{Gamma54.qm}{%
\begin{ensuredisplaymath}
\htuse{Gamma54.gn} = \htuse{Gamma54.td}
\end{ensuredisplaymath}
 & \htuse{Gamma54.qt} & \hfagFitLabel\\
\htuse{CELLO.Gamma54.pub.BEHREND.89B,qt} & \htuse{CELLO.Gamma54.pub.BEHREND.89B,exp} & \htuse{CELLO.Gamma54.pub.BEHREND.89B,ref} \\
\htuse{L3.Gamma54.pub.ADEVA.91F,qt} & \htuse{L3.Gamma54.pub.ADEVA.91F,exp} & \htuse{L3.Gamma54.pub.ADEVA.91F,ref} \\
\htuse{TPC.Gamma54.pub.AIHARA.87B,qt} & \htuse{TPC.Gamma54.pub.AIHARA.87B,exp} & \htuse{TPC.Gamma54.pub.AIHARA.87B,ref}
}%
\htdef{Gamma55.qm}{%
\begin{ensuredisplaymath}
\htuse{Gamma55.gn} = \htuse{Gamma55.td}
\end{ensuredisplaymath}
 & \htuse{Gamma55.qt} & \hfagFitLabel\\
\htuse{L3.Gamma55.pub.ACHARD.01D,qt} & \htuse{L3.Gamma55.pub.ACHARD.01D,exp} & \htuse{L3.Gamma55.pub.ACHARD.01D,ref} \\
\htuse{OPAL.Gamma55.pub.AKERS.95Y,qt} & \htuse{OPAL.Gamma55.pub.AKERS.95Y,exp} & \htuse{OPAL.Gamma55.pub.AKERS.95Y,ref}
}%
\htdef{Gamma56.qm}{%
\begin{ensuredisplaymath}
\htuse{Gamma56.gn} = \htuse{Gamma56.td}
\end{ensuredisplaymath}
 & \htuse{Gamma56.qt} & \hfagFitLabel}%
\htdef{Gamma57.qm}{%
\begin{ensuredisplaymath}
\htuse{Gamma57.gn} = \htuse{Gamma57.td}
\end{ensuredisplaymath}
 & \htuse{Gamma57.qt} & \hfagFitLabel\\
\htuse{CLEO.Gamma57.pub.BALEST.95C,qt} & \htuse{CLEO.Gamma57.pub.BALEST.95C,exp} & \htuse{CLEO.Gamma57.pub.BALEST.95C,ref} \\
\htuse{DELPHI.Gamma57.pub.ABDALLAH.06A,qt} & \htuse{DELPHI.Gamma57.pub.ABDALLAH.06A,exp} & \htuse{DELPHI.Gamma57.pub.ABDALLAH.06A,ref}
}%
\htdef{Gamma57by55.qm}{%
\begin{ensuredisplaymath}
\htuse{Gamma57by55.gn} = \htuse{Gamma57by55.td}
\end{ensuredisplaymath}
 & \htuse{Gamma57by55.qt} & \hfagFitLabel\\
\htuse{OPAL.Gamma57by55.pub.AKERS.95Y,qt} & \htuse{OPAL.Gamma57by55.pub.AKERS.95Y,exp} & \htuse{OPAL.Gamma57by55.pub.AKERS.95Y,ref}
}%
\htdef{Gamma58.qm}{%
\begin{ensuredisplaymath}
\htuse{Gamma58.gn} = \htuse{Gamma58.td}
\end{ensuredisplaymath}
 & \htuse{Gamma58.qt} & \hfagFitLabel\\
\htuse{ALEPH.Gamma58.pub.SCHAEL.05C,qt} & \htuse{ALEPH.Gamma58.pub.SCHAEL.05C,exp} & \htuse{ALEPH.Gamma58.pub.SCHAEL.05C,ref}
}%
\htdef{Gamma59.qm}{%
\begin{ensuredisplaymath}
\htuse{Gamma59.gn} = \htuse{Gamma59.td}
\end{ensuredisplaymath}
 & \htuse{Gamma59.qt} & \hfagFitLabel}%
\htdef{Gamma60.qm}{%
\begin{ensuredisplaymath}
\htuse{Gamma60.gn} = \htuse{Gamma60.td}
\end{ensuredisplaymath}
 & \htuse{Gamma60.qt} & \hfagFitLabel\\
\htuse{BaBar.Gamma60.pub.AUBERT.08,qt} & \htuse{BaBar.Gamma60.pub.AUBERT.08,exp} & \htuse{BaBar.Gamma60.pub.AUBERT.08,ref} \\
\htuse{Belle.Gamma60.pub.LEE.10,qt} & \htuse{Belle.Gamma60.pub.LEE.10,exp} & \htuse{Belle.Gamma60.pub.LEE.10,ref} \\
\htuse{CLEO3.Gamma60.pub.BRIERE.03,qt} & \htuse{CLEO3.Gamma60.pub.BRIERE.03,exp} & \htuse{CLEO3.Gamma60.pub.BRIERE.03,ref}
}%
\htdef{Gamma62.qm}{%
\begin{ensuredisplaymath}
\htuse{Gamma62.gn} = \htuse{Gamma62.td}
\end{ensuredisplaymath}
 & \htuse{Gamma62.qt} & \hfagFitLabel}%
\htdef{Gamma63.qm}{%
\begin{ensuredisplaymath}
\htuse{Gamma63.gn} = \htuse{Gamma63.td}
\end{ensuredisplaymath}
 & \htuse{Gamma63.qt} & \hfagFitLabel}%
\htdef{Gamma64.qm}{%
\begin{ensuredisplaymath}
\htuse{Gamma64.gn} = \htuse{Gamma64.td}
\end{ensuredisplaymath}
 & \htuse{Gamma64.qt} & \hfagFitLabel}%
\htdef{Gamma65.qm}{%
\begin{ensuredisplaymath}
\htuse{Gamma65.gn} = \htuse{Gamma65.td}
\end{ensuredisplaymath}
 & \htuse{Gamma65.qt} & \hfagFitLabel}%
\htdef{Gamma66.qm}{%
\begin{ensuredisplaymath}
\htuse{Gamma66.gn} = \htuse{Gamma66.td}
\end{ensuredisplaymath}
 & \htuse{Gamma66.qt} & \hfagFitLabel\\
\htuse{ALEPH.Gamma66.pub.SCHAEL.05C,qt} & \htuse{ALEPH.Gamma66.pub.SCHAEL.05C,exp} & \htuse{ALEPH.Gamma66.pub.SCHAEL.05C,ref} \\
\htuse{CLEO.Gamma66.pub.BALEST.95C,qt} & \htuse{CLEO.Gamma66.pub.BALEST.95C,exp} & \htuse{CLEO.Gamma66.pub.BALEST.95C,ref} \\
\htuse{DELPHI.Gamma66.pub.ABDALLAH.06A,qt} & \htuse{DELPHI.Gamma66.pub.ABDALLAH.06A,exp} & \htuse{DELPHI.Gamma66.pub.ABDALLAH.06A,ref}
}%
\htdef{Gamma67.qm}{%
\begin{ensuredisplaymath}
\htuse{Gamma67.gn} = \htuse{Gamma67.td}
\end{ensuredisplaymath}
 & \htuse{Gamma67.qt} & \hfagFitLabel}%
\htdef{Gamma68.qm}{%
\begin{ensuredisplaymath}
\htuse{Gamma68.gn} = \htuse{Gamma68.td}
\end{ensuredisplaymath}
 & \htuse{Gamma68.qt} & \hfagFitLabel}%
\htdef{Gamma69.qm}{%
\begin{ensuredisplaymath}
\htuse{Gamma69.gn} = \htuse{Gamma69.td}
\end{ensuredisplaymath}
 & \htuse{Gamma69.qt} & \hfagFitLabel\\
\htuse{CLEO.Gamma69.pub.EDWARDS.00A,qt} & \htuse{CLEO.Gamma69.pub.EDWARDS.00A,exp} & \htuse{CLEO.Gamma69.pub.EDWARDS.00A,ref}
}%
\htdef{Gamma70.qm}{%
\begin{ensuredisplaymath}
\htuse{Gamma70.gn} = \htuse{Gamma70.td}
\end{ensuredisplaymath}
 & \htuse{Gamma70.qt} & \hfagFitLabel}%
\htdef{Gamma74.qm}{%
\begin{ensuredisplaymath}
\htuse{Gamma74.gn} = \htuse{Gamma74.td}
\end{ensuredisplaymath}
 & \htuse{Gamma74.qt} & \hfagFitLabel\\
\htuse{DELPHI.Gamma74.pub.ABDALLAH.06A,qt} & \htuse{DELPHI.Gamma74.pub.ABDALLAH.06A,exp} & \htuse{DELPHI.Gamma74.pub.ABDALLAH.06A,ref}
}%
\htdef{Gamma75.qm}{%
\begin{ensuredisplaymath}
\htuse{Gamma75.gn} = \htuse{Gamma75.td}
\end{ensuredisplaymath}
 & \htuse{Gamma75.qt} & \hfagFitLabel}%
\htdef{Gamma76.qm}{%
\begin{ensuredisplaymath}
\htuse{Gamma76.gn} = \htuse{Gamma76.td}
\end{ensuredisplaymath}
 & \htuse{Gamma76.qt} & \hfagFitLabel\\
\htuse{ALEPH.Gamma76.pub.SCHAEL.05C,qt} & \htuse{ALEPH.Gamma76.pub.SCHAEL.05C,exp} & \htuse{ALEPH.Gamma76.pub.SCHAEL.05C,ref}
}%
\htdef{Gamma76by54.qm}{%
\begin{ensuredisplaymath}
\htuse{Gamma76by54.gn} = \htuse{Gamma76by54.td}
\end{ensuredisplaymath}
 & \htuse{Gamma76by54.qt} & \hfagFitLabel\\
\htuse{CLEO.Gamma76by54.pub.BORTOLETTO.93,qt} & \htuse{CLEO.Gamma76by54.pub.BORTOLETTO.93,exp} & \htuse{CLEO.Gamma76by54.pub.BORTOLETTO.93,ref}
}%
\htdef{Gamma77.qm}{%
\begin{ensuredisplaymath}
\htuse{Gamma77.gn} = \htuse{Gamma77.td}
\end{ensuredisplaymath}
 & \htuse{Gamma77.qt} & \hfagFitLabel}%
\htdef{Gamma78.qm}{%
\begin{ensuredisplaymath}
\htuse{Gamma78.gn} = \htuse{Gamma78.td}
\end{ensuredisplaymath}
 & \htuse{Gamma78.qt} & \hfagFitLabel\\
\htuse{CLEO.Gamma78.pub.ANASTASSOV.01,qt} & \htuse{CLEO.Gamma78.pub.ANASTASSOV.01,exp} & \htuse{CLEO.Gamma78.pub.ANASTASSOV.01,ref}
}%
\htdef{Gamma79.qm}{%
\begin{ensuredisplaymath}
\htuse{Gamma79.gn} = \htuse{Gamma79.td}
\end{ensuredisplaymath}
 & \htuse{Gamma79.qt} & \hfagFitLabel}%
\htdef{Gamma80.qm}{%
\begin{ensuredisplaymath}
\htuse{Gamma80.gn} = \htuse{Gamma80.td}
\end{ensuredisplaymath}
 & \htuse{Gamma80.qt} & \hfagFitLabel}%
\htdef{Gamma80by60.qm}{%
\begin{ensuredisplaymath}
\htuse{Gamma80by60.gn} = \htuse{Gamma80by60.td}
\end{ensuredisplaymath}
 & \htuse{Gamma80by60.qt} & \hfagFitLabel\\
\htuse{CLEO.Gamma80by60.pub.RICHICHI.99,qt} & \htuse{CLEO.Gamma80by60.pub.RICHICHI.99,exp} & \htuse{CLEO.Gamma80by60.pub.RICHICHI.99,ref}
}%
\htdef{Gamma81.qm}{%
\begin{ensuredisplaymath}
\htuse{Gamma81.gn} = \htuse{Gamma81.td}
\end{ensuredisplaymath}
 & \htuse{Gamma81.qt} & \hfagFitLabel}%
\htdef{Gamma81by69.qm}{%
\begin{ensuredisplaymath}
\htuse{Gamma81by69.gn} = \htuse{Gamma81by69.td}
\end{ensuredisplaymath}
 & \htuse{Gamma81by69.qt} & \hfagFitLabel\\
\htuse{CLEO.Gamma81by69.pub.RICHICHI.99,qt} & \htuse{CLEO.Gamma81by69.pub.RICHICHI.99,exp} & \htuse{CLEO.Gamma81by69.pub.RICHICHI.99,ref}
}%
\htdef{Gamma82.qm}{%
\begin{ensuredisplaymath}
\htuse{Gamma82.gn} = \htuse{Gamma82.td}
\end{ensuredisplaymath}
 & \htuse{Gamma82.qt} & \hfagFitLabel\\
\htuse{TPC.Gamma82.pub.BAUER.94,qt} & \htuse{TPC.Gamma82.pub.BAUER.94,exp} & \htuse{TPC.Gamma82.pub.BAUER.94,ref}
}%
\htdef{Gamma83.qm}{%
\begin{ensuredisplaymath}
\htuse{Gamma83.gn} = \htuse{Gamma83.td}
\end{ensuredisplaymath}
 & \htuse{Gamma83.qt} & \hfagFitLabel}%
\htdef{Gamma84.qm}{%
\begin{ensuredisplaymath}
\htuse{Gamma84.gn} = \htuse{Gamma84.td}
\end{ensuredisplaymath}
 & \htuse{Gamma84.qt} & \hfagFitLabel}%
\htdef{Gamma85.qm}{%
\begin{ensuredisplaymath}
\htuse{Gamma85.gn} = \htuse{Gamma85.td}
\end{ensuredisplaymath}
 & \htuse{Gamma85.qt} & \hfagFitLabel\\
\htuse{ALEPH.Gamma85.pub.BARATE.98,qt} & \htuse{ALEPH.Gamma85.pub.BARATE.98,exp} & \htuse{ALEPH.Gamma85.pub.BARATE.98,ref} \\
\htuse{BaBar.Gamma85.pub.AUBERT.08,qt} & \htuse{BaBar.Gamma85.pub.AUBERT.08,exp} & \htuse{BaBar.Gamma85.pub.AUBERT.08,ref} \\
\htuse{Belle.Gamma85.pub.LEE.10,qt} & \htuse{Belle.Gamma85.pub.LEE.10,exp} & \htuse{Belle.Gamma85.pub.LEE.10,ref} \\
\htuse{CLEO3.Gamma85.pub.BRIERE.03,qt} & \htuse{CLEO3.Gamma85.pub.BRIERE.03,exp} & \htuse{CLEO3.Gamma85.pub.BRIERE.03,ref} \\
\htuse{OPAL.Gamma85.pub.ABBIENDI.04J,qt} & \htuse{OPAL.Gamma85.pub.ABBIENDI.04J,exp} & \htuse{OPAL.Gamma85.pub.ABBIENDI.04J,ref}
}%
\htdef{Gamma85by60.qm}{%
\begin{ensuredisplaymath}
\htuse{Gamma85by60.gn} = \htuse{Gamma85by60.td}
\end{ensuredisplaymath}
 & \htuse{Gamma85by60.qt} & \hfagFitLabel}%
\htdef{Gamma87.qm}{%
\begin{ensuredisplaymath}
\htuse{Gamma87.gn} = \htuse{Gamma87.td}
\end{ensuredisplaymath}
 & \htuse{Gamma87.qt} & \hfagFitLabel}%
\htdef{Gamma88.qm}{%
\begin{ensuredisplaymath}
\htuse{Gamma88.gn} = \htuse{Gamma88.td}
\end{ensuredisplaymath}
 & \htuse{Gamma88.qt} & \hfagFitLabel\\
\htuse{ALEPH.Gamma88.pub.BARATE.98,qt} & \htuse{ALEPH.Gamma88.pub.BARATE.98,exp} & \htuse{ALEPH.Gamma88.pub.BARATE.98,ref} \\
\htuse{CLEO3.Gamma88.pub.ARMS.05,qt} & \htuse{CLEO3.Gamma88.pub.ARMS.05,exp} & \htuse{CLEO3.Gamma88.pub.ARMS.05,ref}
}%
\htdef{Gamma89.qm}{%
\begin{ensuredisplaymath}
\htuse{Gamma89.gn} = \htuse{Gamma89.td}
\end{ensuredisplaymath}
 & \htuse{Gamma89.qt} & \hfagFitLabel}%
\htdef{Gamma92.qm}{%
\begin{ensuredisplaymath}
\htuse{Gamma92.gn} = \htuse{Gamma92.td}
\end{ensuredisplaymath}
 & \htuse{Gamma92.qt} & \hfagFitLabel\\
\htuse{OPAL.Gamma92.pub.ABBIENDI.00D,qt} & \htuse{OPAL.Gamma92.pub.ABBIENDI.00D,exp} & \htuse{OPAL.Gamma92.pub.ABBIENDI.00D,ref} \\
\htuse{TPC.Gamma92.pub.BAUER.94,qt} & \htuse{TPC.Gamma92.pub.BAUER.94,exp} & \htuse{TPC.Gamma92.pub.BAUER.94,ref}
}%
\htdef{Gamma93.qm}{%
\begin{ensuredisplaymath}
\htuse{Gamma93.gn} = \htuse{Gamma93.td}
\end{ensuredisplaymath}
 & \htuse{Gamma93.qt} & \hfagFitLabel\\
\htuse{ALEPH.Gamma93.pub.BARATE.98,qt} & \htuse{ALEPH.Gamma93.pub.BARATE.98,exp} & \htuse{ALEPH.Gamma93.pub.BARATE.98,ref} \\
\htuse{BaBar.Gamma93.pub.AUBERT.08,qt} & \htuse{BaBar.Gamma93.pub.AUBERT.08,exp} & \htuse{BaBar.Gamma93.pub.AUBERT.08,ref} \\
\htuse{Belle.Gamma93.pub.LEE.10,qt} & \htuse{Belle.Gamma93.pub.LEE.10,exp} & \htuse{Belle.Gamma93.pub.LEE.10,ref} \\
\htuse{CLEO3.Gamma93.pub.BRIERE.03,qt} & \htuse{CLEO3.Gamma93.pub.BRIERE.03,exp} & \htuse{CLEO3.Gamma93.pub.BRIERE.03,ref}
}%
\htdef{Gamma93by60.qm}{%
\begin{ensuredisplaymath}
\htuse{Gamma93by60.gn} = \htuse{Gamma93by60.td}
\end{ensuredisplaymath}
 & \htuse{Gamma93by60.qt} & \hfagFitLabel\\
\htuse{CLEO.Gamma93by60.pub.RICHICHI.99,qt} & \htuse{CLEO.Gamma93by60.pub.RICHICHI.99,exp} & \htuse{CLEO.Gamma93by60.pub.RICHICHI.99,ref}
}%
\htdef{Gamma94.qm}{%
\begin{ensuredisplaymath}
\htuse{Gamma94.gn} = \htuse{Gamma94.td}
\end{ensuredisplaymath}
 & \htuse{Gamma94.qt} & \hfagFitLabel\\
\htuse{ALEPH.Gamma94.pub.BARATE.98,qt} & \htuse{ALEPH.Gamma94.pub.BARATE.98,exp} & \htuse{ALEPH.Gamma94.pub.BARATE.98,ref} \\
\htuse{CLEO3.Gamma94.pub.ARMS.05,qt} & \htuse{CLEO3.Gamma94.pub.ARMS.05,exp} & \htuse{CLEO3.Gamma94.pub.ARMS.05,ref}
}%
\htdef{Gamma94by69.qm}{%
\begin{ensuredisplaymath}
\htuse{Gamma94by69.gn} = \htuse{Gamma94by69.td}
\end{ensuredisplaymath}
 & \htuse{Gamma94by69.qt} & \hfagFitLabel\\
\htuse{CLEO.Gamma94by69.pub.RICHICHI.99,qt} & \htuse{CLEO.Gamma94by69.pub.RICHICHI.99,exp} & \htuse{CLEO.Gamma94by69.pub.RICHICHI.99,ref}
}%
\htdef{Gamma96.qm}{%
\begin{ensuredisplaymath}
\htuse{Gamma96.gn} = \htuse{Gamma96.td}
\end{ensuredisplaymath}
 & \htuse{Gamma96.qt} & \hfagFitLabel\\
\htuse{BaBar.Gamma96.pub.AUBERT.08,qt} & \htuse{BaBar.Gamma96.pub.AUBERT.08,exp} & \htuse{BaBar.Gamma96.pub.AUBERT.08,ref} \\
\htuse{Belle.Gamma96.pub.LEE.10,qt} & \htuse{Belle.Gamma96.pub.LEE.10,exp} & \htuse{Belle.Gamma96.pub.LEE.10,ref}
}%
\htdef{Gamma102.qm}{%
\begin{ensuredisplaymath}
\htuse{Gamma102.gn} = \htuse{Gamma102.td}
\end{ensuredisplaymath}
 & \htuse{Gamma102.qt} & \hfagFitLabel\\
\htuse{CLEO.Gamma102.pub.GIBAUT.94B,qt} & \htuse{CLEO.Gamma102.pub.GIBAUT.94B,exp} & \htuse{CLEO.Gamma102.pub.GIBAUT.94B,ref} \\
\htuse{HRS.Gamma102.pub.BYLSMA.87,qt} & \htuse{HRS.Gamma102.pub.BYLSMA.87,exp} & \htuse{HRS.Gamma102.pub.BYLSMA.87,ref} \\
\htuse{L3.Gamma102.pub.ACHARD.01D,qt} & \htuse{L3.Gamma102.pub.ACHARD.01D,exp} & \htuse{L3.Gamma102.pub.ACHARD.01D,ref}
}%
\htdef{Gamma103.qm}{%
\begin{ensuredisplaymath}
\htuse{Gamma103.gn} = \htuse{Gamma103.td}
\end{ensuredisplaymath}
 & \htuse{Gamma103.qt} & \hfagFitLabel\\
\htuse{ALEPH.Gamma103.pub.SCHAEL.05C,qt} & \htuse{ALEPH.Gamma103.pub.SCHAEL.05C,exp} & \htuse{ALEPH.Gamma103.pub.SCHAEL.05C,ref} \\
\htuse{ARGUS.Gamma103.pub.ALBRECHT.88B,qt} & \htuse{ARGUS.Gamma103.pub.ALBRECHT.88B,exp} & \htuse{ARGUS.Gamma103.pub.ALBRECHT.88B,ref} \\
\htuse{CLEO.Gamma103.pub.GIBAUT.94B,qt} & \htuse{CLEO.Gamma103.pub.GIBAUT.94B,exp} & \htuse{CLEO.Gamma103.pub.GIBAUT.94B,ref} \\
\htuse{DELPHI.Gamma103.pub.ABDALLAH.06A,qt} & \htuse{DELPHI.Gamma103.pub.ABDALLAH.06A,exp} & \htuse{DELPHI.Gamma103.pub.ABDALLAH.06A,ref} \\
\htuse{HRS.Gamma103.pub.BYLSMA.87,qt} & \htuse{HRS.Gamma103.pub.BYLSMA.87,exp} & \htuse{HRS.Gamma103.pub.BYLSMA.87,ref} \\
\htuse{OPAL.Gamma103.pub.ACKERSTAFF.99E,qt} & \htuse{OPAL.Gamma103.pub.ACKERSTAFF.99E,exp} & \htuse{OPAL.Gamma103.pub.ACKERSTAFF.99E,ref}
}%
\htdef{Gamma104.qm}{%
\begin{ensuredisplaymath}
\htuse{Gamma104.gn} = \htuse{Gamma104.td}
\end{ensuredisplaymath}
 & \htuse{Gamma104.qt} & \hfagFitLabel\\
\htuse{ALEPH.Gamma104.pub.SCHAEL.05C,qt} & \htuse{ALEPH.Gamma104.pub.SCHAEL.05C,exp} & \htuse{ALEPH.Gamma104.pub.SCHAEL.05C,ref} \\
\htuse{CLEO.Gamma104.pub.ANASTASSOV.01,qt} & \htuse{CLEO.Gamma104.pub.ANASTASSOV.01,exp} & \htuse{CLEO.Gamma104.pub.ANASTASSOV.01,ref} \\
\htuse{DELPHI.Gamma104.pub.ABDALLAH.06A,qt} & \htuse{DELPHI.Gamma104.pub.ABDALLAH.06A,exp} & \htuse{DELPHI.Gamma104.pub.ABDALLAH.06A,ref} \\
\htuse{OPAL.Gamma104.pub.ACKERSTAFF.99E,qt} & \htuse{OPAL.Gamma104.pub.ACKERSTAFF.99E,exp} & \htuse{OPAL.Gamma104.pub.ACKERSTAFF.99E,ref}
}%
\htdef{Gamma106.qm}{%
\begin{ensuredisplaymath}
\htuse{Gamma106.gn} = \htuse{Gamma106.td}
\end{ensuredisplaymath}
 & \htuse{Gamma106.qt} & \hfagFitLabel}%
\htdef{Gamma110.qm}{%
\begin{ensuredisplaymath}
\htuse{Gamma110.gn} = \htuse{Gamma110.td}
\end{ensuredisplaymath}
 & \htuse{Gamma110.qt} & \hfagFitLabel}%
\htdef{Gamma126.qm}{%
\begin{ensuredisplaymath}
\htuse{Gamma126.gn} = \htuse{Gamma126.td}
\end{ensuredisplaymath}
 & \htuse{Gamma126.qt} & \hfagFitLabel\\
\htuse{ALEPH.Gamma126.pub.BUSKULIC.97C,qt} & \htuse{ALEPH.Gamma126.pub.BUSKULIC.97C,exp} & \htuse{ALEPH.Gamma126.pub.BUSKULIC.97C,ref} \\
\htuse{Belle.Gamma126.pub.INAMI.09,qt} & \htuse{Belle.Gamma126.pub.INAMI.09,exp} & \htuse{Belle.Gamma126.pub.INAMI.09,ref} \\
\htuse{CLEO.Gamma126.pub.ARTUSO.92,qt} & \htuse{CLEO.Gamma126.pub.ARTUSO.92,exp} & \htuse{CLEO.Gamma126.pub.ARTUSO.92,ref}
}%
\htdef{Gamma128.qm}{%
\begin{ensuredisplaymath}
\htuse{Gamma128.gn} = \htuse{Gamma128.td}
\end{ensuredisplaymath}
 & \htuse{Gamma128.qt} & \hfagFitLabel\\
\htuse{ALEPH.Gamma128.pub.BUSKULIC.97C,qt} & \htuse{ALEPH.Gamma128.pub.BUSKULIC.97C,exp} & \htuse{ALEPH.Gamma128.pub.BUSKULIC.97C,ref} \\
\htuse{BaBar.Gamma128.pub.DEL-AMO-SANCHEZ.11E,qt} & \htuse{BaBar.Gamma128.pub.DEL-AMO-SANCHEZ.11E,exp} & \htuse{BaBar.Gamma128.pub.DEL-AMO-SANCHEZ.11E,ref} \\
\htuse{Belle.Gamma128.pub.INAMI.09,qt} & \htuse{Belle.Gamma128.pub.INAMI.09,exp} & \htuse{Belle.Gamma128.pub.INAMI.09,ref} \\
\htuse{CLEO.Gamma128.pub.BARTELT.96,qt} & \htuse{CLEO.Gamma128.pub.BARTELT.96,exp} & \htuse{CLEO.Gamma128.pub.BARTELT.96,ref}
}%
\htdef{Gamma130.qm}{%
\begin{ensuredisplaymath}
\htuse{Gamma130.gn} = \htuse{Gamma130.td}
\end{ensuredisplaymath}
 & \htuse{Gamma130.qt} & \hfagFitLabel\\
\htuse{Belle.Gamma130.pub.INAMI.09,qt} & \htuse{Belle.Gamma130.pub.INAMI.09,exp} & \htuse{Belle.Gamma130.pub.INAMI.09,ref} \\
\htuse{CLEO.Gamma130.pub.BISHAI.99,qt} & \htuse{CLEO.Gamma130.pub.BISHAI.99,exp} & \htuse{CLEO.Gamma130.pub.BISHAI.99,ref}
}%
\htdef{Gamma132.qm}{%
\begin{ensuredisplaymath}
\htuse{Gamma132.gn} = \htuse{Gamma132.td}
\end{ensuredisplaymath}
 & \htuse{Gamma132.qt} & \hfagFitLabel\\
\htuse{Belle.Gamma132.pub.INAMI.09,qt} & \htuse{Belle.Gamma132.pub.INAMI.09,exp} & \htuse{Belle.Gamma132.pub.INAMI.09,ref} \\
\htuse{CLEO.Gamma132.pub.BISHAI.99,qt} & \htuse{CLEO.Gamma132.pub.BISHAI.99,exp} & \htuse{CLEO.Gamma132.pub.BISHAI.99,ref}
}%
\htdef{Gamma136.qm}{%
\begin{ensuredisplaymath}
\htuse{Gamma136.gn} = \htuse{Gamma136.td}
\end{ensuredisplaymath}
 & \htuse{Gamma136.qt} & \hfagFitLabel}%
\htdef{Gamma149.qm}{%
\begin{ensuredisplaymath}
\htuse{Gamma149.gn} = \htuse{Gamma149.td}
\end{ensuredisplaymath}
 & \htuse{Gamma149.qt} & \hfagFitLabel}%
\htdef{Gamma150.qm}{%
\begin{ensuredisplaymath}
\htuse{Gamma150.gn} = \htuse{Gamma150.td}
\end{ensuredisplaymath}
 & \htuse{Gamma150.qt} & \hfagFitLabel\\
\htuse{ALEPH.Gamma150.pub.BUSKULIC.97C,qt} & \htuse{ALEPH.Gamma150.pub.BUSKULIC.97C,exp} & \htuse{ALEPH.Gamma150.pub.BUSKULIC.97C,ref} \\
\htuse{CLEO.Gamma150.pub.BARINGER.87,qt} & \htuse{CLEO.Gamma150.pub.BARINGER.87,exp} & \htuse{CLEO.Gamma150.pub.BARINGER.87,ref}
}%
\htdef{Gamma150by66.qm}{%
\begin{ensuredisplaymath}
\htuse{Gamma150by66.gn} = \htuse{Gamma150by66.td}
\end{ensuredisplaymath}
 & \htuse{Gamma150by66.qt} & \hfagFitLabel\\
\htuse{ALEPH.Gamma150by66.pub.BUSKULIC.96,qt} & \htuse{ALEPH.Gamma150by66.pub.BUSKULIC.96,exp} & \htuse{ALEPH.Gamma150by66.pub.BUSKULIC.96,ref} \\
\htuse{CLEO.Gamma150by66.pub.BALEST.95C,qt} & \htuse{CLEO.Gamma150by66.pub.BALEST.95C,exp} & \htuse{CLEO.Gamma150by66.pub.BALEST.95C,ref}
}%
\htdef{Gamma151.qm}{%
\begin{ensuredisplaymath}
\htuse{Gamma151.gn} = \htuse{Gamma151.td}
\end{ensuredisplaymath}
 & \htuse{Gamma151.qt} & \hfagFitLabel\\
\htuse{CLEO3.Gamma151.pub.ARMS.05,qt} & \htuse{CLEO3.Gamma151.pub.ARMS.05,exp} & \htuse{CLEO3.Gamma151.pub.ARMS.05,ref}
}%
\htdef{Gamma152.qm}{%
\begin{ensuredisplaymath}
\htuse{Gamma152.gn} = \htuse{Gamma152.td}
\end{ensuredisplaymath}
 & \htuse{Gamma152.qt} & \hfagFitLabel\\
\htuse{ALEPH.Gamma152.pub.BUSKULIC.97C,qt} & \htuse{ALEPH.Gamma152.pub.BUSKULIC.97C,exp} & \htuse{ALEPH.Gamma152.pub.BUSKULIC.97C,ref}
}%
\htdef{Gamma152by54.qm}{%
\begin{ensuredisplaymath}
\htuse{Gamma152by54.gn} = \htuse{Gamma152by54.td}
\end{ensuredisplaymath}
 & \htuse{Gamma152by54.qt} & \hfagFitLabel}%
\htdef{Gamma152by76.qm}{%
\begin{ensuredisplaymath}
\htuse{Gamma152by76.gn} = \htuse{Gamma152by76.td}
\end{ensuredisplaymath}
 & \htuse{Gamma152by76.qt} & \hfagFitLabel\\
\htuse{CLEO.Gamma152by76.pub.BORTOLETTO.93,qt} & \htuse{CLEO.Gamma152by76.pub.BORTOLETTO.93,exp} & \htuse{CLEO.Gamma152by76.pub.BORTOLETTO.93,ref}
}%
\htdef{Gamma167.qm}{%
\begin{ensuredisplaymath}
\htuse{Gamma167.gn} = \htuse{Gamma167.td}
\end{ensuredisplaymath}
 & \htuse{Gamma167.qt} & \hfagFitLabel}%
\htdef{Gamma168.qm}{%
\begin{ensuredisplaymath}
\htuse{Gamma168.gn} = \htuse{Gamma168.td}
\end{ensuredisplaymath}
 & \htuse{Gamma168.qt} & \hfagFitLabel}%
\htdef{Gamma169.qm}{%
\begin{ensuredisplaymath}
\htuse{Gamma169.gn} = \htuse{Gamma169.td}
\end{ensuredisplaymath}
 & \htuse{Gamma169.qt} & \hfagFitLabel}%
\htdef{Gamma800.qm}{%
\begin{ensuredisplaymath}
\htuse{Gamma800.gn} = \htuse{Gamma800.td}
\end{ensuredisplaymath}
 & \htuse{Gamma800.qt} & \hfagFitLabel}%
\htdef{Gamma802.qm}{%
\begin{ensuredisplaymath}
\htuse{Gamma802.gn} = \htuse{Gamma802.td}
\end{ensuredisplaymath}
 & \htuse{Gamma802.qt} & \hfagFitLabel}%
\htdef{Gamma803.qm}{%
\begin{ensuredisplaymath}
\htuse{Gamma803.gn} = \htuse{Gamma803.td}
\end{ensuredisplaymath}
 & \htuse{Gamma803.qt} & \hfagFitLabel}%
\htdef{Gamma804.qm}{%
\begin{ensuredisplaymath}
\htuse{Gamma804.gn} = \htuse{Gamma804.td}
\end{ensuredisplaymath}
 & \htuse{Gamma804.qt} & \hfagFitLabel}%
\htdef{Gamma805.qm}{%
\begin{ensuredisplaymath}
\htuse{Gamma805.gn} = \htuse{Gamma805.td}
\end{ensuredisplaymath}
 & \htuse{Gamma805.qt} & \hfagFitLabel\\
\htuse{ALEPH.Gamma805.pub.SCHAEL.05C,qt} & \htuse{ALEPH.Gamma805.pub.SCHAEL.05C,exp} & \htuse{ALEPH.Gamma805.pub.SCHAEL.05C,ref}
}%
\htdef{Gamma806.qm}{%
\begin{ensuredisplaymath}
\htuse{Gamma806.gn} = \htuse{Gamma806.td}
\end{ensuredisplaymath}
 & \htuse{Gamma806.qt} & \hfagFitLabel}%
\htdef{Gamma810.qm}{%
\begin{ensuredisplaymath}
\htuse{Gamma810.gn} = \htuse{Gamma810.td}
\end{ensuredisplaymath}
 & \htuse{Gamma810.qt} & \hfagFitLabel}%
\htdef{Gamma811.qm}{%
\begin{ensuredisplaymath}
\htuse{Gamma811.gn} = \htuse{Gamma811.td}
\end{ensuredisplaymath}
 & \htuse{Gamma811.qt} & \hfagFitLabel\\
\htuse{BaBar.Gamma811.pub.LEES.12X,qt} & \htuse{BaBar.Gamma811.pub.LEES.12X,exp} & \htuse{BaBar.Gamma811.pub.LEES.12X,ref}
}%
\htdef{Gamma812.qm}{%
\begin{ensuredisplaymath}
\htuse{Gamma812.gn} = \htuse{Gamma812.td}
\end{ensuredisplaymath}
 & \htuse{Gamma812.qt} & \hfagFitLabel\\
\htuse{BaBar.Gamma812.pub.LEES.12X,qt} & \htuse{BaBar.Gamma812.pub.LEES.12X,exp} & \htuse{BaBar.Gamma812.pub.LEES.12X,ref}
}%
\htdef{Gamma820.qm}{%
\begin{ensuredisplaymath}
\htuse{Gamma820.gn} = \htuse{Gamma820.td}
\end{ensuredisplaymath}
 & \htuse{Gamma820.qt} & \hfagFitLabel}%
\htdef{Gamma821.qm}{%
\begin{ensuredisplaymath}
\htuse{Gamma821.gn} = \htuse{Gamma821.td}
\end{ensuredisplaymath}
 & \htuse{Gamma821.qt} & \hfagFitLabel\\
\htuse{BaBar.Gamma821.pub.LEES.12X,qt} & \htuse{BaBar.Gamma821.pub.LEES.12X,exp} & \htuse{BaBar.Gamma821.pub.LEES.12X,ref}
}%
\htdef{Gamma822.qm}{%
\begin{ensuredisplaymath}
\htuse{Gamma822.gn} = \htuse{Gamma822.td}
\end{ensuredisplaymath}
 & \htuse{Gamma822.qt} & \hfagFitLabel\\
\htuse{BaBar.Gamma822.pub.LEES.12X,qt} & \htuse{BaBar.Gamma822.pub.LEES.12X,exp} & \htuse{BaBar.Gamma822.pub.LEES.12X,ref}
}%
\htdef{Gamma830.qm}{%
\begin{ensuredisplaymath}
\htuse{Gamma830.gn} = \htuse{Gamma830.td}
\end{ensuredisplaymath}
 & \htuse{Gamma830.qt} & \hfagFitLabel}%
\htdef{Gamma831.qm}{%
\begin{ensuredisplaymath}
\htuse{Gamma831.gn} = \htuse{Gamma831.td}
\end{ensuredisplaymath}
 & \htuse{Gamma831.qt} & \hfagFitLabel\\
\htuse{BaBar.Gamma831.pub.LEES.12X,qt} & \htuse{BaBar.Gamma831.pub.LEES.12X,exp} & \htuse{BaBar.Gamma831.pub.LEES.12X,ref}
}%
\htdef{Gamma832.qm}{%
\begin{ensuredisplaymath}
\htuse{Gamma832.gn} = \htuse{Gamma832.td}
\end{ensuredisplaymath}
 & \htuse{Gamma832.qt} & \hfagFitLabel\\
\htuse{BaBar.Gamma832.pub.LEES.12X,qt} & \htuse{BaBar.Gamma832.pub.LEES.12X,exp} & \htuse{BaBar.Gamma832.pub.LEES.12X,ref}
}%
\htdef{Gamma833.qm}{%
\begin{ensuredisplaymath}
\htuse{Gamma833.gn} = \htuse{Gamma833.td}
\end{ensuredisplaymath}
 & \htuse{Gamma833.qt} & \hfagFitLabel\\
\htuse{BaBar.Gamma833.pub.LEES.12X,qt} & \htuse{BaBar.Gamma833.pub.LEES.12X,exp} & \htuse{BaBar.Gamma833.pub.LEES.12X,ref}
}%
\htdef{Gamma910.qm}{%
\begin{ensuredisplaymath}
\htuse{Gamma910.gn} = \htuse{Gamma910.td}
\end{ensuredisplaymath}
 & \htuse{Gamma910.qt} & \hfagFitLabel\\
\htuse{BaBar.Gamma910.pub.LEES.12X,qt} & \htuse{BaBar.Gamma910.pub.LEES.12X,exp} & \htuse{BaBar.Gamma910.pub.LEES.12X,ref}
}%
\htdef{Gamma911.qm}{%
\begin{ensuredisplaymath}
\htuse{Gamma911.gn} = \htuse{Gamma911.td}
\end{ensuredisplaymath}
 & \htuse{Gamma911.qt} & \hfagFitLabel\\
\htuse{BaBar.Gamma911.pub.LEES.12X,qt} & \htuse{BaBar.Gamma911.pub.LEES.12X,exp} & \htuse{BaBar.Gamma911.pub.LEES.12X,ref}
}%
\htdef{Gamma920.qm}{%
\begin{ensuredisplaymath}
\htuse{Gamma920.gn} = \htuse{Gamma920.td}
\end{ensuredisplaymath}
 & \htuse{Gamma920.qt} & \hfagFitLabel\\
\htuse{BaBar.Gamma920.pub.LEES.12X,qt} & \htuse{BaBar.Gamma920.pub.LEES.12X,exp} & \htuse{BaBar.Gamma920.pub.LEES.12X,ref}
}%
\htdef{Gamma930.qm}{%
\begin{ensuredisplaymath}
\htuse{Gamma930.gn} = \htuse{Gamma930.td}
\end{ensuredisplaymath}
 & \htuse{Gamma930.qt} & \hfagFitLabel\\
\htuse{BaBar.Gamma930.pub.LEES.12X,qt} & \htuse{BaBar.Gamma930.pub.LEES.12X,exp} & \htuse{BaBar.Gamma930.pub.LEES.12X,ref}
}%
\htdef{Gamma944.qm}{%
\begin{ensuredisplaymath}
\htuse{Gamma944.gn} = \htuse{Gamma944.td}
\end{ensuredisplaymath}
 & \htuse{Gamma944.qt} & \hfagFitLabel\\
\htuse{BaBar.Gamma944.pub.LEES.12X,qt} & \htuse{BaBar.Gamma944.pub.LEES.12X,exp} & \htuse{BaBar.Gamma944.pub.LEES.12X,ref}
}%
\htdef{Gamma945.qm}{%
\begin{ensuredisplaymath}
\htuse{Gamma945.gn} = \htuse{Gamma945.td}
\end{ensuredisplaymath}
 & \htuse{Gamma945.qt} & \hfagFitLabel}%
\htdef{Gamma998.qm}{%
\begin{ensuredisplaymath}
\htuse{Gamma998.gn} = \htuse{Gamma998.td}
\end{ensuredisplaymath}
 & \htuse{Gamma998.qt} & \hfagFitLabel}%
\htdef{BrVal}{%
\htuse{Gamma1.qm} \\
\midrule
\htuse{Gamma2.qm} \\
\midrule
\htuse{Gamma3.qm} \\
\midrule
\htuse{Gamma3by5.qm} \\
\midrule
\htuse{Gamma5.qm} \\
\midrule
\htuse{Gamma7.qm} \\
\midrule
\htuse{Gamma8.qm} \\
\midrule
\htuse{Gamma8by5.qm} \\
\midrule
\htuse{Gamma9.qm} \\
\midrule
\htuse{Gamma9by5.qm} \\
\midrule
\htuse{Gamma10.qm} \\
\midrule
\htuse{Gamma10by5.qm} \\
\midrule
\htuse{Gamma10by9.qm} \\
\midrule
\htuse{Gamma11.qm} \\
\midrule
\htuse{Gamma12.qm} \\
\midrule
\htuse{Gamma13.qm} \\
\midrule
\htuse{Gamma14.qm} \\
\midrule
\htuse{Gamma16.qm} \\
\midrule
\htuse{Gamma17.qm} \\
\midrule
\htuse{Gamma18.qm} \\
\midrule
\htuse{Gamma19.qm} \\
\midrule
\htuse{Gamma19by13.qm} \\
\midrule
\htuse{Gamma20.qm} \\
\midrule
\htuse{Gamma23.qm} \\
\midrule
\htuse{Gamma24.qm} \\
\midrule
\htuse{Gamma25.qm} \\
\midrule
\htuse{Gamma26.qm} \\
\midrule
\htuse{Gamma26by13.qm} \\
\midrule
\htuse{Gamma27.qm} \\
\midrule
\htuse{Gamma28.qm} \\
\midrule
\htuse{Gamma29.qm} \\
\midrule
\htuse{Gamma30.qm} \\
\midrule
\htuse{Gamma31.qm} \\
\midrule
\htuse{Gamma32.qm} \\
\midrule
\htuse{Gamma33.qm} \\
\midrule
\htuse{Gamma34.qm} \\
\midrule
\htuse{Gamma35.qm} \\
\midrule
\htuse{Gamma37.qm} \\
\midrule
\htuse{Gamma38.qm} \\
\midrule
\htuse{Gamma39.qm} \\
\midrule
\htuse{Gamma40.qm} \\
\midrule
\htuse{Gamma42.qm} \\
\midrule
\htuse{Gamma43.qm} \\
\midrule
\htuse{Gamma44.qm} \\
\midrule
\htuse{Gamma46.qm} \\
\midrule
\htuse{Gamma47.qm} \\
\midrule
\htuse{Gamma48.qm} \\
\midrule
\htuse{Gamma49.qm} \\
\midrule
\htuse{Gamma50.qm} \\
\midrule
\htuse{Gamma51.qm} \\
\midrule
\htuse{Gamma53.qm} \\
\midrule
\htuse{Gamma54.qm} \\
\midrule
\htuse{Gamma55.qm} \\
\midrule
\htuse{Gamma56.qm} \\
\midrule
\htuse{Gamma57.qm} \\
\midrule
\htuse{Gamma57by55.qm} \\
\midrule
\htuse{Gamma58.qm} \\
\midrule
\htuse{Gamma59.qm} \\
\midrule
\htuse{Gamma60.qm} \\
\midrule
\htuse{Gamma62.qm} \\
\midrule
\htuse{Gamma63.qm} \\
\midrule
\htuse{Gamma64.qm} \\
\midrule
\htuse{Gamma65.qm} \\
\midrule
\htuse{Gamma66.qm} \\
\midrule
\htuse{Gamma67.qm} \\
\midrule
\htuse{Gamma68.qm} \\
\midrule
\htuse{Gamma69.qm} \\
\midrule
\htuse{Gamma70.qm} \\
\midrule
\htuse{Gamma74.qm} \\
\midrule
\htuse{Gamma75.qm} \\
\midrule
\htuse{Gamma76.qm} \\
\midrule
\htuse{Gamma76by54.qm} \\
\midrule
\htuse{Gamma77.qm} \\
\midrule
\htuse{Gamma78.qm} \\
\midrule
\htuse{Gamma79.qm} \\
\midrule
\htuse{Gamma80.qm} \\
\midrule
\htuse{Gamma80by60.qm} \\
\midrule
\htuse{Gamma81.qm} \\
\midrule
\htuse{Gamma81by69.qm} \\
\midrule
\htuse{Gamma82.qm} \\
\midrule
\htuse{Gamma83.qm} \\
\midrule
\htuse{Gamma84.qm} \\
\midrule
\htuse{Gamma85.qm} \\
\midrule
\htuse{Gamma85by60.qm} \\
\midrule
\htuse{Gamma87.qm} \\
\midrule
\htuse{Gamma88.qm} \\
\midrule
\htuse{Gamma89.qm} \\
\midrule
\htuse{Gamma92.qm} \\
\midrule
\htuse{Gamma93.qm} \\
\midrule
\htuse{Gamma93by60.qm} \\
\midrule
\htuse{Gamma94.qm} \\
\midrule
\htuse{Gamma94by69.qm} \\
\midrule
\htuse{Gamma96.qm} \\
\midrule
\htuse{Gamma102.qm} \\
\midrule
\htuse{Gamma103.qm} \\
\midrule
\htuse{Gamma104.qm} \\
\midrule
\htuse{Gamma106.qm} \\
\midrule
\htuse{Gamma110.qm} \\
\midrule
\htuse{Gamma126.qm} \\
\midrule
\htuse{Gamma128.qm} \\
\midrule
\htuse{Gamma130.qm} \\
\midrule
\htuse{Gamma132.qm} \\
\midrule
\htuse{Gamma136.qm} \\
\midrule
\htuse{Gamma149.qm} \\
\midrule
\htuse{Gamma150.qm} \\
\midrule
\htuse{Gamma150by66.qm} \\
\midrule
\htuse{Gamma151.qm} \\
\midrule
\htuse{Gamma152.qm} \\
\midrule
\htuse{Gamma152by54.qm} \\
\midrule
\htuse{Gamma152by76.qm} \\
\midrule
\htuse{Gamma167.qm} \\
\midrule
\htuse{Gamma168.qm} \\
\midrule
\htuse{Gamma169.qm} \\
\midrule
\htuse{Gamma800.qm} \\
\midrule
\htuse{Gamma802.qm} \\
\midrule
\htuse{Gamma803.qm} \\
\midrule
\htuse{Gamma804.qm} \\
\midrule
\htuse{Gamma805.qm} \\
\midrule
\htuse{Gamma806.qm} \\
\midrule
\htuse{Gamma810.qm} \\
\midrule
\htuse{Gamma811.qm} \\
\midrule
\htuse{Gamma812.qm} \\
\midrule
\htuse{Gamma820.qm} \\
\midrule
\htuse{Gamma821.qm} \\
\midrule
\htuse{Gamma822.qm} \\
\midrule
\htuse{Gamma830.qm} \\
\midrule
\htuse{Gamma831.qm} \\
\midrule
\htuse{Gamma832.qm} \\
\midrule
\htuse{Gamma833.qm} \\
\midrule
\htuse{Gamma910.qm} \\
\midrule
\htuse{Gamma911.qm} \\
\midrule
\htuse{Gamma920.qm} \\
\midrule
\htuse{Gamma930.qm} \\
\midrule
\htuse{Gamma944.qm} \\
\midrule
\htuse{Gamma945.qm} \\
\midrule
\htuse{Gamma998.qm}}%
\htdef{BARATE 98.cite}{\cite{Barate:1997ma}}%
\htdef{BARATE 98.collab}{ALEPH}%
\htdef{BARATE 98.ref}{BARATE 98 (ALEPH) \cite{Barate:1997ma}}%
\htdef{BARATE 98.meas}{%
\begin{ensuredisplaymath}
\htuse{Gamma85.gn} = \htuse{Gamma85.td}
\end{ensuredisplaymath} & \htuse{ALEPH.Gamma85.pub.BARATE.98}
\\
\begin{ensuredisplaymath}
\htuse{Gamma88.gn} = \htuse{Gamma88.td}
\end{ensuredisplaymath} & \htuse{ALEPH.Gamma88.pub.BARATE.98}
\\
\begin{ensuredisplaymath}
\htuse{Gamma93.gn} = \htuse{Gamma93.td}
\end{ensuredisplaymath} & \htuse{ALEPH.Gamma93.pub.BARATE.98}
\\
\begin{ensuredisplaymath}
\htuse{Gamma94.gn} = \htuse{Gamma94.td}
\end{ensuredisplaymath} & \htuse{ALEPH.Gamma94.pub.BARATE.98}}%
\htdef{BARATE 98E.cite}{\cite{Barate:1997tt}}%
\htdef{BARATE 98E.collab}{ALEPH}%
\htdef{BARATE 98E.ref}{BARATE 98E (ALEPH) \cite{Barate:1997tt}}%
\htdef{BARATE 98E.meas}{%
\begin{ensuredisplaymath}
\htuse{Gamma33.gn} = \htuse{Gamma33.td}
\end{ensuredisplaymath} & \htuse{ALEPH.Gamma33.pub.BARATE.98E}
\\
\begin{ensuredisplaymath}
\htuse{Gamma37.gn} = \htuse{Gamma37.td}
\end{ensuredisplaymath} & \htuse{ALEPH.Gamma37.pub.BARATE.98E}
\\
\begin{ensuredisplaymath}
\htuse{Gamma40.gn} = \htuse{Gamma40.td}
\end{ensuredisplaymath} & \htuse{ALEPH.Gamma40.pub.BARATE.98E}
\\
\begin{ensuredisplaymath}
\htuse{Gamma42.gn} = \htuse{Gamma42.td}
\end{ensuredisplaymath} & \htuse{ALEPH.Gamma42.pub.BARATE.98E}
\\
\begin{ensuredisplaymath}
\htuse{Gamma47.gn} = \htuse{Gamma47.td}
\end{ensuredisplaymath} & \htuse{ALEPH.Gamma47.pub.BARATE.98E}
\\
\begin{ensuredisplaymath}
\htuse{Gamma48.gn} = \htuse{Gamma48.td}
\end{ensuredisplaymath} & \htuse{ALEPH.Gamma48.pub.BARATE.98E}
\\
\begin{ensuredisplaymath}
\htuse{Gamma51.gn} = \htuse{Gamma51.td}
\end{ensuredisplaymath} & \htuse{ALEPH.Gamma51.pub.BARATE.98E}
\\
\begin{ensuredisplaymath}
\htuse{Gamma53.gn} = \htuse{Gamma53.td}
\end{ensuredisplaymath} & \htuse{ALEPH.Gamma53.pub.BARATE.98E}}%
\htdef{BARATE 99K.cite}{\cite{Barate:1999hi}}%
\htdef{BARATE 99K.collab}{ALEPH}%
\htdef{BARATE 99K.ref}{BARATE 99K (ALEPH) \cite{Barate:1999hi}}%
\htdef{BARATE 99K.meas}{%
\begin{ensuredisplaymath}
\htuse{Gamma10.gn} = \htuse{Gamma10.td}
\end{ensuredisplaymath} & \htuse{ALEPH.Gamma10.pub.BARATE.99K}
\\
\begin{ensuredisplaymath}
\htuse{Gamma16.gn} = \htuse{Gamma16.td}
\end{ensuredisplaymath} & \htuse{ALEPH.Gamma16.pub.BARATE.99K}
\\
\begin{ensuredisplaymath}
\htuse{Gamma23.gn} = \htuse{Gamma23.td}
\end{ensuredisplaymath} & \htuse{ALEPH.Gamma23.pub.BARATE.99K}
\\
\begin{ensuredisplaymath}
\htuse{Gamma28.gn} = \htuse{Gamma28.td}
\end{ensuredisplaymath} & \htuse{ALEPH.Gamma28.pub.BARATE.99K}
\\
\begin{ensuredisplaymath}
\htuse{Gamma35.gn} = \htuse{Gamma35.td}
\end{ensuredisplaymath} & \htuse{ALEPH.Gamma35.pub.BARATE.99K}
\\
\begin{ensuredisplaymath}
\htuse{Gamma37.gn} = \htuse{Gamma37.td}
\end{ensuredisplaymath} & \htuse{ALEPH.Gamma37.pub.BARATE.99K}
\\
\begin{ensuredisplaymath}
\htuse{Gamma40.gn} = \htuse{Gamma40.td}
\end{ensuredisplaymath} & \htuse{ALEPH.Gamma40.pub.BARATE.99K}
\\
\begin{ensuredisplaymath}
\htuse{Gamma42.gn} = \htuse{Gamma42.td}
\end{ensuredisplaymath} & \htuse{ALEPH.Gamma42.pub.BARATE.99K}}%
\htdef{BARATE 99R.cite}{\cite{Barate:1999hj}}%
\htdef{BARATE 99R.collab}{ALEPH}%
\htdef{BARATE 99R.ref}{BARATE 99R (ALEPH) \cite{Barate:1999hj}}%
\htdef{BARATE 99R.meas}{%
\begin{ensuredisplaymath}
\htuse{Gamma44.gn} = \htuse{Gamma44.td}
\end{ensuredisplaymath} & \htuse{ALEPH.Gamma44.pub.BARATE.99R}}%
\htdef{BUSKULIC 96.cite}{\cite{Buskulic:1995ty}}%
\htdef{BUSKULIC 96.collab}{ALEPH}%
\htdef{BUSKULIC 96.ref}{BUSKULIC 96 (ALEPH) \cite{Buskulic:1995ty}}%
\htdef{BUSKULIC 96.meas}{%
\begin{ensuredisplaymath}
\htuse{Gamma150by66.gn} = \htuse{Gamma150by66.td}
\end{ensuredisplaymath} & \htuse{ALEPH.Gamma150by66.pub.BUSKULIC.96}}%
\htdef{BUSKULIC 97C.cite}{\cite{Buskulic:1996qs}}%
\htdef{BUSKULIC 97C.collab}{ALEPH}%
\htdef{BUSKULIC 97C.ref}{BUSKULIC 97C (ALEPH) \cite{Buskulic:1996qs}}%
\htdef{BUSKULIC 97C.meas}{%
\begin{ensuredisplaymath}
\htuse{Gamma126.gn} = \htuse{Gamma126.td}
\end{ensuredisplaymath} & \htuse{ALEPH.Gamma126.pub.BUSKULIC.97C}
\\
\begin{ensuredisplaymath}
\htuse{Gamma128.gn} = \htuse{Gamma128.td}
\end{ensuredisplaymath} & \htuse{ALEPH.Gamma128.pub.BUSKULIC.97C}
\\
\begin{ensuredisplaymath}
\htuse{Gamma150.gn} = \htuse{Gamma150.td}
\end{ensuredisplaymath} & \htuse{ALEPH.Gamma150.pub.BUSKULIC.97C}
\\
\begin{ensuredisplaymath}
\htuse{Gamma152.gn} = \htuse{Gamma152.td}
\end{ensuredisplaymath} & \htuse{ALEPH.Gamma152.pub.BUSKULIC.97C}}%
\htdef{SCHAEL 05C.cite}{\cite{Schael:2005am}}%
\htdef{SCHAEL 05C.collab}{ALEPH}%
\htdef{SCHAEL 05C.ref}{SCHAEL 05C (ALEPH) \cite{Schael:2005am}}%
\htdef{SCHAEL 05C.meas}{%
\begin{ensuredisplaymath}
\htuse{Gamma3.gn} = \htuse{Gamma3.td}
\end{ensuredisplaymath} & \htuse{ALEPH.Gamma3.pub.SCHAEL.05C}
\\
\begin{ensuredisplaymath}
\htuse{Gamma5.gn} = \htuse{Gamma5.td}
\end{ensuredisplaymath} & \htuse{ALEPH.Gamma5.pub.SCHAEL.05C}
\\
\begin{ensuredisplaymath}
\htuse{Gamma8.gn} = \htuse{Gamma8.td}
\end{ensuredisplaymath} & \htuse{ALEPH.Gamma8.pub.SCHAEL.05C}
\\
\begin{ensuredisplaymath}
\htuse{Gamma13.gn} = \htuse{Gamma13.td}
\end{ensuredisplaymath} & \htuse{ALEPH.Gamma13.pub.SCHAEL.05C}
\\
\begin{ensuredisplaymath}
\htuse{Gamma19.gn} = \htuse{Gamma19.td}
\end{ensuredisplaymath} & \htuse{ALEPH.Gamma19.pub.SCHAEL.05C}
\\
\begin{ensuredisplaymath}
\htuse{Gamma26.gn} = \htuse{Gamma26.td}
\end{ensuredisplaymath} & \htuse{ALEPH.Gamma26.pub.SCHAEL.05C}
\\
\begin{ensuredisplaymath}
\htuse{Gamma30.gn} = \htuse{Gamma30.td}
\end{ensuredisplaymath} & \htuse{ALEPH.Gamma30.pub.SCHAEL.05C}
\\
\begin{ensuredisplaymath}
\htuse{Gamma58.gn} = \htuse{Gamma58.td}
\end{ensuredisplaymath} & \htuse{ALEPH.Gamma58.pub.SCHAEL.05C}
\\
\begin{ensuredisplaymath}
\htuse{Gamma66.gn} = \htuse{Gamma66.td}
\end{ensuredisplaymath} & \htuse{ALEPH.Gamma66.pub.SCHAEL.05C}
\\
\begin{ensuredisplaymath}
\htuse{Gamma76.gn} = \htuse{Gamma76.td}
\end{ensuredisplaymath} & \htuse{ALEPH.Gamma76.pub.SCHAEL.05C}
\\
\begin{ensuredisplaymath}
\htuse{Gamma103.gn} = \htuse{Gamma103.td}
\end{ensuredisplaymath} & \htuse{ALEPH.Gamma103.pub.SCHAEL.05C}
\\
\begin{ensuredisplaymath}
\htuse{Gamma104.gn} = \htuse{Gamma104.td}
\end{ensuredisplaymath} & \htuse{ALEPH.Gamma104.pub.SCHAEL.05C}
\\
\begin{ensuredisplaymath}
\htuse{Gamma805.gn} = \htuse{Gamma805.td}
\end{ensuredisplaymath} & \htuse{ALEPH.Gamma805.pub.SCHAEL.05C}}%
\htdef{ALBRECHT 88B.cite}{\cite{Albrecht:1987zf}}%
\htdef{ALBRECHT 88B.collab}{ARGUS}%
\htdef{ALBRECHT 88B.ref}{ALBRECHT 88B (ARGUS) \cite{Albrecht:1987zf}}%
\htdef{ALBRECHT 88B.meas}{%
\begin{ensuredisplaymath}
\htuse{Gamma103.gn} = \htuse{Gamma103.td}
\end{ensuredisplaymath} & \htuse{ARGUS.Gamma103.pub.ALBRECHT.88B}}%
\htdef{ALBRECHT 92D.cite}{\cite{Albrecht:1991rh}}%
\htdef{ALBRECHT 92D.collab}{ARGUS}%
\htdef{ALBRECHT 92D.ref}{ALBRECHT 92D (ARGUS) \cite{Albrecht:1991rh}}%
\htdef{ALBRECHT 92D.meas}{%
\begin{ensuredisplaymath}
\htuse{Gamma3by5.gn} = \htuse{Gamma3by5.td}
\end{ensuredisplaymath} & \htuse{ARGUS.Gamma3by5.pub.ALBRECHT.92D}}%
\htdef{AUBERT 07AP.cite}{\cite{Aubert:2007jh}}%
\htdef{AUBERT 07AP.collab}{\babar}%
\htdef{AUBERT 07AP.ref}{AUBERT 07AP (\babar) \cite{Aubert:2007jh}}%
\htdef{AUBERT 07AP.meas}{%
\begin{ensuredisplaymath}
\htuse{Gamma16.gn} = \htuse{Gamma16.td}
\end{ensuredisplaymath} & \htuse{BaBar.Gamma16.pub.AUBERT.07AP}}%
\htdef{AUBERT 08.cite}{\cite{Aubert:2007mh}}%
\htdef{AUBERT 08.collab}{\babar}%
\htdef{AUBERT 08.ref}{AUBERT 08 (\babar) \cite{Aubert:2007mh}}%
\htdef{AUBERT 08.meas}{%
\begin{ensuredisplaymath}
\htuse{Gamma60.gn} = \htuse{Gamma60.td}
\end{ensuredisplaymath} & \htuse{BaBar.Gamma60.pub.AUBERT.08}
\\
\begin{ensuredisplaymath}
\htuse{Gamma85.gn} = \htuse{Gamma85.td}
\end{ensuredisplaymath} & \htuse{BaBar.Gamma85.pub.AUBERT.08}
\\
\begin{ensuredisplaymath}
\htuse{Gamma93.gn} = \htuse{Gamma93.td}
\end{ensuredisplaymath} & \htuse{BaBar.Gamma93.pub.AUBERT.08}
\\
\begin{ensuredisplaymath}
\htuse{Gamma96.gn} = \htuse{Gamma96.td}
\end{ensuredisplaymath} & \htuse{BaBar.Gamma96.pub.AUBERT.08}}%
\htdef{AUBERT 10F.cite}{\cite{Aubert:2009qj}}%
\htdef{AUBERT 10F.collab}{\babar}%
\htdef{AUBERT 10F.ref}{AUBERT 10F (\babar) \cite{Aubert:2009qj}}%
\htdef{AUBERT 10F.meas}{%
\begin{ensuredisplaymath}
\htuse{Gamma3by5.gn} = \htuse{Gamma3by5.td}
\end{ensuredisplaymath} & \htuse{BaBar.Gamma3by5.pub.AUBERT.10F}
\\
\begin{ensuredisplaymath}
\htuse{Gamma9by5.gn} = \htuse{Gamma9by5.td}
\end{ensuredisplaymath} & \htuse{BaBar.Gamma9by5.pub.AUBERT.10F}
\\
\begin{ensuredisplaymath}
\htuse{Gamma10by5.gn} = \htuse{Gamma10by5.td}
\end{ensuredisplaymath} & \htuse{BaBar.Gamma10by5.pub.AUBERT.10F}}%
\htdef{DEL-AMO-SANCHEZ 11E.cite}{\cite{delAmoSanchez:2010pc}}%
\htdef{DEL-AMO-SANCHEZ 11E.collab}{\babar}%
\htdef{DEL-AMO-SANCHEZ 11E.ref}{DEL-AMO-SANCHEZ 11E (\babar) \cite{delAmoSanchez:2010pc}}%
\htdef{DEL-AMO-SANCHEZ 11E.meas}{%
\begin{ensuredisplaymath}
\htuse{Gamma128.gn} = \htuse{Gamma128.td}
\end{ensuredisplaymath} & \htuse{BaBar.Gamma128.pub.DEL-AMO-SANCHEZ.11E}}%
\htdef{LEES 12X.cite}{\cite{Lees:2012ks}}%
\htdef{LEES 12X.collab}{\babar}%
\htdef{LEES 12X.ref}{LEES 12X (\babar) \cite{Lees:2012ks}}%
\htdef{LEES 12X.meas}{%
\begin{ensuredisplaymath}
\htuse{Gamma811.gn} = \htuse{Gamma811.td}
\end{ensuredisplaymath} & \htuse{BaBar.Gamma811.pub.LEES.12X}
\\
\begin{ensuredisplaymath}
\htuse{Gamma812.gn} = \htuse{Gamma812.td}
\end{ensuredisplaymath} & \htuse{BaBar.Gamma812.pub.LEES.12X}
\\
\begin{ensuredisplaymath}
\htuse{Gamma821.gn} = \htuse{Gamma821.td}
\end{ensuredisplaymath} & \htuse{BaBar.Gamma821.pub.LEES.12X}
\\
\begin{ensuredisplaymath}
\htuse{Gamma822.gn} = \htuse{Gamma822.td}
\end{ensuredisplaymath} & \htuse{BaBar.Gamma822.pub.LEES.12X}
\\
\begin{ensuredisplaymath}
\htuse{Gamma831.gn} = \htuse{Gamma831.td}
\end{ensuredisplaymath} & \htuse{BaBar.Gamma831.pub.LEES.12X}
\\
\begin{ensuredisplaymath}
\htuse{Gamma832.gn} = \htuse{Gamma832.td}
\end{ensuredisplaymath} & \htuse{BaBar.Gamma832.pub.LEES.12X}
\\
\begin{ensuredisplaymath}
\htuse{Gamma833.gn} = \htuse{Gamma833.td}
\end{ensuredisplaymath} & \htuse{BaBar.Gamma833.pub.LEES.12X}
\\
\begin{ensuredisplaymath}
\htuse{Gamma910.gn} = \htuse{Gamma910.td}
\end{ensuredisplaymath} & \htuse{BaBar.Gamma910.pub.LEES.12X}
\\
\begin{ensuredisplaymath}
\htuse{Gamma911.gn} = \htuse{Gamma911.td}
\end{ensuredisplaymath} & \htuse{BaBar.Gamma911.pub.LEES.12X}
\\
\begin{ensuredisplaymath}
\htuse{Gamma920.gn} = \htuse{Gamma920.td}
\end{ensuredisplaymath} & \htuse{BaBar.Gamma920.pub.LEES.12X}
\\
\begin{ensuredisplaymath}
\htuse{Gamma930.gn} = \htuse{Gamma930.td}
\end{ensuredisplaymath} & \htuse{BaBar.Gamma930.pub.LEES.12X}
\\
\begin{ensuredisplaymath}
\htuse{Gamma944.gn} = \htuse{Gamma944.td}
\end{ensuredisplaymath} & \htuse{BaBar.Gamma944.pub.LEES.12X}}%
\htdef{LEES 12Y.cite}{\cite{Lees:2012de}}%
\htdef{LEES 12Y.collab}{\babar}%
\htdef{LEES 12Y.ref}{LEES 12Y (\babar) \cite{Lees:2012de}}%
\htdef{LEES 12Y.meas}{%
\begin{ensuredisplaymath}
\htuse{Gamma47.gn} = \htuse{Gamma47.td}
\end{ensuredisplaymath} & \htuse{BaBar.Gamma47.pub.LEES.12Y}
\\
\begin{ensuredisplaymath}
\htuse{Gamma50.gn} = \htuse{Gamma50.td}
\end{ensuredisplaymath} & \htuse{BaBar.Gamma50.pub.LEES.12Y}}%
\htdef{FUJIKAWA 08.cite}{\cite{Fujikawa:2008ma}}%
\htdef{FUJIKAWA 08.collab}{Belle}%
\htdef{FUJIKAWA 08.ref}{FUJIKAWA 08 (Belle) \cite{Fujikawa:2008ma}}%
\htdef{FUJIKAWA 08.meas}{%
\begin{ensuredisplaymath}
\htuse{Gamma13.gn} = \htuse{Gamma13.td}
\end{ensuredisplaymath} & \htuse{Belle.Gamma13.pub.FUJIKAWA.08}}%
\htdef{INAMI 09.cite}{\cite{Inami:2008ar}}%
\htdef{INAMI 09.collab}{Belle}%
\htdef{INAMI 09.ref}{INAMI 09 (Belle) \cite{Inami:2008ar}}%
\htdef{INAMI 09.meas}{%
\begin{ensuredisplaymath}
\htuse{Gamma126.gn} = \htuse{Gamma126.td}
\end{ensuredisplaymath} & \htuse{Belle.Gamma126.pub.INAMI.09}
\\
\begin{ensuredisplaymath}
\htuse{Gamma128.gn} = \htuse{Gamma128.td}
\end{ensuredisplaymath} & \htuse{Belle.Gamma128.pub.INAMI.09}
\\
\begin{ensuredisplaymath}
\htuse{Gamma130.gn} = \htuse{Gamma130.td}
\end{ensuredisplaymath} & \htuse{Belle.Gamma130.pub.INAMI.09}
\\
\begin{ensuredisplaymath}
\htuse{Gamma132.gn} = \htuse{Gamma132.td}
\end{ensuredisplaymath} & \htuse{Belle.Gamma132.pub.INAMI.09}}%
\htdef{LEE 10.cite}{\cite{Lee:2010tc}}%
\htdef{LEE 10.collab}{Belle}%
\htdef{LEE 10.ref}{LEE 10 (Belle) \cite{Lee:2010tc}}%
\htdef{LEE 10.meas}{%
\begin{ensuredisplaymath}
\htuse{Gamma60.gn} = \htuse{Gamma60.td}
\end{ensuredisplaymath} & \htuse{Belle.Gamma60.pub.LEE.10}
\\
\begin{ensuredisplaymath}
\htuse{Gamma85.gn} = \htuse{Gamma85.td}
\end{ensuredisplaymath} & \htuse{Belle.Gamma85.pub.LEE.10}
\\
\begin{ensuredisplaymath}
\htuse{Gamma93.gn} = \htuse{Gamma93.td}
\end{ensuredisplaymath} & \htuse{Belle.Gamma93.pub.LEE.10}
\\
\begin{ensuredisplaymath}
\htuse{Gamma96.gn} = \htuse{Gamma96.td}
\end{ensuredisplaymath} & \htuse{Belle.Gamma96.pub.LEE.10}}%
\htdef{RYU 14vpc.cite}{\cite{Ryu:2014vpc}}%
\htdef{RYU 14vpc.collab}{Belle}%
\htdef{RYU 14vpc.ref}{RYU 14vpc (Belle) \cite{Ryu:2014vpc}}%
\htdef{RYU 14vpc.meas}{%
\begin{ensuredisplaymath}
\htuse{Gamma35.gn} = \htuse{Gamma35.td}
\end{ensuredisplaymath} & \htuse{Belle.Gamma35.pub.RYU.14vpc}
\\
\begin{ensuredisplaymath}
\htuse{Gamma37.gn} = \htuse{Gamma37.td}
\end{ensuredisplaymath} & \htuse{Belle.Gamma37.pub.RYU.14vpc}
\\
\begin{ensuredisplaymath}
\htuse{Gamma40.gn} = \htuse{Gamma40.td}
\end{ensuredisplaymath} & \htuse{Belle.Gamma40.pub.RYU.14vpc}
\\
\begin{ensuredisplaymath}
\htuse{Gamma42.gn} = \htuse{Gamma42.td}
\end{ensuredisplaymath} & \htuse{Belle.Gamma42.pub.RYU.14vpc}
\\
\begin{ensuredisplaymath}
\htuse{Gamma47.gn} = \htuse{Gamma47.td}
\end{ensuredisplaymath} & \htuse{Belle.Gamma47.pub.RYU.14vpc}
\\
\begin{ensuredisplaymath}
\htuse{Gamma50.gn} = \htuse{Gamma50.td}
\end{ensuredisplaymath} & \htuse{Belle.Gamma50.pub.RYU.14vpc}}%
\htdef{BEHREND 89B.cite}{\cite{Behrend:1989wc}}%
\htdef{BEHREND 89B.collab}{CELLO}%
\htdef{BEHREND 89B.ref}{BEHREND 89B (CELLO) \cite{Behrend:1989wc}}%
\htdef{BEHREND 89B.meas}{%
\begin{ensuredisplaymath}
\htuse{Gamma54.gn} = \htuse{Gamma54.td}
\end{ensuredisplaymath} & \htuse{CELLO.Gamma54.pub.BEHREND.89B}}%
\htdef{ANASTASSOV 01.cite}{\cite{Anastassov:2000xu}}%
\htdef{ANASTASSOV 01.collab}{CLEO}%
\htdef{ANASTASSOV 01.ref}{ANASTASSOV 01 (CLEO) \cite{Anastassov:2000xu}}%
\htdef{ANASTASSOV 01.meas}{%
\begin{ensuredisplaymath}
\htuse{Gamma78.gn} = \htuse{Gamma78.td}
\end{ensuredisplaymath} & \htuse{CLEO.Gamma78.pub.ANASTASSOV.01}
\\
\begin{ensuredisplaymath}
\htuse{Gamma104.gn} = \htuse{Gamma104.td}
\end{ensuredisplaymath} & \htuse{CLEO.Gamma104.pub.ANASTASSOV.01}}%
\htdef{ANASTASSOV 97.cite}{\cite{Anastassov:1996tc}}%
\htdef{ANASTASSOV 97.collab}{CLEO}%
\htdef{ANASTASSOV 97.ref}{ANASTASSOV 97 (CLEO) \cite{Anastassov:1996tc}}%
\htdef{ANASTASSOV 97.meas}{%
\begin{ensuredisplaymath}
\htuse{Gamma3by5.gn} = \htuse{Gamma3by5.td}
\end{ensuredisplaymath} & \htuse{CLEO.Gamma3by5.pub.ANASTASSOV.97}
\\
\begin{ensuredisplaymath}
\htuse{Gamma5.gn} = \htuse{Gamma5.td}
\end{ensuredisplaymath} & \htuse{CLEO.Gamma5.pub.ANASTASSOV.97}
\\
\begin{ensuredisplaymath}
\htuse{Gamma8.gn} = \htuse{Gamma8.td}
\end{ensuredisplaymath} & \htuse{CLEO.Gamma8.pub.ANASTASSOV.97}}%
\htdef{ARTUSO 92.cite}{\cite{Artuso:1992qu}}%
\htdef{ARTUSO 92.collab}{CLEO}%
\htdef{ARTUSO 92.ref}{ARTUSO 92 (CLEO) \cite{Artuso:1992qu}}%
\htdef{ARTUSO 92.meas}{%
\begin{ensuredisplaymath}
\htuse{Gamma126.gn} = \htuse{Gamma126.td}
\end{ensuredisplaymath} & \htuse{CLEO.Gamma126.pub.ARTUSO.92}}%
\htdef{ARTUSO 94.cite}{\cite{Artuso:1994ii}}%
\htdef{ARTUSO 94.collab}{CLEO}%
\htdef{ARTUSO 94.ref}{ARTUSO 94 (CLEO) \cite{Artuso:1994ii}}%
\htdef{ARTUSO 94.meas}{%
\begin{ensuredisplaymath}
\htuse{Gamma13.gn} = \htuse{Gamma13.td}
\end{ensuredisplaymath} & \htuse{CLEO.Gamma13.pub.ARTUSO.94}}%
\htdef{BALEST 95C.cite}{\cite{Balest:1995kq}}%
\htdef{BALEST 95C.collab}{CLEO}%
\htdef{BALEST 95C.ref}{BALEST 95C (CLEO) \cite{Balest:1995kq}}%
\htdef{BALEST 95C.meas}{%
\begin{ensuredisplaymath}
\htuse{Gamma57.gn} = \htuse{Gamma57.td}
\end{ensuredisplaymath} & \htuse{CLEO.Gamma57.pub.BALEST.95C}
\\
\begin{ensuredisplaymath}
\htuse{Gamma66.gn} = \htuse{Gamma66.td}
\end{ensuredisplaymath} & \htuse{CLEO.Gamma66.pub.BALEST.95C}
\\
\begin{ensuredisplaymath}
\htuse{Gamma150by66.gn} = \htuse{Gamma150by66.td}
\end{ensuredisplaymath} & \htuse{CLEO.Gamma150by66.pub.BALEST.95C}}%
\htdef{BARINGER 87.cite}{\cite{Baringer:1987tr}}%
\htdef{BARINGER 87.collab}{CLEO}%
\htdef{BARINGER 87.ref}{BARINGER 87 (CLEO) \cite{Baringer:1987tr}}%
\htdef{BARINGER 87.meas}{%
\begin{ensuredisplaymath}
\htuse{Gamma150.gn} = \htuse{Gamma150.td}
\end{ensuredisplaymath} & \htuse{CLEO.Gamma150.pub.BARINGER.87}}%
\htdef{BARTELT 96.cite}{\cite{Bartelt:1996iv}}%
\htdef{BARTELT 96.collab}{CLEO}%
\htdef{BARTELT 96.ref}{BARTELT 96 (CLEO) \cite{Bartelt:1996iv}}%
\htdef{BARTELT 96.meas}{%
\begin{ensuredisplaymath}
\htuse{Gamma128.gn} = \htuse{Gamma128.td}
\end{ensuredisplaymath} & \htuse{CLEO.Gamma128.pub.BARTELT.96}}%
\htdef{BATTLE 94.cite}{\cite{Battle:1994by}}%
\htdef{BATTLE 94.collab}{CLEO}%
\htdef{BATTLE 94.ref}{BATTLE 94 (CLEO) \cite{Battle:1994by}}%
\htdef{BATTLE 94.meas}{%
\begin{ensuredisplaymath}
\htuse{Gamma10.gn} = \htuse{Gamma10.td}
\end{ensuredisplaymath} & \htuse{CLEO.Gamma10.pub.BATTLE.94}
\\
\begin{ensuredisplaymath}
\htuse{Gamma16.gn} = \htuse{Gamma16.td}
\end{ensuredisplaymath} & \htuse{CLEO.Gamma16.pub.BATTLE.94}
\\
\begin{ensuredisplaymath}
\htuse{Gamma23.gn} = \htuse{Gamma23.td}
\end{ensuredisplaymath} & \htuse{CLEO.Gamma23.pub.BATTLE.94}
\\
\begin{ensuredisplaymath}
\htuse{Gamma31.gn} = \htuse{Gamma31.td}
\end{ensuredisplaymath} & \htuse{CLEO.Gamma31.pub.BATTLE.94}}%
\htdef{BISHAI 99.cite}{\cite{Bishai:1998gf}}%
\htdef{BISHAI 99.collab}{CLEO}%
\htdef{BISHAI 99.ref}{BISHAI 99 (CLEO) \cite{Bishai:1998gf}}%
\htdef{BISHAI 99.meas}{%
\begin{ensuredisplaymath}
\htuse{Gamma130.gn} = \htuse{Gamma130.td}
\end{ensuredisplaymath} & \htuse{CLEO.Gamma130.pub.BISHAI.99}
\\
\begin{ensuredisplaymath}
\htuse{Gamma132.gn} = \htuse{Gamma132.td}
\end{ensuredisplaymath} & \htuse{CLEO.Gamma132.pub.BISHAI.99}}%
\htdef{BORTOLETTO 93.cite}{\cite{Bortoletto:1993px}}%
\htdef{BORTOLETTO 93.collab}{CLEO}%
\htdef{BORTOLETTO 93.ref}{BORTOLETTO 93 (CLEO) \cite{Bortoletto:1993px}}%
\htdef{BORTOLETTO 93.meas}{%
\begin{ensuredisplaymath}
\htuse{Gamma76by54.gn} = \htuse{Gamma76by54.td}
\end{ensuredisplaymath} & \htuse{CLEO.Gamma76by54.pub.BORTOLETTO.93}
\\
\begin{ensuredisplaymath}
\htuse{Gamma152by76.gn} = \htuse{Gamma152by76.td}
\end{ensuredisplaymath} & \htuse{CLEO.Gamma152by76.pub.BORTOLETTO.93}}%
\htdef{COAN 96.cite}{\cite{Coan:1996iu}}%
\htdef{COAN 96.collab}{CLEO}%
\htdef{COAN 96.ref}{COAN 96 (CLEO) \cite{Coan:1996iu}}%
\htdef{COAN 96.meas}{%
\begin{ensuredisplaymath}
\htuse{Gamma34.gn} = \htuse{Gamma34.td}
\end{ensuredisplaymath} & \htuse{CLEO.Gamma34.pub.COAN.96}
\\
\begin{ensuredisplaymath}
\htuse{Gamma37.gn} = \htuse{Gamma37.td}
\end{ensuredisplaymath} & \htuse{CLEO.Gamma37.pub.COAN.96}
\\
\begin{ensuredisplaymath}
\htuse{Gamma39.gn} = \htuse{Gamma39.td}
\end{ensuredisplaymath} & \htuse{CLEO.Gamma39.pub.COAN.96}
\\
\begin{ensuredisplaymath}
\htuse{Gamma42.gn} = \htuse{Gamma42.td}
\end{ensuredisplaymath} & \htuse{CLEO.Gamma42.pub.COAN.96}
\\
\begin{ensuredisplaymath}
\htuse{Gamma47.gn} = \htuse{Gamma47.td}
\end{ensuredisplaymath} & \htuse{CLEO.Gamma47.pub.COAN.96}}%
\htdef{EDWARDS 00A.cite}{\cite{Edwards:1999fj}}%
\htdef{EDWARDS 00A.collab}{CLEO}%
\htdef{EDWARDS 00A.ref}{EDWARDS 00A (CLEO) \cite{Edwards:1999fj}}%
\htdef{EDWARDS 00A.meas}{%
\begin{ensuredisplaymath}
\htuse{Gamma69.gn} = \htuse{Gamma69.td}
\end{ensuredisplaymath} & \htuse{CLEO.Gamma69.pub.EDWARDS.00A}}%
\htdef{GIBAUT 94B.cite}{\cite{Gibaut:1994ik}}%
\htdef{GIBAUT 94B.collab}{CLEO}%
\htdef{GIBAUT 94B.ref}{GIBAUT 94B (CLEO) \cite{Gibaut:1994ik}}%
\htdef{GIBAUT 94B.meas}{%
\begin{ensuredisplaymath}
\htuse{Gamma102.gn} = \htuse{Gamma102.td}
\end{ensuredisplaymath} & \htuse{CLEO.Gamma102.pub.GIBAUT.94B}
\\
\begin{ensuredisplaymath}
\htuse{Gamma103.gn} = \htuse{Gamma103.td}
\end{ensuredisplaymath} & \htuse{CLEO.Gamma103.pub.GIBAUT.94B}}%
\htdef{PROCARIO 93.cite}{\cite{Procario:1992hd}}%
\htdef{PROCARIO 93.collab}{CLEO}%
\htdef{PROCARIO 93.ref}{PROCARIO 93 (CLEO) \cite{Procario:1992hd}}%
\htdef{PROCARIO 93.meas}{%
\begin{ensuredisplaymath}
\htuse{Gamma19by13.gn} = \htuse{Gamma19by13.td}
\end{ensuredisplaymath} & \htuse{CLEO.Gamma19by13.pub.PROCARIO.93}
\\
\begin{ensuredisplaymath}
\htuse{Gamma26by13.gn} = \htuse{Gamma26by13.td}
\end{ensuredisplaymath} & \htuse{CLEO.Gamma26by13.pub.PROCARIO.93}
\\
\begin{ensuredisplaymath}
\htuse{Gamma29.gn} = \htuse{Gamma29.td}
\end{ensuredisplaymath} & \htuse{CLEO.Gamma29.pub.PROCARIO.93}}%
\htdef{RICHICHI 99.cite}{\cite{Richichi:1998bc}}%
\htdef{RICHICHI 99.collab}{CLEO}%
\htdef{RICHICHI 99.ref}{RICHICHI 99 (CLEO) \cite{Richichi:1998bc}}%
\htdef{RICHICHI 99.meas}{%
\begin{ensuredisplaymath}
\htuse{Gamma80by60.gn} = \htuse{Gamma80by60.td}
\end{ensuredisplaymath} & \htuse{CLEO.Gamma80by60.pub.RICHICHI.99}
\\
\begin{ensuredisplaymath}
\htuse{Gamma81by69.gn} = \htuse{Gamma81by69.td}
\end{ensuredisplaymath} & \htuse{CLEO.Gamma81by69.pub.RICHICHI.99}
\\
\begin{ensuredisplaymath}
\htuse{Gamma93by60.gn} = \htuse{Gamma93by60.td}
\end{ensuredisplaymath} & \htuse{CLEO.Gamma93by60.pub.RICHICHI.99}
\\
\begin{ensuredisplaymath}
\htuse{Gamma94by69.gn} = \htuse{Gamma94by69.td}
\end{ensuredisplaymath} & \htuse{CLEO.Gamma94by69.pub.RICHICHI.99}}%
\htdef{ARMS 05.cite}{\cite{Arms:2005qg}}%
\htdef{ARMS 05.collab}{CLEO3}%
\htdef{ARMS 05.ref}{ARMS 05 (CLEO3) \cite{Arms:2005qg}}%
\htdef{ARMS 05.meas}{%
\begin{ensuredisplaymath}
\htuse{Gamma88.gn} = \htuse{Gamma88.td}
\end{ensuredisplaymath} & \htuse{CLEO3.Gamma88.pub.ARMS.05}
\\
\begin{ensuredisplaymath}
\htuse{Gamma94.gn} = \htuse{Gamma94.td}
\end{ensuredisplaymath} & \htuse{CLEO3.Gamma94.pub.ARMS.05}
\\
\begin{ensuredisplaymath}
\htuse{Gamma151.gn} = \htuse{Gamma151.td}
\end{ensuredisplaymath} & \htuse{CLEO3.Gamma151.pub.ARMS.05}}%
\htdef{BRIERE 03.cite}{\cite{Briere:2003fr}}%
\htdef{BRIERE 03.collab}{CLEO3}%
\htdef{BRIERE 03.ref}{BRIERE 03 (CLEO3) \cite{Briere:2003fr}}%
\htdef{BRIERE 03.meas}{%
\begin{ensuredisplaymath}
\htuse{Gamma60.gn} = \htuse{Gamma60.td}
\end{ensuredisplaymath} & \htuse{CLEO3.Gamma60.pub.BRIERE.03}
\\
\begin{ensuredisplaymath}
\htuse{Gamma85.gn} = \htuse{Gamma85.td}
\end{ensuredisplaymath} & \htuse{CLEO3.Gamma85.pub.BRIERE.03}
\\
\begin{ensuredisplaymath}
\htuse{Gamma93.gn} = \htuse{Gamma93.td}
\end{ensuredisplaymath} & \htuse{CLEO3.Gamma93.pub.BRIERE.03}}%
\htdef{ABDALLAH 06A.cite}{\cite{Abdallah:2003cw}}%
\htdef{ABDALLAH 06A.collab}{DELPHI}%
\htdef{ABDALLAH 06A.ref}{ABDALLAH 06A (DELPHI) \cite{Abdallah:2003cw}}%
\htdef{ABDALLAH 06A.meas}{%
\begin{ensuredisplaymath}
\htuse{Gamma8.gn} = \htuse{Gamma8.td}
\end{ensuredisplaymath} & \htuse{DELPHI.Gamma8.pub.ABDALLAH.06A}
\\
\begin{ensuredisplaymath}
\htuse{Gamma13.gn} = \htuse{Gamma13.td}
\end{ensuredisplaymath} & \htuse{DELPHI.Gamma13.pub.ABDALLAH.06A}
\\
\begin{ensuredisplaymath}
\htuse{Gamma19.gn} = \htuse{Gamma19.td}
\end{ensuredisplaymath} & \htuse{DELPHI.Gamma19.pub.ABDALLAH.06A}
\\
\begin{ensuredisplaymath}
\htuse{Gamma25.gn} = \htuse{Gamma25.td}
\end{ensuredisplaymath} & \htuse{DELPHI.Gamma25.pub.ABDALLAH.06A}
\\
\begin{ensuredisplaymath}
\htuse{Gamma57.gn} = \htuse{Gamma57.td}
\end{ensuredisplaymath} & \htuse{DELPHI.Gamma57.pub.ABDALLAH.06A}
\\
\begin{ensuredisplaymath}
\htuse{Gamma66.gn} = \htuse{Gamma66.td}
\end{ensuredisplaymath} & \htuse{DELPHI.Gamma66.pub.ABDALLAH.06A}
\\
\begin{ensuredisplaymath}
\htuse{Gamma74.gn} = \htuse{Gamma74.td}
\end{ensuredisplaymath} & \htuse{DELPHI.Gamma74.pub.ABDALLAH.06A}
\\
\begin{ensuredisplaymath}
\htuse{Gamma103.gn} = \htuse{Gamma103.td}
\end{ensuredisplaymath} & \htuse{DELPHI.Gamma103.pub.ABDALLAH.06A}
\\
\begin{ensuredisplaymath}
\htuse{Gamma104.gn} = \htuse{Gamma104.td}
\end{ensuredisplaymath} & \htuse{DELPHI.Gamma104.pub.ABDALLAH.06A}}%
\htdef{ABREU 92N.cite}{\cite{Abreu:1992gn}}%
\htdef{ABREU 92N.collab}{DELPHI}%
\htdef{ABREU 92N.ref}{ABREU 92N (DELPHI) \cite{Abreu:1992gn}}%
\htdef{ABREU 92N.meas}{%
\begin{ensuredisplaymath}
\htuse{Gamma7.gn} = \htuse{Gamma7.td}
\end{ensuredisplaymath} & \htuse{DELPHI.Gamma7.pub.ABREU.92N}}%
\htdef{ABREU 94K.cite}{\cite{Abreu:1994fi}}%
\htdef{ABREU 94K.collab}{DELPHI}%
\htdef{ABREU 94K.ref}{ABREU 94K (DELPHI) \cite{Abreu:1994fi}}%
\htdef{ABREU 94K.meas}{%
\begin{ensuredisplaymath}
\htuse{Gamma10.gn} = \htuse{Gamma10.td}
\end{ensuredisplaymath} & \htuse{DELPHI.Gamma10.pub.ABREU.94K}
\\
\begin{ensuredisplaymath}
\htuse{Gamma31.gn} = \htuse{Gamma31.td}
\end{ensuredisplaymath} & \htuse{DELPHI.Gamma31.pub.ABREU.94K}}%
\htdef{ABREU 99X.cite}{\cite{Abreu:1999rb}}%
\htdef{ABREU 99X.collab}{DELPHI}%
\htdef{ABREU 99X.ref}{ABREU 99X (DELPHI) \cite{Abreu:1999rb}}%
\htdef{ABREU 99X.meas}{%
\begin{ensuredisplaymath}
\htuse{Gamma3.gn} = \htuse{Gamma3.td}
\end{ensuredisplaymath} & \htuse{DELPHI.Gamma3.pub.ABREU.99X}
\\
\begin{ensuredisplaymath}
\htuse{Gamma5.gn} = \htuse{Gamma5.td}
\end{ensuredisplaymath} & \htuse{DELPHI.Gamma5.pub.ABREU.99X}}%
\htdef{BYLSMA 87.cite}{\cite{Bylsma:1986zy}}%
\htdef{BYLSMA 87.collab}{HRS}%
\htdef{BYLSMA 87.ref}{BYLSMA 87 (HRS) \cite{Bylsma:1986zy}}%
\htdef{BYLSMA 87.meas}{%
\begin{ensuredisplaymath}
\htuse{Gamma102.gn} = \htuse{Gamma102.td}
\end{ensuredisplaymath} & \htuse{HRS.Gamma102.pub.BYLSMA.87}
\\
\begin{ensuredisplaymath}
\htuse{Gamma103.gn} = \htuse{Gamma103.td}
\end{ensuredisplaymath} & \htuse{HRS.Gamma103.pub.BYLSMA.87}}%
\htdef{ACCIARRI 01F.cite}{\cite{Acciarri:2001sg}}%
\htdef{ACCIARRI 01F.collab}{L3}%
\htdef{ACCIARRI 01F.ref}{ACCIARRI 01F (L3) \cite{Acciarri:2001sg}}%
\htdef{ACCIARRI 01F.meas}{%
\begin{ensuredisplaymath}
\htuse{Gamma3.gn} = \htuse{Gamma3.td}
\end{ensuredisplaymath} & \htuse{L3.Gamma3.pub.ACCIARRI.01F}
\\
\begin{ensuredisplaymath}
\htuse{Gamma5.gn} = \htuse{Gamma5.td}
\end{ensuredisplaymath} & \htuse{L3.Gamma5.pub.ACCIARRI.01F}}%
\htdef{ACCIARRI 95.cite}{\cite{Acciarri:1994vr}}%
\htdef{ACCIARRI 95.collab}{L3}%
\htdef{ACCIARRI 95.ref}{ACCIARRI 95 (L3) \cite{Acciarri:1994vr}}%
\htdef{ACCIARRI 95.meas}{%
\begin{ensuredisplaymath}
\htuse{Gamma7.gn} = \htuse{Gamma7.td}
\end{ensuredisplaymath} & \htuse{L3.Gamma7.pub.ACCIARRI.95}
\\
\begin{ensuredisplaymath}
\htuse{Gamma13.gn} = \htuse{Gamma13.td}
\end{ensuredisplaymath} & \htuse{L3.Gamma13.pub.ACCIARRI.95}
\\
\begin{ensuredisplaymath}
\htuse{Gamma19.gn} = \htuse{Gamma19.td}
\end{ensuredisplaymath} & \htuse{L3.Gamma19.pub.ACCIARRI.95}
\\
\begin{ensuredisplaymath}
\htuse{Gamma26.gn} = \htuse{Gamma26.td}
\end{ensuredisplaymath} & \htuse{L3.Gamma26.pub.ACCIARRI.95}}%
\htdef{ACCIARRI 95F.cite}{\cite{Acciarri:1995kx}}%
\htdef{ACCIARRI 95F.collab}{L3}%
\htdef{ACCIARRI 95F.ref}{ACCIARRI 95F (L3) \cite{Acciarri:1995kx}}%
\htdef{ACCIARRI 95F.meas}{%
\begin{ensuredisplaymath}
\htuse{Gamma35.gn} = \htuse{Gamma35.td}
\end{ensuredisplaymath} & \htuse{L3.Gamma35.pub.ACCIARRI.95F}
\\
\begin{ensuredisplaymath}
\htuse{Gamma40.gn} = \htuse{Gamma40.td}
\end{ensuredisplaymath} & \htuse{L3.Gamma40.pub.ACCIARRI.95F}}%
\htdef{ACHARD 01D.cite}{\cite{Achard:2001pk}}%
\htdef{ACHARD 01D.collab}{L3}%
\htdef{ACHARD 01D.ref}{ACHARD 01D (L3) \cite{Achard:2001pk}}%
\htdef{ACHARD 01D.meas}{%
\begin{ensuredisplaymath}
\htuse{Gamma55.gn} = \htuse{Gamma55.td}
\end{ensuredisplaymath} & \htuse{L3.Gamma55.pub.ACHARD.01D}
\\
\begin{ensuredisplaymath}
\htuse{Gamma102.gn} = \htuse{Gamma102.td}
\end{ensuredisplaymath} & \htuse{L3.Gamma102.pub.ACHARD.01D}}%
\htdef{ADEVA 91F.cite}{\cite{Adeva:1991qq}}%
\htdef{ADEVA 91F.collab}{L3}%
\htdef{ADEVA 91F.ref}{ADEVA 91F (L3) \cite{Adeva:1991qq}}%
\htdef{ADEVA 91F.meas}{%
\begin{ensuredisplaymath}
\htuse{Gamma54.gn} = \htuse{Gamma54.td}
\end{ensuredisplaymath} & \htuse{L3.Gamma54.pub.ADEVA.91F}}%
\htdef{ABBIENDI 00C.cite}{\cite{Abbiendi:1999pm}}%
\htdef{ABBIENDI 00C.collab}{OPAL}%
\htdef{ABBIENDI 00C.ref}{ABBIENDI 00C (OPAL) \cite{Abbiendi:1999pm}}%
\htdef{ABBIENDI 00C.meas}{%
\begin{ensuredisplaymath}
\htuse{Gamma35.gn} = \htuse{Gamma35.td}
\end{ensuredisplaymath} & \htuse{OPAL.Gamma35.pub.ABBIENDI.00C}
\\
\begin{ensuredisplaymath}
\htuse{Gamma38.gn} = \htuse{Gamma38.td}
\end{ensuredisplaymath} & \htuse{OPAL.Gamma38.pub.ABBIENDI.00C}
\\
\begin{ensuredisplaymath}
\htuse{Gamma43.gn} = \htuse{Gamma43.td}
\end{ensuredisplaymath} & \htuse{OPAL.Gamma43.pub.ABBIENDI.00C}}%
\htdef{ABBIENDI 00D.cite}{\cite{Abbiendi:1999cq}}%
\htdef{ABBIENDI 00D.collab}{OPAL}%
\htdef{ABBIENDI 00D.ref}{ABBIENDI 00D (OPAL) \cite{Abbiendi:1999cq}}%
\htdef{ABBIENDI 00D.meas}{%
\begin{ensuredisplaymath}
\htuse{Gamma92.gn} = \htuse{Gamma92.td}
\end{ensuredisplaymath} & \htuse{OPAL.Gamma92.pub.ABBIENDI.00D}}%
\htdef{ABBIENDI 01J.cite}{\cite{Abbiendi:2000ee}}%
\htdef{ABBIENDI 01J.collab}{OPAL}%
\htdef{ABBIENDI 01J.ref}{ABBIENDI 01J (OPAL) \cite{Abbiendi:2000ee}}%
\htdef{ABBIENDI 01J.meas}{%
\begin{ensuredisplaymath}
\htuse{Gamma10.gn} = \htuse{Gamma10.td}
\end{ensuredisplaymath} & \htuse{OPAL.Gamma10.pub.ABBIENDI.01J}
\\
\begin{ensuredisplaymath}
\htuse{Gamma31.gn} = \htuse{Gamma31.td}
\end{ensuredisplaymath} & \htuse{OPAL.Gamma31.pub.ABBIENDI.01J}}%
\htdef{ABBIENDI 03.cite}{\cite{Abbiendi:2002jw}}%
\htdef{ABBIENDI 03.collab}{OPAL}%
\htdef{ABBIENDI 03.ref}{ABBIENDI 03 (OPAL) \cite{Abbiendi:2002jw}}%
\htdef{ABBIENDI 03.meas}{%
\begin{ensuredisplaymath}
\htuse{Gamma3.gn} = \htuse{Gamma3.td}
\end{ensuredisplaymath} & \htuse{OPAL.Gamma3.pub.ABBIENDI.03}}%
\htdef{ABBIENDI 04J.cite}{\cite{Abbiendi:2004xa}}%
\htdef{ABBIENDI 04J.collab}{OPAL}%
\htdef{ABBIENDI 04J.ref}{ABBIENDI 04J (OPAL) \cite{Abbiendi:2004xa}}%
\htdef{ABBIENDI 04J.meas}{%
\begin{ensuredisplaymath}
\htuse{Gamma16.gn} = \htuse{Gamma16.td}
\end{ensuredisplaymath} & \htuse{OPAL.Gamma16.pub.ABBIENDI.04J}
\\
\begin{ensuredisplaymath}
\htuse{Gamma85.gn} = \htuse{Gamma85.td}
\end{ensuredisplaymath} & \htuse{OPAL.Gamma85.pub.ABBIENDI.04J}}%
\htdef{ABBIENDI 99H.cite}{\cite{Abbiendi:1998cx}}%
\htdef{ABBIENDI 99H.collab}{OPAL}%
\htdef{ABBIENDI 99H.ref}{ABBIENDI 99H (OPAL) \cite{Abbiendi:1998cx}}%
\htdef{ABBIENDI 99H.meas}{%
\begin{ensuredisplaymath}
\htuse{Gamma5.gn} = \htuse{Gamma5.td}
\end{ensuredisplaymath} & \htuse{OPAL.Gamma5.pub.ABBIENDI.99H}}%
\htdef{ACKERSTAFF 98M.cite}{\cite{Ackerstaff:1997tx}}%
\htdef{ACKERSTAFF 98M.collab}{OPAL}%
\htdef{ACKERSTAFF 98M.ref}{ACKERSTAFF 98M (OPAL) \cite{Ackerstaff:1997tx}}%
\htdef{ACKERSTAFF 98M.meas}{%
\begin{ensuredisplaymath}
\htuse{Gamma8.gn} = \htuse{Gamma8.td}
\end{ensuredisplaymath} & \htuse{OPAL.Gamma8.pub.ACKERSTAFF.98M}
\\
\begin{ensuredisplaymath}
\htuse{Gamma13.gn} = \htuse{Gamma13.td}
\end{ensuredisplaymath} & \htuse{OPAL.Gamma13.pub.ACKERSTAFF.98M}
\\
\begin{ensuredisplaymath}
\htuse{Gamma17.gn} = \htuse{Gamma17.td}
\end{ensuredisplaymath} & \htuse{OPAL.Gamma17.pub.ACKERSTAFF.98M}}%
\htdef{ACKERSTAFF 99E.cite}{\cite{Ackerstaff:1998ia}}%
\htdef{ACKERSTAFF 99E.collab}{OPAL}%
\htdef{ACKERSTAFF 99E.ref}{ACKERSTAFF 99E (OPAL) \cite{Ackerstaff:1998ia}}%
\htdef{ACKERSTAFF 99E.meas}{%
\begin{ensuredisplaymath}
\htuse{Gamma103.gn} = \htuse{Gamma103.td}
\end{ensuredisplaymath} & \htuse{OPAL.Gamma103.pub.ACKERSTAFF.99E}
\\
\begin{ensuredisplaymath}
\htuse{Gamma104.gn} = \htuse{Gamma104.td}
\end{ensuredisplaymath} & \htuse{OPAL.Gamma104.pub.ACKERSTAFF.99E}}%
\htdef{AKERS 94G.cite}{\cite{Akers:1994td}}%
\htdef{AKERS 94G.collab}{OPAL}%
\htdef{AKERS 94G.ref}{AKERS 94G (OPAL) \cite{Akers:1994td}}%
\htdef{AKERS 94G.meas}{%
\begin{ensuredisplaymath}
\htuse{Gamma33.gn} = \htuse{Gamma33.td}
\end{ensuredisplaymath} & \htuse{OPAL.Gamma33.pub.AKERS.94G}}%
\htdef{AKERS 95Y.cite}{\cite{Akers:1995ry}}%
\htdef{AKERS 95Y.collab}{OPAL}%
\htdef{AKERS 95Y.ref}{AKERS 95Y (OPAL) \cite{Akers:1995ry}}%
\htdef{AKERS 95Y.meas}{%
\begin{ensuredisplaymath}
\htuse{Gamma55.gn} = \htuse{Gamma55.td}
\end{ensuredisplaymath} & \htuse{OPAL.Gamma55.pub.AKERS.95Y}
\\
\begin{ensuredisplaymath}
\htuse{Gamma57by55.gn} = \htuse{Gamma57by55.td}
\end{ensuredisplaymath} & \htuse{OPAL.Gamma57by55.pub.AKERS.95Y}}%
\htdef{ALEXANDER 91D.cite}{\cite{Alexander:1991am}}%
\htdef{ALEXANDER 91D.collab}{OPAL}%
\htdef{ALEXANDER 91D.ref}{ALEXANDER 91D (OPAL) \cite{Alexander:1991am}}%
\htdef{ALEXANDER 91D.meas}{%
\begin{ensuredisplaymath}
\htuse{Gamma7.gn} = \htuse{Gamma7.td}
\end{ensuredisplaymath} & \htuse{OPAL.Gamma7.pub.ALEXANDER.91D}}%
\htdef{AIHARA 87B.cite}{\cite{Aihara:1986mw}}%
\htdef{AIHARA 87B.collab}{TPC}%
\htdef{AIHARA 87B.ref}{AIHARA 87B (TPC) \cite{Aihara:1986mw}}%
\htdef{AIHARA 87B.meas}{%
\begin{ensuredisplaymath}
\htuse{Gamma54.gn} = \htuse{Gamma54.td}
\end{ensuredisplaymath} & \htuse{TPC.Gamma54.pub.AIHARA.87B}}%
\htdef{BAUER 94.cite}{\cite{Bauer:1993wn}}%
\htdef{BAUER 94.collab}{TPC}%
\htdef{BAUER 94.ref}{BAUER 94 (TPC) \cite{Bauer:1993wn}}%
\htdef{BAUER 94.meas}{%
\begin{ensuredisplaymath}
\htuse{Gamma82.gn} = \htuse{Gamma82.td}
\end{ensuredisplaymath} & \htuse{TPC.Gamma82.pub.BAUER.94}
\\
\begin{ensuredisplaymath}
\htuse{Gamma92.gn} = \htuse{Gamma92.td}
\end{ensuredisplaymath} & \htuse{TPC.Gamma92.pub.BAUER.94}}%
\htdef{MeasPaper}{%
\multicolumn{2}{l}{\htuse{BARATE 98.ref}} \\
\htuse{BARATE 98.meas} \\\hline
\multicolumn{2}{l}{\htuse{BARATE 98E.ref}} \\
\htuse{BARATE 98E.meas} \\\hline
\multicolumn{2}{l}{\htuse{BARATE 99K.ref}} \\
\htuse{BARATE 99K.meas} \\\hline
\multicolumn{2}{l}{\htuse{BARATE 99R.ref}} \\
\htuse{BARATE 99R.meas} \\\hline
\multicolumn{2}{l}{\htuse{BUSKULIC 96.ref}} \\
\htuse{BUSKULIC 96.meas} \\\hline
\multicolumn{2}{l}{\htuse{BUSKULIC 97C.ref}} \\
\htuse{BUSKULIC 97C.meas} \\\hline
\multicolumn{2}{l}{\htuse{SCHAEL 05C.ref}} \\
\htuse{SCHAEL 05C.meas} \\\hline
\multicolumn{2}{l}{\htuse{ALBRECHT 88B.ref}} \\
\htuse{ALBRECHT 88B.meas} \\\hline
\multicolumn{2}{l}{\htuse{ALBRECHT 92D.ref}} \\
\htuse{ALBRECHT 92D.meas} \\\hline
\multicolumn{2}{l}{\htuse{AUBERT 07AP.ref}} \\
\htuse{AUBERT 07AP.meas} \\\hline
\multicolumn{2}{l}{\htuse{AUBERT 08.ref}} \\
\htuse{AUBERT 08.meas} \\\hline
\multicolumn{2}{l}{\htuse{AUBERT 10F.ref}} \\
\htuse{AUBERT 10F.meas} \\\hline
\multicolumn{2}{l}{\htuse{DEL-AMO-SANCHEZ 11E.ref}} \\
\htuse{DEL-AMO-SANCHEZ 11E.meas} \\\hline
\multicolumn{2}{l}{\htuse{LEES 12X.ref}} \\
\htuse{LEES 12X.meas} \\\hline
\multicolumn{2}{l}{\htuse{LEES 12Y.ref}} \\
\htuse{LEES 12Y.meas} \\\hline
\multicolumn{2}{l}{\htuse{FUJIKAWA 08.ref}} \\
\htuse{FUJIKAWA 08.meas} \\\hline
\multicolumn{2}{l}{\htuse{INAMI 09.ref}} \\
\htuse{INAMI 09.meas} \\\hline
\multicolumn{2}{l}{\htuse{LEE 10.ref}} \\
\htuse{LEE 10.meas} \\\hline
\multicolumn{2}{l}{\htuse{RYU 14vpc.ref}} \\
\htuse{RYU 14vpc.meas} \\\hline
\multicolumn{2}{l}{\htuse{BEHREND 89B.ref}} \\
\htuse{BEHREND 89B.meas} \\\hline
\multicolumn{2}{l}{\htuse{ANASTASSOV 01.ref}} \\
\htuse{ANASTASSOV 01.meas} \\\hline
\multicolumn{2}{l}{\htuse{ANASTASSOV 97.ref}} \\
\htuse{ANASTASSOV 97.meas} \\\hline
\multicolumn{2}{l}{\htuse{ARTUSO 92.ref}} \\
\htuse{ARTUSO 92.meas} \\\hline
\multicolumn{2}{l}{\htuse{ARTUSO 94.ref}} \\
\htuse{ARTUSO 94.meas} \\\hline
\multicolumn{2}{l}{\htuse{BALEST 95C.ref}} \\
\htuse{BALEST 95C.meas} \\\hline
\multicolumn{2}{l}{\htuse{BARINGER 87.ref}} \\
\htuse{BARINGER 87.meas} \\\hline
\multicolumn{2}{l}{\htuse{BARTELT 96.ref}} \\
\htuse{BARTELT 96.meas} \\\hline
\multicolumn{2}{l}{\htuse{BATTLE 94.ref}} \\
\htuse{BATTLE 94.meas} \\\hline
\multicolumn{2}{l}{\htuse{BISHAI 99.ref}} \\
\htuse{BISHAI 99.meas} \\\hline
\multicolumn{2}{l}{\htuse{BORTOLETTO 93.ref}} \\
\htuse{BORTOLETTO 93.meas} \\\hline
\multicolumn{2}{l}{\htuse{COAN 96.ref}} \\
\htuse{COAN 96.meas} \\\hline
\multicolumn{2}{l}{\htuse{EDWARDS 00A.ref}} \\
\htuse{EDWARDS 00A.meas} \\\hline
\multicolumn{2}{l}{\htuse{GIBAUT 94B.ref}} \\
\htuse{GIBAUT 94B.meas} \\\hline
\multicolumn{2}{l}{\htuse{PROCARIO 93.ref}} \\
\htuse{PROCARIO 93.meas} \\\hline
\multicolumn{2}{l}{\htuse{RICHICHI 99.ref}} \\
\htuse{RICHICHI 99.meas} \\\hline
\multicolumn{2}{l}{\htuse{ARMS 05.ref}} \\
\htuse{ARMS 05.meas} \\\hline
\multicolumn{2}{l}{\htuse{BRIERE 03.ref}} \\
\htuse{BRIERE 03.meas} \\\hline
\multicolumn{2}{l}{\htuse{ABDALLAH 06A.ref}} \\
\htuse{ABDALLAH 06A.meas} \\\hline
\multicolumn{2}{l}{\htuse{ABREU 92N.ref}} \\
\htuse{ABREU 92N.meas} \\\hline
\multicolumn{2}{l}{\htuse{ABREU 94K.ref}} \\
\htuse{ABREU 94K.meas} \\\hline
\multicolumn{2}{l}{\htuse{ABREU 99X.ref}} \\
\htuse{ABREU 99X.meas} \\\hline
\multicolumn{2}{l}{\htuse{BYLSMA 87.ref}} \\
\htuse{BYLSMA 87.meas} \\\hline
\multicolumn{2}{l}{\htuse{ACCIARRI 01F.ref}} \\
\htuse{ACCIARRI 01F.meas} \\\hline
\multicolumn{2}{l}{\htuse{ACCIARRI 95.ref}} \\
\htuse{ACCIARRI 95.meas} \\\hline
\multicolumn{2}{l}{\htuse{ACCIARRI 95F.ref}} \\
\htuse{ACCIARRI 95F.meas} \\\hline
\multicolumn{2}{l}{\htuse{ACHARD 01D.ref}} \\
\htuse{ACHARD 01D.meas} \\\hline
\multicolumn{2}{l}{\htuse{ADEVA 91F.ref}} \\
\htuse{ADEVA 91F.meas} \\\hline
\multicolumn{2}{l}{\htuse{ABBIENDI 00C.ref}} \\
\htuse{ABBIENDI 00C.meas} \\\hline
\multicolumn{2}{l}{\htuse{ABBIENDI 00D.ref}} \\
\htuse{ABBIENDI 00D.meas} \\\hline
\multicolumn{2}{l}{\htuse{ABBIENDI 01J.ref}} \\
\htuse{ABBIENDI 01J.meas} \\\hline
\multicolumn{2}{l}{\htuse{ABBIENDI 03.ref}} \\
\htuse{ABBIENDI 03.meas} \\\hline
\multicolumn{2}{l}{\htuse{ABBIENDI 04J.ref}} \\
\htuse{ABBIENDI 04J.meas} \\\hline
\multicolumn{2}{l}{\htuse{ABBIENDI 99H.ref}} \\
\htuse{ABBIENDI 99H.meas} \\\hline
\multicolumn{2}{l}{\htuse{ACKERSTAFF 98M.ref}} \\
\htuse{ACKERSTAFF 98M.meas} \\\hline
\multicolumn{2}{l}{\htuse{ACKERSTAFF 99E.ref}} \\
\htuse{ACKERSTAFF 99E.meas} \\\hline
\multicolumn{2}{l}{\htuse{AKERS 94G.ref}} \\
\htuse{AKERS 94G.meas} \\\hline
\multicolumn{2}{l}{\htuse{AKERS 95Y.ref}} \\
\htuse{AKERS 95Y.meas} \\\hline
\multicolumn{2}{l}{\htuse{ALEXANDER 91D.ref}} \\
\htuse{ALEXANDER 91D.meas} \\\hline
\multicolumn{2}{l}{\htuse{AIHARA 87B.ref}} \\
\htuse{AIHARA 87B.meas} \\\hline
\multicolumn{2}{l}{\htuse{BAUER 94.ref}} \\
\htuse{BAUER 94.meas}}%
\htdef{BrStrangeVal}{%
\htQuantLine{Gamma10}{0.6960 \pm 0.0096}{-2} 
\htQuantLine{Gamma16}{0.4327 \pm 0.0149}{-2} 
\htQuantLine{Gamma23}{0.0640 \pm 0.0220}{-2} 
\htQuantLine{Gamma28}{0.0428 \pm 0.0216}{-2} 
\htQuantLine{Gamma35}{0.8386 \pm 0.0141}{-2} 
\htQuantLine{Gamma40}{0.3812 \pm 0.0129}{-2} 
\htQuantLine{Gamma44}{0.0234 \pm 0.0231}{-2} 
\htQuantLine{Gamma53}{0.0222 \pm 0.0202}{-2} 
\htQuantLine{Gamma128}{0.0155 \pm 0.0008}{-2} 
\htQuantLine{Gamma130}{0.0048 \pm 0.0012}{-2} 
\htQuantLine{Gamma132}{0.0094 \pm 0.0015}{-2} 
\htQuantLine{Gamma151}{0.0410 \pm 0.0092}{-2} 
\htQuantLine{Gamma168}{0.0022 \pm 0.0008}{-2} 
\htQuantLine{Gamma169}{0.0015 \pm 0.0006}{-2} 
\htQuantLine{Gamma802}{0.2923 \pm 0.0067}{-2} 
\htQuantLine{Gamma803}{0.0410 \pm 0.0143}{-2} 
\htQuantLine{Gamma822}{0.0001 \pm 0.0001}{-2} 
\htQuantLine{Gamma833}{0.0001 \pm 0.0001}{-2}}%
\htdef{BrStrangeTotVal}{%
\htQuantLine{Gamma110}{2.9087 \pm 0.0482}{-2}}%
\htdef{UnitarityQuants}{%
\htConstrLine{Gamma3}{17.3917 \pm 0.0396}{1.0000}{-2}{0} 
\htConstrLine{Gamma5}{17.8162 \pm 0.0410}{1.0000}{-2}{0} 
\htConstrLine{Gamma9}{10.8103 \pm 0.0526}{1.0000}{-2}{0} 
\htConstrLine{Gamma10}{0.6960 \pm 0.0096}{1.0000}{-2}{0} 
\htConstrLine{Gamma14}{25.5023 \pm 0.0918}{1.0000}{-2}{0} 
\htConstrLine{Gamma16}{0.4327 \pm 0.0149}{1.0000}{-2}{0} 
\htConstrLine{Gamma20}{9.2424 \pm 0.0997}{1.0000}{-2}{0} 
\htConstrLine{Gamma23}{0.0640 \pm 0.0220}{1.0000}{-2}{0} 
\htConstrLine{Gamma27}{1.0287 \pm 0.0749}{1.0000}{-2}{0} 
\htConstrLine{Gamma28}{0.0428 \pm 0.0216}{1.0000}{-2}{0} 
\htConstrLine{Gamma30}{0.1099 \pm 0.0391}{1.0000}{-2}{0} 
\htConstrLine{Gamma35}{0.8386 \pm 0.0141}{1.0000}{-2}{0} 
\htConstrLine{Gamma37}{0.1479 \pm 0.0053}{1.0000}{-2}{0} 
\htConstrLine{Gamma40}{0.3812 \pm 0.0129}{1.0000}{-2}{0} 
\htConstrLine{Gamma42}{0.1502 \pm 0.0071}{1.0000}{-2}{0} 
\htConstrLine{Gamma44}{0.0234 \pm 0.0231}{1.0000}{-2}{0} 
\htConstrLine{Gamma47}{0.0233 \pm 0.0007}{2.0000}{-2}{0} 
\htConstrLine{Gamma48}{0.1047 \pm 0.0247}{1.0000}{-2}{0} 
\htConstrLine{Gamma50}{0.0018 \pm 0.0002}{2.0000}{-2}{0} 
\htConstrLine{Gamma51}{0.0318 \pm 0.0119}{1.0000}{-2}{0} 
\htConstrLine{Gamma53}{0.0222 \pm 0.0202}{1.0000}{-2}{0} 
\htConstrLine{Gamma62}{8.9704 \pm 0.0515}{1.0000}{-2}{0} 
\htConstrLine{Gamma70}{2.7694 \pm 0.0711}{1.0000}{-2}{0} 
\htConstrLine{Gamma77}{0.0976 \pm 0.0355}{1.0000}{-2}{0} 
\htConstrLine{Gamma93}{0.1434 \pm 0.0027}{1.0000}{-2}{0} 
\htConstrLine{Gamma94}{0.0061 \pm 0.0018}{1.0000}{-2}{0} 
\htConstrLine{Gamma126}{0.1386 \pm 0.0072}{1.0000}{-2}{0} 
\htConstrLine{Gamma128}{0.0155 \pm 0.0008}{1.0000}{-2}{0} 
\htConstrLine{Gamma130}{0.0048 \pm 0.0012}{1.0000}{-2}{0} 
\htConstrLine{Gamma132}{0.0094 \pm 0.0015}{1.0000}{-2}{0} 
\htConstrLine{Gamma136}{0.0218 \pm 0.0013}{1.0000}{-2}{0} 
\htConstrLine{Gamma151}{0.0410 \pm 0.0092}{1.0000}{-2}{0} 
\htConstrLine{Gamma152}{0.4058 \pm 0.0419}{1.0000}{-2}{0} 
\htConstrLine{Gamma167}{0.0044 \pm 0.0016}{0.8310}{-2}{0} 
\htConstrLine{Gamma800}{1.9544 \pm 0.0647}{1.0000}{-2}{0} 
\htConstrLine{Gamma802}{0.2923 \pm 0.0067}{1.0000}{-2}{0} 
\htConstrLine{Gamma803}{0.0410 \pm 0.0143}{1.0000}{-2}{0} 
\htConstrLine{Gamma805}{0.0400 \pm 0.0200}{1.0000}{-2}{0} 
\htConstrLine{Gamma811}{0.0071 \pm 0.0016}{1.0000}{-2}{0} 
\htConstrLine{Gamma812}{0.0013 \pm 0.0027}{1.0000}{-2}{0} 
\htConstrLine{Gamma821}{0.0768 \pm 0.0030}{1.0000}{-2}{0} 
\htConstrLine{Gamma822}{0.0001 \pm 0.0001}{1.0000}{-2}{0} 
\htConstrLine{Gamma831}{0.0084 \pm 0.0006}{1.0000}{-2}{0} 
\htConstrLine{Gamma832}{0.0038 \pm 0.0009}{1.0000}{-2}{0} 
\htConstrLine{Gamma833}{0.0001 \pm 0.0001}{1.0000}{-2}{0} 
\htConstrLine{Gamma920}{0.0052 \pm 0.0004}{1.0000}{-2}{0} 
\htConstrLine{Gamma945}{0.0193 \pm 0.0038}{1.0000}{-2}{0} 
\htConstrLine{Gamma998}{0.0355 \pm 0.1031}{1.0000}{-2}{0}}%
\htdef{BaseQuants}{%
\htQuantLine{Gamma3}{17.3917 \pm 0.0396}{-2} 
\htQuantLine{Gamma5}{17.8162 \pm 0.0410}{-2} 
\htQuantLine{Gamma9}{10.8103 \pm 0.0526}{-2} 
\htQuantLine{Gamma10}{0.6960 \pm 0.0096}{-2} 
\htQuantLine{Gamma14}{25.5023 \pm 0.0918}{-2} 
\htQuantLine{Gamma16}{0.4327 \pm 0.0149}{-2} 
\htQuantLine{Gamma20}{9.2424 \pm 0.0997}{-2} 
\htQuantLine{Gamma23}{0.0640 \pm 0.0220}{-2} 
\htQuantLine{Gamma27}{1.0287 \pm 0.0749}{-2} 
\htQuantLine{Gamma28}{0.0428 \pm 0.0216}{-2} 
\htQuantLine{Gamma30}{0.1099 \pm 0.0391}{-2} 
\htQuantLine{Gamma35}{0.8386 \pm 0.0141}{-2} 
\htQuantLine{Gamma37}{0.1479 \pm 0.0053}{-2} 
\htQuantLine{Gamma40}{0.3812 \pm 0.0129}{-2} 
\htQuantLine{Gamma42}{0.1502 \pm 0.0071}{-2} 
\htQuantLine{Gamma44}{0.0234 \pm 0.0231}{-2} 
\htQuantLine{Gamma47}{0.0233 \pm 0.0007}{-2} 
\htQuantLine{Gamma48}{0.1047 \pm 0.0247}{-2} 
\htQuantLine{Gamma50}{0.0018 \pm 0.0002}{-2} 
\htQuantLine{Gamma51}{0.0318 \pm 0.0119}{-2} 
\htQuantLine{Gamma53}{0.0222 \pm 0.0202}{-2} 
\htQuantLine{Gamma62}{8.9704 \pm 0.0515}{-2} 
\htQuantLine{Gamma70}{2.7694 \pm 0.0711}{-2} 
\htQuantLine{Gamma77}{0.0976 \pm 0.0355}{-2} 
\htQuantLine{Gamma93}{0.1434 \pm 0.0027}{-2} 
\htQuantLine{Gamma94}{0.0061 \pm 0.0018}{-2} 
\htQuantLine{Gamma126}{0.1386 \pm 0.0072}{-2} 
\htQuantLine{Gamma128}{0.0155 \pm 0.0008}{-2} 
\htQuantLine{Gamma130}{0.0048 \pm 0.0012}{-2} 
\htQuantLine{Gamma132}{0.0094 \pm 0.0015}{-2} 
\htQuantLine{Gamma136}{0.0218 \pm 0.0013}{-2} 
\htQuantLine{Gamma151}{0.0410 \pm 0.0092}{-2} 
\htQuantLine{Gamma152}{0.4058 \pm 0.0419}{-2} 
\htQuantLine{Gamma167}{0.0044 \pm 0.0016}{-2} 
\htQuantLine{Gamma800}{1.9544 \pm 0.0647}{-2} 
\htQuantLine{Gamma802}{0.2923 \pm 0.0067}{-2} 
\htQuantLine{Gamma803}{0.0410 \pm 0.0143}{-2} 
\htQuantLine{Gamma805}{0.0400 \pm 0.0200}{-2} 
\htQuantLine{Gamma811}{0.0071 \pm 0.0016}{-2} 
\htQuantLine{Gamma812}{0.0013 \pm 0.0027}{-2} 
\htQuantLine{Gamma821}{0.0768 \pm 0.0030}{-2} 
\htQuantLine{Gamma822}{0.0001 \pm 0.0001}{-2} 
\htQuantLine{Gamma831}{0.0084 \pm 0.0006}{-2} 
\htQuantLine{Gamma832}{0.0038 \pm 0.0009}{-2} 
\htQuantLine{Gamma833}{0.0001 \pm 0.0001}{-2} 
\htQuantLine{Gamma920}{0.0052 \pm 0.0004}{-2} 
\htQuantLine{Gamma945}{0.0193 \pm 0.0038}{-2}}%
\htdef{BrCorr}{%
\ifhevea\begin{table}\fi
\begin{center}
\ifhevea
\caption{Basis quantities correlation coefficients in percent, subtable 1.\label{tab:tau:br-fit-corr1}}%
\else
\begin{minipage}{\linewidth}
\begin{center}
\captionof{table}{Basis quantities correlation coefficients in percent, subtable 1.}\label{tab:tau:br-fit-corr1}%
\fi
\begin{envsmall}
\begin{center}
\begin{tabular}{rrrrrrrrrrrrrrr}
\hline
\( \Gamma_{5} \) &   23 &  &  &  &  &  &  &  &  &  &  &  &  &  \\
\( \Gamma_{9} \) &    7 &    5 &  &  &  &  &  &  &  &  &  &  &  &  \\
\( \Gamma_{10} \) &    3 &    5 &    1 &  &  &  &  &  &  &  &  &  &  &  \\
\( \Gamma_{14} \) &  -13 &  -14 &  -12 &   -3 &  &  &  &  &  &  &  &  &  &  \\
\( \Gamma_{16} \) &    0 &   -1 &    2 &   -1 &  -16 &  &  &  &  &  &  &  &  &  \\
\( \Gamma_{20} \) &   -5 &   -5 &   -7 &   -1 &  -40 &    2 &  &  &  &  &  &  &  &  \\
\( \Gamma_{23} \) &    0 &    0 &    0 &   -2 &    2 &  -13 &  -22 &  &  &  &  &  &  &  \\
\( \Gamma_{27} \) &   -4 &   -3 &   -8 &   -1 &    0 &    3 &  -36 &    6 &  &  &  &  &  &  \\
\( \Gamma_{28} \) &    0 &    0 &    0 &   -2 &    2 &  -13 &    5 &  -21 &  -29 &  &  &  &  &  \\
\( \Gamma_{30} \) &   -5 &   -4 &  -11 &   -2 &   -9 &    0 &    6 &    0 &  -42 &    0 &  &  &  &  \\
\( \Gamma_{35} \) &    0 &    0 &    0 &    0 &    0 &    0 &    0 &    1 &    0 &    1 &    0 &  &  &  \\
\( \Gamma_{37} \) &    0 &    0 &    0 &    0 &    0 &   -2 &    1 &   -3 &    1 &   -3 &    0 &  -22 &  &  \\
\( \Gamma_{40} \) &    0 &    0 &    0 &    0 &    0 &    1 &    0 &    1 &   -2 &    1 &    0 &  -12 &    4 &  \\
 & \( \Gamma_{3} \) & \( \Gamma_{5} \) & \( \Gamma_{9} \) & \( \Gamma_{10} \) & \( \Gamma_{14} \) & \( \Gamma_{16} \) & \( \Gamma_{20} \) & \( \Gamma_{23} \) & \( \Gamma_{27} \) & \( \Gamma_{28} \) & \( \Gamma_{30} \) & \( \Gamma_{35} \) & \( \Gamma_{37} \) & \( \Gamma_{40} \)
\\\hline
\end{tabular}
\end{center}
\end{envsmall}
\ifhevea\else
\end{center}
\end{minipage}
\fi
\end{center}
\ifhevea\end{table}\fi
\ifhevea\begin{table}\fi
\begin{center}
\ifhevea
\caption{Basis quantities correlation coefficients in percent, subtable 2.\label{tab:tau:br-fit-corr2}}%
\else
\begin{minipage}{\linewidth}
\begin{center}
\captionof{table}{Basis quantities correlation coefficients in percent, subtable 2.}\label{tab:tau:br-fit-corr2}%
\fi
\begin{envsmall}
\begin{center}
\begin{tabular}{rrrrrrrrrrrrrrr}
\hline
\( \Gamma_{42} \) &    0 &    0 &    0 &    0 &    1 &   -3 &    1 &   -5 &    1 &   -5 &    0 &    2 &  -21 &  -20 \\
\( \Gamma_{44} \) &    0 &    0 &    0 &    0 &    0 &    0 &    0 &    0 &    0 &    0 &    0 &   -1 &    0 &   -4 \\
\( \Gamma_{47} \) &    0 &    0 &    0 &    0 &    0 &    0 &    0 &    0 &    0 &    0 &    0 &   -1 &    1 &   -4 \\
\( \Gamma_{48} \) &    0 &    0 &    0 &    0 &    0 &    0 &    0 &    0 &    0 &    0 &    0 &   -3 &    0 &   -2 \\
\( \Gamma_{50} \) &    0 &    0 &    0 &    0 &    0 &    0 &    0 &   -1 &    0 &   -1 &    0 &    0 &    7 &    0 \\
\( \Gamma_{51} \) &    0 &    0 &    0 &    0 &    0 &    0 &    0 &    0 &    0 &    0 &    0 &   -1 &    0 &   -1 \\
\( \Gamma_{53} \) &    0 &    0 &    0 &    0 &    0 &    0 &    0 &    0 &    0 &    0 &    0 &    0 &    0 &    0 \\
\( \Gamma_{62} \) &   -3 &   -5 &    8 &    0 &   -4 &    5 &   -7 &   -1 &   -5 &   -1 &   -5 &    0 &    0 &    0 \\
\( \Gamma_{70} \) &   -6 &   -6 &   -7 &   -1 &   -8 &   -1 &   -1 &    0 &   -1 &    0 &    3 &    0 &    0 &    0 \\
\( \Gamma_{77} \) &   -1 &    0 &   -3 &   -1 &   -2 &    0 &    0 &    0 &    2 &    0 &    2 &    0 &    0 &    0 \\
\( \Gamma_{93} \) &   -1 &   -1 &    3 &    0 &   -1 &    2 &   -1 &    0 &   -1 &    0 &   -1 &    0 &    0 &    0 \\
\( \Gamma_{94} \) &    0 &    0 &    0 &    0 &    0 &    0 &    0 &    0 &    0 &    0 &    0 &    0 &    0 &    0 \\
\( \Gamma_{126} \) &    0 &    0 &    0 &    0 &    0 &    0 &   -1 &    0 &    0 &    0 &   -2 &    0 &    0 &    0 \\
\( \Gamma_{128} \) &    0 &    0 &    1 &    0 &    0 &    1 &    0 &   -1 &    0 &   -1 &    0 &    0 &    0 &    0 \\
 & \( \Gamma_{3} \) & \( \Gamma_{5} \) & \( \Gamma_{9} \) & \( \Gamma_{10} \) & \( \Gamma_{14} \) & \( \Gamma_{16} \) & \( \Gamma_{20} \) & \( \Gamma_{23} \) & \( \Gamma_{27} \) & \( \Gamma_{28} \) & \( \Gamma_{30} \) & \( \Gamma_{35} \) & \( \Gamma_{37} \) & \( \Gamma_{40} \)
\\\hline
\end{tabular}
\end{center}
\end{envsmall}
\ifhevea\else
\end{center}
\end{minipage}
\fi
\end{center}
\ifhevea\end{table}\fi
\ifhevea\begin{table}\fi
\begin{center}
\ifhevea
\caption{Basis quantities correlation coefficients in percent, subtable 3.\label{tab:tau:br-fit-corr3}}%
\else
\begin{minipage}{\linewidth}
\begin{center}
\captionof{table}{Basis quantities correlation coefficients in percent, subtable 3.}\label{tab:tau:br-fit-corr3}%
\fi
\begin{envsmall}
\begin{center}
\begin{tabular}{rrrrrrrrrrrrrrr}
\hline
\( \Gamma_{130} \) &    0 &    0 &    0 &    0 &    0 &    0 &    0 &    0 &    0 &    0 &    0 &    0 &    0 &    0 \\
\( \Gamma_{132} \) &    0 &    0 &    0 &    0 &    0 &    0 &    0 &    0 &    0 &    0 &    0 &    0 &    0 &    0 \\
\( \Gamma_{136} \) &    0 &    0 &    0 &    0 &    0 &    0 &    0 &    0 &    0 &    0 &    0 &    0 &    0 &    0 \\
\( \Gamma_{151} \) &    0 &    0 &    0 &    0 &    0 &    0 &    0 &    0 &    0 &    0 &    0 &    0 &    0 &    0 \\
\( \Gamma_{152} \) &   -1 &    0 &   -3 &   -1 &   -2 &    0 &   -1 &    0 &    2 &    0 &    2 &    0 &    0 &    0 \\
\( \Gamma_{167} \) &    0 &    0 &    0 &    0 &    0 &    0 &    0 &    0 &    0 &    0 &    0 &    0 &    0 &    0 \\
\( \Gamma_{800} \) &   -2 &   -2 &   -2 &    0 &   -3 &    0 &    0 &    0 &    0 &    0 &    1 &    0 &    0 &    0 \\
\( \Gamma_{802} \) &   -1 &   -1 &    0 &    0 &   -1 &    0 &   -2 &    0 &   -2 &    0 &   -1 &    0 &    0 &    0 \\
\( \Gamma_{803} \) &    0 &    0 &    0 &    0 &    0 &    0 &    0 &    0 &    0 &    0 &    0 &    0 &    0 &    0 \\
\( \Gamma_{805} \) &    0 &    0 &    0 &    0 &    0 &    0 &    0 &    0 &    0 &    0 &    0 &    0 &    0 &    0 \\
\( \Gamma_{811} \) &    0 &    0 &    0 &    0 &    0 &    0 &    0 &    0 &    0 &    0 &    0 &    0 &    0 &    0 \\
\( \Gamma_{812} \) &    0 &    1 &    0 &    0 &    0 &    0 &    0 &    0 &    0 &    0 &    0 &    0 &    0 &    0 \\
\( \Gamma_{821} \) &    0 &    0 &    1 &    0 &    0 &    0 &   -1 &    0 &    0 &    0 &   -1 &    0 &    0 &    0 \\
\( \Gamma_{822} \) &    0 &    0 &    0 &    0 &    0 &    0 &    0 &    0 &    0 &    0 &    0 &    0 &    0 &    0 \\
 & \( \Gamma_{3} \) & \( \Gamma_{5} \) & \( \Gamma_{9} \) & \( \Gamma_{10} \) & \( \Gamma_{14} \) & \( \Gamma_{16} \) & \( \Gamma_{20} \) & \( \Gamma_{23} \) & \( \Gamma_{27} \) & \( \Gamma_{28} \) & \( \Gamma_{30} \) & \( \Gamma_{35} \) & \( \Gamma_{37} \) & \( \Gamma_{40} \)
\\\hline
\end{tabular}
\end{center}
\end{envsmall}
\ifhevea\else
\end{center}
\end{minipage}
\fi
\end{center}
\ifhevea\end{table}\fi
\ifhevea\begin{table}\fi
\begin{center}
\ifhevea
\caption{Basis quantities correlation coefficients in percent, subtable 4.\label{tab:tau:br-fit-corr4}}%
\else
\begin{minipage}{\linewidth}
\begin{center}
\captionof{table}{Basis quantities correlation coefficients in percent, subtable 4.}\label{tab:tau:br-fit-corr4}%
\fi
\begin{envsmall}
\begin{center}
\begin{tabular}{rrrrrrrrrrrrrrr}
\hline
\( \Gamma_{831} \) &    0 &    0 &    0 &    0 &    0 &    0 &    0 &    0 &    0 &    0 &    0 &    0 &    0 &    0 \\
\( \Gamma_{832} \) &    0 &    0 &    0 &    0 &    0 &    0 &    0 &    0 &    0 &    0 &    0 &    0 &    0 &    0 \\
\( \Gamma_{833} \) &    0 &    0 &    0 &    0 &    0 &    0 &    0 &    0 &    0 &    0 &    0 &    0 &    0 &    0 \\
\( \Gamma_{920} \) &    0 &    0 &    0 &    0 &    0 &    0 &    0 &    0 &    0 &    0 &    0 &    0 &    0 &    0 \\
\( \Gamma_{945} \) &    0 &    0 &    0 &    0 &    0 &    0 &    0 &    0 &    0 &    0 &    0 &    0 &    0 &    0 \\
 & \( \Gamma_{3} \) & \( \Gamma_{5} \) & \( \Gamma_{9} \) & \( \Gamma_{10} \) & \( \Gamma_{14} \) & \( \Gamma_{16} \) & \( \Gamma_{20} \) & \( \Gamma_{23} \) & \( \Gamma_{27} \) & \( \Gamma_{28} \) & \( \Gamma_{30} \) & \( \Gamma_{35} \) & \( \Gamma_{37} \) & \( \Gamma_{40} \)
\\\hline
\end{tabular}
\end{center}
\end{envsmall}
\ifhevea\else
\end{center}
\end{minipage}
\fi
\end{center}
\ifhevea\end{table}\fi
\ifhevea\begin{table}\fi
\begin{center}
\ifhevea
\caption{Basis quantities correlation coefficients in percent, subtable 5.\label{tab:tau:br-fit-corr5}}%
\else
\begin{minipage}{\linewidth}
\begin{center}
\captionof{table}{Basis quantities correlation coefficients in percent, subtable 5.}\label{tab:tau:br-fit-corr5}%
\fi
\begin{envsmall}
\begin{center}
\begin{tabular}{rrrrrrrrrrrrrrr}
\hline
\( \Gamma_{44} \) &    0 &  &  &  &  &  &  &  &  &  &  &  &  &  \\
\( \Gamma_{47} \) &    1 &    0 &  &  &  &  &  &  &  &  &  &  &  &  \\
\( \Gamma_{48} \) &   -1 &   -6 &    0 &  &  &  &  &  &  &  &  &  &  &  \\
\( \Gamma_{50} \) &    5 &    0 &   -7 &    0 &  &  &  &  &  &  &  &  &  &  \\
\( \Gamma_{51} \) &    0 &   -3 &    0 &   -6 &    0 &  &  &  &  &  &  &  &  &  \\
\( \Gamma_{53} \) &    0 &    0 &    0 &    0 &    0 &    0 &  &  &  &  &  &  &  &  \\
\( \Gamma_{62} \) &    0 &    0 &    1 &    0 &    0 &    0 &    0 &  &  &  &  &  &  &  \\
\( \Gamma_{70} \) &    0 &    0 &    0 &    0 &    0 &    0 &    0 &  -20 &  &  &  &  &  &  \\
\( \Gamma_{77} \) &    0 &    0 &    0 &    0 &    0 &    0 &    0 &   -1 &   -7 &  &  &  &  &  \\
\( \Gamma_{93} \) &    0 &    0 &    0 &    0 &    0 &    0 &    0 &   14 &   -4 &    0 &  &  &  &  \\
\( \Gamma_{94} \) &    0 &    0 &    0 &    0 &    0 &    0 &    0 &    0 &   -2 &    0 &    0 &  &  &  \\
\( \Gamma_{126} \) &    0 &    0 &    1 &    0 &    0 &    0 &    0 &    1 &    0 &   -5 &    0 &    0 &  &  \\
\( \Gamma_{128} \) &    0 &    0 &    1 &    0 &    0 &    0 &    0 &    2 &    0 &    0 &    1 &    0 &    4 &  \\
 & \( \Gamma_{42} \) & \( \Gamma_{44} \) & \( \Gamma_{47} \) & \( \Gamma_{48} \) & \( \Gamma_{50} \) & \( \Gamma_{51} \) & \( \Gamma_{53} \) & \( \Gamma_{62} \) & \( \Gamma_{70} \) & \( \Gamma_{77} \) & \( \Gamma_{93} \) & \( \Gamma_{94} \) & \( \Gamma_{126} \) & \( \Gamma_{128} \)
\\\hline
\end{tabular}
\end{center}
\end{envsmall}
\ifhevea\else
\end{center}
\end{minipage}
\fi
\end{center}
\ifhevea\end{table}\fi
\ifhevea\begin{table}\fi
\begin{center}
\ifhevea
\caption{Basis quantities correlation coefficients in percent, subtable 6.\label{tab:tau:br-fit-corr6}}%
\else
\begin{minipage}{\linewidth}
\begin{center}
\captionof{table}{Basis quantities correlation coefficients in percent, subtable 6.}\label{tab:tau:br-fit-corr6}%
\fi
\begin{envsmall}
\begin{center}
\begin{tabular}{rrrrrrrrrrrrrrr}
\hline
\( \Gamma_{130} \) &    0 &    0 &    0 &    0 &    0 &    0 &    0 &    0 &    0 &   -1 &    0 &    0 &    1 &    1 \\
\( \Gamma_{132} \) &    0 &    0 &    0 &    0 &    0 &    0 &    0 &    0 &    0 &    0 &    0 &    0 &    2 &    1 \\
\( \Gamma_{136} \) &    0 &    0 &    0 &    0 &    0 &    0 &    0 &    0 &   -1 &    0 &    0 &    0 &    0 &    0 \\
\( \Gamma_{151} \) &    0 &    0 &    0 &    0 &    0 &    0 &    0 &    0 &   12 &    0 &    0 &    0 &    0 &    0 \\
\( \Gamma_{152} \) &    0 &    0 &    0 &    0 &    0 &    0 &    0 &   -1 &  -11 &  -64 &    0 &    0 &    0 &    0 \\
\( \Gamma_{167} \) &    0 &    0 &    0 &    0 &    0 &    0 &    0 &   -1 &    0 &    0 &    1 &    0 &    0 &    0 \\
\( \Gamma_{800} \) &    0 &    0 &    0 &    0 &    0 &    0 &    0 &   -8 &  -69 &   -2 &   -1 &    0 &    0 &    0 \\
\( \Gamma_{802} \) &    0 &    0 &    0 &    0 &    0 &    0 &    0 &   16 &   -6 &    0 &    0 &    0 &    0 &    0 \\
\( \Gamma_{803} \) &    0 &    0 &    0 &    0 &    0 &    0 &    0 &   -1 &  -19 &    0 &    0 &   -2 &    0 &   -1 \\
\( \Gamma_{805} \) &    0 &    0 &    0 &    0 &    0 &    0 &    0 &    0 &    0 &    0 &    0 &    0 &    0 &    0 \\
\( \Gamma_{811} \) &    0 &    0 &    0 &    0 &    0 &    0 &    0 &    0 &    0 &    0 &    0 &    0 &    0 &    0 \\
\( \Gamma_{812} \) &    0 &    0 &    0 &    0 &   -1 &    0 &    0 &    0 &   -1 &    0 &    0 &    0 &    0 &    0 \\
\( \Gamma_{821} \) &    0 &    0 &    0 &    0 &    0 &    0 &    0 &    0 &   -1 &    0 &    0 &    0 &    0 &    0 \\
\( \Gamma_{822} \) &    0 &    0 &    0 &    0 &    0 &    0 &    0 &    0 &    0 &    0 &    0 &    0 &    0 &    0 \\
 & \( \Gamma_{42} \) & \( \Gamma_{44} \) & \( \Gamma_{47} \) & \( \Gamma_{48} \) & \( \Gamma_{50} \) & \( \Gamma_{51} \) & \( \Gamma_{53} \) & \( \Gamma_{62} \) & \( \Gamma_{70} \) & \( \Gamma_{77} \) & \( \Gamma_{93} \) & \( \Gamma_{94} \) & \( \Gamma_{126} \) & \( \Gamma_{128} \)
\\\hline
\end{tabular}
\end{center}
\end{envsmall}
\ifhevea\else
\end{center}
\end{minipage}
\fi
\end{center}
\ifhevea\end{table}\fi
\ifhevea\begin{table}\fi
\begin{center}
\ifhevea
\caption{Basis quantities correlation coefficients in percent, subtable 7.\label{tab:tau:br-fit-corr7}}%
\else
\begin{minipage}{\linewidth}
\begin{center}
\captionof{table}{Basis quantities correlation coefficients in percent, subtable 7.}\label{tab:tau:br-fit-corr7}%
\fi
\begin{envsmall}
\begin{center}
\begin{tabular}{rrrrrrrrrrrrrrr}
\hline
\( \Gamma_{831} \) &    0 &    0 &    0 &    0 &    0 &    0 &    0 &    0 &    0 &    0 &    0 &    0 &    0 &    0 \\
\( \Gamma_{832} \) &    0 &    0 &    0 &    0 &    0 &    0 &    0 &    0 &    0 &    0 &    0 &    0 &    0 &    0 \\
\( \Gamma_{833} \) &    0 &    0 &    0 &    0 &    0 &    0 &    0 &    0 &    0 &    0 &    0 &    0 &    0 &    0 \\
\( \Gamma_{920} \) &    0 &    0 &    0 &    0 &    0 &    0 &    0 &    0 &    0 &    0 &    0 &    0 &    0 &    0 \\
\( \Gamma_{945} \) &    0 &    0 &    0 &    0 &    0 &    0 &    0 &    0 &    0 &    0 &    0 &    0 &    0 &    0 \\
 & \( \Gamma_{42} \) & \( \Gamma_{44} \) & \( \Gamma_{47} \) & \( \Gamma_{48} \) & \( \Gamma_{50} \) & \( \Gamma_{51} \) & \( \Gamma_{53} \) & \( \Gamma_{62} \) & \( \Gamma_{70} \) & \( \Gamma_{77} \) & \( \Gamma_{93} \) & \( \Gamma_{94} \) & \( \Gamma_{126} \) & \( \Gamma_{128} \)
\\\hline
\end{tabular}
\end{center}
\end{envsmall}
\ifhevea\else
\end{center}
\end{minipage}
\fi
\end{center}
\ifhevea\end{table}\fi
\ifhevea\begin{table}\fi
\begin{center}
\ifhevea
\caption{Basis quantities correlation coefficients in percent, subtable 8.\label{tab:tau:br-fit-corr8}}%
\else
\begin{minipage}{\linewidth}
\begin{center}
\captionof{table}{Basis quantities correlation coefficients in percent, subtable 8.}\label{tab:tau:br-fit-corr8}%
\fi
\begin{envsmall}
\begin{center}
\begin{tabular}{rrrrrrrrrrrrrrr}
\hline
\( \Gamma_{132} \) &    0 &  &  &  &  &  &  &  &  &  &  &  &  &  \\
\( \Gamma_{136} \) &    0 &    0 &  &  &  &  &  &  &  &  &  &  &  &  \\
\( \Gamma_{151} \) &    0 &    0 &    0 &  &  &  &  &  &  &  &  &  &  &  \\
\( \Gamma_{152} \) &    0 &    0 &    0 &    0 &  &  &  &  &  &  &  &  &  &  \\
\( \Gamma_{167} \) &    0 &    0 &    0 &    0 &    0 &  &  &  &  &  &  &  &  &  \\
\( \Gamma_{800} \) &    0 &    0 &    0 &  -14 &   -3 &    0 &  &  &  &  &  &  &  &  \\
\( \Gamma_{802} \) &    0 &    0 &    0 &   -2 &    0 &    1 &   -1 &  &  &  &  &  &  &  \\
\( \Gamma_{803} \) &    0 &    0 &    0 &  -58 &    0 &    0 &    9 &    1 &  &  &  &  &  &  \\
\( \Gamma_{805} \) &    0 &    0 &    0 &    0 &    0 &    0 &    0 &    0 &    0 &  &  &  &  &  \\
\( \Gamma_{811} \) &    0 &   -1 &   20 &    0 &    0 &    0 &    0 &    0 &    0 &    0 &  &  &  &  \\
\( \Gamma_{812} \) &    0 &   -2 &   -8 &    0 &    0 &    0 &    0 &    0 &    0 &    0 &  -16 &  &  &  \\
\( \Gamma_{821} \) &    0 &    0 &   47 &    0 &    0 &    0 &    0 &    0 &    0 &    0 &    8 &   -4 &  &  \\
\( \Gamma_{822} \) &    0 &    0 &   -1 &    0 &    0 &    0 &    0 &    0 &    0 &    0 &    0 &    0 &   -1 &  \\
 & \( \Gamma_{130} \) & \( \Gamma_{132} \) & \( \Gamma_{136} \) & \( \Gamma_{151} \) & \( \Gamma_{152} \) & \( \Gamma_{167} \) & \( \Gamma_{800} \) & \( \Gamma_{802} \) & \( \Gamma_{803} \) & \( \Gamma_{805} \) & \( \Gamma_{811} \) & \( \Gamma_{812} \) & \( \Gamma_{821} \) & \( \Gamma_{822} \)
\\\hline
\end{tabular}
\end{center}
\end{envsmall}
\ifhevea\else
\end{center}
\end{minipage}
\fi
\end{center}
\ifhevea\end{table}\fi
\ifhevea\begin{table}\fi
\begin{center}
\ifhevea
\caption{Basis quantities correlation coefficients in percent, subtable 9.\label{tab:tau:br-fit-corr9}}%
\else
\begin{minipage}{\linewidth}
\begin{center}
\captionof{table}{Basis quantities correlation coefficients in percent, subtable 9.}\label{tab:tau:br-fit-corr9}%
\fi
\begin{envsmall}
\begin{center}
\begin{tabular}{rrrrrrrrrrrrrrr}
\hline
\( \Gamma_{831} \) &    0 &    0 &   39 &    0 &    0 &    0 &    0 &    0 &    0 &    0 &   14 &   -4 &   39 &   -1 \\
\( \Gamma_{832} \) &    0 &    0 &    3 &    0 &    0 &    0 &    0 &    0 &    0 &    0 &    2 &    0 &    3 &    0 \\
\( \Gamma_{833} \) &    0 &    0 &   -1 &    0 &    0 &    0 &    0 &    0 &    0 &    0 &    0 &    0 &   -1 &    0 \\
\( \Gamma_{920} \) &    0 &    0 &   21 &    0 &    0 &    0 &    0 &    0 &    0 &    0 &    3 &   -2 &   35 &   -1 \\
\( \Gamma_{945} \) &    0 &   -1 &   25 &    0 &    0 &    0 &    0 &    0 &    0 &    0 &   10 &  -11 &   10 &    0 \\
 & \( \Gamma_{130} \) & \( \Gamma_{132} \) & \( \Gamma_{136} \) & \( \Gamma_{151} \) & \( \Gamma_{152} \) & \( \Gamma_{167} \) & \( \Gamma_{800} \) & \( \Gamma_{802} \) & \( \Gamma_{803} \) & \( \Gamma_{805} \) & \( \Gamma_{811} \) & \( \Gamma_{812} \) & \( \Gamma_{821} \) & \( \Gamma_{822} \)
\\\hline
\end{tabular}
\end{center}
\end{envsmall}
\ifhevea\else
\end{center}
\end{minipage}
\fi
\end{center}
\ifhevea\end{table}\fi
\ifhevea\begin{table}\fi
\begin{center}
\ifhevea
\caption{Basis quantities correlation coefficients in percent, subtable 10.\label{tab:tau:br-fit-corr10}}%
\else
\begin{minipage}{\linewidth}
\begin{center}
\captionof{table}{Basis quantities correlation coefficients in percent, subtable 10.}\label{tab:tau:br-fit-corr10}%
\fi
\begin{envsmall}
\begin{center}
\begin{tabular}{rrrrrr}
\hline
\( \Gamma_{832} \) &   -2 &  &  &  &  \\
\( \Gamma_{833} \) &   -1 &   -1 &  &  &  \\
\( \Gamma_{920} \) &   17 &    1 &    0 &  &  \\
\( \Gamma_{945} \) &   17 &    2 &    0 &    4 &  \\
 & \( \Gamma_{831} \) & \( \Gamma_{832} \) & \( \Gamma_{833} \) & \( \Gamma_{920} \) & \( \Gamma_{945} \)
\\\hline
\end{tabular}
\end{center}
\end{envsmall}
\ifhevea\else
\end{center}
\end{minipage}
\fi
\end{center}
\ifhevea\end{table}\fi}%
\htconstrdef{Gamma1.c}{\Gamma_{1}}{\Gamma_{3} + \Gamma_{5} + \Gamma_{9} + \Gamma_{10} + \Gamma_{14} + \Gamma_{16} + \Gamma_{20} + \Gamma_{23} + \Gamma_{27} + \Gamma_{28} + \Gamma_{30} + \Gamma_{35} + \Gamma_{40} + \Gamma_{44} + \Gamma_{37} + \Gamma_{42} + \Gamma_{47} + \Gamma_{48} + \Gamma_{804} + \Gamma_{50} + \Gamma_{51} + \Gamma_{806} + \Gamma_{126}\cdot{}\Gamma_{\eta\to\text{neutral}} + \Gamma_{128}\cdot{}\Gamma_{\eta\to\text{neutral}} + \Gamma_{130}\cdot{}\Gamma_{\eta\to\text{neutral}} + \Gamma_{132}\cdot{}\Gamma_{\eta\to\text{neutral}} + \Gamma_{800}\cdot{}\Gamma_{\omega\to\pi^0\gamma} + \Gamma_{151}\cdot{}\Gamma_{\omega\to\pi^0\gamma} + \Gamma_{152}\cdot{}\Gamma_{\omega\to\pi^0\gamma} + \Gamma_{167}\cdot{}\Gamma_{\phi\to K_S K_L}}{\Gamma_{3} + \Gamma_{5} + \Gamma_{9} + \Gamma_{10} + \Gamma_{14} + \Gamma_{16}  \\ 
  {}& + \Gamma_{20} + \Gamma_{23} + \Gamma_{27} + \Gamma_{28} + \Gamma_{30} + \Gamma_{35}  \\ 
  {}& + \Gamma_{40} + \Gamma_{44} + \Gamma_{37} + \Gamma_{42} + \Gamma_{47} + \Gamma_{48}  \\ 
  {}& + \Gamma_{804} + \Gamma_{50} + \Gamma_{51} + \Gamma_{806} + \Gamma_{126}\cdot{}\Gamma_{\eta\to\text{neutral}}  \\ 
  {}& + \Gamma_{128}\cdot{}\Gamma_{\eta\to\text{neutral}} + \Gamma_{130}\cdot{}\Gamma_{\eta\to\text{neutral}} + \Gamma_{132}\cdot{}\Gamma_{\eta\to\text{neutral}}  \\ 
  {}& + \Gamma_{800}\cdot{}\Gamma_{\omega\to\pi^0\gamma} + \Gamma_{151}\cdot{}\Gamma_{\omega\to\pi^0\gamma} + \Gamma_{152}\cdot{}\Gamma_{\omega\to\pi^0\gamma}  \\ 
  {}& + \Gamma_{167}\cdot{}\Gamma_{\phi\to K_S K_L}}%
\htconstrdef{Gamma2.c}{\Gamma_{2}}{\Gamma_{3} + \Gamma_{5} + \Gamma_{9} + \Gamma_{10} + \Gamma_{14} + \Gamma_{16} + \Gamma_{20} + \Gamma_{23} + \Gamma_{27} + \Gamma_{28} + \Gamma_{30} + \Gamma_{35}\cdot{}(\Gamma_{<\bar{K}^0|K_S>}\cdot{}\Gamma_{K_S\to\pi^0\pi^0}+\Gamma_{<\bar{K}^0|K_L>}) + \Gamma_{40}\cdot{}(\Gamma_{<\bar{K}^0|K_S>}\cdot{}\Gamma_{K_S\to\pi^0\pi^0}+\Gamma_{<\bar{K}^0|K_L>}) + \Gamma_{44}\cdot{}(\Gamma_{<\bar{K}^0|K_S>}\cdot{}\Gamma_{K_S\to\pi^0\pi^0}+\Gamma_{<\bar{K}^0|K_L>}) + \Gamma_{37}\cdot{}(\Gamma_{<\bar{K}^0|K_S>}\cdot{}\Gamma_{K_S\to\pi^0\pi^0}+\Gamma_{<\bar{K}^0|K_L>}) + \Gamma_{42}\cdot{}(\Gamma_{<\bar{K}^0|K_S>}\cdot{}\Gamma_{K_S\to\pi^0\pi^0}+\Gamma_{<\bar{K}^0|K_L>}) + \Gamma_{47}\cdot{}(\Gamma_{K_S\to\pi^0\pi^0}\cdot{}\Gamma_{K_S\to\pi^0\pi^0}) + \Gamma_{48}\cdot{}\Gamma_{K_S\to\pi^0\pi^0} + \Gamma_{804} + \Gamma_{50}\cdot{}(\Gamma_{K_S\to\pi^0\pi^0}\cdot{}\Gamma_{K_S\to\pi^0\pi^0}) + \Gamma_{51}\cdot{}\Gamma_{K_S\to\pi^0\pi^0} + \Gamma_{806} + \Gamma_{126}\cdot{}\Gamma_{\eta\to\text{neutral}} + \Gamma_{128}\cdot{}\Gamma_{\eta\to\text{neutral}} + \Gamma_{130}\cdot{}\Gamma_{\eta\to\text{neutral}} + \Gamma_{132}\cdot{}(\Gamma_{\eta\to\text{neutral}}\cdot{}(\Gamma_{<\bar{K}^0|K_S>}\cdot{}\Gamma_{K_S\to\pi^0\pi^0}+\Gamma_{<\bar{K}^0|K_L>})) + \Gamma_{800}\cdot{}\Gamma_{\omega\to\pi^0\gamma} + \Gamma_{151}\cdot{}\Gamma_{\omega\to\pi^0\gamma} + \Gamma_{152}\cdot{}\Gamma_{\omega\to\pi^0\gamma} + \Gamma_{167}\cdot{}(\Gamma_{\phi\to K_S K_L}\cdot{}\Gamma_{K_S\to\pi^0\pi^0})}{\Gamma_{3} + \Gamma_{5} + \Gamma_{9} + \Gamma_{10} + \Gamma_{14} + \Gamma_{16}  \\ 
  {}& + \Gamma_{20} + \Gamma_{23} + \Gamma_{27} + \Gamma_{28} + \Gamma_{30} + \Gamma_{35}\cdot{}(\Gamma_{<\bar{K}^0|K_S>}\cdot{}\Gamma_{K_S\to\pi^0\pi^0} \\ 
  {}& +\Gamma_{<\bar{K}^0|K_L>}) + \Gamma_{40}\cdot{}(\Gamma_{<\bar{K}^0|K_S>}\cdot{}\Gamma_{K_S\to\pi^0\pi^0}+\Gamma_{<\bar{K}^0|K_L>}) + \Gamma_{44}\cdot{}(\Gamma_{<\bar{K}^0|K_S>}\cdot{}\Gamma_{K_S\to\pi^0\pi^0} \\ 
  {}& +\Gamma_{<\bar{K}^0|K_L>}) + \Gamma_{37}\cdot{}(\Gamma_{<\bar{K}^0|K_S>}\cdot{}\Gamma_{K_S\to\pi^0\pi^0}+\Gamma_{<\bar{K}^0|K_L>}) + \Gamma_{42}\cdot{}(\Gamma_{<\bar{K}^0|K_S>}\cdot{}\Gamma_{K_S\to\pi^0\pi^0} \\ 
  {}& +\Gamma_{<\bar{K}^0|K_L>}) + \Gamma_{47}\cdot{}(\Gamma_{K_S\to\pi^0\pi^0}\cdot{}\Gamma_{K_S\to\pi^0\pi^0}) + \Gamma_{48}\cdot{}\Gamma_{K_S\to\pi^0\pi^0}  \\ 
  {}& + \Gamma_{804} + \Gamma_{50}\cdot{}(\Gamma_{K_S\to\pi^0\pi^0}\cdot{}\Gamma_{K_S\to\pi^0\pi^0}) + \Gamma_{51}\cdot{}\Gamma_{K_S\to\pi^0\pi^0}  \\ 
  {}& + \Gamma_{806} + \Gamma_{126}\cdot{}\Gamma_{\eta\to\text{neutral}} + \Gamma_{128}\cdot{}\Gamma_{\eta\to\text{neutral}} + \Gamma_{130}\cdot{}\Gamma_{\eta\to\text{neutral}}  \\ 
  {}& + \Gamma_{132}\cdot{}(\Gamma_{\eta\to\text{neutral}}\cdot{}(\Gamma_{<\bar{K}^0|K_S>}\cdot{}\Gamma_{K_S\to\pi^0\pi^0}+\Gamma_{<\bar{K}^0|K_L>})) + \Gamma_{800}\cdot{}\Gamma_{\omega\to\pi^0\gamma}  \\ 
  {}& + \Gamma_{151}\cdot{}\Gamma_{\omega\to\pi^0\gamma} + \Gamma_{152}\cdot{}\Gamma_{\omega\to\pi^0\gamma} + \Gamma_{167}\cdot{}(\Gamma_{\phi\to K_S K_L}\cdot{}\Gamma_{K_S\to\pi^0\pi^0})}%
\htconstrdef{Gamma3by5.c}{\frac{\Gamma_{3}}{\Gamma_{5}}}{\frac{\Gamma_{3}}{\Gamma_{5}}}{\frac{\Gamma_{3}}{\Gamma_{5}}}%
\htconstrdef{Gamma7.c}{\Gamma_{7}}{\Gamma_{35}\cdot{}\Gamma_{<\bar{K}^0|K_L>} + \Gamma_{9} + \Gamma_{804} + \Gamma_{37}\cdot{}\Gamma_{<K^0|K_L>} + \Gamma_{10}}{\Gamma_{35}\cdot{}\Gamma_{<\bar{K}^0|K_L>} + \Gamma_{9} + \Gamma_{804} + \Gamma_{37}\cdot{}\Gamma_{<K^0|K_L>}  \\ 
  {}& + \Gamma_{10}}%
\htconstrdef{Gamma8.c}{\Gamma_{8}}{\Gamma_{9} + \Gamma_{10}}{\Gamma_{9} + \Gamma_{10}}%
\htconstrdef{Gamma8by5.c}{\frac{\Gamma_{8}}{\Gamma_{5}}}{\frac{\Gamma_{8}}{\Gamma_{5}}}{\frac{\Gamma_{8}}{\Gamma_{5}}}%
\htconstrdef{Gamma9by5.c}{\frac{\Gamma_{9}}{\Gamma_{5}}}{\frac{\Gamma_{9}}{\Gamma_{5}}}{\frac{\Gamma_{9}}{\Gamma_{5}}}%
\htconstrdef{Gamma10by5.c}{\frac{\Gamma_{10}}{\Gamma_{5}}}{\frac{\Gamma_{10}}{\Gamma_{5}}}{\frac{\Gamma_{10}}{\Gamma_{5}}}%
\htconstrdef{Gamma10by9.c}{\frac{\Gamma_{10}}{\Gamma_{9}}}{\frac{\Gamma_{10}}{\Gamma_{9}}}{\frac{\Gamma_{10}}{\Gamma_{9}}}%
\htconstrdef{Gamma11.c}{\Gamma_{11}}{\Gamma_{14} + \Gamma_{16} + \Gamma_{20} + \Gamma_{23} + \Gamma_{27} + \Gamma_{28} + \Gamma_{30} + \Gamma_{35}\cdot{}(\Gamma_{<K^0|K_S>}\cdot{}\Gamma_{K_S\to\pi^0\pi^0}) + \Gamma_{37}\cdot{}(\Gamma_{<K^0|K_S>}\cdot{}\Gamma_{K_S\to\pi^0\pi^0}) + \Gamma_{40}\cdot{}(\Gamma_{<K^0|K_S>}\cdot{}\Gamma_{K_S\to\pi^0\pi^0}) + \Gamma_{42}\cdot{}(\Gamma_{<K^0|K_S>}\cdot{}\Gamma_{K_S\to\pi^0\pi^0}) + \Gamma_{47}\cdot{}(\Gamma_{K_S\to\pi^0\pi^0}\cdot{}\Gamma_{K_S\to\pi^0\pi^0}) + \Gamma_{50}\cdot{}(\Gamma_{K_S\to\pi^0\pi^0}\cdot{}\Gamma_{K_S\to\pi^0\pi^0}) + \Gamma_{126}\cdot{}\Gamma_{\eta\to\text{neutral}} + \Gamma_{128}\cdot{}\Gamma_{\eta\to\text{neutral}} + \Gamma_{130}\cdot{}\Gamma_{\eta\to\text{neutral}} + \Gamma_{132}\cdot{}(\Gamma_{<K^0|K_S>}\cdot{}\Gamma_{K_S\to\pi^0\pi^0}\cdot{}\Gamma_{\eta\to\text{neutral}}) + \Gamma_{151}\cdot{}\Gamma_{\omega\to\pi^0\gamma} + \Gamma_{152}\cdot{}\Gamma_{\omega\to\pi^0\gamma} + \Gamma_{800}\cdot{}\Gamma_{\omega\to\pi^0\gamma}}{\Gamma_{14} + \Gamma_{16} + \Gamma_{20} + \Gamma_{23} + \Gamma_{27} + \Gamma_{28}  \\ 
  {}& + \Gamma_{30} + \Gamma_{35}\cdot{}(\Gamma_{<K^0|K_S>}\cdot{}\Gamma_{K_S\to\pi^0\pi^0}) + \Gamma_{37}\cdot{}(\Gamma_{<K^0|K_S>}\cdot{}\Gamma_{K_S\to\pi^0\pi^0})  \\ 
  {}& + \Gamma_{40}\cdot{}(\Gamma_{<K^0|K_S>}\cdot{}\Gamma_{K_S\to\pi^0\pi^0}) + \Gamma_{42}\cdot{}(\Gamma_{<K^0|K_S>}\cdot{}\Gamma_{K_S\to\pi^0\pi^0})  \\ 
  {}& + \Gamma_{47}\cdot{}(\Gamma_{K_S\to\pi^0\pi^0}\cdot{}\Gamma_{K_S\to\pi^0\pi^0}) + \Gamma_{50}\cdot{}(\Gamma_{K_S\to\pi^0\pi^0}\cdot{}\Gamma_{K_S\to\pi^0\pi^0})  \\ 
  {}& + \Gamma_{126}\cdot{}\Gamma_{\eta\to\text{neutral}} + \Gamma_{128}\cdot{}\Gamma_{\eta\to\text{neutral}} + \Gamma_{130}\cdot{}\Gamma_{\eta\to\text{neutral}}  \\ 
  {}& + \Gamma_{132}\cdot{}(\Gamma_{<K^0|K_S>}\cdot{}\Gamma_{K_S\to\pi^0\pi^0}\cdot{}\Gamma_{\eta\to\text{neutral}}) + \Gamma_{151}\cdot{}\Gamma_{\omega\to\pi^0\gamma}  \\ 
  {}& + \Gamma_{152}\cdot{}\Gamma_{\omega\to\pi^0\gamma} + \Gamma_{800}\cdot{}\Gamma_{\omega\to\pi^0\gamma}}%
\htconstrdef{Gamma12.c}{\Gamma_{12}}{\Gamma_{128}\cdot{}\Gamma_{\eta\to3\pi^0} + \Gamma_{30} + \Gamma_{23} + \Gamma_{28} + \Gamma_{14} + \Gamma_{16} + \Gamma_{20} + \Gamma_{27} + \Gamma_{126}\cdot{}\Gamma_{\eta\to3\pi^0} + \Gamma_{130}\cdot{}\Gamma_{\eta\to3\pi^0}}{\Gamma_{128}\cdot{}\Gamma_{\eta\to3\pi^0} + \Gamma_{30} + \Gamma_{23} + \Gamma_{28} + \Gamma_{14}  \\ 
  {}& + \Gamma_{16} + \Gamma_{20} + \Gamma_{27} + \Gamma_{126}\cdot{}\Gamma_{\eta\to3\pi^0} + \Gamma_{130}\cdot{}\Gamma_{\eta\to3\pi^0}}%
\htconstrdef{Gamma13.c}{\Gamma_{13}}{\Gamma_{14} + \Gamma_{16}}{\Gamma_{14} + \Gamma_{16}}%
\htconstrdef{Gamma17.c}{\Gamma_{17}}{\Gamma_{128}\cdot{}\Gamma_{\eta\to3\pi^0} + \Gamma_{30} + \Gamma_{23} + \Gamma_{28} + \Gamma_{35}\cdot{}(\Gamma_{<K^0|K_S>}\cdot{}\Gamma_{K_S\to\pi^0\pi^0}) + \Gamma_{40}\cdot{}(\Gamma_{<K^0|K_S>}\cdot{}\Gamma_{K_S\to\pi^0\pi^0}) + \Gamma_{42}\cdot{}(\Gamma_{<K^0|K_S>}\cdot{}\Gamma_{K_S\to\pi^0\pi^0}) + \Gamma_{20} + \Gamma_{27} + \Gamma_{47}\cdot{}(\Gamma_{K_S\to\pi^0\pi^0}\cdot{}\Gamma_{K_S\to\pi^0\pi^0}) + \Gamma_{50}\cdot{}(\Gamma_{K_S\to\pi^0\pi^0}\cdot{}\Gamma_{K_S\to\pi^0\pi^0}) + \Gamma_{126}\cdot{}\Gamma_{\eta\to3\pi^0} + \Gamma_{37}\cdot{}(\Gamma_{<K^0|K_S>}\cdot{}\Gamma_{K_S\to\pi^0\pi^0}) + \Gamma_{130}\cdot{}\Gamma_{\eta\to3\pi^0}}{\Gamma_{128}\cdot{}\Gamma_{\eta\to3\pi^0} + \Gamma_{30} + \Gamma_{23} + \Gamma_{28} + \Gamma_{35}\cdot{}(\Gamma_{<K^0|K_S>}\cdot{}\Gamma_{K_S\to\pi^0\pi^0})  \\ 
  {}& + \Gamma_{40}\cdot{}(\Gamma_{<K^0|K_S>}\cdot{}\Gamma_{K_S\to\pi^0\pi^0}) + \Gamma_{42}\cdot{}(\Gamma_{<K^0|K_S>}\cdot{}\Gamma_{K_S\to\pi^0\pi^0})  \\ 
  {}& + \Gamma_{20} + \Gamma_{27} + \Gamma_{47}\cdot{}(\Gamma_{K_S\to\pi^0\pi^0}\cdot{}\Gamma_{K_S\to\pi^0\pi^0}) + \Gamma_{50}\cdot{}(\Gamma_{K_S\to\pi^0\pi^0}\cdot{}\Gamma_{K_S\to\pi^0\pi^0})  \\ 
  {}& + \Gamma_{126}\cdot{}\Gamma_{\eta\to3\pi^0} + \Gamma_{37}\cdot{}(\Gamma_{<K^0|K_S>}\cdot{}\Gamma_{K_S\to\pi^0\pi^0}) + \Gamma_{130}\cdot{}\Gamma_{\eta\to3\pi^0}}%
\htconstrdef{Gamma18.c}{\Gamma_{18}}{\Gamma_{23} + \Gamma_{35}\cdot{}(\Gamma_{<K^0|K_S>}\cdot{}\Gamma_{K_S\to\pi^0\pi^0}) + \Gamma_{20} + \Gamma_{37}\cdot{}(\Gamma_{<K^0|K_S>}\cdot{}\Gamma_{K_S\to\pi^0\pi^0})}{\Gamma_{23} + \Gamma_{35}\cdot{}(\Gamma_{<K^0|K_S>}\cdot{}\Gamma_{K_S\to\pi^0\pi^0}) + \Gamma_{20} + \Gamma_{37}\cdot{}(\Gamma_{<K^0|K_S>}\cdot{}\Gamma_{K_S\to\pi^0\pi^0})}%
\htconstrdef{Gamma19.c}{\Gamma_{19}}{\Gamma_{23} + \Gamma_{20}}{\Gamma_{23} + \Gamma_{20}}%
\htconstrdef{Gamma19by13.c}{\frac{\Gamma_{19}}{\Gamma_{13}}}{\frac{\Gamma_{19}}{\Gamma_{13}}}{\frac{\Gamma_{19}}{\Gamma_{13}}}%
\htconstrdef{Gamma24.c}{\Gamma_{24}}{\Gamma_{27} + \Gamma_{28} + \Gamma_{30} + \Gamma_{40}\cdot{}(\Gamma_{<K^0|K_S>}\cdot{}\Gamma_{K_S\to\pi^0\pi^0}) + \Gamma_{42}\cdot{}(\Gamma_{<K^0|K_S>}\cdot{}\Gamma_{K_S\to\pi^0\pi^0}) + \Gamma_{47}\cdot{}(\Gamma_{K_S\to\pi^0\pi^0}\cdot{}\Gamma_{K_S\to\pi^0\pi^0}) + \Gamma_{50}\cdot{}(\Gamma_{K_S\to\pi^0\pi^0}\cdot{}\Gamma_{K_S\to\pi^0\pi^0}) + \Gamma_{126}\cdot{}\Gamma_{\eta\to3\pi^0} + \Gamma_{128}\cdot{}\Gamma_{\eta\to3\pi^0} + \Gamma_{130}\cdot{}\Gamma_{\eta\to3\pi^0} + \Gamma_{132}\cdot{}(\Gamma_{<K^0|K_S>}\cdot{}\Gamma_{K_S\to\pi^0\pi^0}\cdot{}\Gamma_{\eta\to3\pi^0})}{\Gamma_{27} + \Gamma_{28} + \Gamma_{30} + \Gamma_{40}\cdot{}(\Gamma_{<K^0|K_S>}\cdot{}\Gamma_{K_S\to\pi^0\pi^0})  \\ 
  {}& + \Gamma_{42}\cdot{}(\Gamma_{<K^0|K_S>}\cdot{}\Gamma_{K_S\to\pi^0\pi^0}) + \Gamma_{47}\cdot{}(\Gamma_{K_S\to\pi^0\pi^0}\cdot{}\Gamma_{K_S\to\pi^0\pi^0})  \\ 
  {}& + \Gamma_{50}\cdot{}(\Gamma_{K_S\to\pi^0\pi^0}\cdot{}\Gamma_{K_S\to\pi^0\pi^0}) + \Gamma_{126}\cdot{}\Gamma_{\eta\to3\pi^0} + \Gamma_{128}\cdot{}\Gamma_{\eta\to3\pi^0}  \\ 
  {}& + \Gamma_{130}\cdot{}\Gamma_{\eta\to3\pi^0} + \Gamma_{132}\cdot{}(\Gamma_{<K^0|K_S>}\cdot{}\Gamma_{K_S\to\pi^0\pi^0}\cdot{}\Gamma_{\eta\to3\pi^0})}%
\htconstrdef{Gamma25.c}{\Gamma_{25}}{\Gamma_{128}\cdot{}\Gamma_{\eta\to3\pi^0} + \Gamma_{30} + \Gamma_{28} + \Gamma_{27} + \Gamma_{126}\cdot{}\Gamma_{\eta\to3\pi^0} + \Gamma_{130}\cdot{}\Gamma_{\eta\to3\pi^0}}{\Gamma_{128}\cdot{}\Gamma_{\eta\to3\pi^0} + \Gamma_{30} + \Gamma_{28} + \Gamma_{27} + \Gamma_{126}\cdot{}\Gamma_{\eta\to3\pi^0}  \\ 
  {}& + \Gamma_{130}\cdot{}\Gamma_{\eta\to3\pi^0}}%
\htconstrdef{Gamma26.c}{\Gamma_{26}}{\Gamma_{128}\cdot{}\Gamma_{\eta\to3\pi^0} + \Gamma_{28} + \Gamma_{40}\cdot{}(\Gamma_{<K^0|K_S>}\cdot{}\Gamma_{K_S\to\pi^0\pi^0}) + \Gamma_{42}\cdot{}(\Gamma_{<K^0|K_S>}\cdot{}\Gamma_{K_S\to\pi^0\pi^0}) + \Gamma_{27}}{\Gamma_{128}\cdot{}\Gamma_{\eta\to3\pi^0} + \Gamma_{28} + \Gamma_{40}\cdot{}(\Gamma_{<K^0|K_S>}\cdot{}\Gamma_{K_S\to\pi^0\pi^0})  \\ 
  {}& + \Gamma_{42}\cdot{}(\Gamma_{<K^0|K_S>}\cdot{}\Gamma_{K_S\to\pi^0\pi^0}) + \Gamma_{27}}%
\htconstrdef{Gamma26by13.c}{\frac{\Gamma_{26}}{\Gamma_{13}}}{\frac{\Gamma_{26}}{\Gamma_{13}}}{\frac{\Gamma_{26}}{\Gamma_{13}}}%
\htconstrdef{Gamma29.c}{\Gamma_{29}}{\Gamma_{30} + \Gamma_{126}\cdot{}\Gamma_{\eta\to3\pi^0} + \Gamma_{130}\cdot{}\Gamma_{\eta\to3\pi^0}}{\Gamma_{30} + \Gamma_{126}\cdot{}\Gamma_{\eta\to3\pi^0} + \Gamma_{130}\cdot{}\Gamma_{\eta\to3\pi^0}}%
\htconstrdef{Gamma31.c}{\Gamma_{31}}{\Gamma_{128}\cdot{}\Gamma_{\eta\to\text{neutral}} + \Gamma_{23} + \Gamma_{28} + \Gamma_{42} + \Gamma_{16} + \Gamma_{37} + \Gamma_{10} + \Gamma_{167}\cdot{}(\Gamma_{\phi\to K_S K_L}\cdot{}\Gamma_{K_S\to\pi^0\pi^0})}{\Gamma_{128}\cdot{}\Gamma_{\eta\to\text{neutral}} + \Gamma_{23} + \Gamma_{28} + \Gamma_{42} + \Gamma_{16}  \\ 
  {}& + \Gamma_{37} + \Gamma_{10} + \Gamma_{167}\cdot{}(\Gamma_{\phi\to K_S K_L}\cdot{}\Gamma_{K_S\to\pi^0\pi^0})}%
\htconstrdef{Gamma32.c}{\Gamma_{32}}{\Gamma_{16} + \Gamma_{23} + \Gamma_{28} + \Gamma_{37} + \Gamma_{42} + \Gamma_{128}\cdot{}\Gamma_{\eta\to\text{neutral}} + \Gamma_{130}\cdot{}\Gamma_{\eta\to\text{neutral}} + \Gamma_{167}\cdot{}(\Gamma_{\phi\to K_S K_L}\cdot{}\Gamma_{K_S\to\pi^0\pi^0})}{\Gamma_{16} + \Gamma_{23} + \Gamma_{28} + \Gamma_{37} + \Gamma_{42} + \Gamma_{128}\cdot{}\Gamma_{\eta\to\text{neutral}}  \\ 
  {}& + \Gamma_{130}\cdot{}\Gamma_{\eta\to\text{neutral}} + \Gamma_{167}\cdot{}(\Gamma_{\phi\to K_S K_L}\cdot{}\Gamma_{K_S\to\pi^0\pi^0})}%
\htconstrdef{Gamma33.c}{\Gamma_{33}}{\Gamma_{35}\cdot{}\Gamma_{<\bar{K}^0|K_S>} + \Gamma_{40}\cdot{}\Gamma_{<\bar{K}^0|K_S>} + \Gamma_{42}\cdot{}\Gamma_{<K^0|K_S>} + \Gamma_{47} + \Gamma_{48} + \Gamma_{50} + \Gamma_{51} + \Gamma_{37}\cdot{}\Gamma_{<K^0|K_S>} + \Gamma_{132}\cdot{}(\Gamma_{<\bar{K}^0|K_S>}\cdot{}\Gamma_{\eta\to\text{neutral}}) + \Gamma_{44}\cdot{}\Gamma_{<\bar{K}^0|K_S>} + \Gamma_{167}\cdot{}\Gamma_{\phi\to K_S K_L}}{\Gamma_{35}\cdot{}\Gamma_{<\bar{K}^0|K_S>} + \Gamma_{40}\cdot{}\Gamma_{<\bar{K}^0|K_S>} + \Gamma_{42}\cdot{}\Gamma_{<K^0|K_S>}  \\ 
  {}& + \Gamma_{47} + \Gamma_{48} + \Gamma_{50} + \Gamma_{51} + \Gamma_{37}\cdot{}\Gamma_{<K^0|K_S>}  \\ 
  {}& + \Gamma_{132}\cdot{}(\Gamma_{<\bar{K}^0|K_S>}\cdot{}\Gamma_{\eta\to\text{neutral}}) + \Gamma_{44}\cdot{}\Gamma_{<\bar{K}^0|K_S>} + \Gamma_{167}\cdot{}\Gamma_{\phi\to K_S K_L}}%
\htconstrdef{Gamma34.c}{\Gamma_{34}}{\Gamma_{35} + \Gamma_{37}}{\Gamma_{35} + \Gamma_{37}}%
\htconstrdef{Gamma38.c}{\Gamma_{38}}{\Gamma_{42} + \Gamma_{37}}{\Gamma_{42} + \Gamma_{37}}%
\htconstrdef{Gamma39.c}{\Gamma_{39}}{\Gamma_{40} + \Gamma_{42}}{\Gamma_{40} + \Gamma_{42}}%
\htconstrdef{Gamma43.c}{\Gamma_{43}}{\Gamma_{40} + \Gamma_{44}}{\Gamma_{40} + \Gamma_{44}}%
\htconstrdef{Gamma46.c}{\Gamma_{46}}{\Gamma_{48} + \Gamma_{47} + \Gamma_{804}}{\Gamma_{48} + \Gamma_{47} + \Gamma_{804}}%
\htconstrdef{Gamma49.c}{\Gamma_{49}}{\Gamma_{50} + \Gamma_{51} + \Gamma_{806}}{\Gamma_{50} + \Gamma_{51} + \Gamma_{806}}%
\htconstrdef{Gamma54.c}{\Gamma_{54}}{\Gamma_{35}\cdot{}(\Gamma_{<K^0|K_S>}\cdot{}\Gamma_{K_S\to\pi^+\pi^-}) + \Gamma_{37}\cdot{}(\Gamma_{<K^0|K_S>}\cdot{}\Gamma_{K_S\to\pi^+\pi^-}) + \Gamma_{40}\cdot{}(\Gamma_{<K^0|K_S>}\cdot{}\Gamma_{K_S\to\pi^+\pi^-}) + \Gamma_{42}\cdot{}(\Gamma_{<K^0|K_S>}\cdot{}\Gamma_{K_S\to\pi^+\pi^-}) + \Gamma_{47}\cdot{}(2\cdot{}\Gamma_{K_S\to\pi^+\pi^-}\cdot{}\Gamma_{K_S\to\pi^0\pi^0}) + \Gamma_{48}\cdot{}\Gamma_{K_S\to\pi^+\pi^-} + \Gamma_{50}\cdot{}(2\cdot{}\Gamma_{K_S\to\pi^+\pi^-}\cdot{}\Gamma_{K_S\to\pi^0\pi^0}) + \Gamma_{51}\cdot{}\Gamma_{K_S\to\pi^+\pi^-} + \Gamma_{53}\cdot{}(\Gamma_{<\bar{K}^0|K_S>}\cdot{}\Gamma_{K_S\to\pi^0\pi^0}+\Gamma_{<\bar{K}^0|K_L>}) + \Gamma_{62} + \Gamma_{70} + \Gamma_{77} + \Gamma_{78} + \Gamma_{93} + \Gamma_{94} + \Gamma_{126}\cdot{}\Gamma_{\eta\to\text{charged}} + \Gamma_{128}\cdot{}\Gamma_{\eta\to\text{charged}} + \Gamma_{130}\cdot{}\Gamma_{\eta\to\text{charged}} + \Gamma_{132}\cdot{}(\Gamma_{<\bar{K}^0|K_L>}\cdot{}\Gamma_{\eta\to\pi^+\pi^-\pi^0} + \Gamma_{<\bar{K}^0|K_S>}\cdot{}\Gamma_{K_S\to\pi^0\pi^0}\cdot{}\Gamma_{\eta\to\pi^+\pi^-\pi^0} + \Gamma_{<\bar{K}^0|K_S>}\cdot{}\Gamma_{K_S\to\pi^+\pi^-}\cdot{}\Gamma_{\eta\to3\pi^0}) + \Gamma_{151}\cdot{}(\Gamma_{\omega\to\pi^+\pi^-\pi^0}+\Gamma_{\omega\to\pi^+\pi^-}) + \Gamma_{152}\cdot{}(\Gamma_{\omega\to\pi^+\pi^-\pi^0}+\Gamma_{\omega\to\pi^+\pi^-}) + \Gamma_{167}\cdot{}(\Gamma_{\phi\to K^+K^-} + \Gamma_{\phi\to K_S K_L}\cdot{}\Gamma_{K_S\to\pi^+\pi^-}) + \Gamma_{802} + \Gamma_{803} + \Gamma_{800}\cdot{}(\Gamma_{\omega\to\pi^+\pi^-\pi^0}+\Gamma_{\omega\to\pi^+\pi^-})}{\Gamma_{35}\cdot{}(\Gamma_{<K^0|K_S>}\cdot{}\Gamma_{K_S\to\pi^+\pi^-}) + \Gamma_{37}\cdot{}(\Gamma_{<K^0|K_S>}\cdot{}\Gamma_{K_S\to\pi^+\pi^-})  \\ 
  {}& + \Gamma_{40}\cdot{}(\Gamma_{<K^0|K_S>}\cdot{}\Gamma_{K_S\to\pi^+\pi^-}) + \Gamma_{42}\cdot{}(\Gamma_{<K^0|K_S>}\cdot{}\Gamma_{K_S\to\pi^+\pi^-})  \\ 
  {}& + \Gamma_{47}\cdot{}(2\cdot{}\Gamma_{K_S\to\pi^+\pi^-}\cdot{}\Gamma_{K_S\to\pi^0\pi^0}) + \Gamma_{48}\cdot{}\Gamma_{K_S\to\pi^+\pi^-}  \\ 
  {}& + \Gamma_{50}\cdot{}(2\cdot{}\Gamma_{K_S\to\pi^+\pi^-}\cdot{}\Gamma_{K_S\to\pi^0\pi^0}) + \Gamma_{51}\cdot{}\Gamma_{K_S\to\pi^+\pi^-}  \\ 
  {}& + \Gamma_{53}\cdot{}(\Gamma_{<\bar{K}^0|K_S>}\cdot{}\Gamma_{K_S\to\pi^0\pi^0}+\Gamma_{<\bar{K}^0|K_L>}) + \Gamma_{62} + \Gamma_{70}  \\ 
  {}& + \Gamma_{77} + \Gamma_{78} + \Gamma_{93} + \Gamma_{94} + \Gamma_{126}\cdot{}\Gamma_{\eta\to\text{charged}}  \\ 
  {}& + \Gamma_{128}\cdot{}\Gamma_{\eta\to\text{charged}} + \Gamma_{130}\cdot{}\Gamma_{\eta\to\text{charged}} + \Gamma_{132}\cdot{}(\Gamma_{<\bar{K}^0|K_L>}\cdot{}\Gamma_{\eta\to\pi^+\pi^-\pi^0}  \\ 
  {}& + \Gamma_{<\bar{K}^0|K_S>}\cdot{}\Gamma_{K_S\to\pi^0\pi^0}\cdot{}\Gamma_{\eta\to\pi^+\pi^-\pi^0} + \Gamma_{<\bar{K}^0|K_S>}\cdot{}\Gamma_{K_S\to\pi^+\pi^-}\cdot{}\Gamma_{\eta\to3\pi^0})  \\ 
  {}& + \Gamma_{151}\cdot{}(\Gamma_{\omega\to\pi^+\pi^-\pi^0}+\Gamma_{\omega\to\pi^+\pi^-}) + \Gamma_{152}\cdot{}(\Gamma_{\omega\to\pi^+\pi^-\pi^0}+\Gamma_{\omega\to\pi^+\pi^-})  \\ 
  {}& + \Gamma_{167}\cdot{}(\Gamma_{\phi\to K^+K^-} + \Gamma_{\phi\to K_S K_L}\cdot{}\Gamma_{K_S\to\pi^+\pi^-}) + \Gamma_{802} + \Gamma_{803}  \\ 
  {}& + \Gamma_{800}\cdot{}(\Gamma_{\omega\to\pi^+\pi^-\pi^0}+\Gamma_{\omega\to\pi^+\pi^-})}%
\htconstrdef{Gamma55.c}{\Gamma_{55}}{\Gamma_{128}\cdot{}\Gamma_{\eta\to\text{charged}} + \Gamma_{152}\cdot{}(\Gamma_{\omega\to\pi^+\pi^-\pi^0}+\Gamma_{\omega\to\pi^+\pi^-}) + \Gamma_{78} + \Gamma_{77} + \Gamma_{94} + \Gamma_{62} + \Gamma_{70} + \Gamma_{93} + \Gamma_{126}\cdot{}\Gamma_{\eta\to\text{charged}} + \Gamma_{802} + \Gamma_{803} + \Gamma_{800}\cdot{}(\Gamma_{\omega\to\pi^+\pi^-\pi^0}+\Gamma_{\omega\to\pi^+\pi^-}) + \Gamma_{151}\cdot{}(\Gamma_{\omega\to\pi^+\pi^-\pi^0}+\Gamma_{\omega\to\pi^+\pi^-}) + \Gamma_{130}\cdot{}\Gamma_{\eta\to\text{charged}} + \Gamma_{168}}{\Gamma_{128}\cdot{}\Gamma_{\eta\to\text{charged}} + \Gamma_{152}\cdot{}(\Gamma_{\omega\to\pi^+\pi^-\pi^0}+\Gamma_{\omega\to\pi^+\pi^-}) + \Gamma_{78}  \\ 
  {}& + \Gamma_{77} + \Gamma_{94} + \Gamma_{62} + \Gamma_{70} + \Gamma_{93} + \Gamma_{126}\cdot{}\Gamma_{\eta\to\text{charged}}  \\ 
  {}& + \Gamma_{802} + \Gamma_{803} + \Gamma_{800}\cdot{}(\Gamma_{\omega\to\pi^+\pi^-\pi^0}+\Gamma_{\omega\to\pi^+\pi^-}) + \Gamma_{151}\cdot{}(\Gamma_{\omega\to\pi^+\pi^-\pi^0} \\ 
  {}& +\Gamma_{\omega\to\pi^+\pi^-}) + \Gamma_{130}\cdot{}\Gamma_{\eta\to\text{charged}} + \Gamma_{168}}%
\htconstrdef{Gamma56.c}{\Gamma_{56}}{\Gamma_{35}\cdot{}(\Gamma_{<K^0|K_S>}\cdot{}\Gamma_{K_S\to\pi^+\pi^-}) + \Gamma_{62} + \Gamma_{93} + \Gamma_{37}\cdot{}(\Gamma_{<K^0|K_S>}\cdot{}\Gamma_{K_S\to\pi^+\pi^-}) + \Gamma_{802} + \Gamma_{800}\cdot{}\Gamma_{\omega\to\pi^+\pi^-} + \Gamma_{151}\cdot{}\Gamma_{\omega\to\pi^+\pi^-} + \Gamma_{168}}{\Gamma_{35}\cdot{}(\Gamma_{<K^0|K_S>}\cdot{}\Gamma_{K_S\to\pi^+\pi^-}) + \Gamma_{62} + \Gamma_{93} + \Gamma_{37}\cdot{}(\Gamma_{<K^0|K_S>}\cdot{}\Gamma_{K_S\to\pi^+\pi^-})  \\ 
  {}& + \Gamma_{802} + \Gamma_{800}\cdot{}\Gamma_{\omega\to\pi^+\pi^-} + \Gamma_{151}\cdot{}\Gamma_{\omega\to\pi^+\pi^-} + \Gamma_{168}}%
\htconstrdef{Gamma57.c}{\Gamma_{57}}{\Gamma_{62} + \Gamma_{93} + \Gamma_{802} + \Gamma_{800}\cdot{}\Gamma_{\omega\to\pi^+\pi^-} + \Gamma_{151}\cdot{}\Gamma_{\omega\to\pi^+\pi^-} + \Gamma_{167}\cdot{}\Gamma_{\phi\to K^+K^-}}{\Gamma_{62} + \Gamma_{93} + \Gamma_{802} + \Gamma_{800}\cdot{}\Gamma_{\omega\to\pi^+\pi^-} + \Gamma_{151}\cdot{}\Gamma_{\omega\to\pi^+\pi^-}  \\ 
  {}& + \Gamma_{167}\cdot{}\Gamma_{\phi\to K^+K^-}}%
\htconstrdef{Gamma57by55.c}{\frac{\Gamma_{57}}{\Gamma_{55}}}{\frac{\Gamma_{57}}{\Gamma_{55}}}{\frac{\Gamma_{57}}{\Gamma_{55}}}%
\htconstrdef{Gamma58.c}{\Gamma_{58}}{\Gamma_{62} + \Gamma_{93} + \Gamma_{802} + \Gamma_{167}\cdot{}\Gamma_{\phi\to K^+K^-}}{\Gamma_{62} + \Gamma_{93} + \Gamma_{802} + \Gamma_{167}\cdot{}\Gamma_{\phi\to K^+K^-}}%
\htconstrdef{Gamma59.c}{\Gamma_{59}}{\Gamma_{35}\cdot{}(\Gamma_{<K^0|K_S>}\cdot{}\Gamma_{K_S\to\pi^+\pi^-}) + \Gamma_{62} + \Gamma_{800}\cdot{}\Gamma_{\omega\to\pi^+\pi^-}}{\Gamma_{35}\cdot{}(\Gamma_{<K^0|K_S>}\cdot{}\Gamma_{K_S\to\pi^+\pi^-}) + \Gamma_{62} + \Gamma_{800}\cdot{}\Gamma_{\omega\to\pi^+\pi^-}}%
\htconstrdef{Gamma60.c}{\Gamma_{60}}{\Gamma_{62} + \Gamma_{800}\cdot{}\Gamma_{\omega\to\pi^+\pi^-}}{\Gamma_{62} + \Gamma_{800}\cdot{}\Gamma_{\omega\to\pi^+\pi^-}}%
\htconstrdef{Gamma63.c}{\Gamma_{63}}{\Gamma_{40}\cdot{}(\Gamma_{<K^0|K_S>}\cdot{}\Gamma_{K_S\to\pi^+\pi^-}) + \Gamma_{42}\cdot{}(\Gamma_{<K^0|K_S>}\cdot{}\Gamma_{K_S\to\pi^+\pi^-}) + \Gamma_{47}\cdot{}(2\cdot{}\Gamma_{K_S\to\pi^+\pi^-}\cdot{}\Gamma_{K_S\to\pi^0\pi^0}) + \Gamma_{50}\cdot{}(2\cdot{}\Gamma_{K_S\to\pi^+\pi^-}\cdot{}\Gamma_{K_S\to\pi^0\pi^0}) + \Gamma_{70} + \Gamma_{77} + \Gamma_{78} + \Gamma_{94} + \Gamma_{126}\cdot{}\Gamma_{\eta\to\text{charged}} + \Gamma_{128}\cdot{}\Gamma_{\eta\to\text{charged}} + \Gamma_{130}\cdot{}\Gamma_{\eta\to\text{charged}} + \Gamma_{132}\cdot{}(\Gamma_{<\bar{K}^0|K_S>}\cdot{}\Gamma_{K_S\to\pi^+\pi^-}\cdot{}\Gamma_{\eta\to\text{neutral}} + \Gamma_{<\bar{K}^0|K_S>}\cdot{}\Gamma_{K_S\to\pi^0\pi^0}\cdot{}\Gamma_{\eta\to\text{charged}}) + \Gamma_{151}\cdot{}\Gamma_{\omega\to\pi^+\pi^-\pi^0} + \Gamma_{152}\cdot{}(\Gamma_{\omega\to\pi^+\pi^-\pi^0}+\Gamma_{\omega\to\pi^+\pi^-}) + \Gamma_{800}\cdot{}\Gamma_{\omega\to\pi^+\pi^-\pi^0} + \Gamma_{803}}{\Gamma_{40}\cdot{}(\Gamma_{<K^0|K_S>}\cdot{}\Gamma_{K_S\to\pi^+\pi^-}) + \Gamma_{42}\cdot{}(\Gamma_{<K^0|K_S>}\cdot{}\Gamma_{K_S\to\pi^+\pi^-})  \\ 
  {}& + \Gamma_{47}\cdot{}(2\cdot{}\Gamma_{K_S\to\pi^+\pi^-}\cdot{}\Gamma_{K_S\to\pi^0\pi^0}) + \Gamma_{50}\cdot{}(2\cdot{}\Gamma_{K_S\to\pi^+\pi^-}\cdot{}\Gamma_{K_S\to\pi^0\pi^0})  \\ 
  {}& + \Gamma_{70} + \Gamma_{77} + \Gamma_{78} + \Gamma_{94} + \Gamma_{126}\cdot{}\Gamma_{\eta\to\text{charged}}  \\ 
  {}& + \Gamma_{128}\cdot{}\Gamma_{\eta\to\text{charged}} + \Gamma_{130}\cdot{}\Gamma_{\eta\to\text{charged}} + \Gamma_{132}\cdot{}(\Gamma_{<\bar{K}^0|K_S>}\cdot{}\Gamma_{K_S\to\pi^+\pi^-}\cdot{}\Gamma_{\eta\to\text{neutral}}  \\ 
  {}& + \Gamma_{<\bar{K}^0|K_S>}\cdot{}\Gamma_{K_S\to\pi^0\pi^0}\cdot{}\Gamma_{\eta\to\text{charged}}) + \Gamma_{151}\cdot{}\Gamma_{\omega\to\pi^+\pi^-\pi^0} + \Gamma_{152}\cdot{}(\Gamma_{\omega\to\pi^+\pi^-\pi^0} \\ 
  {}& +\Gamma_{\omega\to\pi^+\pi^-}) + \Gamma_{800}\cdot{}\Gamma_{\omega\to\pi^+\pi^-\pi^0} + \Gamma_{803}}%
\htconstrdef{Gamma64.c}{\Gamma_{64}}{\Gamma_{78} + \Gamma_{77} + \Gamma_{94} + \Gamma_{70} + \Gamma_{126}\cdot{}\Gamma_{\eta\to\pi^+\pi^-\pi^0} + \Gamma_{128}\cdot{}\Gamma_{\eta\to\pi^+\pi^-\pi^0} + \Gamma_{130}\cdot{}\Gamma_{\eta\to\pi^+\pi^-\pi^0} + \Gamma_{800}\cdot{}\Gamma_{\omega\to\pi^+\pi^-\pi^0} + \Gamma_{151}\cdot{}\Gamma_{\omega\to\pi^+\pi^-\pi^0} + \Gamma_{152}\cdot{}(\Gamma_{\omega\to\pi^+\pi^-\pi^0}+\Gamma_{\omega\to\pi^+\pi^-}) + \Gamma_{803}}{\Gamma_{78} + \Gamma_{77} + \Gamma_{94} + \Gamma_{70} + \Gamma_{126}\cdot{}\Gamma_{\eta\to\pi^+\pi^-\pi^0}  \\ 
  {}& + \Gamma_{128}\cdot{}\Gamma_{\eta\to\pi^+\pi^-\pi^0} + \Gamma_{130}\cdot{}\Gamma_{\eta\to\pi^+\pi^-\pi^0} + \Gamma_{800}\cdot{}\Gamma_{\omega\to\pi^+\pi^-\pi^0}  \\ 
  {}& + \Gamma_{151}\cdot{}\Gamma_{\omega\to\pi^+\pi^-\pi^0} + \Gamma_{152}\cdot{}(\Gamma_{\omega\to\pi^+\pi^-\pi^0}+\Gamma_{\omega\to\pi^+\pi^-}) + \Gamma_{803}}%
\htconstrdef{Gamma65.c}{\Gamma_{65}}{\Gamma_{40}\cdot{}(\Gamma_{<K^0|K_S>}\cdot{}\Gamma_{K_S\to\pi^+\pi^-}) + \Gamma_{42}\cdot{}(\Gamma_{<K^0|K_S>}\cdot{}\Gamma_{K_S\to\pi^+\pi^-}) + \Gamma_{70} + \Gamma_{94} + \Gamma_{128}\cdot{}\Gamma_{\eta\to\pi^+\pi^-\pi^0} + \Gamma_{151}\cdot{}\Gamma_{\omega\to\pi^+\pi^-\pi^0} + \Gamma_{152}\cdot{}\Gamma_{\omega\to\pi^+\pi^-} + \Gamma_{800}\cdot{}\Gamma_{\omega\to\pi^+\pi^-\pi^0} + \Gamma_{803}}{\Gamma_{40}\cdot{}(\Gamma_{<K^0|K_S>}\cdot{}\Gamma_{K_S\to\pi^+\pi^-}) + \Gamma_{42}\cdot{}(\Gamma_{<K^0|K_S>}\cdot{}\Gamma_{K_S\to\pi^+\pi^-})  \\ 
  {}& + \Gamma_{70} + \Gamma_{94} + \Gamma_{128}\cdot{}\Gamma_{\eta\to\pi^+\pi^-\pi^0} + \Gamma_{151}\cdot{}\Gamma_{\omega\to\pi^+\pi^-\pi^0}  \\ 
  {}& + \Gamma_{152}\cdot{}\Gamma_{\omega\to\pi^+\pi^-} + \Gamma_{800}\cdot{}\Gamma_{\omega\to\pi^+\pi^-\pi^0} + \Gamma_{803}}%
\htconstrdef{Gamma66.c}{\Gamma_{66}}{\Gamma_{70} + \Gamma_{94} + \Gamma_{128}\cdot{}\Gamma_{\eta\to\pi^+\pi^-\pi^0} + \Gamma_{151}\cdot{}\Gamma_{\omega\to\pi^+\pi^-\pi^0} + \Gamma_{152}\cdot{}\Gamma_{\omega\to\pi^+\pi^-} + \Gamma_{800}\cdot{}\Gamma_{\omega\to\pi^+\pi^-\pi^0} + \Gamma_{803}}{\Gamma_{70} + \Gamma_{94} + \Gamma_{128}\cdot{}\Gamma_{\eta\to\pi^+\pi^-\pi^0} + \Gamma_{151}\cdot{}\Gamma_{\omega\to\pi^+\pi^-\pi^0}  \\ 
  {}& + \Gamma_{152}\cdot{}\Gamma_{\omega\to\pi^+\pi^-} + \Gamma_{800}\cdot{}\Gamma_{\omega\to\pi^+\pi^-\pi^0} + \Gamma_{803}}%
\htconstrdef{Gamma67.c}{\Gamma_{67}}{\Gamma_{70} + \Gamma_{94} + \Gamma_{128}\cdot{}\Gamma_{\eta\to\pi^+\pi^-\pi^0} + \Gamma_{803}}{\Gamma_{70} + \Gamma_{94} + \Gamma_{128}\cdot{}\Gamma_{\eta\to\pi^+\pi^-\pi^0} + \Gamma_{803}}%
\htconstrdef{Gamma68.c}{\Gamma_{68}}{\Gamma_{40}\cdot{}(\Gamma_{<K^0|K_S>}\cdot{}\Gamma_{K_S\to\pi^+\pi^-}) + \Gamma_{70} + \Gamma_{152}\cdot{}\Gamma_{\omega\to\pi^+\pi^-} + \Gamma_{800}\cdot{}\Gamma_{\omega\to\pi^+\pi^-\pi^0}}{\Gamma_{40}\cdot{}(\Gamma_{<K^0|K_S>}\cdot{}\Gamma_{K_S\to\pi^+\pi^-}) + \Gamma_{70} + \Gamma_{152}\cdot{}\Gamma_{\omega\to\pi^+\pi^-}  \\ 
  {}& + \Gamma_{800}\cdot{}\Gamma_{\omega\to\pi^+\pi^-\pi^0}}%
\htconstrdef{Gamma69.c}{\Gamma_{69}}{\Gamma_{152}\cdot{}\Gamma_{\omega\to\pi^+\pi^-} + \Gamma_{70} + \Gamma_{800}\cdot{}\Gamma_{\omega\to\pi^+\pi^-\pi^0}}{\Gamma_{152}\cdot{}\Gamma_{\omega\to\pi^+\pi^-} + \Gamma_{70} + \Gamma_{800}\cdot{}\Gamma_{\omega\to\pi^+\pi^-\pi^0}}%
\htconstrdef{Gamma74.c}{\Gamma_{74}}{\Gamma_{152}\cdot{}\Gamma_{\omega\to\pi^+\pi^-\pi^0} + \Gamma_{78} + \Gamma_{77} + \Gamma_{126}\cdot{}\Gamma_{\eta\to\pi^+\pi^-\pi^0} + \Gamma_{130}\cdot{}\Gamma_{\eta\to\pi^+\pi^-\pi^0}}{\Gamma_{152}\cdot{}\Gamma_{\omega\to\pi^+\pi^-\pi^0} + \Gamma_{78} + \Gamma_{77} + \Gamma_{126}\cdot{}\Gamma_{\eta\to\pi^+\pi^-\pi^0}  \\ 
  {}& + \Gamma_{130}\cdot{}\Gamma_{\eta\to\pi^+\pi^-\pi^0}}%
\htconstrdef{Gamma75.c}{\Gamma_{75}}{\Gamma_{152}\cdot{}\Gamma_{\omega\to\pi^+\pi^-\pi^0} + \Gamma_{47}\cdot{}(2\cdot{}\Gamma_{K_S\to\pi^+\pi^-}\cdot{}\Gamma_{K_S\to\pi^0\pi^0}) + \Gamma_{77} + \Gamma_{126}\cdot{}\Gamma_{\eta\to\pi^+\pi^-\pi^0} + \Gamma_{130}\cdot{}\Gamma_{\eta\to\pi^+\pi^-\pi^0}}{\Gamma_{152}\cdot{}\Gamma_{\omega\to\pi^+\pi^-\pi^0} + \Gamma_{47}\cdot{}(2\cdot{}\Gamma_{K_S\to\pi^+\pi^-}\cdot{}\Gamma_{K_S\to\pi^0\pi^0})  \\ 
  {}& + \Gamma_{77} + \Gamma_{126}\cdot{}\Gamma_{\eta\to\pi^+\pi^-\pi^0} + \Gamma_{130}\cdot{}\Gamma_{\eta\to\pi^+\pi^-\pi^0}}%
\htconstrdef{Gamma76.c}{\Gamma_{76}}{\Gamma_{152}\cdot{}\Gamma_{\omega\to\pi^+\pi^-\pi^0} + \Gamma_{77} + \Gamma_{126}\cdot{}\Gamma_{\eta\to\pi^+\pi^-\pi^0} + \Gamma_{130}\cdot{}\Gamma_{\eta\to\pi^+\pi^-\pi^0}}{\Gamma_{152}\cdot{}\Gamma_{\omega\to\pi^+\pi^-\pi^0} + \Gamma_{77} + \Gamma_{126}\cdot{}\Gamma_{\eta\to\pi^+\pi^-\pi^0} + \Gamma_{130}\cdot{}\Gamma_{\eta\to\pi^+\pi^-\pi^0}}%
\htconstrdef{Gamma76by54.c}{\frac{\Gamma_{76}}{\Gamma_{54}}}{\frac{\Gamma_{76}}{\Gamma_{54}}}{\frac{\Gamma_{76}}{\Gamma_{54}}}%
\htconstrdef{Gamma78.c}{\Gamma_{78}}{\Gamma_{810} + \Gamma_{50}\cdot{}(2\cdot{}\Gamma_{K_S\to\pi^+\pi^-}\cdot{}\Gamma_{K_S\to\pi^0\pi^0}) + \Gamma_{132}\cdot{}(\Gamma_{<\bar{K}^0|K_S>}\cdot{}\Gamma_{K_S\to\pi^+\pi^-}\cdot{}\Gamma_{\eta\to3\pi^0})}{\Gamma_{810} + \Gamma_{50}\cdot{}(2\cdot{}\Gamma_{K_S\to\pi^+\pi^-}\cdot{}\Gamma_{K_S\to\pi^0\pi^0}) + \Gamma_{132}\cdot{}(\Gamma_{<\bar{K}^0|K_S>}\cdot{}\Gamma_{K_S\to\pi^+\pi^-}\cdot{}\Gamma_{\eta\to3\pi^0})}%
\htconstrdef{Gamma79.c}{\Gamma_{79}}{\Gamma_{37}\cdot{}(\Gamma_{<K^0|K_S>}\cdot{}\Gamma_{K_S\to\pi^+\pi^-}) + \Gamma_{42}\cdot{}(\Gamma_{<K^0|K_S>}\cdot{}\Gamma_{K_S\to\pi^+\pi^-}) + \Gamma_{93} + \Gamma_{94} + \Gamma_{128}\cdot{}\Gamma_{\eta\to\text{charged}} + \Gamma_{151}\cdot{}(\Gamma_{\omega\to\pi^+\pi^-\pi^0}+\Gamma_{\omega\to\pi^+\pi^-}) + \Gamma_{168} + \Gamma_{802} + \Gamma_{803}}{\Gamma_{37}\cdot{}(\Gamma_{<K^0|K_S>}\cdot{}\Gamma_{K_S\to\pi^+\pi^-}) + \Gamma_{42}\cdot{}(\Gamma_{<K^0|K_S>}\cdot{}\Gamma_{K_S\to\pi^+\pi^-})  \\ 
  {}& + \Gamma_{93} + \Gamma_{94} + \Gamma_{128}\cdot{}\Gamma_{\eta\to\text{charged}} + \Gamma_{151}\cdot{}(\Gamma_{\omega\to\pi^+\pi^-\pi^0} \\ 
  {}& +\Gamma_{\omega\to\pi^+\pi^-}) + \Gamma_{168} + \Gamma_{802} + \Gamma_{803}}%
\htconstrdef{Gamma80.c}{\Gamma_{80}}{\Gamma_{93} + \Gamma_{802} + \Gamma_{151}\cdot{}\Gamma_{\omega\to\pi^+\pi^-}}{\Gamma_{93} + \Gamma_{802} + \Gamma_{151}\cdot{}\Gamma_{\omega\to\pi^+\pi^-}}%
\htconstrdef{Gamma80by60.c}{\frac{\Gamma_{80}}{\Gamma_{60}}}{\frac{\Gamma_{80}}{\Gamma_{60}}}{\frac{\Gamma_{80}}{\Gamma_{60}}}%
\htconstrdef{Gamma81.c}{\Gamma_{81}}{\Gamma_{128}\cdot{}\Gamma_{\eta\to\pi^+\pi^-\pi^0} + \Gamma_{94} + \Gamma_{803} + \Gamma_{151}\cdot{}\Gamma_{\omega\to\pi^+\pi^-\pi^0}}{\Gamma_{128}\cdot{}\Gamma_{\eta\to\pi^+\pi^-\pi^0} + \Gamma_{94} + \Gamma_{803} + \Gamma_{151}\cdot{}\Gamma_{\omega\to\pi^+\pi^-\pi^0}}%
\htconstrdef{Gamma81by69.c}{\frac{\Gamma_{81}}{\Gamma_{69}}}{\frac{\Gamma_{81}}{\Gamma_{69}}}{\frac{\Gamma_{81}}{\Gamma_{69}}}%
\htconstrdef{Gamma82.c}{\Gamma_{82}}{\Gamma_{128}\cdot{}\Gamma_{\eta\to\text{charged}} + \Gamma_{42}\cdot{}(\Gamma_{<K^0|K_S>}\cdot{}\Gamma_{K_S\to\pi^+\pi^-}) + \Gamma_{802} + \Gamma_{803} + \Gamma_{151}\cdot{}(\Gamma_{\omega\to\pi^+\pi^-\pi^0}+\Gamma_{\omega\to\pi^+\pi^-}) + \Gamma_{37}\cdot{}(\Gamma_{<K^0|K_S>}\cdot{}\Gamma_{K_S\to\pi^+\pi^-})}{\Gamma_{128}\cdot{}\Gamma_{\eta\to\text{charged}} + \Gamma_{42}\cdot{}(\Gamma_{<K^0|K_S>}\cdot{}\Gamma_{K_S\to\pi^+\pi^-}) + \Gamma_{802}  \\ 
  {}& + \Gamma_{803} + \Gamma_{151}\cdot{}(\Gamma_{\omega\to\pi^+\pi^-\pi^0}+\Gamma_{\omega\to\pi^+\pi^-}) + \Gamma_{37}\cdot{}(\Gamma_{<K^0|K_S>}\cdot{}\Gamma_{K_S\to\pi^+\pi^-})}%
\htconstrdef{Gamma83.c}{\Gamma_{83}}{\Gamma_{128}\cdot{}\Gamma_{\eta\to\pi^+\pi^-\pi^0} + \Gamma_{802} + \Gamma_{803} + \Gamma_{151}\cdot{}(\Gamma_{\omega\to\pi^+\pi^-\pi^0}+\Gamma_{\omega\to\pi^+\pi^-})}{\Gamma_{128}\cdot{}\Gamma_{\eta\to\pi^+\pi^-\pi^0} + \Gamma_{802} + \Gamma_{803} + \Gamma_{151}\cdot{}(\Gamma_{\omega\to\pi^+\pi^-\pi^0} \\ 
  {}& +\Gamma_{\omega\to\pi^+\pi^-})}%
\htconstrdef{Gamma84.c}{\Gamma_{84}}{\Gamma_{802} + \Gamma_{151}\cdot{}\Gamma_{\omega\to\pi^+\pi^-} + \Gamma_{37}\cdot{}(\Gamma_{<K^0|K_S>}\cdot{}\Gamma_{K_S\to\pi^+\pi^-})}{\Gamma_{802} + \Gamma_{151}\cdot{}\Gamma_{\omega\to\pi^+\pi^-} + \Gamma_{37}\cdot{}(\Gamma_{<K^0|K_S>}\cdot{}\Gamma_{K_S\to\pi^+\pi^-})}%
\htconstrdef{Gamma85.c}{\Gamma_{85}}{\Gamma_{802} + \Gamma_{151}\cdot{}\Gamma_{\omega\to\pi^+\pi^-}}{\Gamma_{802} + \Gamma_{151}\cdot{}\Gamma_{\omega\to\pi^+\pi^-}}%
\htconstrdef{Gamma85by60.c}{\frac{\Gamma_{85}}{\Gamma_{60}}}{\frac{\Gamma_{85}}{\Gamma_{60}}}{\frac{\Gamma_{85}}{\Gamma_{60}}}%
\htconstrdef{Gamma87.c}{\Gamma_{87}}{\Gamma_{42}\cdot{}(\Gamma_{<K^0|K_S>}\cdot{}\Gamma_{K_S\to\pi^+\pi^-}) + \Gamma_{128}\cdot{}\Gamma_{\eta\to\pi^+\pi^-\pi^0} + \Gamma_{151}\cdot{}\Gamma_{\omega\to\pi^+\pi^-\pi^0} + \Gamma_{803}}{\Gamma_{42}\cdot{}(\Gamma_{<K^0|K_S>}\cdot{}\Gamma_{K_S\to\pi^+\pi^-}) + \Gamma_{128}\cdot{}\Gamma_{\eta\to\pi^+\pi^-\pi^0} + \Gamma_{151}\cdot{}\Gamma_{\omega\to\pi^+\pi^-\pi^0}  \\ 
  {}& + \Gamma_{803}}%
\htconstrdef{Gamma88.c}{\Gamma_{88}}{\Gamma_{128}\cdot{}\Gamma_{\eta\to\pi^+\pi^-\pi^0} + \Gamma_{803} + \Gamma_{151}\cdot{}\Gamma_{\omega\to\pi^+\pi^-\pi^0}}{\Gamma_{128}\cdot{}\Gamma_{\eta\to\pi^+\pi^-\pi^0} + \Gamma_{803} + \Gamma_{151}\cdot{}\Gamma_{\omega\to\pi^+\pi^-\pi^0}}%
\htconstrdef{Gamma89.c}{\Gamma_{89}}{\Gamma_{803} + \Gamma_{151}\cdot{}\Gamma_{\omega\to\pi^+\pi^-\pi^0}}{\Gamma_{803} + \Gamma_{151}\cdot{}\Gamma_{\omega\to\pi^+\pi^-\pi^0}}%
\htconstrdef{Gamma92.c}{\Gamma_{92}}{\Gamma_{94} + \Gamma_{93}}{\Gamma_{94} + \Gamma_{93}}%
\htconstrdef{Gamma93by60.c}{\frac{\Gamma_{93}}{\Gamma_{60}}}{\frac{\Gamma_{93}}{\Gamma_{60}}}{\frac{\Gamma_{93}}{\Gamma_{60}}}%
\htconstrdef{Gamma94by69.c}{\frac{\Gamma_{94}}{\Gamma_{69}}}{\frac{\Gamma_{94}}{\Gamma_{69}}}{\frac{\Gamma_{94}}{\Gamma_{69}}}%
\htconstrdef{Gamma96.c}{\Gamma_{96}}{\Gamma_{167}\cdot{}\Gamma_{\phi\to K^+K^-}}{\Gamma_{167}\cdot{}\Gamma_{\phi\to K^+K^-}}%
\htconstrdef{Gamma102.c}{\Gamma_{102}}{\Gamma_{103} + \Gamma_{104}}{\Gamma_{103} + \Gamma_{104}}%
\htconstrdef{Gamma103.c}{\Gamma_{103}}{\Gamma_{820} + \Gamma_{822} + \Gamma_{831}\cdot{}\Gamma_{\omega\to\pi^+\pi^-}}{\Gamma_{820} + \Gamma_{822} + \Gamma_{831}\cdot{}\Gamma_{\omega\to\pi^+\pi^-}}%
\htconstrdef{Gamma104.c}{\Gamma_{104}}{\Gamma_{830} + \Gamma_{833}}{\Gamma_{830} + \Gamma_{833}}%
\htconstrdef{Gamma106.c}{\Gamma_{106}}{\Gamma_{30} + \Gamma_{44}\cdot{}\Gamma_{<\bar{K}^0|K_S>} + \Gamma_{47} + \Gamma_{53}\cdot{}\Gamma_{<K^0|K_S>} + \Gamma_{77} + \Gamma_{103} + \Gamma_{126}\cdot{}(\Gamma_{\eta\to3\pi^0}+\Gamma_{\eta\to\pi^+\pi^-\pi^0}) + \Gamma_{152}\cdot{}\Gamma_{\omega\to\pi^+\pi^-\pi^0}}{\Gamma_{30} + \Gamma_{44}\cdot{}\Gamma_{<\bar{K}^0|K_S>} + \Gamma_{47} + \Gamma_{53}\cdot{}\Gamma_{<K^0|K_S>}  \\ 
  {}& + \Gamma_{77} + \Gamma_{103} + \Gamma_{126}\cdot{}(\Gamma_{\eta\to3\pi^0}+\Gamma_{\eta\to\pi^+\pi^-\pi^0}) + \Gamma_{152}\cdot{}\Gamma_{\omega\to\pi^+\pi^-\pi^0}}%
\htconstrdef{Gamma110.c}{\Gamma_{110}}{\Gamma_{10} + \Gamma_{16} + \Gamma_{23} + \Gamma_{28} + \Gamma_{35} + \Gamma_{40} + \Gamma_{128} + \Gamma_{802} + \Gamma_{803} + \Gamma_{151} + \Gamma_{130} + \Gamma_{132} + \Gamma_{44} + \Gamma_{53} + \Gamma_{168} + \Gamma_{169} + \Gamma_{822} + \Gamma_{833}}{\Gamma_{10} + \Gamma_{16} + \Gamma_{23} + \Gamma_{28} + \Gamma_{35} + \Gamma_{40}  \\ 
  {}& + \Gamma_{128} + \Gamma_{802} + \Gamma_{803} + \Gamma_{151} + \Gamma_{130} + \Gamma_{132}  \\ 
  {}& + \Gamma_{44} + \Gamma_{53} + \Gamma_{168} + \Gamma_{169} + \Gamma_{822} + \Gamma_{833}}%
\htconstrdef{Gamma149.c}{\Gamma_{149}}{\Gamma_{152} + \Gamma_{800} + \Gamma_{151}}{\Gamma_{152} + \Gamma_{800} + \Gamma_{151}}%
\htconstrdef{Gamma150.c}{\Gamma_{150}}{\Gamma_{800} + \Gamma_{151}}{\Gamma_{800} + \Gamma_{151}}%
\htconstrdef{Gamma150by66.c}{\frac{\Gamma_{150}}{\Gamma_{66}}}{\frac{\Gamma_{150}}{\Gamma_{66}}}{\frac{\Gamma_{150}}{\Gamma_{66}}}%
\htconstrdef{Gamma152by54.c}{\frac{\Gamma_{152}}{\Gamma_{54}}}{\frac{\Gamma_{152}}{\Gamma_{54}}}{\frac{\Gamma_{152}}{\Gamma_{54}}}%
\htconstrdef{Gamma152by76.c}{\frac{\Gamma_{152}}{\Gamma_{76}}}{\frac{\Gamma_{152}}{\Gamma_{76}}}{\frac{\Gamma_{152}}{\Gamma_{76}}}%
\htconstrdef{Gamma168.c}{\Gamma_{168}}{\Gamma_{167}\cdot{}\Gamma_{\phi\to K^+K^-}}{\Gamma_{167}\cdot{}\Gamma_{\phi\to K^+K^-}}%
\htconstrdef{Gamma169.c}{\Gamma_{169}}{\Gamma_{167}\cdot{}\Gamma_{\phi\to K_S K_L}}{\Gamma_{167}\cdot{}\Gamma_{\phi\to K_S K_L}}%
\htconstrdef{Gamma804.c}{\Gamma_{804}}{\Gamma_{47} \cdot{} ((\Gamma_{<K^0|K_L>}\cdot{}\Gamma_{<\bar{K}^0|K_L>}) / (\Gamma_{<K^0|K_S>}\cdot{}\Gamma_{<\bar{K}^0|K_S>}))}{\Gamma_{47} \cdot{} ((\Gamma_{<K^0|K_L>}\cdot{}\Gamma_{<\bar{K}^0|K_L>}) / (\Gamma_{<K^0|K_S>}\cdot{}\Gamma_{<\bar{K}^0|K_S>}))}%
\htconstrdef{Gamma806.c}{\Gamma_{806}}{\Gamma_{50} \cdot{} ((\Gamma_{<K^0|K_L>}\cdot{}\Gamma_{<\bar{K}^0|K_L>}) / (\Gamma_{<K^0|K_S>}\cdot{}\Gamma_{<\bar{K}^0|K_S>}))}{\Gamma_{50} \cdot{} ((\Gamma_{<K^0|K_L>}\cdot{}\Gamma_{<\bar{K}^0|K_L>}) / (\Gamma_{<K^0|K_S>}\cdot{}\Gamma_{<\bar{K}^0|K_S>}))}%
\htconstrdef{Gamma810.c}{\Gamma_{810}}{\Gamma_{910} + \Gamma_{911} + \Gamma_{811}\cdot{}\Gamma_{\omega\to\pi^+\pi^-\pi^0} + \Gamma_{812}}{\Gamma_{910} + \Gamma_{911} + \Gamma_{811}\cdot{}\Gamma_{\omega\to\pi^+\pi^-\pi^0} + \Gamma_{812}}%
\htconstrdef{Gamma820.c}{\Gamma_{820}}{\Gamma_{920} + \Gamma_{821}}{\Gamma_{920} + \Gamma_{821}}%
\htconstrdef{Gamma830.c}{\Gamma_{830}}{\Gamma_{930} + \Gamma_{831}\cdot{}\Gamma_{\omega\to\pi^+\pi^-\pi^0} + \Gamma_{832}}{\Gamma_{930} + \Gamma_{831}\cdot{}\Gamma_{\omega\to\pi^+\pi^-\pi^0} + \Gamma_{832}}%
\htconstrdef{Gamma910.c}{\Gamma_{910}}{\Gamma_{136}\cdot{}\Gamma_{\eta\to3\pi^0}}{\Gamma_{136}\cdot{}\Gamma_{\eta\to3\pi^0}}%
\htconstrdef{Gamma911.c}{\Gamma_{911}}{\Gamma_{945}\cdot{}\Gamma_{\eta\to\pi^+\pi^-\pi^0}}{\Gamma_{945}\cdot{}\Gamma_{\eta\to\pi^+\pi^-\pi^0}}%
\htconstrdef{Gamma930.c}{\Gamma_{930}}{\Gamma_{136}\cdot{}\Gamma_{\eta\to\pi^+\pi^-\pi^0}}{\Gamma_{136}\cdot{}\Gamma_{\eta\to\pi^+\pi^-\pi^0}}%
\htconstrdef{Gamma944.c}{\Gamma_{944}}{\Gamma_{136}\cdot{}\Gamma_{\eta\to\gamma\gamma}}{\Gamma_{136}\cdot{}\Gamma_{\eta\to\gamma\gamma}}%
\htconstrdef{GammaAll.c}{\Gamma_{\text{All}}}{\Gamma_{3} + \Gamma_{5} + \Gamma_{9} + \Gamma_{10} + \Gamma_{14} + \Gamma_{16} + \Gamma_{20} + \Gamma_{23} + \Gamma_{27} + \Gamma_{28} + \Gamma_{30} + \Gamma_{35} + \Gamma_{37} + \Gamma_{40} + \Gamma_{42} + \Gamma_{47}\cdot{}(1 + ((\Gamma_{<K^0|K_L>}\cdot{}\Gamma_{<\bar{K}^0|K_L>}) / (\Gamma_{<K^0|K_S>}\cdot{}\Gamma_{<\bar{K}^0|K_S>}))) + \Gamma_{48} + \Gamma_{62} + \Gamma_{70} + \Gamma_{77} + \Gamma_{811} + \Gamma_{812} + \Gamma_{93} + \Gamma_{94} + \Gamma_{832} + \Gamma_{833} + \Gamma_{126} + \Gamma_{128} + \Gamma_{802} + \Gamma_{803} + \Gamma_{800} + \Gamma_{151} + \Gamma_{130} + \Gamma_{132} + \Gamma_{44} + \Gamma_{53} + \Gamma_{50}\cdot{}(1 + ((\Gamma_{<K^0|K_L>}\cdot{}\Gamma_{<\bar{K}^0|K_L>}) / (\Gamma_{<K^0|K_S>}\cdot{}\Gamma_{<\bar{K}^0|K_S>}))) + \Gamma_{51} + \Gamma_{167}\cdot{}(\Gamma_{\phi\to K^+K^-}+\Gamma_{\phi\to K_S K_L}) + \Gamma_{152} + \Gamma_{920} + \Gamma_{821} + \Gamma_{822} + \Gamma_{831} + \Gamma_{136} + \Gamma_{945} + \Gamma_{805}}{\Gamma_{3} + \Gamma_{5} + \Gamma_{9} + \Gamma_{10} + \Gamma_{14} + \Gamma_{16}  \\ 
  {}& + \Gamma_{20} + \Gamma_{23} + \Gamma_{27} + \Gamma_{28} + \Gamma_{30} + \Gamma_{35}  \\ 
  {}& + \Gamma_{37} + \Gamma_{40} + \Gamma_{42} + \Gamma_{47}\cdot{}(1 + ((\Gamma_{<K^0|K_L>}\cdot{}\Gamma_{<\bar{K}^0|K_L>}) / (\Gamma_{<K^0|K_S>}\cdot{}\Gamma_{<\bar{K}^0|K_S>})))  \\ 
  {}& + \Gamma_{48} + \Gamma_{62} + \Gamma_{70} + \Gamma_{77} + \Gamma_{811} + \Gamma_{812}  \\ 
  {}& + \Gamma_{93} + \Gamma_{94} + \Gamma_{832} + \Gamma_{833} + \Gamma_{126} + \Gamma_{128}  \\ 
  {}& + \Gamma_{802} + \Gamma_{803} + \Gamma_{800} + \Gamma_{151} + \Gamma_{130} + \Gamma_{132}  \\ 
  {}& + \Gamma_{44} + \Gamma_{53} + \Gamma_{50}\cdot{}(1 + ((\Gamma_{<K^0|K_L>}\cdot{}\Gamma_{<\bar{K}^0|K_L>}) / (\Gamma_{<K^0|K_S>}\cdot{}\Gamma_{<\bar{K}^0|K_S>})))  \\ 
  {}& + \Gamma_{51} + \Gamma_{167}\cdot{}(\Gamma_{\phi\to K^+K^-}+\Gamma_{\phi\to K_S K_L}) + \Gamma_{152} + \Gamma_{920}  \\ 
  {}& + \Gamma_{821} + \Gamma_{822} + \Gamma_{831} + \Gamma_{136} + \Gamma_{945} + \Gamma_{805}}%
\htconstrdef{Unitarity}{1}{\Gamma_{\text{All}} + \Gamma_{998}}{\Gamma_{\text{All}} + \Gamma_{998}}%
\htdef{ConstrEqs}{%
\begin{align*}
\htuse{Gamma1.c.left} ={}& \htuse{Gamma1.c.right.split}
\end{align*}
\begin{align*}
\htuse{Gamma2.c.left} ={}& \htuse{Gamma2.c.right.split}
\end{align*}
\begin{align*}
\htuse{Gamma7.c.left} ={}& \htuse{Gamma7.c.right.split}
\end{align*}
\begin{align*}
\htuse{Gamma8.c.left} ={}& \htuse{Gamma8.c.right.split}
\end{align*}
\begin{align*}
\htuse{Gamma11.c.left} ={}& \htuse{Gamma11.c.right.split}
\end{align*}
\begin{align*}
\htuse{Gamma12.c.left} ={}& \htuse{Gamma12.c.right.split}
\end{align*}
\begin{align*}
\htuse{Gamma13.c.left} ={}& \htuse{Gamma13.c.right.split}
\end{align*}
\begin{align*}
\htuse{Gamma17.c.left} ={}& \htuse{Gamma17.c.right.split}
\end{align*}
\begin{align*}
\htuse{Gamma18.c.left} ={}& \htuse{Gamma18.c.right.split}
\end{align*}
\begin{align*}
\htuse{Gamma19.c.left} ={}& \htuse{Gamma19.c.right.split}
\end{align*}
\begin{align*}
\htuse{Gamma24.c.left} ={}& \htuse{Gamma24.c.right.split}
\end{align*}
\begin{align*}
\htuse{Gamma25.c.left} ={}& \htuse{Gamma25.c.right.split}
\end{align*}
\begin{align*}
\htuse{Gamma26.c.left} ={}& \htuse{Gamma26.c.right.split}
\end{align*}
\begin{align*}
\htuse{Gamma29.c.left} ={}& \htuse{Gamma29.c.right.split}
\end{align*}
\begin{align*}
\htuse{Gamma31.c.left} ={}& \htuse{Gamma31.c.right.split}
\end{align*}
\begin{align*}
\htuse{Gamma32.c.left} ={}& \htuse{Gamma32.c.right.split}
\end{align*}
\begin{align*}
\htuse{Gamma33.c.left} ={}& \htuse{Gamma33.c.right.split}
\end{align*}
\begin{align*}
\htuse{Gamma34.c.left} ={}& \htuse{Gamma34.c.right.split}
\end{align*}
\begin{align*}
\htuse{Gamma38.c.left} ={}& \htuse{Gamma38.c.right.split}
\end{align*}
\begin{align*}
\htuse{Gamma39.c.left} ={}& \htuse{Gamma39.c.right.split}
\end{align*}
\begin{align*}
\htuse{Gamma43.c.left} ={}& \htuse{Gamma43.c.right.split}
\end{align*}
\begin{align*}
\htuse{Gamma46.c.left} ={}& \htuse{Gamma46.c.right.split}
\end{align*}
\begin{align*}
\htuse{Gamma49.c.left} ={}& \htuse{Gamma49.c.right.split}
\end{align*}
\begin{align*}
\htuse{Gamma54.c.left} ={}& \htuse{Gamma54.c.right.split}
\end{align*}
\begin{align*}
\htuse{Gamma55.c.left} ={}& \htuse{Gamma55.c.right.split}
\end{align*}
\begin{align*}
\htuse{Gamma56.c.left} ={}& \htuse{Gamma56.c.right.split}
\end{align*}
\begin{align*}
\htuse{Gamma57.c.left} ={}& \htuse{Gamma57.c.right.split}
\end{align*}
\begin{align*}
\htuse{Gamma58.c.left} ={}& \htuse{Gamma58.c.right.split}
\end{align*}
\begin{align*}
\htuse{Gamma59.c.left} ={}& \htuse{Gamma59.c.right.split}
\end{align*}
\begin{align*}
\htuse{Gamma60.c.left} ={}& \htuse{Gamma60.c.right.split}
\end{align*}
\begin{align*}
\htuse{Gamma63.c.left} ={}& \htuse{Gamma63.c.right.split}
\end{align*}
\begin{align*}
\htuse{Gamma64.c.left} ={}& \htuse{Gamma64.c.right.split}
\end{align*}
\begin{align*}
\htuse{Gamma65.c.left} ={}& \htuse{Gamma65.c.right.split}
\end{align*}
\begin{align*}
\htuse{Gamma66.c.left} ={}& \htuse{Gamma66.c.right.split}
\end{align*}
\begin{align*}
\htuse{Gamma67.c.left} ={}& \htuse{Gamma67.c.right.split}
\end{align*}
\begin{align*}
\htuse{Gamma68.c.left} ={}& \htuse{Gamma68.c.right.split}
\end{align*}
\begin{align*}
\htuse{Gamma69.c.left} ={}& \htuse{Gamma69.c.right.split}
\end{align*}
\begin{align*}
\htuse{Gamma74.c.left} ={}& \htuse{Gamma74.c.right.split}
\end{align*}
\begin{align*}
\htuse{Gamma75.c.left} ={}& \htuse{Gamma75.c.right.split}
\end{align*}
\begin{align*}
\htuse{Gamma76.c.left} ={}& \htuse{Gamma76.c.right.split}
\end{align*}
\begin{align*}
\htuse{Gamma78.c.left} ={}& \htuse{Gamma78.c.right.split}
\end{align*}
\begin{align*}
\htuse{Gamma79.c.left} ={}& \htuse{Gamma79.c.right.split}
\end{align*}
\begin{align*}
\htuse{Gamma80.c.left} ={}& \htuse{Gamma80.c.right.split}
\end{align*}
\begin{align*}
\htuse{Gamma81.c.left} ={}& \htuse{Gamma81.c.right.split}
\end{align*}
\begin{align*}
\htuse{Gamma82.c.left} ={}& \htuse{Gamma82.c.right.split}
\end{align*}
\begin{align*}
\htuse{Gamma83.c.left} ={}& \htuse{Gamma83.c.right.split}
\end{align*}
\begin{align*}
\htuse{Gamma84.c.left} ={}& \htuse{Gamma84.c.right.split}
\end{align*}
\begin{align*}
\htuse{Gamma85.c.left} ={}& \htuse{Gamma85.c.right.split}
\end{align*}
\begin{align*}
\htuse{Gamma87.c.left} ={}& \htuse{Gamma87.c.right.split}
\end{align*}
\begin{align*}
\htuse{Gamma88.c.left} ={}& \htuse{Gamma88.c.right.split}
\end{align*}
\begin{align*}
\htuse{Gamma89.c.left} ={}& \htuse{Gamma89.c.right.split}
\end{align*}
\begin{align*}
\htuse{Gamma92.c.left} ={}& \htuse{Gamma92.c.right.split}
\end{align*}
\begin{align*}
\htuse{Gamma96.c.left} ={}& \htuse{Gamma96.c.right.split}
\end{align*}
\begin{align*}
\htuse{Gamma102.c.left} ={}& \htuse{Gamma102.c.right.split}
\end{align*}
\begin{align*}
\htuse{Gamma103.c.left} ={}& \htuse{Gamma103.c.right.split}
\end{align*}
\begin{align*}
\htuse{Gamma104.c.left} ={}& \htuse{Gamma104.c.right.split}
\end{align*}
\begin{align*}
\htuse{Gamma106.c.left} ={}& \htuse{Gamma106.c.right.split}
\end{align*}
\begin{align*}
\htuse{Gamma110.c.left} ={}& \htuse{Gamma110.c.right.split}
\end{align*}
\begin{align*}
\htuse{Gamma149.c.left} ={}& \htuse{Gamma149.c.right.split}
\end{align*}
\begin{align*}
\htuse{Gamma150.c.left} ={}& \htuse{Gamma150.c.right.split}
\end{align*}
\begin{align*}
\htuse{Gamma168.c.left} ={}& \htuse{Gamma168.c.right.split}
\end{align*}
\begin{align*}
\htuse{Gamma169.c.left} ={}& \htuse{Gamma169.c.right.split}
\end{align*}
\begin{align*}
\htuse{Gamma804.c.left} ={}& \htuse{Gamma804.c.right.split}
\end{align*}
\begin{align*}
\htuse{Gamma806.c.left} ={}& \htuse{Gamma806.c.right.split}
\end{align*}
\begin{align*}
\htuse{Gamma810.c.left} ={}& \htuse{Gamma810.c.right.split}
\end{align*}
\begin{align*}
\htuse{Gamma820.c.left} ={}& \htuse{Gamma820.c.right.split}
\end{align*}
\begin{align*}
\htuse{Gamma830.c.left} ={}& \htuse{Gamma830.c.right.split}
\end{align*}
\begin{align*}
\htuse{Gamma910.c.left} ={}& \htuse{Gamma910.c.right.split}
\end{align*}
\begin{align*}
\htuse{Gamma911.c.left} ={}& \htuse{Gamma911.c.right.split}
\end{align*}
\begin{align*}
\htuse{Gamma930.c.left} ={}& \htuse{Gamma930.c.right.split}
\end{align*}
\begin{align*}
\htuse{Gamma944.c.left} ={}& \htuse{Gamma944.c.right.split}
\end{align*}
\begin{align*}
\htuse{GammaAll.c.left} ={}& \htuse{GammaAll.c.right.split}
\end{align*}}%
\htdef{NumMeasALEPH}{39}%
\htdef{NumMeasARGUS}{2}%
\htdef{NumMeasBaBar}{23}%
\htdef{NumMeasBelle}{15}%
\htdef{NumMeasCELLO}{1}%
\htdef{NumMeasCLEO}{35}%
\htdef{NumMeasCLEO3}{6}%
\htdef{NumMeasDELPHI}{14}%
\htdef{NumMeasHRS}{2}%
\htdef{NumMeasL3}{11}%
\htdef{NumMeasOPAL}{19}%
\htdef{NumMeasTPC}{3}%

%% file: tau-br-fit-hfag-2016-elab-vadirect.tex
\htquantdef{B_tau_had_fit}{B_tau_had_fit}{}{64.76 \pm 0.10}{64.76}{0.10}%
\htquantdef{B_tau_s_fit}{B_tau_s_fit}{}{2.909 \pm 0.048}{2.909}{0.048}%
\htquantdef{B_tau_s_unitarity}{B_tau_s_unitarity}{}{(2.944 \pm 0.103) \cdot 10^{-2}}{2.944\cdot 10^{-2}}{0.103\cdot 10^{-2}}%
\htquantdef{B_tau_VA}{B_tau_VA}{}{0.6185 \pm 0.0010}{0.6185}{0.0010}%
\htquantdef{B_tau_VA_fit}{B_tau_VA_fit}{}{61.85 \pm 0.10}{61.85}{0.10}%
\htquantdef{B_tau_VA_unitarity}{B_tau_VA_unitarity}{}{0.61883 \pm 0.00080}{0.61883}{0.00080}%
\htquantdef{Be_fit}{Be_fit}{}{0.17816 \pm 0.00041}{0.17816}{0.00041}%
\htquantdef{Be_from_Bmu}{Be_from_Bmu}{}{0.17882 \pm 0.00041}{0.17882}{0.00041}%
\htquantdef{Be_from_taulife}{Be_from_taulife}{}{0.17780 \pm 0.00032}{0.17780}{0.00032}%
\htquantdef{Be_lept}{Be_lept}{}{17.850 \pm 0.032}{17.850}{0.032}%
\htquantdef{Be_unitarity}{Be_unitarity}{}{0.1785 \pm 0.0010}{0.1785}{0.0010}%
\htquantdef{Be_univ}{Be_univ}{}{17.815 \pm 0.023}{17.815}{0.023}%
\htquantdef{Bmu_by_Be_th}{Bmu_by_Be_th}{}{0.9725606 \pm 0.0000036}{0.9725606}{0.0000036}%
\htquantdef{Bmu_fit}{Bmu_fit}{}{0.17392 \pm 0.00040}{0.17392}{0.00040}%
\htquantdef{Bmu_from_taulife}{Bmu_from_taulife}{}{0.17292 \pm 0.00032}{0.17292}{0.00032}%
\htquantdef{Bmu_unitarity}{Bmu_unitarity}{}{0.1743 \pm 0.0010}{0.1743}{0.0010}%
\htquantdef{BR_a1_pigamma}{BR_a1_pigamma}{}{0.2100\cdot 10^{-2}}{0.2100\cdot 10^{-2}}{0}%
\htquantdef{BR_eta_2gam}{BR_eta_2gam}{}{0.3941}{0.3941}{0}%
\htquantdef{BR_eta_3piz}{BR_eta_3piz}{}{0.3268}{0.3268}{0}%
\htquantdef{BR_eta_charged}{BR_eta_charged}{}{0.2810}{0.2810}{0}%
\htquantdef{BR_eta_neutral}{BR_eta_neutral}{}{0.7212}{0.7212}{0}%
\htquantdef{BR_eta_pimpipgamma}{BR_eta_pimpipgamma}{}{4.220\cdot 10^{-2}}{4.220\cdot 10^{-2}}{0}%
\htquantdef{BR_eta_pimpippiz}{BR_eta_pimpippiz}{}{0.2292}{0.2292}{0}%
\htquantdef{BR_f1_2pip2pim}{BR_f1_2pip2pim}{}{0.1100}{0.1100}{0}%
\htquantdef{BR_f1_2pizpippim}{BR_f1_2pizpippim}{}{0.2200}{0.2200}{0}%
\htquantdef{BR_KS_2piz}{BR_KS_2piz}{}{0.3069}{0.3069}{0}%
\htquantdef{BR_KS_pimpip}{BR_KS_pimpip}{}{0.6920}{0.6920}{0}%
\htquantdef{BR_om_pimpip}{BR_om_pimpip}{}{1.530\cdot 10^{-2}}{1.530\cdot 10^{-2}}{0}%
\htquantdef{BR_om_pimpippiz}{BR_om_pimpippiz}{}{0.8920}{0.8920}{0}%
\htquantdef{BR_om_pizgamma}{BR_om_pizgamma}{}{8.280\cdot 10^{-2}}{8.280\cdot 10^{-2}}{0}%
\htquantdef{BR_phi_KmKp}{BR_phi_KmKp}{}{0.4890}{0.4890}{0}%
\htquantdef{BR_phi_KSKL}{BR_phi_KSKL}{}{0.3420}{0.3420}{0}%
\htquantdef{BRA_Kz_KL_KET}{BRA_Kz_KL_KET}{}{0.5000}{0.5000}{0}%
\htquantdef{BRA_Kz_KS_KET}{BRA_Kz_KS_KET}{}{0.5000}{0.5000}{0}%
\htquantdef{BRA_Kzbar_KL_KET}{BRA_Kzbar_KL_KET}{}{0.5000}{0.5000}{0}%
\htquantdef{BRA_Kzbar_KS_KET}{BRA_Kzbar_KS_KET}{}{0.5000}{0.5000}{0}%
\htquantdef{delta_mu_gamma}{delta_mu_gamma}{}{0.9958}{0.9958}{0}%
\htquantdef{delta_mu_W}{delta_mu_W}{}{1.00000103667 \pm 0.00000000039}{1.00000103667}{0.00000000039}%
\htquantdef{delta_tau_gamma}{delta_tau_gamma}{}{0.9957}{0.9957}{0}%
\htquantdef{delta_tau_W}{delta_tau_W}{}{1.00029627 \pm 0.00000012}{1.00029627}{0.00000012}%
\htquantdef{deltaR_su3break}{deltaR_su3break}{}{0.242 \pm 0.032}{0.242}{0.032}%
\htquantdef{deltaR_su3break_d2pert}{deltaR_su3break_d2pert}{}{9.300 \pm 3.400}{9.300}{3.400}%
\htquantdef{deltaR_su3break_pheno}{deltaR_su3break_pheno}{}{0.1544 \pm 0.0037}{0.1544}{0.0037}%
\htquantdef{deltaR_su3break_remain}{deltaR_su3break_remain}{}{(0.3400 \pm 0.2800) \cdot 10^{-2}}{0.3400\cdot 10^{-2}}{0.2800\cdot 10^{-2}}%
\htquantdef{dRrad_k_munu}{dRrad_k_munu}{}{1.30 \pm 0.20}{1.30}{0.20}%
\htquantdef{dRrad_kmunu_by_pimunu}{dRrad_kmunu_by_pimunu}{}{-1.13 \pm 0.23}{-1.13}{0.23}%
\htquantdef{dRrad_tauK_by_Kmu}{dRrad_tauK_by_Kmu}{}{0.90 \pm 0.22}{0.90}{0.22}%
\htquantdef{dRrad_taupi_by_pimu}{dRrad_taupi_by_pimu}{}{0.16 \pm 0.14}{0.16}{0.14}%
\htquantdef{EmNuNumb}{EmNuNumb}{}{0.1783}{0.1783}{0}%
\htquantdef{f_K}{f_K}{}{155.6 \pm 0.4}{155.6}{0.4}%
\htquantdef{f_K_by_f_pi}{f_K_by_f_pi}{}{1.1930 \pm 0.0030}{1.1930}{0.0030}%
\htquantdef{fp0_Kpi}{fp0_Kpi}{}{0.9677 \pm 0.0027}{0.9677}{0.0027}%
\htquantdef{G_F_by_hcut3_c3}{G_F_by_hcut3_c3}{}{(1.16637870 \pm 0.00000060) \cdot 10^{-11}}{1.16637870\cdot 10^{-11}}{0.00000060\cdot 10^{-11}}%
\htquantdef{Gamma1}{\Gamma_{1}}{\BRF{\tau^-}{(\text{particles})^- \ge{} 0\, \text{neutrals} \ge{} 0\,  K^0\, \nu_\tau}}{0.8519 \pm 0.0011}{0.8519}{0.0011}%
\htquantdef{Gamma10}{\Gamma_{10}}{\BRF{\tau^-}{K^- \nu_\tau}}{(0.6960 \pm 0.0096) \cdot 10^{-2}}{0.6960\cdot 10^{-2}}{0.0096\cdot 10^{-2}}%
\htquantdef{Gamma102}{\Gamma_{102}}{\BRF{\tau^-}{3h^- 2h^+ \ge{} 0\,  \text{neutrals}\, \nu_\tau\;(\text{ex.~} K^0)}}{(9.850 \pm 0.370) \cdot 10^{-4}}{9.850\cdot 10^{-4}}{0.370\cdot 10^{-4}}%
\htquantdef{Gamma103}{\Gamma_{103}}{\BRF{\tau^-}{3h^- 2h^+ \nu_\tau ~(\text{ex.~}K^0)}}{(8.216 \pm 0.316) \cdot 10^{-4}}{8.216\cdot 10^{-4}}{0.316\cdot 10^{-4}}%
\htquantdef{Gamma104}{\Gamma_{104}}{\BRF{\tau^-}{3h^- 2h^+ \pi^0 \nu_\tau ~(\text{ex.~}K^0)}}{(1.634 \pm 0.114) \cdot 10^{-4}}{1.634\cdot 10^{-4}}{0.114\cdot 10^{-4}}%
\htquantdef{Gamma106}{\Gamma_{106}}{\BRF{\tau^-}{(5\pi)^- \nu_\tau}}{(0.7748 \pm 0.0534) \cdot 10^{-2}}{0.7748\cdot 10^{-2}}{0.0534\cdot 10^{-2}}%
\htquantdef{Gamma10by5}{\frac{\Gamma_{10}}{\Gamma_{5}}}{\frac{\BRF{\tau^-}{K^- \nu_\tau}}{\BRF{\tau^-}{e^- \bar{\nu}_e \nu_\tau}}}{(3.906 \pm 0.054) \cdot 10^{-2}}{3.906\cdot 10^{-2}}{0.054\cdot 10^{-2}}%
\htquantdef{Gamma10by9}{\frac{\Gamma_{10}}{\Gamma_{9}}}{\frac{\BRF{\tau^-}{K^- \nu_\tau}}{\BRF{\tau^-}{\pi^- \nu_\tau}}}{(6.438 \pm 0.094) \cdot 10^{-2}}{6.438\cdot 10^{-2}}{0.094\cdot 10^{-2}}%
\htquantdef{Gamma11}{\Gamma_{11}}{\BRF{\tau^-}{h^- \ge{} 1\,  \text{neutrals}\, \nu_\tau}}{0.36973 \pm 0.00097}{0.36973}{0.00097}%
\htquantdef{Gamma110}{\Gamma_{110}}{\BRF{\tau^-}{X_s^- \nu_\tau}}{(2.909 \pm 0.048) \cdot 10^{-2}}{2.909\cdot 10^{-2}}{0.048\cdot 10^{-2}}%
\htquantdef{Gamma110_pdg09}{\Gamma_{110}_pdg09}{}{(2.841 \pm 0.038) \cdot 10^{-2}}{2.841\cdot 10^{-2}}{0.038\cdot 10^{-2}}%
\htquantdef{Gamma12}{\Gamma_{12}}{\BRF{\tau^-}{h^- \ge{} 1\, \pi^0\, \nu_\tau\;(\text{ex.~} K^0)}}{0.36475 \pm 0.00097}{0.36475}{0.00097}%
\htquantdef{Gamma126}{\Gamma_{126}}{\BRF{\tau^-}{\pi^- \pi^0 \eta \nu_\tau}}{(0.1386 \pm 0.0072) \cdot 10^{-2}}{0.1386\cdot 10^{-2}}{0.0072\cdot 10^{-2}}%
\htquantdef{Gamma128}{\Gamma_{128}}{\BRF{\tau^-}{K^- \eta \nu_\tau}}{(1.547 \pm 0.080) \cdot 10^{-4}}{1.547\cdot 10^{-4}}{0.080\cdot 10^{-4}}%
\htquantdef{Gamma13}{\Gamma_{13}}{\BRF{\tau^-}{h^- \pi^0 \nu_\tau}}{0.25935 \pm 0.00091}{0.25935}{0.00091}%
\htquantdef{Gamma130}{\Gamma_{130}}{\BRF{\tau^-}{K^- \pi^0 \eta \nu_\tau}}{(4.827 \pm 1.161) \cdot 10^{-5}}{4.827\cdot 10^{-5}}{1.161\cdot 10^{-5}}%
\htquantdef{Gamma132}{\Gamma_{132}}{\BRF{\tau^-}{\pi^- \bar{K}^0 \eta \nu_\tau}}{(9.371 \pm 1.491) \cdot 10^{-5}}{9.371\cdot 10^{-5}}{1.491\cdot 10^{-5}}%
\htquantdef{Gamma136}{\Gamma_{136}}{\BRF{\tau^-}{\pi^- \pi^+ \pi^- \eta \nu_\tau\;(\text{ex.~} K^0)}}{(2.184 \pm 0.130) \cdot 10^{-4}}{2.184\cdot 10^{-4}}{0.130\cdot 10^{-4}}%
\htquantdef{Gamma14}{\Gamma_{14}}{\BRF{\tau^-}{\pi^- \pi^0 \nu_\tau}}{0.25502 \pm 0.00092}{0.25502}{0.00092}%
\htquantdef{Gamma149}{\Gamma_{149}}{\BRF{\tau^-}{h^- \omega \ge{} 0\,  \text{neutrals}\, \nu_\tau}}{(2.401 \pm 0.075) \cdot 10^{-2}}{2.401\cdot 10^{-2}}{0.075\cdot 10^{-2}}%
\htquantdef{Gamma150}{\Gamma_{150}}{\BRF{\tau^-}{h^- \omega \nu_\tau}}{(1.995 \pm 0.064) \cdot 10^{-2}}{1.995\cdot 10^{-2}}{0.064\cdot 10^{-2}}%
\htquantdef{Gamma150by66}{\frac{\Gamma_{150}}{\Gamma_{66}}}{\frac{\BRF{\tau^-}{h^- \omega \nu_\tau}}{\BRF{\tau^-}{h^- h^- h^+ \pi^0 \nu_\tau\;(\text{ex.~} K^0)}}}{0.4332 \pm 0.0139}{0.4332}{0.0139}%
\htquantdef{Gamma151}{\Gamma_{151}}{\BRF{\tau^-}{K^- \omega \nu_\tau}}{(4.100 \pm 0.922) \cdot 10^{-4}}{4.100\cdot 10^{-4}}{0.922\cdot 10^{-4}}%
\htquantdef{Gamma152}{\Gamma_{152}}{\BRF{\tau^-}{h^- \pi^0 \omega \nu_\tau}}{(0.4058 \pm 0.0419) \cdot 10^{-2}}{0.4058\cdot 10^{-2}}{0.0419\cdot 10^{-2}}%
\htquantdef{Gamma152by54}{\frac{\Gamma_{152}}{\Gamma_{54}}}{\frac{\BRF{\tau^-}{h^- \omega \pi^0 \nu_\tau}}{\BRF{\tau^-}{h^- h^- h^+ \ge{} 0\, \text{neutrals} \ge{} 0\,  K_L^0\, \nu_\tau}}}{(2.667 \pm 0.275) \cdot 10^{-2}}{2.667\cdot 10^{-2}}{0.275\cdot 10^{-2}}%
\htquantdef{Gamma152by76}{\frac{\Gamma_{152}}{\Gamma_{76}}}{\frac{\BRF{\tau^-}{h^- \omega \pi^0 \nu_\tau}}{\BRF{\tau^-}{h^- h^- h^+ 2\pi^0 \nu_\tau\;(\text{ex.~} K^0)}}}{0.8241 \pm 0.0757}{0.8241}{0.0757}%
\htquantdef{Gamma16}{\Gamma_{16}}{\BRF{\tau^-}{K^- \pi^0 \nu_\tau}}{(0.4327 \pm 0.0149) \cdot 10^{-2}}{0.4327\cdot 10^{-2}}{0.0149\cdot 10^{-2}}%
\htquantdef{Gamma167}{\Gamma_{167}}{\BRF{\tau^-}{K^- \phi \nu_\tau}}{(4.445 \pm 1.636) \cdot 10^{-5}}{4.445\cdot 10^{-5}}{1.636\cdot 10^{-5}}%
\htquantdef{Gamma168}{\Gamma_{168}}{\BRF{\tau^-}{K^- \phi \nu_\tau ~(\phi \to K^+ K^-)}}{(2.174 \pm 0.800) \cdot 10^{-5}}{2.174\cdot 10^{-5}}{0.800\cdot 10^{-5}}%
\htquantdef{Gamma169}{\Gamma_{169}}{\BRF{\tau^-}{K^- \phi \nu_\tau ~(\phi \to K_S^0 K_L^0)}}{(1.520 \pm 0.560) \cdot 10^{-5}}{1.520\cdot 10^{-5}}{0.560\cdot 10^{-5}}%
\htquantdef{Gamma17}{\Gamma_{17}}{\BRF{\tau^-}{h^- \ge{} 2\,  \pi^0\, \nu_\tau}}{0.10775 \pm 0.00095}{0.10775}{0.00095}%
\htquantdef{Gamma18}{\Gamma_{18}}{\BRF{\tau^-}{h^- 2\pi^0 \nu_\tau}}{(9.458 \pm 0.097) \cdot 10^{-2}}{9.458\cdot 10^{-2}}{0.097\cdot 10^{-2}}%
\htquantdef{Gamma19}{\Gamma_{19}}{\BRF{\tau^-}{h^- 2\pi^0 \nu_\tau\;(\text{ex.~} K^0)}}{(9.306 \pm 0.097) \cdot 10^{-2}}{9.306\cdot 10^{-2}}{0.097\cdot 10^{-2}}%
\htquantdef{Gamma19by13}{\frac{\Gamma_{19}}{\Gamma_{13}}}{\frac{\BRF{\tau^-}{h^- 2\pi^0 \nu_\tau\;(\text{ex.~} K^0)}}{\BRF{\tau^-}{h^- \pi^0 \nu_\tau}}}{0.3588 \pm 0.0044}{0.3588}{0.0044}%
\htquantdef{Gamma2}{\Gamma_{2}}{\BRF{\tau^-}{(\text{particles})^- \ge{} 0\, \text{neutrals} \ge{} 0\,  K_L^0\, \nu_\tau}}{0.8453 \pm 0.0010}{0.8453}{0.0010}%
\htquantdef{Gamma20}{\Gamma_{20}}{\BRF{\tau^-}{\pi^- 2\pi^0 \nu_\tau ~(\text{ex.~}K^0)}}{(9.242 \pm 0.100) \cdot 10^{-2}}{9.242\cdot 10^{-2}}{0.100\cdot 10^{-2}}%
\htquantdef{Gamma23}{\Gamma_{23}}{\BRF{\tau^-}{K^- 2\pi^0 \nu_\tau ~(\text{ex.~}K^0)}}{(6.398 \pm 2.204) \cdot 10^{-4}}{6.398\cdot 10^{-4}}{2.204\cdot 10^{-4}}%
\htquantdef{Gamma24}{\Gamma_{24}}{\BRF{\tau^-}{h^- \ge{} 3\, \pi^0\, \nu_\tau}}{(1.318 \pm 0.065) \cdot 10^{-2}}{1.318\cdot 10^{-2}}{0.065\cdot 10^{-2}}%
\htquantdef{Gamma25}{\Gamma_{25}}{\BRF{\tau^-}{h^- \ge{} 3\, \pi^0\, \nu_\tau\;(\text{ex.~} K^0)}}{(1.233 \pm 0.065) \cdot 10^{-2}}{1.233\cdot 10^{-2}}{0.065\cdot 10^{-2}}%
\htquantdef{Gamma26}{\Gamma_{26}}{\BRF{\tau^-}{h^- 3\pi^0 \nu_\tau}}{(1.158 \pm 0.072) \cdot 10^{-2}}{1.158\cdot 10^{-2}}{0.072\cdot 10^{-2}}%
\htquantdef{Gamma26by13}{\frac{\Gamma_{26}}{\Gamma_{13}}}{\frac{\BRF{\tau^-}{h^- 3\pi^0 \nu_\tau}}{\BRF{\tau^-}{h^- \pi^0 \nu_\tau}}}{(4.465 \pm 0.277) \cdot 10^{-2}}{4.465\cdot 10^{-2}}{0.277\cdot 10^{-2}}%
\htquantdef{Gamma27}{\Gamma_{27}}{\BRF{\tau^-}{\pi^- 3\pi^0 \nu_\tau ~(\text{ex.~}K^0)}}{(1.029 \pm 0.075) \cdot 10^{-2}}{1.029\cdot 10^{-2}}{0.075\cdot 10^{-2}}%
\htquantdef{Gamma28}{\Gamma_{28}}{\BRF{\tau^-}{K^- 3\pi^0 \nu_\tau ~(\text{ex.~}K^0,\eta)}}{(4.283 \pm 2.161) \cdot 10^{-4}}{4.283\cdot 10^{-4}}{2.161\cdot 10^{-4}}%
\htquantdef{Gamma29}{\Gamma_{29}}{\BRF{\tau^-}{h^- 4\pi^0 \nu_\tau\;(\text{ex.~} K^0)}}{(0.1568 \pm 0.0391) \cdot 10^{-2}}{0.1568\cdot 10^{-2}}{0.0391\cdot 10^{-2}}%
\htquantdef{Gamma3}{\Gamma_{3}}{\BRF{\tau^-}{\mu^- \bar{\nu}_\mu \nu_\tau}}{0.17392 \pm 0.00040}{0.17392}{0.00040}%
\htquantdef{Gamma30}{\Gamma_{30}}{\BRF{\tau^-}{h^- 4\pi^0 \nu_\tau ~(\text{ex.~}K^0,\eta)}}{(0.1099 \pm 0.0391) \cdot 10^{-2}}{0.1099\cdot 10^{-2}}{0.0391\cdot 10^{-2}}%
\htquantdef{Gamma31}{\Gamma_{31}}{\BRF{\tau^-}{K^- \ge{} 0\, \pi^0 \ge{} 0\, K^0 \ge{} 0\, \gamma \nu_\tau}}{(1.545 \pm 0.030) \cdot 10^{-2}}{1.545\cdot 10^{-2}}{0.030\cdot 10^{-2}}%
\htquantdef{Gamma32}{\Gamma_{32}}{\BRF{\tau^-}{K^- \ge{} 1\, (\pi^0\,\text{or}\,K^0\,\text{or}\,\gamma) \nu_\tau}}{(0.8528 \pm 0.0286) \cdot 10^{-2}}{0.8528\cdot 10^{-2}}{0.0286\cdot 10^{-2}}%
\htquantdef{Gamma33}{\Gamma_{33}}{\BRF{\tau^-}{K_S^0 (\text{particles})^- \nu_\tau}}{(0.9372 \pm 0.0292) \cdot 10^{-2}}{0.9372\cdot 10^{-2}}{0.0292\cdot 10^{-2}}%
\htquantdef{Gamma34}{\Gamma_{34}}{\BRF{\tau^-}{h^- \bar{K}^0 \nu_\tau}}{(0.9865 \pm 0.0139) \cdot 10^{-2}}{0.9865\cdot 10^{-2}}{0.0139\cdot 10^{-2}}%
\htquantdef{Gamma35}{\Gamma_{35}}{\BRF{\tau^-}{\pi^- \bar{K}^0 \nu_\tau}}{(0.8386 \pm 0.0141) \cdot 10^{-2}}{0.8386\cdot 10^{-2}}{0.0141\cdot 10^{-2}}%
\htquantdef{Gamma37}{\Gamma_{37}}{\BRF{\tau^-}{K^- K^0 \nu_\tau}}{(0.1479 \pm 0.0053) \cdot 10^{-2}}{0.1479\cdot 10^{-2}}{0.0053\cdot 10^{-2}}%
\htquantdef{Gamma38}{\Gamma_{38}}{\BRF{\tau^-}{K^- K^0 \ge{} 0\,  \pi^0\, \nu_\tau}}{(0.2982 \pm 0.0079) \cdot 10^{-2}}{0.2982\cdot 10^{-2}}{0.0079\cdot 10^{-2}}%
\htquantdef{Gamma39}{\Gamma_{39}}{\BRF{\tau^-}{h^- \bar{K}^0 \pi^0 \nu_\tau}}{(0.5314 \pm 0.0134) \cdot 10^{-2}}{0.5314\cdot 10^{-2}}{0.0134\cdot 10^{-2}}%
\htquantdef{Gamma3by5}{\frac{\Gamma_{3}}{\Gamma_{5}}}{\frac{\BRF{\tau^-}{\mu^- \bar{\nu}_\mu \nu_\tau}}{\BRF{\tau^-}{e^- \bar{\nu}_e \nu_\tau}}}{0.9762 \pm 0.0028}{0.9762}{0.0028}%
\htquantdef{Gamma40}{\Gamma_{40}}{\BRF{\tau^-}{\pi^- \bar{K}^0 \pi^0 \nu_\tau}}{(0.3812 \pm 0.0129) \cdot 10^{-2}}{0.3812\cdot 10^{-2}}{0.0129\cdot 10^{-2}}%
\htquantdef{Gamma42}{\Gamma_{42}}{\BRF{\tau^-}{K^- \pi^0 K^0 \nu_\tau}}{(0.1502 \pm 0.0071) \cdot 10^{-2}}{0.1502\cdot 10^{-2}}{0.0071\cdot 10^{-2}}%
\htquantdef{Gamma43}{\Gamma_{43}}{\BRF{\tau^-}{\pi^- \bar{K}^0 \ge{} 1\,  \pi^0\, \nu_\tau}}{(0.4046 \pm 0.0260) \cdot 10^{-2}}{0.4046\cdot 10^{-2}}{0.0260\cdot 10^{-2}}%
\htquantdef{Gamma44}{\Gamma_{44}}{\BRF{\tau^-}{\pi^- \bar{K}^0 \pi^0 \pi^0 \nu_\tau ~(\text{ex.~}K^0)}}{(2.340 \pm 2.306) \cdot 10^{-4}}{2.340\cdot 10^{-4}}{2.306\cdot 10^{-4}}%
\htquantdef{Gamma46}{\Gamma_{46}}{\BRF{\tau^-}{\pi^- K^0 \bar{K}^0 \nu_\tau}}{(0.1513 \pm 0.0247) \cdot 10^{-2}}{0.1513\cdot 10^{-2}}{0.0247\cdot 10^{-2}}%
\htquantdef{Gamma47}{\Gamma_{47}}{\BRF{\tau^-}{\pi^- K_S^0 K_S^0 \nu_\tau}}{(2.332 \pm 0.065) \cdot 10^{-4}}{2.332\cdot 10^{-4}}{0.065\cdot 10^{-4}}%
\htquantdef{Gamma48}{\Gamma_{48}}{\BRF{\tau^-}{\pi^- K_S^0 K_L^0 \nu_\tau}}{(0.1047 \pm 0.0247) \cdot 10^{-2}}{0.1047\cdot 10^{-2}}{0.0247\cdot 10^{-2}}%
\htquantdef{Gamma49}{\Gamma_{49}}{\BRF{\tau^-}{\pi^- K^0 \bar{K}^0 \pi^0 \nu_\tau}}{(3.540 \pm 1.193) \cdot 10^{-4}}{3.540\cdot 10^{-4}}{1.193\cdot 10^{-4}}%
\htquantdef{Gamma5}{\Gamma_{5}}{\BRF{\tau^-}{e^- \bar{\nu}_e \nu_\tau}}{0.17816 \pm 0.00041}{0.17816}{0.00041}%
\htquantdef{Gamma50}{\Gamma_{50}}{\BRF{\tau^-}{\pi^- \pi^0 K_S^0 K_S^0 \nu_\tau}}{(1.815 \pm 0.207) \cdot 10^{-5}}{1.815\cdot 10^{-5}}{0.207\cdot 10^{-5}}%
\htquantdef{Gamma51}{\Gamma_{51}}{\BRF{\tau^-}{\pi^- \pi^0 K_S^0 K_L^0 \nu_\tau}}{(3.177 \pm 1.192) \cdot 10^{-4}}{3.177\cdot 10^{-4}}{1.192\cdot 10^{-4}}%
\htquantdef{Gamma53}{\Gamma_{53}}{\BRF{\tau^-}{\bar{K}^0 h^- h^- h^+ \nu_\tau}}{(2.218 \pm 2.024) \cdot 10^{-4}}{2.218\cdot 10^{-4}}{2.024\cdot 10^{-4}}%
\htquantdef{Gamma54}{\Gamma_{54}}{\BRF{\tau^-}{h^- h^- h^+ \ge{} 0\, \text{neutrals} \ge{} 0\,  K_L^0\, \nu_\tau}}{0.15215 \pm 0.00061}{0.15215}{0.00061}%
\htquantdef{Gamma55}{\Gamma_{55}}{\BRF{\tau^-}{h^- h^- h^+ \ge{} 0\,  \text{neutrals}\, \nu_\tau\;(\text{ex.~} K^0)}}{0.14567 \pm 0.00057}{0.14567}{0.00057}%
\htquantdef{Gamma56}{\Gamma_{56}}{\BRF{\tau^-}{h^- h^- h^+ \nu_\tau}}{(9.780 \pm 0.054) \cdot 10^{-2}}{9.780\cdot 10^{-2}}{0.054\cdot 10^{-2}}%
\htquantdef{Gamma57}{\Gamma_{57}}{\BRF{\tau^-}{h^- h^- h^+ \nu_\tau\;(\text{ex.~} K^0)}}{(9.439 \pm 0.053) \cdot 10^{-2}}{9.439\cdot 10^{-2}}{0.053\cdot 10^{-2}}%
\htquantdef{Gamma57by55}{\frac{\Gamma_{57}}{\Gamma_{55}}}{\frac{\BRF{\tau^-}{h^- h^- h^+ \nu_\tau\;(\text{ex.~} K^0)}}{\BRF{\tau^-}{h^- h^- h^+ \ge{} 0\,  \text{neutrals}\, \nu_\tau\;(\text{ex.~} K^0)}}}{0.6480 \pm 0.0030}{0.6480}{0.0030}%
\htquantdef{Gamma58}{\Gamma_{58}}{\BRF{\tau^-}{h^- h^- h^+ \nu_\tau\;(\text{ex.~} K^0, \omega)}}{(9.408 \pm 0.053) \cdot 10^{-2}}{9.408\cdot 10^{-2}}{0.053\cdot 10^{-2}}%
\htquantdef{Gamma59}{\Gamma_{59}}{\BRF{\tau^-}{\pi^- \pi^+ \pi^- \nu_\tau}}{(9.290 \pm 0.052) \cdot 10^{-2}}{9.290\cdot 10^{-2}}{0.052\cdot 10^{-2}}%
\htquantdef{Gamma60}{\Gamma_{60}}{\BRF{\tau^-}{\pi^- \pi^+ \pi^- \nu_\tau\;(\text{ex.~} K^0)}}{(9.000 \pm 0.051) \cdot 10^{-2}}{9.000\cdot 10^{-2}}{0.051\cdot 10^{-2}}%
\htquantdef{Gamma62}{\Gamma_{62}}{\BRF{\tau^-}{\pi^- \pi^- \pi^+ \nu_\tau ~(\text{ex.~}K^0,\omega)}}{(8.970 \pm 0.052) \cdot 10^{-2}}{8.970\cdot 10^{-2}}{0.052\cdot 10^{-2}}%
\htquantdef{Gamma63}{\Gamma_{63}}{\BRF{\tau^-}{h^- h^- h^+ \ge{} 1\,  \text{neutrals}\, \nu_\tau}}{(5.325 \pm 0.050) \cdot 10^{-2}}{5.325\cdot 10^{-2}}{0.050\cdot 10^{-2}}%
\htquantdef{Gamma64}{\Gamma_{64}}{\BRF{\tau^-}{h^- h^- h^+ \ge{} 1\,  \pi^0\, \nu_\tau\;(\text{ex.~} K^0)}}{(5.120 \pm 0.049) \cdot 10^{-2}}{5.120\cdot 10^{-2}}{0.049\cdot 10^{-2}}%
\htquantdef{Gamma65}{\Gamma_{65}}{\BRF{\tau^-}{h^- h^- h^+ \pi^0 \nu_\tau}}{(4.790 \pm 0.052) \cdot 10^{-2}}{4.790\cdot 10^{-2}}{0.052\cdot 10^{-2}}%
\htquantdef{Gamma66}{\Gamma_{66}}{\BRF{\tau^-}{h^- h^- h^+ \pi^0 \nu_\tau\;(\text{ex.~} K^0)}}{(4.606 \pm 0.051) \cdot 10^{-2}}{4.606\cdot 10^{-2}}{0.051\cdot 10^{-2}}%
\htquantdef{Gamma67}{\Gamma_{67}}{\BRF{\tau^-}{h^- h^- h^+ \pi^0 \nu_\tau\;(\text{ex.~} K^0, \omega)}}{(2.820 \pm 0.070) \cdot 10^{-2}}{2.820\cdot 10^{-2}}{0.070\cdot 10^{-2}}%
\htquantdef{Gamma68}{\Gamma_{68}}{\BRF{\tau^-}{\pi^- \pi^+ \pi^- \pi^0 \nu_\tau}}{(4.651 \pm 0.053) \cdot 10^{-2}}{4.651\cdot 10^{-2}}{0.053\cdot 10^{-2}}%
\htquantdef{Gamma69}{\Gamma_{69}}{\BRF{\tau^-}{\pi^- \pi^+ \pi^- \pi^0 \nu_\tau\;(\text{ex.~} K^0)}}{(4.519 \pm 0.052) \cdot 10^{-2}}{4.519\cdot 10^{-2}}{0.052\cdot 10^{-2}}%
\htquantdef{Gamma7}{\Gamma_{7}}{\BRF{\tau^-}{h^- \ge{} 0\,  K_L^0\, \nu_\tau}}{0.12023 \pm 0.00054}{0.12023}{0.00054}%
\htquantdef{Gamma70}{\Gamma_{70}}{\BRF{\tau^-}{\pi^- \pi^- \pi^+ \pi^0 \nu_\tau ~(\text{ex.~}K^0,\omega)}}{(2.769 \pm 0.071) \cdot 10^{-2}}{2.769\cdot 10^{-2}}{0.071\cdot 10^{-2}}%
\htquantdef{Gamma74}{\Gamma_{74}}{\BRF{\tau^-}{h^- h^- h^+ \ge{} 2\, \pi^0\, \nu_\tau\;(\text{ex.~} K^0)}}{(0.5135 \pm 0.0312) \cdot 10^{-2}}{0.5135\cdot 10^{-2}}{0.0312\cdot 10^{-2}}%
\htquantdef{Gamma75}{\Gamma_{75}}{\BRF{\tau^-}{h^- h^- h^+ 2\pi^0 \nu_\tau}}{(0.5024 \pm 0.0310) \cdot 10^{-2}}{0.5024\cdot 10^{-2}}{0.0310\cdot 10^{-2}}%
\htquantdef{Gamma76}{\Gamma_{76}}{\BRF{\tau^-}{h^- h^- h^+ 2\pi^0 \nu_\tau\;(\text{ex.~} K^0)}}{(0.4925 \pm 0.0310) \cdot 10^{-2}}{0.4925\cdot 10^{-2}}{0.0310\cdot 10^{-2}}%
\htquantdef{Gamma76by54}{\frac{\Gamma_{76}}{\Gamma_{54}}}{\frac{\BRF{\tau^-}{h^- h^- h^+ 2\pi^0 \nu_\tau\;(\text{ex.~} K^0)}}{\BRF{\tau^-}{h^- h^- h^+ \ge{} 0\, \text{neutrals} \ge{} 0\,  K_L^0\, \nu_\tau}}}{(3.237 \pm 0.202) \cdot 10^{-2}}{3.237\cdot 10^{-2}}{0.202\cdot 10^{-2}}%
\htquantdef{Gamma77}{\Gamma_{77}}{\BRF{\tau^-}{h^- h^- h^+ 2\pi^0 \nu_\tau ~(\text{ex.~}K^0,\omega,\eta)}}{(9.759 \pm 3.550) \cdot 10^{-4}}{9.759\cdot 10^{-4}}{3.550\cdot 10^{-4}}%
\htquantdef{Gamma78}{\Gamma_{78}}{\BRF{\tau^-}{h^- h^- h^+ 3\pi^0 \nu_\tau}}{(2.107 \pm 0.299) \cdot 10^{-4}}{2.107\cdot 10^{-4}}{0.299\cdot 10^{-4}}%
\htquantdef{Gamma79}{\Gamma_{79}}{\BRF{\tau^-}{K^- h^- h^+ \ge{} 0\,  \text{neutrals}\, \nu_\tau}}{(0.6297 \pm 0.0141) \cdot 10^{-2}}{0.6297\cdot 10^{-2}}{0.0141\cdot 10^{-2}}%
\htquantdef{Gamma8}{\Gamma_{8}}{\BRF{\tau^-}{h^- \nu_\tau}}{0.11506 \pm 0.00054}{0.11506}{0.00054}%
\htquantdef{Gamma80}{\Gamma_{80}}{\BRF{\tau^-}{K^- \pi^- h^+ \nu_\tau\;(\text{ex.~} K^0)}}{(0.4363 \pm 0.0073) \cdot 10^{-2}}{0.4363\cdot 10^{-2}}{0.0073\cdot 10^{-2}}%
\htquantdef{Gamma800}{\Gamma_{800}}{\BRF{\tau^-}{\pi^- \omega \nu_\tau}}{(1.954 \pm 0.065) \cdot 10^{-2}}{1.954\cdot 10^{-2}}{0.065\cdot 10^{-2}}%
\htquantdef{Gamma802}{\Gamma_{802}}{\BRF{\tau^-}{K^- \pi^- \pi^+ \nu_\tau ~(\text{ex.~}K^0,\omega)}}{(0.2923 \pm 0.0067) \cdot 10^{-2}}{0.2923\cdot 10^{-2}}{0.0067\cdot 10^{-2}}%
\htquantdef{Gamma803}{\Gamma_{803}}{\BRF{\tau^-}{K^- \pi^- \pi^+ \pi^0 \nu_\tau ~(\text{ex.~}K^0,\omega,\eta)}}{(4.103 \pm 1.429) \cdot 10^{-4}}{4.103\cdot 10^{-4}}{1.429\cdot 10^{-4}}%
\htquantdef{Gamma804}{\Gamma_{804}}{\BRF{\tau^-}{\pi^- K_L^0 K_L^0 \nu_\tau}}{(2.332 \pm 0.065) \cdot 10^{-4}}{2.332\cdot 10^{-4}}{0.065\cdot 10^{-4}}%
\htquantdef{Gamma805}{\Gamma_{805}}{\BRF{\tau^-}{a_1^- (\to \pi^- \gamma) \nu_\tau}}{(4.000 \pm 2.000) \cdot 10^{-4}}{4.000\cdot 10^{-4}}{2.000\cdot 10^{-4}}%
\htquantdef{Gamma806}{\Gamma_{806}}{\BRF{\tau^-}{\pi^- \pi^0 K_L^0 K_L^0 \nu_\tau}}{(1.815 \pm 0.207) \cdot 10^{-5}}{1.815\cdot 10^{-5}}{0.207\cdot 10^{-5}}%
\htquantdef{Gamma80by60}{\frac{\Gamma_{80}}{\Gamma_{60}}}{\frac{\BRF{\tau^-}{K^- \pi^- h^+ \nu_\tau\;(\text{ex.~} K^0)}}{\BRF{\tau^-}{\pi^- \pi^+ \pi^- \nu_\tau\;(\text{ex.~} K^0)}}}{(4.847 \pm 0.080) \cdot 10^{-2}}{4.847\cdot 10^{-2}}{0.080\cdot 10^{-2}}%
\htquantdef{Gamma81}{\Gamma_{81}}{\BRF{\tau^-}{K^- \pi^- h^+ \pi^0 \nu_\tau\;(\text{ex.~} K^0)}}{(8.726 \pm 1.177) \cdot 10^{-4}}{8.726\cdot 10^{-4}}{1.177\cdot 10^{-4}}%
\htquantdef{Gamma810}{\Gamma_{810}}{\BRF{\tau^-}{2\pi^- \pi^+ 3\pi^0 \nu_\tau ~(\text{ex.~}K^0)}}{(1.924 \pm 0.298) \cdot 10^{-4}}{1.924\cdot 10^{-4}}{0.298\cdot 10^{-4}}%
\htquantdef{Gamma811}{\Gamma_{811}}{\BRF{\tau^-}{\pi^- 2\pi^0 \omega \nu_\tau ~(\text{ex.~}K^0)}}{(7.105 \pm 1.586) \cdot 10^{-5}}{7.105\cdot 10^{-5}}{1.586\cdot 10^{-5}}%
\htquantdef{Gamma812}{\Gamma_{812}}{\BRF{\tau^-}{2\pi^- \pi^+ 3\pi^0 \nu_\tau ~(\text{ex.~}K^0, \eta, \omega, f_1)}}{(1.344 \pm 2.683) \cdot 10^{-5}}{1.344\cdot 10^{-5}}{2.683\cdot 10^{-5}}%
\htquantdef{Gamma81by69}{\frac{\Gamma_{81}}{\Gamma_{69}}}{\frac{\BRF{\tau^-}{K^- \pi^- h^+ \pi^0 \nu_\tau\;(\text{ex.~} K^0)}}{\BRF{\tau^-}{\pi^- \pi^+ \pi^- \pi^0 \nu_\tau\;(\text{ex.~} K^0)}}}{(1.931 \pm 0.266) \cdot 10^{-2}}{1.931\cdot 10^{-2}}{0.266\cdot 10^{-2}}%
\htquantdef{Gamma82}{\Gamma_{82}}{\BRF{\tau^-}{K^- \pi^- \pi^+ \ge{} 0\,  \text{neutrals}\, \nu_\tau}}{(0.4780 \pm 0.0137) \cdot 10^{-2}}{0.4780\cdot 10^{-2}}{0.0137\cdot 10^{-2}}%
\htquantdef{Gamma820}{\Gamma_{820}}{\BRF{\tau^-}{3\pi^- 2\pi^+ \nu_\tau ~(\text{ex.~}K^0, \omega)}}{(8.197 \pm 0.315) \cdot 10^{-4}}{8.197\cdot 10^{-4}}{0.315\cdot 10^{-4}}%
\htquantdef{Gamma821}{\Gamma_{821}}{\BRF{\tau^-}{3\pi^- 2\pi^+ \nu_\tau ~(\text{ex.~}K^0, \omega, f_1)}}{(7.677 \pm 0.297) \cdot 10^{-4}}{7.677\cdot 10^{-4}}{0.297\cdot 10^{-4}}%
\htquantdef{Gamma822}{\Gamma_{822}}{\BRF{\tau^-}{K^- 2\pi^- 2\pi^+ \nu_\tau ~(\text{ex.~}K^0)}}{(0.596 \pm 1.208) \cdot 10^{-6}}{0.596\cdot 10^{-6}}{1.208\cdot 10^{-6}}%
\htquantdef{Gamma83}{\Gamma_{83}}{\BRF{\tau^-}{K^- \pi^- \pi^+ \ge{} 0\,  \pi^0\, \nu_\tau\;(\text{ex.~} K^0)}}{(0.3741 \pm 0.0135) \cdot 10^{-2}}{0.3741\cdot 10^{-2}}{0.0135\cdot 10^{-2}}%
\htquantdef{Gamma830}{\Gamma_{830}}{\BRF{\tau^-}{3\pi^- 2\pi^+ \pi^0 \nu_\tau ~(\text{ex.~}K^0)}}{(1.623 \pm 0.114) \cdot 10^{-4}}{1.623\cdot 10^{-4}}{0.114\cdot 10^{-4}}%
\htquantdef{Gamma831}{\Gamma_{831}}{\BRF{\tau^-}{2\pi^- \pi^+ \omega \nu_\tau ~(\text{ex.~}K^0)}}{(8.359 \pm 0.626) \cdot 10^{-5}}{8.359\cdot 10^{-5}}{0.626\cdot 10^{-5}}%
\htquantdef{Gamma832}{\Gamma_{832}}{\BRF{\tau^-}{3\pi^- 2\pi^+ \pi^0 \nu_\tau ~(\text{ex.~}K^0, \eta, \omega, f_1)}}{(3.771 \pm 0.875) \cdot 10^{-5}}{3.771\cdot 10^{-5}}{0.875\cdot 10^{-5}}%
\htquantdef{Gamma833}{\Gamma_{833}}{\BRF{\tau^-}{K^- 2\pi^- 2\pi^+ \pi^0 \nu_\tau ~(\text{ex.~}K^0)}}{(1.108 \pm 0.566) \cdot 10^{-6}}{1.108\cdot 10^{-6}}{0.566\cdot 10^{-6}}%
\htquantdef{Gamma84}{\Gamma_{84}}{\BRF{\tau^-}{K^- \pi^- \pi^+ \nu_\tau}}{(0.3441 \pm 0.0070) \cdot 10^{-2}}{0.3441\cdot 10^{-2}}{0.0070\cdot 10^{-2}}%
\htquantdef{Gamma85}{\Gamma_{85}}{\BRF{\tau^-}{K^- \pi^+ \pi^- \nu_\tau\;(\text{ex.~} K^0)}}{(0.2929 \pm 0.0067) \cdot 10^{-2}}{0.2929\cdot 10^{-2}}{0.0067\cdot 10^{-2}}%
\htquantdef{Gamma85by60}{\frac{\Gamma_{85}}{\Gamma_{60}}}{\frac{\BRF{\tau^-}{K^- \pi^+ \pi^- \nu_\tau\;(\text{ex.~}K^0)}}{\BRF{\tau^-}{\pi^- \pi^+ \pi^- \nu_\tau\;(\text{ex.~}K^0)}}}{(3.254 \pm 0.074) \cdot 10^{-2}}{3.254\cdot 10^{-2}}{0.074\cdot 10^{-2}}%
\htquantdef{Gamma87}{\Gamma_{87}}{\BRF{\tau^-}{K^- \pi^- \pi^+ \pi^0 \nu_\tau}}{(0.1331 \pm 0.0119) \cdot 10^{-2}}{0.1331\cdot 10^{-2}}{0.0119\cdot 10^{-2}}%
\htquantdef{Gamma88}{\Gamma_{88}}{\BRF{\tau^-}{K^- \pi^- \pi^+ \pi^0 \nu_\tau\;(\text{ex.~} K^0)}}{(8.115 \pm 1.168) \cdot 10^{-4}}{8.115\cdot 10^{-4}}{1.168\cdot 10^{-4}}%
\htquantdef{Gamma89}{\Gamma_{89}}{\BRF{\tau^-}{K^- \pi^- \pi^+ \pi^0 \nu_\tau\;(\text{ex.~} K^0, \eta)}}{(7.761 \pm 1.168) \cdot 10^{-4}}{7.761\cdot 10^{-4}}{1.168\cdot 10^{-4}}%
\htquantdef{Gamma8by5}{\frac{\Gamma_{8}}{\Gamma_{5}}}{\frac{\BRF{\tau^-}{h^- \nu_\tau}}{\BRF{\tau^-}{e^- \bar{\nu}_e \nu_\tau}}}{0.6458 \pm 0.0033}{0.6458}{0.0033}%
\htquantdef{Gamma9}{\Gamma_{9}}{\BRF{\tau^-}{\pi^- \nu_\tau}}{0.10810 \pm 0.00053}{0.10810}{0.00053}%
\htquantdef{Gamma910}{\Gamma_{910}}{\BRF{\tau^-}{2\pi^- \pi^+ \eta \nu_\tau ~(\eta \to 3\pi^0) ~(\text{ex.~}K^0)}}{(7.136 \pm 0.424) \cdot 10^{-5}}{7.136\cdot 10^{-5}}{0.424\cdot 10^{-5}}%
\htquantdef{Gamma911}{\Gamma_{911}}{\BRF{\tau^-}{\pi^- 2\pi^0 \eta \nu_\tau ~(\eta \to \pi^+ \pi^- \pi^0) ~(\text{ex.~}K^0)}}{(4.420 \pm 0.867) \cdot 10^{-5}}{4.420\cdot 10^{-5}}{0.867\cdot 10^{-5}}%
\htquantdef{Gamma92}{\Gamma_{92}}{\BRF{\tau^-}{\pi^- K^- K^+ \ge{} 0\,  \text{neutrals}\, \nu_\tau}}{(0.1495 \pm 0.0033) \cdot 10^{-2}}{0.1495\cdot 10^{-2}}{0.0033\cdot 10^{-2}}%
\htquantdef{Gamma920}{\Gamma_{920}}{\BRF{\tau^-}{\pi^- f_1 \nu_\tau ~(f_1 \to 2\pi^- 2\pi^+)}}{(5.197 \pm 0.444) \cdot 10^{-5}}{5.197\cdot 10^{-5}}{0.444\cdot 10^{-5}}%
\htquantdef{Gamma93}{\Gamma_{93}}{\BRF{\tau^-}{\pi^- K^- K^+ \nu_\tau}}{(0.1434 \pm 0.0027) \cdot 10^{-2}}{0.1434\cdot 10^{-2}}{0.0027\cdot 10^{-2}}%
\htquantdef{Gamma930}{\Gamma_{930}}{\BRF{\tau^-}{2\pi^- \pi^+ \eta \nu_\tau ~(\eta \to \pi^+\pi^-\pi^0) ~(\text{ex.~}K^0)}}{(5.005 \pm 0.297) \cdot 10^{-5}}{5.005\cdot 10^{-5}}{0.297\cdot 10^{-5}}%
\htquantdef{Gamma93by60}{\frac{\Gamma_{93}}{\Gamma_{60}}}{\frac{\BRF{\tau^-}{\pi^- K^- K^+ \nu_\tau}}{\BRF{\tau^-}{\pi^- \pi^+ \pi^- \nu_\tau\;(\text{ex.~} K^0)}}}{(1.593 \pm 0.030) \cdot 10^{-2}}{1.593\cdot 10^{-2}}{0.030\cdot 10^{-2}}%
\htquantdef{Gamma94}{\Gamma_{94}}{\BRF{\tau^-}{\pi^- K^- K^+ \pi^0 \nu_\tau}}{(6.113 \pm 1.829) \cdot 10^{-5}}{6.113\cdot 10^{-5}}{1.829\cdot 10^{-5}}%
\htquantdef{Gamma944}{\Gamma_{944}}{\BRF{\tau^-}{2\pi^- \pi^+ \eta \nu_\tau ~(\eta \to \gamma\gamma) ~(\text{ex.~}K^0)}}{(8.606 \pm 0.511) \cdot 10^{-5}}{8.606\cdot 10^{-5}}{0.511\cdot 10^{-5}}%
\htquantdef{Gamma945}{\Gamma_{945}}{\BRF{\tau^-}{\pi^- 2\pi^0 \eta \nu_\tau}}{(1.929 \pm 0.378) \cdot 10^{-4}}{1.929\cdot 10^{-4}}{0.378\cdot 10^{-4}}%
\htquantdef{Gamma94by69}{\frac{\Gamma_{94}}{\Gamma_{69}}}{\frac{\BRF{\tau^-}{\pi^- K^- K^+ \pi^0 \nu_\tau}}{\BRF{\tau^-}{\pi^- \pi^+ \pi^- \pi^0 \nu_\tau\;(\text{ex.~} K^0)}}}{(0.1353 \pm 0.0405) \cdot 10^{-2}}{0.1353\cdot 10^{-2}}{0.0405\cdot 10^{-2}}%
\htquantdef{Gamma96}{\Gamma_{96}}{\BRF{\tau^-}{K^- K^- K^+ \nu_\tau}}{(2.174 \pm 0.800) \cdot 10^{-5}}{2.174\cdot 10^{-5}}{0.800\cdot 10^{-5}}%
\htquantdef{Gamma998}{\Gamma_{998}}{1 - \Gamma_{\text{All}}}{(0.0355 \pm 0.1031) \cdot 10^{-2}}{0.0355\cdot 10^{-2}}{0.1031\cdot 10^{-2}}%
\htquantdef{Gamma9by5}{\frac{\Gamma_{9}}{\Gamma_{5}}}{\frac{\BRF{\tau^-}{\pi^- \nu_\tau}}{\BRF{\tau^-}{e^- \bar{\nu}_e \nu_\tau}}}{0.6068 \pm 0.0032}{0.6068}{0.0032}%
\htquantdef{GammaAll}{\Gamma_{\text{All}}}{\Gamma_{\text{All}}}{0.9996 \pm 0.0010}{0.9996}{0.0010}%
\htquantdef{gmubyge_tau}{gmubyge_tau}{}{1.0019 \pm 0.0014}{1.0019}{0.0014}%
\htquantdef{gtaubyge_tau}{gtaubyge_tau}{}{1.0029 \pm 0.0015}{1.0029}{0.0015}%
\htquantdef{gtaubygmu_fit}{gtaubygmu_fit}{}{1.0000 \pm 0.0014}{1.0000}{0.0014}%
\htquantdef{gtaubygmu_K}{gtaubygmu_K}{}{0.9860 \pm 0.0070}{0.9860}{0.0070}%
\htquantdef{gtaubygmu_pi}{gtaubygmu_pi}{}{0.9961 \pm 0.0027}{0.9961}{0.0027}%
\htquantdef{gtaubygmu_tau}{gtaubygmu_tau}{}{1.0010 \pm 0.0015}{1.0010}{0.0015}%
\htquantdef{hcut}{hcut}{}{(6.582119514 \pm 0.000000040) \cdot 10^{-22}}{6.582119514\cdot 10^{-22}}{0.000000040\cdot 10^{-22}}%
\htquantdef{KmKzsNu}{KmKzsNu}{}{7.450\cdot 10^{-4}}{7.450\cdot 10^{-4}}{0}%
\htquantdef{KmPizKzsNu}{KmPizKzsNu}{}{7.550\cdot 10^{-4}}{7.550\cdot 10^{-4}}{0}%
\htquantdef{KtoENu}{KtoENu}{}{(1.5820 \pm 0.0070) \cdot 10^{-5}}{1.5820\cdot 10^{-5}}{0.0070\cdot 10^{-5}}%
\htquantdef{KtoMuNu}{KtoMuNu}{}{0.6356 \pm 0.0011}{0.6356}{0.0011}%
\htquantdef{m_e}{m_e}{}{0.510998928 \pm 0.000000011}{0.510998928}{0.000000011}%
\htquantdef{m_K}{m_K}{}{493.677 \pm 0.016}{493.677}{0.016}%
\htquantdef{m_mu}{m_mu}{}{105.6583715 \pm 0.0000035}{105.6583715}{0.0000035}%
\htquantdef{m_pi}{m_pi}{}{139.57018 \pm 0.00035}{139.57018}{0.00035}%
\htquantdef{m_s}{m_s}{}{95.00 \pm 5.00}{95.00}{5.00}%
\htquantdef{m_tau}{m_tau}{}{(1.77686 \pm 0.00012) \cdot 10^{3}}{1.77686\cdot 10^{3}}{0.00012\cdot 10^{3}}%
\htquantdef{m_W}{m_W}{}{(8.0385 \pm 0.0015) \cdot 10^{4}}{8.0385\cdot 10^{4}}{0.0015\cdot 10^{4}}%
\htquantdef{MumNuNumb}{MumNuNumb}{}{0.1741}{0.1741}{0}%
\htquantdef{phspf_mebymmu}{phspf_mebymmu}{}{0.999812949174 \pm 0.000000000015}{0.999812949174}{0.000000000015}%
\htquantdef{phspf_mebymtau}{phspf_mebymtau}{}{0.999999338359 \pm 0.000000000089}{0.999999338359}{0.000000000089}%
\htquantdef{phspf_mmubymtau}{phspf_mmubymtau}{}{0.9725600 \pm 0.0000036}{0.9725600}{0.0000036}%
\htquantdef{pi}{pi}{}{3.142}{3.142}{0}%
\htquantdef{PimKmKpNu}{PimKmKpNu}{}{0.1440\cdot 10^{-2}}{0.1440\cdot 10^{-2}}{0}%
\htquantdef{PimKmPipNu}{PimKmPipNu}{}{0.2940\cdot 10^{-2}}{0.2940\cdot 10^{-2}}{0}%
\htquantdef{PimKzsKzlNu}{PimKzsKzlNu}{}{0.1200\cdot 10^{-2}}{0.1200\cdot 10^{-2}}{0}%
\htquantdef{PimKzsKzsNu}{PimKzsKzsNu}{}{2.320\cdot 10^{-4}}{2.320\cdot 10^{-4}}{0}%
\htquantdef{PimPimPipNu}{PimPimPipNu}{}{8.990\cdot 10^{-2}}{8.990\cdot 10^{-2}}{0}%
\htquantdef{PimPimPipPizNu}{PimPimPipPizNu}{}{4.610\cdot 10^{-2}}{4.610\cdot 10^{-2}}{0}%
\htquantdef{PimPizKzsNu}{PimPizKzsNu}{}{0.1940\cdot 10^{-2}}{0.1940\cdot 10^{-2}}{0}%
\htquantdef{PimPizNu}{PimPizNu}{}{0.2552}{0.2552}{0}%
\htquantdef{pitoENu}{pitoENu}{}{(1.2300 \pm 0.0040) \cdot 10^{-4}}{1.2300\cdot 10^{-4}}{0.0040\cdot 10^{-4}}%
\htquantdef{pitoMuNu}{pitoMuNu}{}{0.99987700 \pm 0.00000040}{0.99987700}{0.00000040}%
\htquantdef{R_tau}{R_tau}{}{3.6349 \pm 0.0082}{3.6349}{0.0082}%
\htquantdef{R_tau_s}{R_tau_s}{}{0.1633 \pm 0.0027}{0.1633}{0.0027}%
\htquantdef{R_tau_VA}{R_tau_VA}{}{3.4717 \pm 0.0081}{3.4717}{0.0081}%
\htquantdef{Rrad_kmunu_by_pimunu}{Rrad_kmunu_by_pimunu}{}{0.9930 \pm 0.0035}{0.9930}{0.0035}%
\htquantdef{Rrad_SEW_tau_Knu}{Rrad_SEW_tau_Knu}{}{1.02010 \pm 0.00030}{1.02010}{0.00030}%
\htquantdef{Rrad_tauK_by_taupi}{Rrad_tauK_by_taupi}{}{1.00 \pm 0.00}{1.00}{0.00}%
\htquantdef{sigmataupmy4s}{sigmataupmy4s}{}{0.9190}{0.9190}{0}%
\htquantdef{tau_K}{tau_K}{}{(1.2380 \pm 0.0020) \cdot 10^{-8}}{1.2380\cdot 10^{-8}}{0.0020\cdot 10^{-8}}%
\htquantdef{tau_mu}{tau_mu}{}{(2.196981 \pm 0.000022) \cdot 10^{-6}}{2.196981\cdot 10^{-6}}{0.000022\cdot 10^{-6}}%
\htquantdef{tau_pi}{tau_pi}{}{(2.60330 \pm 0.00050) \cdot 10^{-8}}{2.60330\cdot 10^{-8}}{0.00050\cdot 10^{-8}}%
\htquantdef{tau_tau}{tau_tau}{}{290.3 \pm 0.5}{290.3}{0.5}%
\htquantdef{Vud}{Vud}{}{0.97417 \pm 0.00021}{0.97417}{0.00021}%
\htquantdef{Vud_moulson_ckm14}{Vud_moulson_ckm14}{}{0.97417 \pm 0.00021}{0.97417}{0.00021}%
\htquantdef{Vus}{Vus}{}{0.2186 \pm 0.0021}{0.2186}{0.0021}%
\htquantdef{Vus_err_exp}{Vus_err_exp}{}{0.1869\cdot 10^{-2}}{0.1869\cdot 10^{-2}}{0}%
\htquantdef{Vus_err_exp_perc}{Vus_err_exp_perc}{}{0.8547}{0.8547}{0}%
\htquantdef{Vus_err_perc}{Vus_err_perc}{}{0.9765}{0.9765}{0}%
\htquantdef{Vus_err_th}{Vus_err_th}{}{0.1032\cdot 10^{-2}}{0.1032\cdot 10^{-2}}{0}%
\htquantdef{Vus_err_th_perc}{Vus_err_th_perc}{}{0.47}{0.47}{0}%
\htquantdef{Vus_kl2_maltman_mainz16}{Vus_kl2_maltman_mainz16}{}{0.22500 \pm 0.00098}{0.22500}{0.00098}%
\htquantdef{Vus_kl2_moulson_ckm14}{Vus_kl2_moulson_ckm14}{}{0.22484 \pm 0.00059}{0.22484}{0.00059}%
\htquantdef{Vus_kl3_maltman_mainz16}{Vus_kl3_maltman_mainz16}{}{0.22310 \pm 0.00081}{0.22310}{0.00081}%
\htquantdef{Vus_kl3_moulson_ckm14}{Vus_kl3_moulson_ckm14}{}{0.22320 \pm 0.00090}{0.22320}{0.00090}%
\htquantdef{Vus_maltman_mainz16}{Vus_maltman_mainz16}{}{0.2228 \pm 0.0024}{0.2228}{0.0024}%
\htquantdef{Vus_mism}{Vus_mism}{}{(-0.7206 \pm 0.2334) \cdot 10^{-2}}{-0.7206\cdot 10^{-2}}{0.2334\cdot 10^{-2}}%
\htquantdef{Vus_mism_sigma}{Vus_mism_sigma}{}{-3.1}{-3.1}{0}%
\htquantdef{Vus_mism_sigma_abs}{Vus_mism_sigma_abs}{}{3.1}{3.1}{0}%
\htquantdef{Vus_tau}{Vus_tau}{}{0.2216 \pm 0.0015}{0.2216}{0.0015}%
\htquantdef{Vus_tau_mism}{Vus_tau_mism}{}{(-0.4171 \pm 0.1730) \cdot 10^{-2}}{-0.4171\cdot 10^{-2}}{0.1730\cdot 10^{-2}}%
\htquantdef{Vus_tau_mism_sigma}{Vus_tau_mism_sigma}{}{-2.4}{-2.4}{0}%
\htquantdef{Vus_tau_mism_sigma_abs}{Vus_tau_mism_sigma_abs}{}{2.4}{2.4}{0}%
\htquantdef{Vus_tauKnu}{Vus_tauKnu}{}{0.2221 \pm 0.0017}{0.2221}{0.0017}%
\htquantdef{Vus_tauKnu_err_th_perc}{Vus_tauKnu_err_th_perc}{}{0.2962}{0.2962}{0}%
\htquantdef{Vus_tauKnu_mism}{Vus_tauKnu_mism}{}{(-0.3736 \pm 0.1899) \cdot 10^{-2}}{-0.3736\cdot 10^{-2}}{0.1899\cdot 10^{-2}}%
\htquantdef{Vus_tauKnu_mism_sigma}{Vus_tauKnu_mism_sigma}{}{-2.0}{-2.0}{0}%
\htquantdef{Vus_tauKnu_mism_sigma_abs}{Vus_tauKnu_mism_sigma_abs}{}{2.0}{2.0}{0}%
\htquantdef{Vus_tauKpi}{Vus_tauKpi}{}{0.2236 \pm 0.0018}{0.2236}{0.0018}%
\htquantdef{Vus_tauKpi_err_th_perc}{Vus_tauKpi_err_th_perc}{}{0.3058}{0.3058}{0}%
\htquantdef{Vus_tauKpi_err_th_perc_dRrad_kmunu_by_pimunu}{Vus_tauKpi_err_th_perc_dRrad_kmunu_by_pimunu}{}{-0.1163}{-0.1163}{0}%
\htquantdef{Vus_tauKpi_err_th_perc_dRrad_tauK_by_Kmu}{Vus_tauKpi_err_th_perc_dRrad_tauK_by_Kmu}{}{-0.1090}{-0.1090}{0}%
\htquantdef{Vus_tauKpi_err_th_perc_dRrad_taupi_by_pimu}{Vus_tauKpi_err_th_perc_dRrad_taupi_by_pimu}{}{6.989\cdot 10^{-2}}{6.989\cdot 10^{-2}}{0}%
\htquantdef{Vus_tauKpi_err_th_perc_f_K_by_f_pi}{Vus_tauKpi_err_th_perc_f_K_by_f_pi}{}{-0.2515}{-0.2515}{0}%
\htquantdef{Vus_tauKpi_mism}{Vus_tauKpi_mism}{}{(-0.2231 \pm 0.1998) \cdot 10^{-2}}{-0.2231\cdot 10^{-2}}{0.1998\cdot 10^{-2}}%
\htquantdef{Vus_tauKpi_mism_sigma}{Vus_tauKpi_mism_sigma}{}{-1.1}{-1.1}{0}%
\htquantdef{Vus_tauKpi_mism_sigma_abs}{Vus_tauKpi_mism_sigma_abs}{}{1.1}{1.1}{0}%
\htquantdef{Vus_uni}{Vus_uni}{}{0.22582 \pm 0.00089}{0.22582}{0.00089}%
\htquantdef{VusbyVud_moulson_ckm14}{VusbyVud_moulson_ckm14}{}{0.23080 \pm 0.00060}{0.23080}{0.00060}%
\htdef{couplingsCorr}{%
$\left( \frac{g_\tau}{g_e} \right)$ &   53\\
$\left( \frac{g_\mu}{g_e} \right)$ &  -49 &   48\\
$\left( \frac{g_\tau}{g_\mu} \right)_\pi$ &   24 &   26 &    2\\
$\left( \frac{g_\tau}{g_\mu} \right)_K$ &   11 &   10 &   -1 &    6\\
 & $\left( \frac{g_\tau}{g_\mu} \right)$ & $\left( \frac{g_\tau}{g_e} \right)$ & $\left( \frac{g_\mu}{g_e} \right)$ & $\left( \frac{g_\tau}{g_\mu} \right)_\pi$}%